\documentclass[a4paper]{article}

\usepackage{feynmf}
\usepackage{amsmath}
\usepackage{amssymb}
\usepackage{graphicx}
\usepackage{slashed}
\usepackage{wrapfig}
\usepackage[english]{babel}
\usepackage{array}

\DeclareMathAlphabet{\mathpzc}{OT1}{pzc}{m}{it}

\DeclareMathSizes{10}{10}{6}{4}

\numberwithin{equation}{section}
\numberwithin{table}{section}
\numberwithin{figure}{section}

\long\def\@makecaption#1#2{%
  \vskip\abovecaptionskip
  \sbox\@tempboxa{#1: #2}%
  \ifdim \wd\@tempboxa >\hsize
    #1: #2\par
  \else
    \global \@minipagefalse
    \hb@xt@\hsize{\box\@tempboxa\hfil}%
  \fi
  \vskip\belowcaptionskip}
\makeatother

\newcommand{\eq}[1]{(\ref{#1})}
\newcommand{\Tr}{\mathrm{Tr}}
\newcommand{\Mhalf}{M_{\scriptscriptstyle 1\hspace*{-1.3pt}\protect\raisebox{-0.2ex}{$\scriptscriptstyle /$}\hspace*{-1.3pt}\protect\raisebox{-0.7ex}{$\scriptscriptstyle 2$}}}

\makeatletter  
\renewcommand{\verbatim@font}{%
  \ttfamily\small\catcode`\<=\active\catcode`\>=\active%
}
\makeatother   

\begin{document}

\setlength{\unitlength}{1mm}
\setlength{\extrarowheight}{1mm}

\vspace{3cm}
\begin{center}{\Large Supersymmetric Phenomenology in the mSUGRA Parameter Space}
\vspace{24pt}

\emph{Master thesis at the Institute of Mathematics, Astrophysics and Particle Physics at the Radboud University Nijmegen}
\vspace{12pt}

Irene Niessen
\vspace{12pt}

Supervisors: Wim Beenakker and Nicolo de Groot

\vspace{3cm}

{\large Abstract}\end{center}

\noindent In this master thesis the possible supersymmetric phenomenologies associated with low-mass mSUGRA are investigated. The main characteristics of the supersymmetric mass spectrum are explained and a systematic method is presented to predict phenomenology directly from the mass spectrum. The resulting phenomenologies are compared to the current ATLAS benchmark points. It is found that some phenomenological scenarios are not covered by these benchmark points.

\thispagestyle{empty}
\newpage
\mbox{}
\thispagestyle{empty}
\newpage

\setcounter{page}{1}
\section*{Acknowledgements}

I occasionally say that this thesis is the result of pure stubbornness, but that is not quite true. I could not have written it without the help 
and support of the people around me. Firstly, I would like to thank everyone from the high energy physics department, especially Folkert 
Koetsveld, Lisa Hartgring and Lucian Ancu for proofreading my thesis and their help in both physics and personal matters, Frank Filthaut for the 
insightful discussions, and Gemma Koppers, Marjo van Wees and Annelies Oosterhof for making the department the great place it is. And of course 
my supervisors. I would like to thank Nicolo de Groot for teaching me to keep in mind the applicability of my research. And 
Wim Beenakker, for always taking the time to help and guide me, and most importantly, helping me rediscover the scientist in me.

Finally, I would like to thank my friends and family, especially my parents and sister for their unconditional support over the past few years.

\tableofcontents

\section{Introduction}

The field of High Energy Physics deals with the smallest particles, usually elementary particles, and their interactions. Ironically, the tiniest particles are studied with the largest machines in the world. In particle accelerators small particles are given enormous energies before being collided. The products of these collisions provide information on the building blocks of the universe.

At the moment, the most powerful accelerator to date is being finished. In the Large Hadron Collider (LHC), protons with an energy of 7 TeV each will be collided. These collisions will be studied by four detectors. It is expected that they will find the famous Higgs boson, which is predicted by the Standard Model of particle physics. 

If the Higgs boson is found, the Standard Model is complete. However, no one believes that nature is completely described by the Standard Model. Therefore the LHC also aims to search for new physics. Two of the four detectors at the LHC are multi-purpose detectors, designed to detect a variety of signals, in particular indications for new physics. The multi-purpose detector that will be mentioned specifically in this thesis is ATLAS (A Toroidal LHC Apparatus).

One of the most popular extensions of the Standard Model is supersymmetry, and the search for supersymmetry is an important physics goal for ATLAS. Unfortunately supersymmetry is not characterized by a single signal. There are many different possible phenomenologies depending on the supersymmetric model and the parameters within the model.

Since a different phenomenology yields a different signal at the LHC, it is important to identify the possible phenomenologies in order to recognize supersymmetry. In this thesis, the possible phenomenologies associated with minimal supergravity breaking (mSUGRA) are investigated.

To this end, we first discuss some key concepts of the Standard Model in section \ref{s:standardmodel}. In section \ref{s:supersymmetry}, supersymmetry and mSUGRA are introduced. The methods and programs used in this research are presented in section \ref{s:methods}. In section \ref{s:massspectrum}, the mass spectrum following from mSUGRA is explained. The results are applied in section \ref{s:phenomenology}, where we develop a systematic method to investigate the possibilities for supersymmetric phenomenology in the context of mSUGRA.

\subsection{Notation and Conventions}

Natural units are used throughout this thesis, so $\hbar=c=1$. Also, the Einstein summation convention is used, so repeated indices are summed over unless explicitly stated otherwise. All other conventions are introduced when they are needed.

\begin{fmffile}{feynstmod}
\fmfset{dot_size}{1thick}
\fmfset{curly_len}{1.5mm}
\fmfset{arrow_len}{3mm}
\fmfpen{thin}

\fmfcmd{%
style_def wiggly_arrow expr p =
cdraw (wiggly p);
shrink (1);
cfill (arrow p);
endshrink;
enddef;}

\section{The Standard Model}\label{s:standardmodel}

The Standard Model is one of the most succesful theories ever. The match between theoretical predictions and actual measurements is astounding. Also, the entire theory can be derived from symmetry arguments. The fact that such a simple concept yields such a successful theory is truly amazing. The purpose of this section is to give a short overview of some important concepts that are used later in this thesis. For a rigorous treatment of quantum field theory, see for instance \cite{peskinschroeder}, \cite{itzyksonzuber} or \cite{chengli}.

\subsection{Gauge Invariance}\label{s:gauge}

When we look at nature, we see certain symmetries. Or more precise, we see that physics is invariant under certain transformations. These transformations are described by group theory. In this sense the Standard Model `is described by the group $SU(3)\times SU(2)\times U(1)$': it is derived by requiring invariance under this group. So how do we do that?

In this section, we start with a Lagrangian for spin 1/2 fermions and introduce the postulate of local gauge invariance. We first encounter this postulate in an Abelian (commuting) theory and find that it reproduces the theory of electrodynamics. We then apply it to non-Abelian theories and obtain the Standard Model gauge group.
\newline

\textsc{Abelian Gauge Symmetries.} Spin 1/2 fermions are the most commonly observed particles. They are described by spinors $\psi$ that have four degrees of freedom: two for the particle and two for the corresponding antiparticle. We can split a fermion into a left-handed part $\psi_L$ and a right-handed part $\psi_R$ using the helicity operator, which is the projection of the spin onto the direction of motion. Massive fermions are always a linear combination of $\psi_L$ and $\psi_R$ since the helicity can be flipped with a Lorentz boost\footnote{For massive particles, an appropriate Lorentz boost flips the direction of motion while the spin is unchanged, so the helicity changes sign.}. Massless fermions, that travel with the speed of light, are pure helicity eigenstates. Therefore the spinors that describe them only have two degrees of freedom. The basic relativistic Dirac Lagrangian for a massless fermion\footnote{For massive fermions we also need a mass term of the form $m\bar\psi\psi$, but we will omit that for now.} consists of kinetic terms of the form:
\begin{equation}\label{U1lagrangian}\mathcal{L}=i\bar\psi\gamma^\mu\partial_\mu\psi\end{equation}
where the $\gamma^\mu$ are the Dirac gamma matrices. The question we ask ourselves is what symmetries this Lagrangian has. According to Noether's theorem, a continuous symmetry gives rise to a conserved charge and current, so conserved quantities can guide us to an underlying symmetry. The simplest example is electric charge. This is the conserved charge generated by the $U(1)$-symmetry of the fermion Lagrangian \eq{U1lagrangian} under the constant phase transformation:
\begin{equation}\label{U1transformation}\psi\to e^{i\varphi}\psi\end{equation}

This type of symmetry is called global, since it is the same for every point in spacetime. The trick is to require the symmetry to be \emph{local}, so $\varphi$ can depend on place and time. This makes sense, for instance because gravity is inherently local. However, the most important reason to introduce the postulate of local gauge symmetry is that it works. As we will see in a moment, local gauge invariance yields very nice results. So let us study how the theory is changed if we require the symmetry to be local.

We still want our Lagrangian to be invariant under equation \eq{U1transformation}, even if $\varphi$ depends on place and time, so we replace the derivative $\partial_\mu$ with a covariant derivative
\begin{equation}D_\mu=\partial_\mu-ig_1A_\mu\label{U1covariantderivative}\end{equation}
where $A_\mu$ is a vector field (gauge field) and $g_1$ is the coupling strength. In case of $U(1)$-symmetry, the covariant derivative is invariant under the local gauge transformation \eq{U1transformation} provided that $A_\mu$ transforms as:
\begin{equation}
A_\mu\to A_\mu+\frac{1}{g_1}\partial_\mu\varphi\label{gaugefreedomA}
\end{equation}

Introducing the covariant derivative is identical to what is known in electrodynamics as minimal substitution, which describes interactions between the electromagnetic field $A_\mu$ and charged particles. Also, we recognise equation \eq{gaugefreedomA} as the gauge freedom in electrodynamics.

We complete the Lagrangian by adding kinetic terms for the vector field $A_\mu$. They need to contain derivatives of the field $A_\mu$ and should leave the theory gauge invariant. The simplest object we can use is the commutator of two covariant derivatives. In case of $U(1)$-symmetry this gives the electromagnetic field tensor:
\[\left[D_\mu,D_\nu\right]=-ig_1\left(\partial_\mu A_\nu-\partial_\nu A_\mu\right)=-ig_1F_{\mu\nu}\]
The kinetic term for a vector boson is then given by $-\frac{1}{4}F^{\mu\nu}F_{\mu\nu}$. In case of electrodynamics the equations of motion for the complete Lagrangian, including the kinetic terms, reproduce Maxwell's equations. The field $A_\mu$ describes the massless spin 1 gauge bosons that are more commonly known as photons. So the theory of electrodynamics can be derived from a simple symmetry argument!
\newline

\textsc{Non-Abelian Gauge Symmetries.} Encouraged by this success, we try to generalize this approach. Electric charge is not the only conserved quantity. Another example is conservation of electron number. An electron can decay to a $W$ boson and an electron-neutrino, but not to for instance a muon-neutrino. This can be described mathematically by saying that the electron and the electron-neutrino are grouped together in a doublet and that they are linked by a unitary transformation. The underlying symmetry group turns out to be $SU(2)$.

Another example is quantum chromodynamics (QCD). Quarks have a three-valued conserved quantum number called `colour'. The underlying symmetry group is $SU(3)$. Since free coloured particles have never been observed, we do not see this symmetry directly, but we can infer it from the multiplet structure we observe in hadrons \cite{griffiths}.

The underlying symmetries of non-Abelian gauge theories are associated with additional degrees of freedom. If they are labeled by indices, the Lagrangian is invariant under a unitary transformation:
\begin{equation}\psi^j\to U^{jk}\psi^k\label{intdoftransformation}\end{equation}
Any unitary matrix can be written as the exponential function of a Hermitian matrix that can in turn be written as a linear combination of the generators $T^a$ of the Lie-algebra of the symmetry group, so equation \eq{intdoftransformation} can be written as:
\begin{equation}\label{SUNtransformation}
\psi\to U\psi=e^{i\theta^a T^a}\psi\equiv e^{i\vec\theta\cdot\vec T}\psi\end{equation}
In case of $SU(2)$ the index $a$ can be 1, 2 or 3, while for $SU(3)$ the index can run from 1 to 8. The generators $T^a$ do not commute, so equation \eq{SUNtransformation} is a generalization of equation \eq{U1transformation} to non-Abelian gauge transformations. The arguments used to construct the Lagrangian are the same, but the resulting terms are more complicated. For instance, the covariant derivative that follows from $SU(2)$ gauge invariance is given by:
\begin{equation}D_\mu=\partial_\mu-ig_2T^aA_\mu^a\label{SU2covariantderivative}\end{equation}
with $T^a$ the $SU(2)$ generators, $A_\mu^a$ the gauge fields and $g_2$ the coupling constant. The kinetic terms are constructed in the same way as in the $U(1)$ case. This introduces the commutator of two covariant derivatives that have the form of equation \eq{SU2covariantderivative}. The generators $T^a$ do not commute, so this implies self-interactions between the gauge bosons. Working out the details shows that the number of gauge bosons that is obtained with $SU(3)$, $SU(2)$ and $U(1)$ symmetry is indeed the number that is observed experimentally.
\newline

\textsc{Group Theory.} In group theory language we say that particles are in certain representations that are characterized by their dimension \cite{sternberg}. If we say that quarks are in an $SU(3)$ triplet, we mean that they are in a three dimensional representation of $SU(3)$, because a quark can have three different colours\footnote{To get a feeling for representations, it is useful to realize that the Standard Model fermions are in the fundamental representation. In familiar terms that means they can be viewed as vectors. E.g. a doublet is a 2D vector, a singlet is a number. If $\psi$ is a column vector, $\bar\psi$ is a row vector, so $\bar\psi\psi$ can be contracted to form a singlet. The gauge bosons are in the adjoint representation, which means they can be seen as matrices.}. Table~\ref{standardmodelparticles} lists the Standard Model particles and the $SU(3)\times SU(2)\times U(1)$ representations they are in. Fermion representations are split into a left-handed doublet and a right-handed singlet representation since experiments show that weak interactions only couple to the left-handed part.

\begin{table}[!h]
\centering{
\begin{tabular}{|l|l|c|}
\hline
Name & Symbol & $(SU(3),SU(2),U(1))$\\
\hline
\multicolumn{3}{|l|}{Quarks} \\
\hline
Left-handed doublet & $(u,d)_L\equiv Q_L$ & $(3,2,\frac{1}{6})_L$\\
Right-handed up-type singlet & $u_R$ & $(3,1,\frac{2}{3})_R$\\
Right-handed down-type singlet & $d_R$ & $(3,1,-\frac{1}{3})_R$\\
\hline
\multicolumn{3}{|l|}{Leptons} \\
\hline
Left-handed leptons & $(\nu_e,e^-)_L\equiv L_L$ & $(1,2,-\frac{1}{2})_L$\\
Right-handed charged leptons& $e_R^-$ & $(1,1,-1)_R$\\
\hline
\multicolumn{3}{|l|}{Gauge bosons} \\
\hline
Gluons (strong force)& $g$ & $(8,1,0)$\\
W and Z bosons (weak force)\footnotemark& $W^\pm,Z$ & $(1,3,0)$\\
Photon (electromagnetic force)\footnotemark[\value{footnote}]& $\gamma$ & $(1,1,0)$\\
\hline
\multicolumn{3}{|l|}{Scalar (see section~\ref{s:higgs})} \\
\hline
Higgs boson & $H$ & $(1,2,\frac{1}{2})$\\
\hline
\end{tabular}
\caption[Standard Model particle content]{Particle content of the Standard Model. Of the three generations of quarks and leptons only the first is shown. Note that left-handed fermions are doublets under $SU(2)$ while right-handed fermions are singlets. $SU(2)$ and $SU(3)$ representations are labeled by their dimension while $U(1)$ representations are labeled by the eigenvalue of the $U(1)$ hypercharge generator $Y$, which is related to the electric charge $Q$ and the third component of isospin $T_3$ by the Gell-Mann-Nishijima formula $Q=T_3+Y$}\label{standardmodelparticles}}
\end{table}
\footnotetext[\value{footnote}]{The $W$ and $Z$ bosons and the photon are actually mixtures of the unbroken representations listed here. They emerge after spontaneous symmetry breaking (see section~\ref{s:higgs}).}

\subsection{The Higgs Mechanism}\label{s:higgs}

Obtaining the Standard Model gauge bosons from group theory is certainly a nice result, but there is one problem: all particles in our theory are massless. This is fine for the photon and the gluons and perhaps for the neutrinos, but we know all other particles have mass. The simplest way to solve this problem is by adding mass terms to the Lagrangian. However, if we do that, we run into problems.

Mass terms for fermions break chiral symmetry, under which left-handed and right-handed fermions transform independently. A fermion mass term has the form $m\bar\psi\psi=m(\bar\psi_L\psi_R+\bar\psi_R\psi_L)$ and couples a left-handed fermion, which is an $SU(2)$ doublet, to a right-handed fermion, an $SU(2)$ singlet. That means that we will end up with a Lagrangian that is a doublet under $SU(2)$. This is not gauge invariant and therefore unacceptable.

Adding a mass term for gauge bosons, which is of the form $\frac{1}{2}m^2A_\mu A^\mu$, breaks the gauge invariance we based the theory on. This essentially leaves us without a theory, so that is not an option either. It is clear that we need a proper way to generate mass.

Let us start with fermion masses. The simplest way to add a mass term without breaking chiral symmetry is by introducing another $SU(2)$ doublet $\phi$. This can be used to couple a left-handed doublet to a right-handed singlet with a Yukawa interaction:
\begin{equation}\mathcal{L}_{y}=-y\bar\psi_L\phi\psi_R=-y\left(\bar\psi_{u},\bar\psi_{d}\right)_L\left(\begin{array}{c}\phi_u\\\phi_d\end{array}\right)\psi_R=-y\phi_u\bar\psi_{uL}\psi_R-y\phi_d\bar\psi_{dL}\psi_R\label{higgsfermioncoupling}\end{equation}
Here $u$ and $d$ denote the up and down components of $SU(2)$ doublets, so they should not be summed over, and the coupling $y$ is different for fermions with different masses. Although equation \eq{higgsfermioncoupling} couples left-handed fermions to right-handed fermions in a gauge invariant way, it does not automatically generate a mass term. But if $\phi$ has a vacuum expectation value (vev), it can be written as:
\begin{equation}
\phi=\langle\phi\rangle+\eta\label{higgsexpansion}
\end{equation}
where the vev $\langle\phi\rangle$ is the first order expansion of $\phi$, and $\eta$ contains the higher order terms (quantum fluctuations). The coupling to $\eta$ is an interaction between three fields, but the coupling to the constant $\langle\phi\rangle$ has the form of a mass term. That means that an additional $SU(2)$ doublet can generate a mass term if it has a non-zero vev.

\begin{figure}[!ht]
\begin{minipage}[t]{0.5\linewidth}
\caption[Higgs potential for $\mu^2>0$]{$\mu^2>0$}\label{scalarpotential}
\begin{minipage}[t]{0.95\linewidth}
\begin{minipage}[t]{0.59\linewidth}
\includegraphics[width=\linewidth]{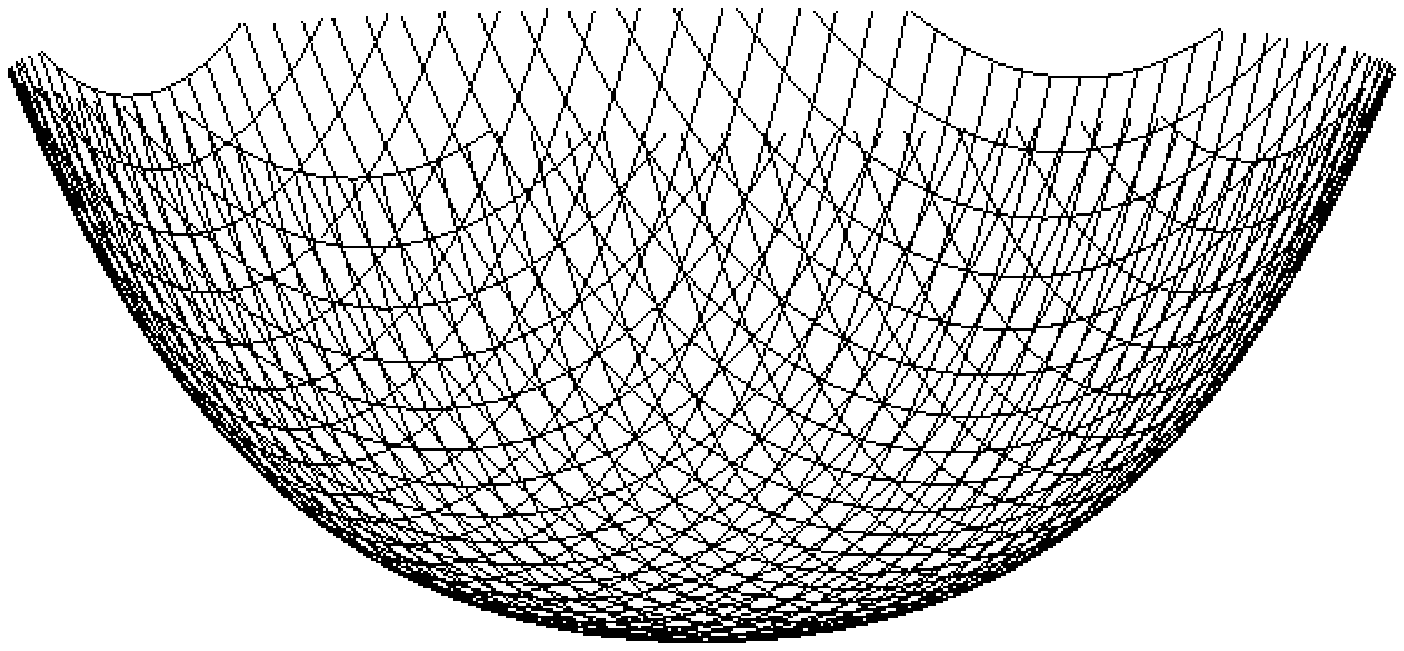}
\end{minipage}
\begin{minipage}[t]{0.39\linewidth}
\includegraphics[width=\linewidth]{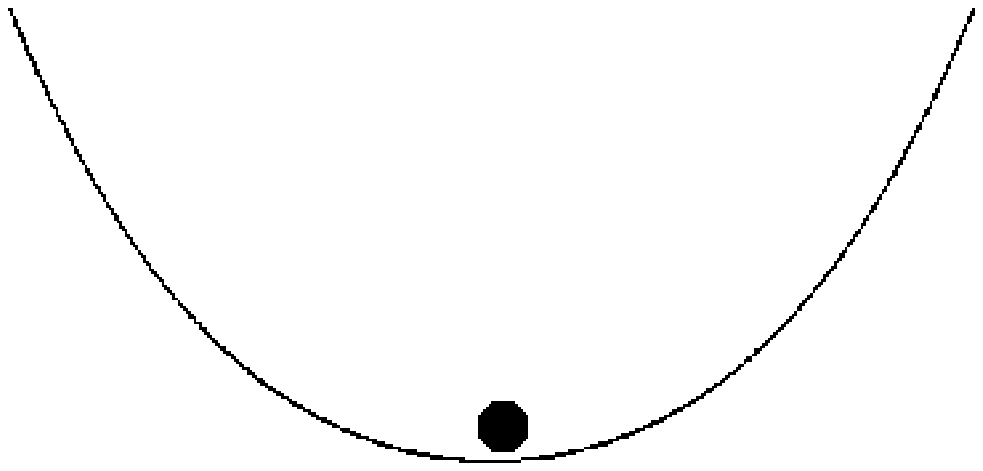}
\end{minipage}
\end{minipage}
\end{minipage}
\hfill
\begin{minipage}[t]{0.5\linewidth}
\caption[Higgs potential for $\mu^2<0$]{$\mu^2<0$}\label{mexicanhat}
\begin{minipage}[t]{0.95\linewidth}
\begin{minipage}[t]{0.59\linewidth}
\includegraphics[width=\linewidth]{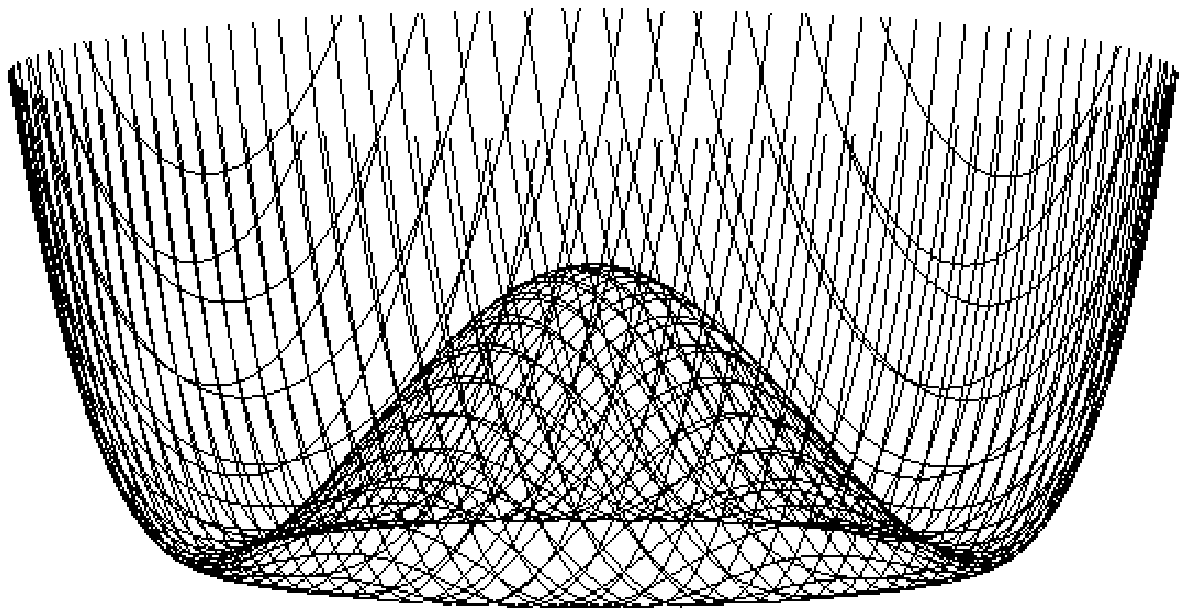}
\end{minipage}
\begin{minipage}[t]{0.39\linewidth}
\includegraphics[width=\linewidth]{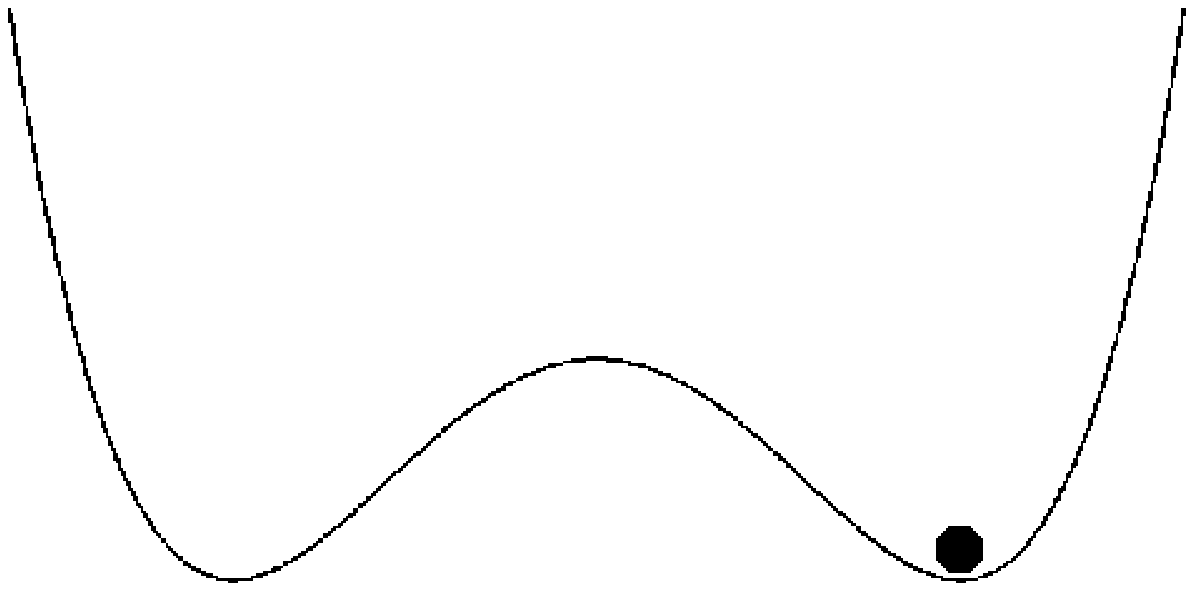}
\end{minipage}
\end{minipage}
\end{minipage}
\hfill
\vspace{4pt}

Figure~\ref{scalarpotential} and~\ref{mexicanhat}: The scalar Lagrangian from equation \eq{scalarlagrangian} gives rise to two possible shapes of the potential. If $\mu^2>0$, the potential looks like the graph in figure~\ref{scalarpotential}. If $\mu^2<0$, the potential looks like the graph in figure~\ref{mexicanhat}.
\end{figure}

The particles we encountered so far do not have a vev and we cannot add it without breaking either gauge or Lorentz invariance. However, there is a field that can obtain a vev. For a complex scalar, the most general Lagrangian that is gauge invariant and renormalizable (see section~\ref{s:renormalization}) is:
\begin{equation}\mathcal{L}_\phi=(D_\mu\phi)^\dagger(D^\mu\phi)-V=(D_\mu\phi)^\dagger(D^\mu\phi)-\mu^2\phi^\dagger\phi-\frac{1}{4}\lambda(\phi^\dagger\phi)^2\label{scalarlagrangian}\end{equation}
with $\mu^2,\lambda\in\mathbb{R}$. The covariant derivative in equation~\ref{scalarlagrangian} follows from $U(1)$ and $SU(2)$ gauge invariance and is found by combining equations \eq{U1covariantderivative} and \eq{SU2covariantderivative}:
\[D_\mu=\partial_\mu-\frac{i}{2}g_1B_\mu-ig_2T^aA_\mu^a\]
where $B_\mu$ is the $U(1)$ hypercharge gauge boson and $A_\mu^a$ are the $SU(2)$ gauge bosons. The potential $V$ needs to be bounded from below, so $\lambda$ must be positive. For $\mu^2$ we can distinguish two cases, that are shown in figures~\ref{scalarpotential} and~\ref{mexicanhat}. In figure~\ref{scalarpotential} the minimum of the potential is zero and nothing interesting happens. In contrast, in figure~\ref{mexicanhat} the minimum is non-zero and $\phi$ obtains a vev.

Now we simplify equation \eq{higgsfermioncoupling} by fixing the gauge. First we use $SU(2)$ rotations to bring $\langle\phi\rangle$ to the form
\begin{equation}\langle\phi\rangle=\frac{1}{\sqrt{2}}\left(\begin{array}{c}0\\v\end{array}\right)\label{higgsvev}\end{equation}
and then we make $v$ real with a $U(1)$ phase transformation. Inserting this in equation \eq{higgsfermioncoupling} indeed yields a mass-term for down-type fermions. A mass-term for the up-type fermions is generated by the charge conjugate\footnote{Charge conjugation is essentially a transformation between matter and antimatter. The charge conjugate of a complex $SU(2)$ doublet is given by $\phi^c=i\sigma^2\phi^*$ where the Pauli matrix $\sigma^2$ is responsible for bringing equation \eq{higgsvev} into the right form to generate mass for the up-type fermions.} of $\phi$. That means our first problem is solved: we now have massive fermions.

It turns out that $\phi$ generates masses for the gauge bosons as well. We can see this by expanding the Lagrangian \eq{scalarlagrangian} around the new minimum \eq{higgsvev}. This is an exercise for a rainy Sunday afternoon, but writing out the Lagrangian to first order shows that $\langle\phi\rangle$ gives rise to a term for the $U(1)$ and $SU(2)$ bosons $(A^1_\mu,A^2_\mu,A^3_\mu,B_\mu)\equiv V_\mu$ that is given by: \[\mathcal{L}_{m}=\frac{1}{2}V_\mu M^2V^\mu\]
where the mass matrix $M^2$ is:
\[M^2=\frac{v^2}{4}\left(\begin{array}{cccc}g_2^2&0&0&0\\0&g_2^2&0&0\\0&0&g_2^2&-g_1g_2\\0&0&-g_1g_2&g_1^2\end{array}\right)\]
The off-diagonal terms in $M^2$ mix the neutral $SU(2)$ boson $A^3_\mu$ and the $U(1)$ boson $B_\mu$. To obtain the mass-eigenstates, we have to diagonalize $M^2$. This results in massive $W$ and $Z$ bosons and a massless photon, exactly as observed in nature.

The next order expansion of $\phi$, which is the field $\eta$ from equation \eq{higgsexpansion}, behaves like a scalar field, describing the famous Higgs particle that will hopefully be discovered at the LHC. That covers all four real degrees of freedom of $\phi$: one scalar particle and three longitudinal components of the now massive vector bosons\footnote{A massive vector boson has three degrees of freedom while a massless vector boson only has two. Three real degrees of freedom of the Higgs doublet are `eaten' by the vector bosons $W^+,W^-$ and $Z$, which acquire mass.}. Higher order  terms in the Lagrangian give rise to interactions among the Higgs and gauge bosons. For instance it follows that a photon couples to a $W$ boson, as it should since $W$ bosons are charged particles.

This way of generating mass is called the Higgs mechanism, after one of its inventors. It is also known as spontaneous symmetry breaking. The last name can be understood from figures~\ref{scalarpotential} and~\ref{mexicanhat}. Both potentials have the same $SU(2)$ rotation symmetry, but seen from a minimum of figure~\ref{mexicanhat} the potential does not seem to be symmetric. In that case the symmetry is said to be `spontaneously broken'. Although the symmetry \emph{appears} to be broken, the theory is still intact, and so are most of its properties.

\subsection{Feynman Rules}\label{s:feynman}

Now that we have a Lagrangian, we can use the Euler-Lagrange equations to obtain the equations of motion. That is a good check on the validity of our Lagrangian, but experiments do not measure equations of motion, they measure cross sections. We can calculate those from the Lagrangian as follows:
\begin{enumerate}
\item Go from the Lagrangian to the Hamiltonian in the usual way.
\item Insert the Hamiltonian in the quantum mechanical wave equation to get the time evolution of the system. Using the interaction picture it follows that the time evolution between initial time $t_i$ and final time $t_f$ is given by: \[U(t_f,t_i)=\mathrm{T}\exp\left(-i\int_{t_i}^{t_f}H_Idt\right)\]
where $H_I$ is the interaction Hamiltonion and T stands for time-ordering.
\item Define the $S$-matrix as $S=\displaystyle\lim_{t\to\infty}U(t,-t)$. Then the transition amplitude for a particular scattering process follows from the expression:
\begin{equation}\left<\mbox{final state}|S|\mbox{initial state}\right>=\left<\mbox{final state}\left|\mathrm{T}e^{-i\int_{-\infty}^{\infty}H_Idt}\right|\mbox{initial state}\right>\label{transitionamplitude}\end{equation}
\item Combining the transition amplitude with relativistic kinematics gives the cross section.
\end{enumerate}

This is quite a lot of work if we have to repeat these steps for every single cross section. To make matters worse, the matrix element \eq{transitionamplitude} seems impossible to calculate. Luckily perturbation theory comes to the rescue. A Taylor expansion of the exponential function gives a power series that can be represented graphically as propagators, that denote the propagating fields, and vertices, that represent interactions. Propagators and vertices are associated with Feynman rules. Once these rules are defined, any process can be calculated by drawing all possible arrangements of propagators and vertices with the given initial and final state. The formulas can then simply be read off from the resulting Feynman graphs. For $\varphi^3$-theory, which describes a real scalar with the Lagrangian $\mathcal{L}=\frac{1}{2}\partial_\mu\varphi\partial^\mu\varphi-\frac{1}{2}m^2\varphi^2-\frac{1}{3!}g\varphi^3$, the Feynman rules are given by:
\begin{align}
\parbox{15mm}{\begin{fmfgraph*}(12,5)
\fmfleft{q1}
\fmfright{q2}
\fmf{dashes,label=$p$}{q1,q2}
\fmfdot{q1}
\fmfdot{q2}
\end{fmfgraph*}} & = \frac{i}{p^2-m^2+i\varepsilon}\equiv G_2\label{feynexampleinteraction}\\
\parbox{11mm}{\begin{fmfgraph}(8,10)
\fmfleft{q1}
\fmfright{q2,q3}
\fmf{dashes}{q1,v}
\fmf{dashes}{v,q2}
\fmf{dashes}{v,q3}
\fmfdot{v}
\end{fmfgraph}} & = -ig
\end{align}
where $p$ is the momentum. For fields with internal degrees of freedom, interactions contain the generators of the Lie-algebra of the corresponding symmetry group. In that case one has to pay more attention to the order of the building blocks in the matrix element, but the concept is the same. Fermion propagators and their vertices also contain Dirac matrices. Combining several indices is also possible. For example, a coupling of two quarks to a gluon is:
\[\parbox{13mm}{\begin{fmfgraph*}(10,10)
\fmfleft{q1,q2}
\fmfright{q3}
\fmf{fermion}{q1,v}
\fmf{fermion}{v,q2}
\fmf{gluon,label=$a,,\mu$,label.side=right}{v,q3}
\fmfdot{v}
\end{fmfgraph*}} = ig_S\gamma^\mu T^a\label{quarkquarkgluon}\]
A list of all relevant Feynman rules used in this thesis can be found in appendix~\ref{feynmanrules}. Apart from rules for propagators and vertices, there are also rules for Feynman graphs as a whole, such as combinatorial factors for interchangeable lines, integration over internal momenta in loops and a minus sign for every fermion loop \cite{mandlshaw}.

\subsection{Renormalization}\label{s:renormalization}

Now that we know how to calculate cross sections, the theory seems to be complete. Unfortunately there is a complication. Take as an example the $\phi^3$ theory from section~\ref{s:feynman}. The value of the bare propagator $G_2$ is given by equation \eq{feynexampleinteraction}, but this is not the propagator we measure. If we include higher order corrections from the perturbative Taylor expansion, the effective propagator also has contributions from loop diagrams such as:
\begin{align}
\parbox{33mm}{
\begin{fmfgraph*}(30,15)
\fmfleft{in}
\fmfright{out}
\fmf{dashes_arrow,label=$p$,label.side=right}{in,v1}
\fmf{dashes_arrow,label=$p$}{v2,out}
\fmf{dashes_arrow,right,tension=0.7,label=$k$}{v2,v1}
\fmf{dashes_arrow,right,tension=0.7,label=$k+p$}{v1,v2}
\fmfdot{v1}
\fmfdot{v2}
\fmfdot{in}
\fmfdot{out}
\end{fmfgraph*}}
&=G_2^2\,\times\,\frac{g^2}{2}\int_{-\infty}^\infty\frac{d^4k}{(2\pi)^4}\frac{1}{(k^2-m^2+i\varepsilon)((p+k)^2-m^2+i\varepsilon)}\label{loop}
\end{align}

The momentum inside the loop is not determined so we have to integrate over all possible momenta. That is a problem, because this integral diverges. That means that the value of graph \eq{loop}, and therefore the effective propagator, is not well-defined. The solution is renormalization: redefine masses, coupling constants and fields to absorb infinities in a consistent manner.

Renormalization consists of several parts. The first step is to label the infinities in the theory. This is called regularization. Second we decide how to deal with these infinities. That is the actual renormalization. Finally, the freedom in both regularization and renormalization prescription leads to renormalization group equations (RGEs), which predict that the masses and couplings change with the energy scale, as observed in experiments. The rest of this section illustrates these concepts using $\phi^3$ theory and loosely sketches the consequences for the Standard Model. For a thorough review of the topic, see for instance \cite{collins1984}.

\subsubsection{Regularization}\label{s:regularization}

The first step in dealing with infinities is to label them. There are many ways to do this, but we will use a cut-off procedure. This regularization scheme is rarely used in practice, but it illustrates the basic concepts.

Equation \eq{loop} diverges if we integrate over all possible momenta, but doing so implies that we can extend the Standard Model to infinite momenta. It is quite unrealistic to assume physics is unchanged up to infinite energies. Even in the unlikely case that gravity is the only thing missing from the Standard Model, something is bound to happen at the Planck scale. To take this into account, we introduce a cut-off $\Lambda$ in all four components of the loop momentum $k$ and define the one-loop self-energy as:
\begin{align*}\Sigma_{1}(p^2,m^2,\Lambda)\equiv i\times
\parbox{22mm}{
\begin{fmfgraph}(20,5)
\fmfleft{in}
\fmfright{out}
\fmf{dashes}{in,v1}
\fmf{dashes}{v2,out}
\fmf{dashes,left,tension=0.7}{v1,v2}
\fmf{dashes,right,tension=0.7}{v1,v2}
\fmfdot{v1}
\fmfdot{v2}
\end{fmfgraph}
}
=\frac{ig^2}{2(2\pi)^4}\int^\Lambda_{-\Lambda}\frac{d^4k}{(k^2-m^2+i\varepsilon)((p+k)^2-m^2+i\varepsilon)}\end{align*}
To make this more precise we split the self-energy in a finite part that contains information about the process and a regulator part that contains the divergences. The problem occurs for $\Lambda\to\infty$, so the natural way to do this is by subtracting a term with the same asymptotic behaviour as $\Sigma_1$:
\begin{align}
\Sigma_1(p^2,m^2,\Lambda)=\,&\frac{ig^2}{32\pi^4}\int^\Lambda_{-\Lambda} d^4k\left[\frac{1}{[k^2-m^2+i\epsilon][(p+k)^2-m^2+i\epsilon]}-\frac{1}{(k^2-\mu^2+i\epsilon)^2}\right]\nonumber\\
&+\frac{ig^2}{32\pi^4}\int^\Lambda_{-\Lambda} d^4k\frac{1}{(k^2-\mu^2+i\epsilon)^2}\nonumber\\
\equiv\,&\Sigma_{1fin}(p^2,m^2,\mu^2,\Lambda)+\Sigma_{1div}(\mu^2,\Lambda)\label{selfenergysplit}
\end{align}
This introduces an additional parameter $\mu$ that we can choose any way we want. A specific choice of $\mu$ is a renormalization prescription. The divergence is now absorbed in the regulator part $\Sigma_{1div}$, while the external momentum $p$ and mass $m$ of the original propagator only occur in the finite part $\Sigma_{1fin}$.

Graph \eq{loop} is not the only contribution to the propagator. The effective propagator at one-loop level also has higher order contributions and the total one-loop correction is:
\begin{align}
G_{2,eff}&=\, \parbox{8mm}{\begin{fmfgraph}(8,5)
\fmfleft{q1}
\fmfright{q2}
\fmf{dashes}{q1,q2}
\fmfdot{q1}
\fmfdot{q2}
\end{fmfgraph}}+\, \parbox{20mm}{\begin{fmfgraph}(20,5)
\fmfleft{in}
\fmfright{out}
\fmf{dashes}{in,v1}
\fmf{dashes}{v2,out}
\fmf{dashes,left,tension=0.7}{v1,v2}
\fmf{dashes,right,tension=0.7}{v1,v2}
\fmfdot{v1}
\fmfdot{v2}
\fmfdot{in}
\fmfdot{out}
\end{fmfgraph}}+\,
\parbox{30mm}{\begin{fmfgraph}(30,5)
\fmfleft{in}
\fmfright{out}
\fmf{dashes}{in,v1}
\fmf{dashes}{v4,out}
\fmf{dashes,left,tension=0.7}{v1,v2}
\fmf{dashes,right,tension=0.7}{v1,v2}
\fmf{dashes}{v2,v3}
\fmf{dashes,left,tension=0.7}{v3,v4}
\fmf{dashes,right,tension=0.7}{v3,v4}
\fmfdot{v1}
\fmfdot{v2}
\fmfdot{v3}
\fmfdot{v4}
\fmfdot{in}
\fmfdot{out}
\end{fmfgraph}}+\cdots=\nonumber\\
&=G_2+G_2(-i\Sigma_1)G_2+G_2(-i\Sigma_1)G_2(-i\Sigma_1)G_2+\cdots=\frac{i}{p^2-m^2-\Sigma_1+i\varepsilon}\label{resummation}\end{align}
This looks familiar. If we can absorb the self-energy in some $m_{eff}^2$, equation~\ref{resummation} is the usual propagator, just with a different mass. Something similar occurs for the coupling constants appearing in the interaction vertices. In this case the correction is of the form:

\begin{align}
\parbox{35mm}{\begin{fmfgraph*}(25,15)
\fmfleft{q1}
\fmfright{q2,q3}
\fmf{dashes_arrow}{q1,v}
\fmf{dashes_arrow,label=$p+k$}{v,v1}
\fmf{dashes_arrow}{v1,q2}
\fmf{dashes_arrow,label=$k$}{v2,v}
\fmf{dashes_arrow}{v2,q3}
\fmflabel{$p$}{q1}
\fmflabel{$p-q$}{q2}
\fmflabel{$q$}{q3}
\fmffreeze
\fmf{dashes_arrow,label=$q+k$,l.side=right}{v1,v2}
\fmfdot{v}
\fmfdot{v1}
\fmfdot{v2}
\end{fmfgraph*}}&=ig^3\int\frac{d^4k}{(2\pi)^4}G_2(p+k)G_2(k)G_2(q+k)\label{scalarcouplingcorrection}\\
\nonumber
\end{align}
For the $\varphi^3$ theory diagram \eq{scalarcouplingcorrection} is finite, but for more complex theories this does not need to be the case. If a contribution to the vertex diverges, we need to renormalize the coupling constant. If all corrections have the same form as terms that are already present in the Lagrangian, a theory is called renormalizable. In that case all diagrams can be dealt with by a redefinition of masses and couplings. Non-renormalizable theories yield divergent higher order corrections with a form that is different from any term in the Lagrangian. For such theories the predictive power of the perturbative expansion is lost since additional interactions have to be introduced at every next order to absorb all divergences. An example is the theory that results from quantizing general relativity, which is why gravity is not incorporated in the Standard Model. An easy way to check if a theory is renormalizable is by power counting. The total mass dimension of the Lagrangian must be four. If this cannot be accomplished without negative dimensional couplings, higher order corrections will introduce terms with a different structure and a theory is non-renormalizable.

\subsubsection{Renormalization}\label{s:renormalizationsub}

Now that we have labeled our infinities, we need to decide how to deal with them. Equation \eq{selfenergysplit} splits the self-energy into a finite and a diverging part, but the value of both parts depends on the parameter $\mu$. In principle we can pick any value for $\mu$, since an arbitrary parameter should cancel in the final result: a change in renormalization prescription should not change physics. However, if a finite number of terms in the perturbative expansion is taken into account, the result does depend on $\mu$. In that case it is important that the perturbative expansion converges as fast as possible, so some values of $\mu$ are more natural than others. In the physical scheme, $\mu$ is chosen in such a way that the pole of the propagator resides at the physical mass $m_{ph}$. To this end we rewrite $m^2$ in the Lagrangian as:
\begin{align}m^2=m_{ph}^2+\delta m^2\label{renormalizationcounterterm}\end{align}
The resulting $\delta m^2$-dependent term in the Lagrangian is called a `counterterm' since it compensates infinities. Then we define the finite one-loop renormalized self-energy $\Sigma_{1R}$ as:
\[\Sigma_{1R}(p^2)\equiv\left[\Sigma_1(p^2,m_{ph}^2,\Lambda)+\delta m^2\right]_{\Lambda\to\infty}=\Sigma_{1fin}(p^2,m_{ph}^2,\mu^2)+[\delta m^2+\Sigma_{1div}(\mu^2)]\]

If a theory is renormalizable, it is always possible to find a $\delta m^2$ with $\delta m^2+\Sigma_{1div}<\infty$. Now we can tune the free parameter $\mu$ in such a way that $\Sigma_{1R}(m_{ph}^2)=0$. That ensures that the pole of the effective propagator resides at the physical mass. Finding the actual function $\Sigma_{1R}(p^2)$ requires some tedious calculation, but the upshot is that the propagator in the physical scheme is:
\[G_{2,ph}=\frac{i}{p^2-m_{ph}^2-\Sigma_{1R}(p^2)+i\varepsilon}\]

Since $\Sigma_{1R}(m_{ph}^2)=0$, the mass that we measure is indeed $m_{ph}$. The function $\Sigma_{1R}(p^2)$ determines the shape of the resonance. The residue of $G_{2,ph}$ at $p^2=m_{ph}^2$ changes the normalization of the propagator. In general it does not need to be finite. If it diverges, the field itself has to be renormalized by rewriting the field $\phi$ as $\phi=Z^{1/2}\phi_R$. This results in a rescaling of the propagator and is called the wave function renormalization. Of course $Z$ needs to be tuned to produce the correct normalization.

Finally we need to renormalize the coupling constant for corrections such as diagram \eq{scalarcouplingcorrection}. In the physical scheme this means that we choose the counterterms and free parameters such that the renormalized coupling constant coincides with the physical coupling constant. The end result is a consistent theory with redefined masses, fields and couplings.

The renormalization procedure seems a bit strange: we have expressions that yield infinity, then we subtract infinity, and we end up with well-defined quantities. How can that be? The answer is that we do not measure the bare quantities that occur in the original Lagrangian. It is impossible to measure a propagator without its loop corrections. That is also the reason we can add counterterms to the Lagrangian. As long as they have the same form as terms that are already present, we can never measure them anyway. The renormalization procedure redefines the theory in terms of quantities that have a physical meaning. It does not matter that the counterterms diverge. We can only get information about the theory through measurements, and they involve the physical quantities.

Of course the story does not end with one-loop corrections, there are also higher order contributions. The parameters of the theory need to be retuned for every additional loop correction one takes into account and calculating higher order corrections is a difficult and painstaking process. However, it can be shown for a number of theories that they are renormalizable to all orders. This is for instance the case for the Standard Model.

\subsubsection{Renormalization in the Standard Model}\label{s:renormalizationsm}

Since in the Standard Model there are a lot more possible loop diagrams than in the $\varphi^3$ theory, at first glance it is not even clear if the Standard Model is renormalizable. This is where symmetry becomes important: gauge theories are renormalizable \cite{thooftgauge} even if the symmetry is spontaneously broken \cite{thooftbrokensymmetry}. We will not prove this here, but we will show some examples of renormalization in the Standard Model using the Feynman rules listed in appendix~\ref{feynmanrules}.
\newline

\textsc{Natural Smallness of Parameters.} Apart from being renormalizable, the Standard Model has another nice property. For most cases it ensures that parameters that start out small stay naturally small even after renormalization. As an example, consider the fermion self-energy generated by the following graph:

\begin{align}\label{fermionselfenergy}
i\times\parbox{43mm}{\begin{fmfchar*}(40,17)
\fmfleft{in} \fmfright{out}
\fmfv{l=$\nu$,l.a=120,l.d=3thick}{bc}
\fmfv{l=$\mu$,l.a=60,l.d=3thick}{cs}
\fmf{plain_arrow,label=$p$}{in,bc}
\fmf{plain_arrow,label=$k$}{bc,cs}
\fmf{plain_arrow,label=$p$}{cs,out}
\fmffreeze
\fmf{wiggly_arrow,right, tension=0,label=$k-p$}{cs,bc}
\fmfdot{bc,cs}
\end{fmfchar*}}&=i(iQe)^2\int \frac{d^4k}{(2\pi)^4}\frac{-i^2\gamma^\mu(\slashed{k}+m)\gamma^\nu g_{\mu\nu}}{(k^2-m^2+i\varepsilon)((k-p)^2+i\varepsilon)}\nonumber\\
&=\frac{2iQ^2e^2}{(2\pi)^4}\int d^4k\frac{\slashed{k}-2m}{(k^2-m^2+i\varepsilon)((k-p)^2+i\varepsilon)}\end{align}
To quantify the divergence, we introduce a cut-off $\Lambda$ and do a Taylor expansion in $p$ of the photon propagator. Omitting the $i\varepsilon$ terms, the one loop self-energy is:
\[\Sigma_1=\frac{2iQ^2e^2}{(2\pi)^4}\int^\Lambda_{-\Lambda} d^4k\frac{\slashed{k}-2m}{k^2-m^2}\frac{1}{k^2}\left(1+\frac{2k\cdot p}{k^2}+\cdots\right)\]
The most divergent part is the first term in the Taylor expansion combined with $\slashed{k}$. However, this term is odd under $k\to-k$, so it integrates to zero. That means that the leading order terms of diagram \eq{fermionselfenergy} are:
\begin{equation}\Sigma_1\approx\frac{4iQ^2e^2}{(2\pi)^4}\int^\Lambda_{-\Lambda} d^4k\left[\frac{p\cdot k\slashed{k}}{k^4(k^2-m^2)}-\frac{m}{k^2(k^2-m^2)}\right]\label{fermionselfenergyleadingorder}\end{equation}
We can rewrite this by using the following identity, which holds for any even function $f(k)$:
\begin{equation}\int^\Lambda_{-\Lambda} d^4k\; \frac{p\cdot k\slashed{k}}{f(k)}=p_\mu\gamma_\nu\int^\Lambda_{-\Lambda} d^4k\; \frac{k^\mu k^\nu}{f(k)}=p_\mu\gamma_\nu\int^\Lambda_{-\Lambda} d^4k\; \frac{g^{\mu\nu}k^2}{4f(k)}=\slashed{p}\int^\Lambda_{-\Lambda} d^4k\; \frac{1}{4}\frac{k^2}{f(k)}\label{kmuknutrick}\end{equation}
Here we used that for $\mu\ne\nu$ the function $k^\mu k^\nu$ is odd. Since the cut-off is symmetric in all components of $k$, integrating these terms yields zero. The denominators in equation \eq{fermionselfenergyleadingorder} are even functions, so using equation \eq{kmuknutrick} the leading order terms of the self-energy yield:
\[\Sigma_1\propto\left(\slashed{p}-4m\right)\log\Lambda=-3m\log\Lambda+\left(\slashed{p}-m\right)\log\Lambda\]
The first term has to be removed by mass renormalization, the second term by wave function renormalization. We conclude that the mass term generated by graph \eq{fermionselfenergy} is logarithmically dependent on $\Lambda$ and proportional to the lowest order mass of the fermion $m$. This explains why fermion masses are naturally small. If the lowest order fermion mass is small, even choosing a large value for the renormalization parameter $\mu$ will not yield a high fermion mass at higher loop order. It turns out that this holds for higher orders as well. Thus the Standard Model preserves the smallness of the fermion mass in a natural way.
\newline

\textsc{Symmetry and Dimensional Regularization.} An example where this is less clear is the photon self-energy. Here we have an additional constraint, since the photon mass needs to remain strictly zero. However, the correction of a fermion loop to the photon self-energy is:
\begin{align}\label{photonselfenergycutoff}
i\times\parbox{33mm}{
\begin{fmfgraph*}(30,15)
\fmfleft{in}
\fmfright{out}
\fmf{photon,label=$p,,\nu$,label.side=right}{in,v1}
\fmf{photon,label=$p,,\mu$}{v2,out}
\fmf{fermion,right,tension=0.7,label=$k+p$}{v1,v2}
\fmf{fermion,right,tension=0.7,label=$k$}{v2,v1}
\fmfdot{v1}
\fmfdot{v2}
\end{fmfgraph*}}&=-i(iQe)^2\int \frac{d^4k}{(2\pi)^4}\frac{i^2\Tr\left(\gamma^\mu\left(\slashed{k}+\slashed{p}+m\right)\gamma^\nu\left(\slashed{k}+m\right)\right)}{\left((k+p)^2-m^2+i\varepsilon\right)\left(k^2-m^2+i\varepsilon\right)}\nonumber\\
&=-\frac{4ie^2Q^2}{(2\pi)^4}\int d^4k\frac{(k^\nu+p^\nu)k^\mu+(k^\mu+p^\mu)k^\nu-(k^2+p\cdot k-m^2)g^{\mu\nu}}{((k+p)^2-m^2+i\varepsilon)(k^2-m^2+i\varepsilon)}
\end{align}
Using the same procedure as for the fermion self-energy gives for the most divergent term:
\[\Sigma_1^{\mu\nu}\approx-\frac{4ie^2Q^2}{(2\pi)^4}\int^\Lambda_{-\Lambda} d^4k\frac{2k^\mu k^\nu-k^2g^{\mu\nu}}{((k+p)^2-m^2+i\varepsilon)(k^2-m^2+i\varepsilon)}\propto -g^{\mu\nu}\Lambda^2\]
The effective propagator follows in the same way as in equation \eq{resummation}. Absorbing all constants in $\Lambda^2$, the result is:
\[G_{2,eff}=i\frac{-g_{\mu\nu}}{p^2-\Lambda^2}\]

We cannot absorb $\Lambda^2$ in any term in the Lagrangian, so we now have a photon with an effective mass $\Lambda^2$. Clearly this is not correct. The reason that the cut-off procedure does not work is that the newly introduced momentum $\Lambda$ is not Lorentz invariant. This breaks the gauge symmetry of the theory. Without the cut-off, it is possible to shift $k$ in such a way that the integral has the required gauge symmetric form $\propto p^\mu p^\nu-g^{\mu\nu}p^2$, leaving the photon massless. After introducing $\Lambda$, a Lorentz boost does not leave the integral invariant, so $k$ cannot be shifted. It turns out that a regularization scheme that breaks a symmetry often fails to produce the correct results.

A regularization scheme that does respect Standard Model symmetries is dimensional regularization. Take a $D$-dimensional integral of the form that is encountered in loop diagrams:
\[\Sigma=\int d^Dk\frac{1}{(k^2-m^2+i\varepsilon)^2}\]
If $D\to4$ this integral diverges, but for dimensions smaller than four it is finite. In dimensional regularization we use this to our advantage by taking the dimension as regulator. The usual renormalization scheme used with dimensional regularization is minimal subtraction, where the counterterms are pure poles proportional to $1/(D-4)$.
 
It turns out that almost all Standard Model propagators and vertices have a logarithmic divergence that is proportional to the original parameters of the theory, similar to graph \eq{fermionselfenergy}. This ensures that these parameters will stay small if the tree-level parameters are small. The underlying reason is symmetry. Symmetries `protect' parameters, preventing them from becoming large, even if the symmetry is broken by the Higgs mechanism described in section \ref{s:higgs}. Gauge boson masses are protected by gauge symmetry, whereas fermion masses are protected by chiral symmetry. This also explains that regularization schemes have to respect the symmetries of the theory in order to give correct results.
\newline

\textsc{The Hierarchy Problem.} The only particle in the Standard Model that is not protected by any symmetry is the Higgs boson. This means we can safely use a cut-off for its regularization. The Higgs boson has many possible loop corrections, but we will pay particular attention to the fermion loop:
\begin{equation}
i\times\parbox{23mm}{\begin{fmfgraph}(20,10)
\fmfleft{in}
\fmfright{out}
\fmf{dashes}{in,v1}
\fmf{dashes}{v2,out}
\fmf{fermion,right,tension=0.6}{v1,v2}
\fmf{fermion,right,tension=0.6}{v2,v1}
\fmfdot{v1}
\fmfdot{v2}
\end{fmfgraph}}
=-\frac{i|y|^2}{(2\pi)^4}\int^\Lambda_{-\Lambda}d^4k\frac{\Tr\left((\slashed{k}+m)(\slashed{k}+\slashed{p}+m)\right)}{(k^2-m^2+i\varepsilon)((k+p)^2-m^2+i\varepsilon)}\propto-|y|^2\Lambda^2\label{hierarchyproblem}
\end{equation}
In the Standard Model, the Yukawa coupling $y$ is not related to any other interactions, so there are no diagrams that can cancel this divergence. Equation \eq{hierarchyproblem} implies that the smallness of the Higgs mass does not survive renormalization if the renormalization parameter $\mu$ is chosen to be large. Yet all predictions show that the Higgs mass must be relatively low. This discrepancy is called the hierarchy problem and it is one of the unsolved riddles in the Standard Model. We will come back to the hierarchy problem in the context of supersymmetry.

\subsubsection{Renormalization-Group Equations}\label{s:RGE}

The physical scheme mentioned in section \ref{s:renormalizationsub} is not the only possible renormalization prescription. Every possible value of $\mu$ is in principle a valid solution. If we also take different regularization schemes into account, the number of possibilities is even greater, since every regulator comes with its own collection of renormalization prescriptions. For the purpose of clarity, the discussion will be limited to a change in the parameter $\mu$ introduced in section~\ref{s:regularization}.

One could raise the question why we would want to use a different renormalization prescription than the physical scheme. The most important reason is to study high-energy behaviour. Recall from section~\ref{s:feynman} that Feynman diagrams are a perturbative expansion. If we want to get precise results at some high energy, we either have to do the expansion around some value close to the desired energy scale or compute high-order corrections. Looking at the simple one-loop examples from the previous section and realizing that both the number and the complexity of loop diagrams increases at higher order, it is clear that the first option is more attractive. That means we have to choose $\mu$ to have the same order of magnitude as the energy scale associated with the proces we want to study. The price we pay for this is that our masses and couplings change. This change is governed by the renormalization-group equations.

This approach is radically different from the physical scheme introduced in section~\ref{s:renormalization}. There the behaviour at the physical mass pole is calculated completely. Here we just want to obtain the general energy dependence of masses and couplings. That is most easily done by using minimal subtraction, taking pure poles as counterterms. We will not go into that much detail here, but it is useful to keep in mind that the goal in this section is to obtain large-scale behaviour only.

First recall how our theory is affected by a change in renormalization prescription. In a theory with just one bare mass $m_0$ and coupling $g_0$, we describe physics in terms of the renormalized masses and couplings $m(\mu)$ and $g(\mu)$, where $g$ is made dimensionless by multiplying with the appropriate power of $\mu$. We also have the wave function renormalization $Z$, but we will omit that for now since it is not important to the main argument. Changing $\mu$ does not affect physical observables such as the $S$-matrix, but it does change the effective masses and couplings. The set of all possible masses and couplings for all values of $\mu$ forms the renormalization-group (RG). 
If we have the RG, we can use it to translate from one energy scale to another, so our next task is to find the RG. For this we use the fact that the $S$-matrix is invariant under a change in variables $\left(\mu,m(\mu),g(\mu)\right)\to\left(\mu',m(\mu'),g(\mu')\right)$. We can rewrite this statement as:
\[\mu\frac{dS}{d\mu}=\left[\mu\frac{\partial}{\partial\mu}+\mu\frac{dg(\mu)}{d\mu}\frac{\partial}{\partial g}+\mu\frac{dm^2(\mu)}{d\mu}\frac{\partial}{\partial m^2}\right]S=0\]
From this equation we define the renormalization-group coefficients:
\begin{align}
\beta&=\mu\frac{dg(\mu)}{d\mu}\\
\gamma_m&=-\frac{\mu}{m^2}\frac{dm^2(\mu)}{d\mu}
\end{align}

\begin{figure}[!h]
\centering{
\includegraphics[width=9cm]{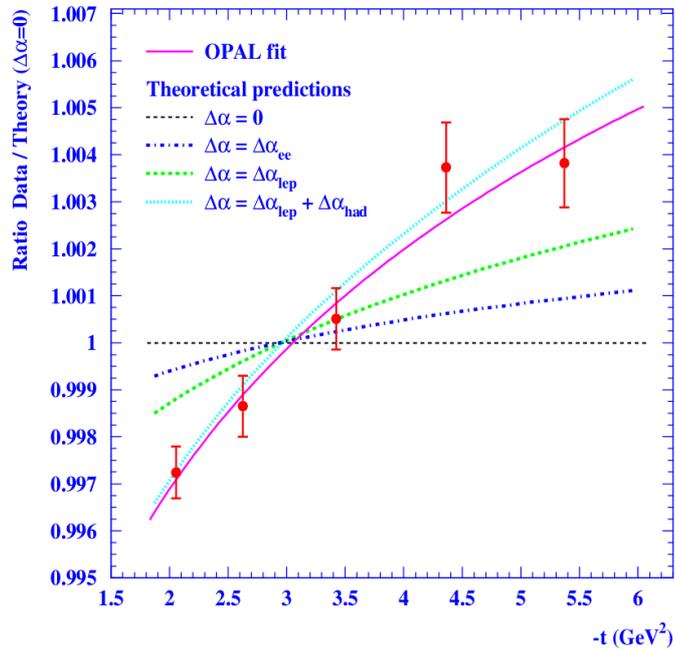}
\caption[Running QED coupling]{Measurement of the running QED coupling in Bhabha scattering by the OPAL collaboration \cite{Abbiendi:2005rx}. The flat line corresponds to a fixed coupling, the other dotted lines are predictions that include different types of loop contributions. The uppermost line includes all possible loops, while the solid line is the best fit to the data.}\label{f:runningcoupling}}
\end{figure}

If we know the RG coefficients, we have coupled differential equations for $g(\mu)$ and $m^2(\mu)$ so we can compute the effective couplings and masses at any energy scale $\mu$. These differential equations are the renormalization-group equations (RGEs).
\newpage
Our problem is now reduced to obtaining the RG coefficients, which can be done perturbatively. From equation \eq{renormalizationcounterterm} we know how to write the bare masses and couplings $m_0$ and $g_0$ perturbatively in terms of the effective masses and couplings $m^2(\mu)$ and $g(\mu)$
\begin{align}
g_0&=g(\mu)\left(1+\frac{[\delta g]_{1}}{g}+\frac{[\delta g]_{2}}{g}+\cdots\right)\label{g0expansion}\\
m^2_0&=m^2(\mu)\left(1+\frac{[\delta m^2]_{1}}{m^2}+\frac{[\delta m^2]_{2}}{m^2}+\cdots\right)\label{m^2expansion}
\end{align}
where $[\delta g]_j$ and $[\delta m^2]_j$ are the perturbative corrections of order $j$, which are proportional to $g^{2j}$. Calculating these terms is complicated, but if we have the loop corrections to a certain order, we can use them to calculate the RG coefficients by realizing that the bare masses and couplings do not depend on $\mu$. This gives us the differential equations:
\begin{align*}
\mu\frac{dg_0}{d\mu}&=\left[\mu\frac{\partial}{\partial\mu}+\beta(g)\frac{\partial}{\partial g}-\gamma_m(g)m^2\frac{\partial}{\partial m^2}\right]g_0=0\\
\mu\frac{dm^2_0}{d\mu}&=\left[\mu\frac{\partial}{\partial\mu}+\beta(g)\frac{\partial}{\partial g}-\gamma_m(g)m^2\frac{\partial}{\partial m^2}\right]m^2_0=0
\end{align*}
that we can solve for $\beta$ and $\gamma_m$ by inserting \eq{g0expansion} and \eq{m^2expansion}. So if we calculate equations \eq{g0expansion} and \eq{m^2expansion} up to a certain order, we can obtain the RGEs to the same order. The RGEs can then be used to predict high-energy phenemenology. They determine how masses and couplings change with the energy scale. These so-called `running' masses and couplings have been confirmed by experiment, as can be seen in figure~\ref{f:runningcoupling} for the QED coupling.
\vspace{0.5cm}
\newpage
Obviously the most difficult part of this procedure is obtaining the RGEs, since this involves calculating loop diagrams. One may wonder why we replace one type of loop calculation by another. The answer is that even though we still need to calculate some loop diagrams, we can afford to calculate them to a much lower order. We essentially rearrange the perturbation series to take into account the $\mu$-dependence, which becomes extremely important when the difference between the renormalization scale and the energy scale of the process is large. This way we get a good precision using relatively low-order results, even for extremely high energies. The two-loop RGEs have been calculated for many theories and can be used to obtain precise predictions for high-energy behaviour from low-energy measurements.

\end{fmffile}
\begin{fmffile}{feynsusy}
\fmfset{dot_size}{1thick}
\fmfset{curly_len}{1.5mm}
\fmfset{arrow_len}{3mm}
\fmfpen{thin}

\fmfcmd{%
vardef cross_bar (expr p, len, ang) =
 ((-len/2,0)--(len/2,0))
 rotated (ang + angle direction length(p)/2 of p)
 shifted point length(p)/2 of p
 enddef;
 style_def phantom_crossed expr p=
 ccutdraw cross_bar (p, 3mm, 45);
 ccutdraw cross_bar (p, 3mm, -45);
enddef;
  style_def gaugino expr p =
  cdraw (wiggly p);
  cdraw p;
 enddef;}

\section{Supersymmetry}\label{s:supersymmetry}

The Standard Model covers almost all experimental observations of particles. An exception might be neutrino masses \cite{Fukuda:1998mi}, but they can easily be incorporated \cite{Akhmedov:1999tm}. However, although the agreement with experiment is very good, there is a number of theoretical problems and unanswered questions. Some examples are:
\begin{itemize}
\item Gravity is not described by the Standard Model and as mentioned in section~\ref{s:renormalizationsub}, attempts to incorporate it lead to a non-renormalizable theory.
\item Symmetry is a beautiful concept to build a theory on, but why is the Standard Model described by $SU(3)\times SU(2)\times U(1)$ and not by some other group?
\item Why do tree-level masses and coupling constants have the values they have? There are a lot of free parameters in the Standard Model and the mass differences between for instance the fermion masses are quite big. Even worse, compared to the Planck scale all masses are ridiculously small. A theory that explains this would be more compelling.
\item As mentioned in section \ref{s:regularization}, it is unlikely that there is no physics beyond the Standard Model. If the Standard Model Higgs interacts with the particles associated with this new physics, these high-mass particles contribute to loop corrections. The Standard Model does not protect the mass of the Higgs boson against such high-scale corrections, leading to the hierarchy problem described in section~\ref{s:renormalizationsm}. An additional symmetry could protect the Higgs mass in the same way the other Standard Model masses are protected from high-scale corrections.
\item The Standard Model accounts for only 4\% of the universe. From astrophysical data we know that an additional 20\% consists of so-called dark matter \cite{Bertone:2004pz}\cite{Spergel:2006hy}. We cannot observe it directly so we do not know any of its properties. The remaining 76\% is the even more mysterious dark energy. Clearly a theoretical framework is needed to explain all this\label{susyanddarkmatter}.
\end{itemize}
This list is not exhaustive \cite{Mohapatra:1990jc}\cite{schellekensbtsm}, but it gives enough reasons to look beyond the Standard Model. Many theories have been proposed to address these problems. Some examples are string theory \cite{zwiebach2004}, which is the only theory that successfully includes gravity, Grand Unified Theories (GUTs, see Appendix~\ref{app:gut}), which partially address the group structure and the small tree-level masses in the Standard Model, and supersymmetry, which is the topic of this thesis.

As we will see in this section, supersymmetry protects the Higgs mass from high-scale corrections and thus provides a solution for the hierarchy problem. In addition, most supersymmetric theories predict that the lightest supersymmetric particle is stable and nearly undetectable. This makes it a good candidate for dark matter, so supersymmetry explains both the hierarchy problem and dark matter.

Supersymmetry is also interesting in the context of other theories. For instance, the minimal GUT is not consistent with the Standard Model, while it is with supersymmetry (see Appendix~\ref{app:gut}). In addition, most string theories predict supersymmetry. Thus supersymmetry is a promising theory for both experimental and theoretical reasons.

\subsection{What is Supersymmetry?}

We have seen in section~\ref{s:gauge} that symmetry arguments are a powerful tool in theory building. Therefore a natural way to extend the Standard Model would be to find an additional symmetry. Supersymmetry is a symmetry between bosons and fermions, so a transformation $Q$ with:
\begin{equation}Q|\mathrm{boson}\rangle=|\mathrm{fermion}\rangle\quad\quad Q|\mathrm{fermion}\rangle=|\mathrm{boson}\rangle\label{susytransformation}\end{equation}
Supersymmetry is usually treated as a global symmetry, so $\partial_\mu Q=0$. From equation \eq{susytransformation} it follows that $Q$ is fermionic, which leads to an algebra of anticommutation relations. It turns out that the anticommutator is proportional to the momentum \cite{aitchison2005}:
\[\{Q,Q^\dagger\}\sim p^\mu\]
Since $Q$ commutes with both $p^2$ and the gauge transformations, any particles that are in the same supermultiplet have equal masses and gauge charges. Furthermore, within a supermultiplet the numbers of bosonic and fermionic degrees of freedom need to be equal.

These properties have important phenomenological consequences. Looking at the Standard Model particles in table~\ref{standardmodelparticles}, we see that none of them can be each others superpartners. That means we have to introduce new particles and that we will end up with twice as many particles as we have now. Also, since the masses of these particles are predicted to be the same as those of their Standard Model partners, they should have been measured long ago. This apparent contradiction with experiment is addressed by supersymmetry breaking in section~\ref{s:susybreaking}.

A technical detail of supersymmetry is that fermions are usually described by left-handed Weyl spinors. The Weyl representation splits a fermion field into a left-handed and a right-handed part, which is more natural since weak interactions treat left-handed particles different than right-handed particles. A four-component Dirac spinor can be written in terms of two Weyl spinors\footnote{A Dirac spinor written in terms of Weyl spinors has the form
$\Psi_D=\left( \xi_\alpha,\chi^{\dagger\dot\alpha}\right)$ where $\xi$ is a left-handed Weyl spinor and $\chi^\dagger$ a right-handed Weyl spinor. The independent indices $\alpha$ and $\dot\alpha$ can take the values 1 and 2. The undotted indices denote the first two components of the Dirac spinor and the dotted indices the last two components. Weyl spinors are contracted using the anti-symmetric $\epsilon$ tensor: $\epsilon^{12}=-\epsilon^{21}=\epsilon_{21}=-\epsilon_{12}=1,\epsilon^{11}=\epsilon^{22}=\epsilon_{11}=\epsilon_{22}=0$. This notation is very useful in proving that a Lagrangian is supersymmetric, but it can be quite confusing. We will not explicitly prove supersymmetry for our Lagrangians, but cite the results from \cite{martin2006}.\label{epsilonfootnote}} that each have two complex degrees of freedom. Using charge conjugation, a right-handed Weyl spinor can be written as a left-handed one with opposite hypercharge. In working with Lagrangians, the following notation will be used:
\begin{equation}f\equiv f_L\qquad\bar f^\dagger\equiv f_R\label{bardefinition}\end{equation}
This notation only applies to fields. When discussing particles, the labels $L$ and $R$ are used and a bar is used to label an antiparticle. The meaning of a bar should be clear from the context.

Finally a few words on nomenclature. Supersymmetric partners of fermions have the name of their Standard Model counterparts preceded by an `s'. For partners of bosons, we add `ino' to the end of their name. This leads to silly names like winos and squarks, but it is unambiguous. The symbol for a supersymmetric particle is usually the Standard Model symbol with a tilde, so a top squark (`stop') is denoted as $\tilde t$. An exception is the chargino and neutralino sector\footnote{A chargino is a linear combination of a Wino and a charged Higgsino. A neutralino contains the photino, the Zino and the neutral Higgsinos.}, where the gauge eigenstates do not correspond to the mass eigenstates due to mixing effects.

\subsection{Supersymmetric Lagrangians}\label{s:susylagrangians}

We want supersymmetry to be an extension of the Standard Model, a theory with spin 0, spin 1/2 and spin 1 particles. The simplest way to accomplish that is by using two kinds of supermultiplets:
\begin{itemize}
\item a \textit{chiral supermultiplet}, which combines a Weyl fermion with a complex scalar.
\item a \textit{gauge supermultiplet}, which combines a gauge boson with a Weyl fermion.
\end{itemize}
These supermultiplets are the building blocks of our supersymmetric theory. However, they are incomplete. A complex scalar has two degrees of freedom, while an off-shell\footnote{A particle that is on-shell satisfies the `classical' equation of motion, while an off-shell particle does not. Particles that correspond to internal lines in Feynman diagrams are in general off-shell. External lines are on-shell.} Weyl fermion has four. This mismatch within the chiral supermultiplet can be solved by introducing an auxiliary field $F$, a complex scalar field that enters the Lagrangian as:
\begin{equation}
\mathcal{L}_{F}=F^*F\label{auxfieldF}
\end{equation}
The gauge multiplet needs a similar adjustment, since an off-shell gauge boson only has three degrees of freedom instead of the necessary four to match the Weyl fermion. Here we introduce a real bosonic auxiliary field $D^a$ with a corresponding term in the Lagrangian:
\begin{equation}
\mathcal{L}_{D}=\frac{1}{2}D^aD^a\label{auxfieldD}
\end{equation}
The auxiliary fields $D$ and $F$ do not have any kinetic terms, since they are not propagating fields. This can be understood by the fact that the mismatch occurs off-shell only. On-shell, all fields involved have two real degrees of freedom and the $F$ and $D$ terms vanish. Since the auxiliary fields are not physical, they are fixed by their equations of motion, which follow from the Lagrangians of the chiral and gauge supermultiplets.

Now that we have the building blocks, the next step is to obtain a supersymmetric Lagrangian. The basic concepts are the same as in the Standard Model case of section~\ref{s:gauge}. We start with the chiral supermultiplets and their interactions. These are specified by the superpotential. Then we add the gauge supermultiplets and their interactions by requiring gauge invariance. The resulting Lagrangian will have many terms, but they all follow from the superpotential and the gauge group.
\newline

\textsc{The Superpotential.} In the Standard Model our starting point was a fermion Lagrangian, so let us write down the most general Lagrangian for a collection of chiral supermultiplets labeled by indices $i$ and $j$. This involves kinetic terms for the scalars $\phi_i$ and the fermions $\psi_i$, the Lagrangian \eq{auxfieldF} for the auxiliary fields $F_i$ and interactions between $\phi_i$, $\psi_i$ and $F_i$. The possible interaction terms are limited by demanding the Lagrangian to be renormalizable, so the mass dimension cannot be larger than four (see section~\ref{s:renormalizationsub}), and supersymmetric, so it is invariant under $Q$. This leads to an interaction Lagrangian of the form \cite{martin2006}:
\begin{equation}\mathcal{L}_{int}=\left(-\frac{1}{2}W^{ij}\psi_i\psi_j+W^iF_i\right)+c.c.\label{generalinteractionlagrangian}\end{equation}
where $W^{ij}$ and $W^i$ are polynomials in the scalar fields that are related to each other as:
\[W^{ij}=\frac{\delta^2W}{\delta\phi_i\delta\phi_j}\quad\quad\quad W^i=\frac{\delta W}{\delta\phi_i}\]
with $W$ an analytic function in the scalar fields $\phi_i$ that is called the \textit{superpotential}:
\begin{equation}
W=L^i\phi_i+\frac{1}{2}M^{ij}\phi_i\phi_j+\frac{1}{6}y^{ijk}\phi_i\phi_j\phi_k\label{generalsuperpotential}
\end{equation}
Inserting equation \eq{generalsuperpotential} in the first term of the Lagrangian \eq{generalinteractionlagrangian}, we can recognise a Yukawa interaction similar to equation \eq{higgsfermioncoupling} in the term containing $y^{ijk}$. The term with $M^{ij}$ contains a fermion mass term, which is allowed in a general theory. The linear term is only allowed if $\phi_i$ is a gauge singlet, so it does not occur in the Standard Model. We will discuss the second term in equation \eq{generalinteractionlagrangian} later.
\newline

\textsc{Gauge Invariance.} Having specified the Lagrangian for the chiral supermultiplet, we require it to be gauge invariant. The recipe is essentially the same as in section~\ref{s:gauge}, except that we are dealing with supermultiplets. In the Standard Model, gauge invariance results in gauge bosons. In a supersymmetric theory we cannot introduce new particles without introducing their supersymmetric partners. That means gauge invariance yields gauge supermultiplets that contain not only gauge bosons, but also gauginos and $D$ terms. And of course all these fields interact with each other.

Since the interactions are supersymmetric, any Standard Model gauge coupling will have a supersymmetric counterpart. In other words, we have to take into account all interactions between the gauge and the chiral supermultiplet as well. In order to complete the Lagrangian, we write down all possible interaction terms that are renormalizable and yield a supersymmetric Lagrangian.

The $F$ and $D$ terms are then fixed by their equations of motion. It turns out that \cite{martin2006}:
\begin{equation}
F_i=-W_i^*\qquad F^{*i}=-W^i\qquad D^a=-g\phi^*T^a\phi\label{FandDfixed}
\end{equation}
where $T^a$ is the generator of the gauge group and $g$ the gauge coupling. Inserting this into equations \eq{auxfieldF} and \eq{auxfieldD}, we can see that the $F$ and the $D$ terms introduce trilinear and quartic scalar interactions into the Lagrangian.
\newpage
\noindent We now have the complete Lagrangian that contains:
\begin{itemize}
\setlength{\itemsep}{2pt}
\setlength{\parskip}{0pt}
\setlength{\parsep}{0pt}
\item Interaction terms within the chiral supermultiplet, specified by the superpotential.
\item Kinetic terms for the chiral supermultiplet.
\item Interaction terms between the chiral supermultiplet and the gauge supermultiplet.
\item Kinetic terms for the gauge supermultiplet.
\item Interaction terms within the gauge supermultiplet.
\item Trilinear and quartic scalar interactions coming from the $F$ and $D$ terms.
\end{itemize}

In spite of the plethora of terms, the Lagrangian is completely determined by the superpotential and the gauge group. The superpotential specifies the chiral supermultiplets and their interactions. By requiring gauge invariance we can derive the gauge supermultiplets and all other interactions. Thus the entire theory is specified by a single equation and the gauge group.

\subsection{The Minimal Supersymmetric Standard Model}\label{s:mssm}

Although we now have the basic form of a supersymmetric Lagrangian, this still leaves a lot of freedom in constructing an actual theory. One can introduce any number of chiral supermultiplets and gauge symmetries or even have several supersymmetry generators. Some of these theories have interesting characteristics\footnote{For supersymmetric gauge theories with four supersymmetry generators, it can be shown that non-renormalization theorems apply, which state that parameters do not have to be renormalized or that higher loop corrections do not contribute \cite{Dine:1999jq}. This spurred the hope for a renormalizable theory of gravity using supersymmetry. So far such a theory has not been found.}, but we will restrict ourselves to the simplest extension of the Standard Model, with one supersymmetric generator and a minimal particle content. The Minimal Supersymmetric Standard Model (MSSM) is specified by the superpotential:
\begin{equation}\label{W_MSSM}
W=\tilde{\bar u}\mathbf{y}_u\tilde QH_2-\tilde{\bar d}\mathbf{y}_d\tilde QH_1-\tilde{\bar e}\mathbf{y}_e\tilde LH_1+\mu H_2H_1
\end{equation}
The Yukawa couplings $\mathbf{y}_i$ are $3\times3$ matrices in the family space of generations and all gauge and family indices have been suppressed. The sfermion fields correspond to their Standard Model counterparts in the notation of table~\ref{standardmodelparticles}. Equation \eq{W_MSSM} has some interesting implications.
\newline

\textsc{Two Higgs Doublets.} The first three terms resemble the Standard Model Yukawa couplings, but in supersymmetry we need two Higgs doublets. Remember from section~\ref{s:higgs} that the charge conjugate of the Higgs doublet is used to give mass to the up-type fermions. However, the charge conjugate of a scalar field is not an analytic function in that field, so it cannot appear in the superpotential. Therefore a second Higgs doublet with opposite hypercharge is introduced to give mass to the up-type fermions after spontaneous symmetry breaking\footnote{A second Higgs doublet is also needed for anomaly cancellations \cite{Haber:1993wf}.}. Explicitly, the Higgs doublets are:
\[H_1=\left(\begin{array}{c}H_1^0\\H_1^-\end{array}\right)\qquad H_2=\left(\begin{array}{c}H_2^+\\H_2^0\end{array}\right)\]
Only the neutral components $H_i^0$ can get a vev, for otherwise the photon would couple to the charged vev and acquire a mass. So it would seem that the $H_2$ doublet gives mass to the down-type fermions. However, contraction using an $\epsilon$-tensor (see footnote~\ref{epsilonfootnote}) is implicitly understood in equation \eq{W_MSSM}, so the lower component of $H_2$ combines with the up-type fermions and the upper component of $H_1$ with the down-type fermions.
\newline

\textsc{The Hierarchy Problem.} In the Standard Model, the Yukawa coupling $y$ is not related to any other interaction. This is not the case in supersymmetry. It follows from equations \eq{auxfieldF} and \eq{FandDfixed} that $y$ also occurs quadratically in quartic interactions coming from the $F$ terms that have the form $|y_f|^2H_iH_i^*\tilde f\tilde f^*$ or $|y_f|^2H_i^*H_i\tilde{\bar f}^*\tilde{\bar f}$, with $H_i$ ($i=1,2$) the appropriate Higgs field for the sfermion $\tilde f$. They give rise to radiative corrections to the Higgs propagator:
\newline

\begin{equation}i\times\parbox{15mm}{\begin{fmfchar*}(15,7)
\fmfforce{sw}{in}
\fmfforce{se}{out}
\fmf{dashes}{in,v,out}
\fmf{dashes_arrow,tension=0.6,label=$\tilde{f}_L,,\tilde{f}_R$}{v,v}
\fmfdot{v}
\end{fmfchar*}}=\frac{i|y_f|^2}{(2\pi)^4}\int^\Lambda_{-\Lambda} d^4k\frac{1}{k^2-m_{\tilde f}^2+i\epsilon}\propto |y_f|^2\Lambda^2\end{equation}
Taking into account the combinatorial factors, it turns out that the quadratic divergences in these diagrams exactly cancel the quadratic divergences from the fermion loop correction in section~\ref{s:renormalizationsm}. Thus supersymmetry solves the hierarchy problem. If the supersymmetric particles are too heavy, a new hierarchy problem is introduced by the next order logarithmic term, which depends quadratically on the mass difference between the superpartners. This indicates that supersymmetric particles cannot be too heavy. Usually it is assumed that the masses of the lightest superpartners cannot be much larger than 1 TeV.
\newline

\textsc{$R$-Parity and Dark Matter.} The superpotential \eq{W_MSSM} reproduces the Standard Model particles and interactions, but it is not the most general superpotential for the MSSM chiral supermultiplets. Other terms that are allowed by supersymmetry and gauge invariance do not conserve baryon and lepton number, so experiment puts severe limits on the magnitude of such couplings \cite{Yao:2006px}. To explain the absence of these terms we introduce a new symmetry with a multiplicative quantum number called $R$-parity $P_R=(-1)^{3(B-L)-2s}$ with $B$ the baryon number, $L$ the lepton number and $s$ the spin of the particle. For Standard Model particles $P_R=1$. For their supersymmetric partners $P_R=-1$, since their spin is shifted by $\pm1/2$. Conservation of $R$-parity implies that in collider experiments supersymmetric particles are always produced in pairs as the initial state consists of Standard Model particles. It also means that the Lightest Supersymmetric Particle (LSP) is stable, since it cannot decay to Standard Model particles alone. In many supersymmetric models, the LSP is a neutralino. This particle is almost impossible to detect, which makes it a good candidate for dark matter (see page~\pageref{susyanddarkmatter}).
\newline

\textsc{Masses and Spontaneous Symmetry Breaking.} The MSSM Lagrangian has all Standard Model symmetries, so particles that are massless in the unbroken Standard Model do not have a mass here either. As a result only the Higgs supermultiplets have the equivalent of a mass term, which is the last term in the superpotential \eq{W_MSSM}. It contains the only new parameter in the theory $\mu$. The Higgs potential is completely determined by the $F$ and the $D$ terms and is given by:
\begin{equation}
V=|\mu|^2\left(|H_2^0|^2+|H_2^+|^2+|H_1^0|^2+|H_1^-
|^2\right)+\mbox{``quartic terms''}\geq 0
\end{equation}
This potential is positive with one global minimum at the origin, so it does not give rise to spontaneous symmetry breaking. That means that the Higgs mechanism, which generates mass in the Standard Model (see section~\ref{s:higgs}), does not work in unbroken supersymmetry. Since the MSSM superpotential only contains mass terms for the Higgs fields, that implies that all particles except the Higgs are massless! This is another reason that supersymmetry cannot be exact.

\subsection{Supersymmetry Breaking}\label{s:susybreaking}

Unbroken supersymmetry predicts that there are no mass differences within supermultiplets and the MSSM does not explain mass at all. Clearly we need to find a way to break supersymmetry. We have seen in section~\ref{s:higgs} how to break a symmetry while preserving key aspects of the theory. However, for supersymmetry this is not easy. Several methods have been proposed, but none of them are able to derive a broken supersymmetric Lagrangian from first principles. We will first shortly discuss some generalities of supersymmetry breaking and then look at how supersymmetry is broken in practice.

\subsubsection{Breaking Mechanisms}\label{s:breakingmechanisms}

The only way we know to break a symmetry while maintaining the characteristics of a theory is spontaneous symmetry breaking. So our only hope is to do something similar and change our theory in such a way that the vacuum is not invariant under the supersymmetry transformation $Q$. There are two terms in the supersymmetry Lagrangian that can acquire a vev without breaking other symmetries: the $F$ terms and the $D$ terms.

Fayet-Iliopoulos breaking \cite{Fayet:1974jb} adds a linear term to the auxiliary field $D$. Since we want our theory to remain gauge invariant, this is only possible for the $D$ term of the $U(1)_Y$ gauge supermultiplet. But even in that case it does not work. The extra term changes the equations of motion for the $D$ field and in order to make the field vanish on-shell, some of the squarks and charged sleptons have to acquire a vev. This would break not only supersymmetry, but also the gauge symmetries and is therefore unacceptable.

The other option is that the $F$ field acquires a vev. This is called O'Raifeartaigh breaking \cite{O'Raifeartaigh:1975pr} and is equivalent to keeping the linear term in equation \eq{generalsuperpotential}. As mentioned in section~\ref{s:susylagrangians}, we do not have a gauge singlet in the Standard Model that could provide this term. Therefore one usually assumes that this type of breaking originates in some hidden sector that contains particles that couple very weakly to the particles we observe. These small interactions would be responsible for symmetry breaking through radiative corrections.

There are many possible scenarios for this hidden sector and the way it couples to the visible sector. A few examples are gauge-mediated supersymmetry breaking \cite{Giudice:1998bp}, anomaly-mediated supersymmetry breaking \cite{Feng:1999hg} and extra-dimensional mediated supersymmetry breaking \cite{Kaplan:2000av}. Probably the most popular scenario is Planck-scale-mediated supersymmetry breaking \cite{Cremmer:1982vy}. This assumes that the breaking sector couples to the visible sector through gravitational interactions. Since general relativity is inherently local, this makes it necessary to treat supersymmetry as a local symmetry, just like the gauge symmetries. This is mathematically complicated. In addition, we do not have a quantum theory of gravity, so it is impossible to derive low-energy phenomenology from first principles. However, as we will see in section~\ref{s:msugra}, gravity-mediated supersymmetry breaking can put important constraints on the effective form of supersymmetry breaking.

\subsubsection{Soft Breaking Terms}\label{s:softbreakingterms}

The lack of a supersymmetry breaking mechanism does not seem very promising for supersymmetry, but in practice we ignore our ignorance by manually introducing breaking terms. We describe the theory by an effective Lagrangian that includes all possible breaking terms that leave the basic properties of the theory intact. In other words, the Lagrangian has to remain renormalizable and the breaking terms should not introduce quadratic divergences that would give rise to a new hierarchy problem. This is called soft breaking. The possible soft breaking terms are \cite{martin2006}:
\begin{itemize}
\item Gaugino masses for each gauge group: $-\frac{1}{2}(M_3\tilde g\tilde g+M_2\tilde W\tilde W+M_1\tilde B\tilde B+c.c.)$ with $M_i\in\mathbb{C}$
\item Sfermion masses: $-\tilde Q^\dagger\mathbf{m}_{\tilde Q}^2\tilde Q-\tilde L^\dagger\mathbf{m}_{\tilde L}^2\tilde L-\tilde{\bar u}\mathbf{m}_{\tilde{\bar u}}^2\tilde{\bar u}^\dagger-\tilde{\bar d}\mathbf{m}_{\tilde{\bar d}}^2\tilde{\bar d}^\dagger-\tilde{\bar e}\mathbf{m}_{\tilde{\bar e}}^2\tilde{\bar e}^\dagger$ where the mass matrices are $3\times3$ hermitian matrices in family space
\item Higgs masses and mixing: $-m_{H_2}^2H_2^\dagger H_2-m_{H_1}^2H_1^\dagger H_1-(bH_2 H_1+c.c.)$ with $m_{H_i}^2\in\mathbb{R},b\in\mathbb{C}$
\item Triple scalar couplings: $-(\tilde{\bar u}\mathbf{a}_u\tilde Q H_2-\tilde{\bar d}\mathbf{a}_d\tilde Q H_1-\tilde{\bar e}\mathbf{a}_e\tilde L H_1+c.c.)$ where the scalar couplings are complex $3\times3$ matrices in family space.
\end{itemize}
These terms break supersymmetry, but they conserve $R$-parity. Also, in contrast to the equivalent mass terms for the Standard Model particles, the gaugino and sfermion mass terms respect chiral and gauge symmetry. In that sense the mass difference between Standard Model particles and their supersymmetric partners is quite natural.

\subsubsection{Spontaneous Symmetry Breaking}\label{s:susyhiggsmech}

For spontaneous symmetry breaking the Higgs potential needs to have minima outside the origin. This requirement is not met in unbroken supersymmetry, but the soft breaking terms change the potential. Writing out the Higgs fields in their components, the Higgs potential with the breaking terms included is:
\begin{align*}
V=&\left(|\mu|^2+m_{H_2}^2\right)\left(|H_2^0|^2+|H_2^+|^2\right)+\left(|\mu|^2+m_{H_1}^2\right)\left(|H_1^0|^2+|H_1^-|^2)+[b(H_2^+H_1^--H_2^0H_1^0\right)+c.c.]\\
&+\frac{1}{8}\left(g^2+g'^2\right)\left(|H_2^0|^2+|H_2^+|^2-|H_1^0|^2-|H_1^-|^2\right)^2+\frac{1}{2}g^2\left|H_2^+H_1^{0*}+H_2^0H_1^{-*}\right|^2
\end{align*}
where the breaking parameters $b$ and $m_{H_i}$ can take complex values. Just like in section~\ref{s:higgs}, we can simplify this expression by fixing the gauge. First we use $SU(2)$ rotations to choose $H_2^+=0$. Since in a minimum we must have $\partial V/\partial H^+_2=0$, this also implies $H_1^-=0$. Then we use that the $b$ term is the only term that depends on the phase of the Higgs fields. Using a $U(1)$ phase rotation to make $b$ real and positive, it follows that in a minimum of the potential $H_2^0H_1^0$ must be real and positive as well. In other words: the Higgses have opposite phases. Since they have opposite hypercharge, we can choose both Higgses to be real and positive. In this gauge the equation takes a more manageable form:
\[V=(|\mu|^2+m_{H_2}^2)|H_2^0|^2+(|\mu|^2+m_{H_1}^2)|H_1^0|^2-[bH_2^0H_1^0+c.c.]+\frac{1}{8}(g^2+g'^2)(|H_2^0|^2-|H_1^0|^2)^2\]
The quartic terms ensure that the potential is bounded from below for almost all values of $H_2^0$ and $H_1^0$. However for $H_2^0=H_1^0$ the quartic terms vanish and the potential will not be bounded from below unless:
\begin{equation}2b<\left(|\mu|^2+m_{H_2}^2\right)+\left(|\mu|^2+m_{H_1}^2\right)\label{spontaneoussymmetrybreakinglowerlimit}\end{equation}
This implies that not both $|\mu|^2+m_{H_1}^2$ and $|\mu|^2+m_{H_2}^2$ can be negative, so we can only have a minimum outside the origin if the origin itself is a saddle point, which is the case if:
\begin{equation}b^2>(|\mu|^2+m_{H_2}^2)(|\mu|^2+m_{H_1}^2)\label{spontaneoussymmetrybreakingupperlimit}\end{equation}
If equations \eq{spontaneoussymmetrybreakinglowerlimit} and \eq{spontaneoussymmetrybreakingupperlimit} are met, supersymmetry breaking results in spontaneous symmetry breaking. The ratio of the vevs of the neutral Higgs components is usually written as:
\begin{equation}
 \tan\beta=\frac{\langle H_2\rangle}{\langle H_1\rangle}\label{tanbeta}
\end{equation}
Just like in the Standard Model, the Higgs vevs are related to the mass of the $W$ and $Z$ bosons $m_W$ and $m_Z$. The minimum of the potential satisfies $\partial V/\partial H_1=\partial V/\partial H_2=0$, which gives us two equations we can use to express $b$ and $|\mu|^2$ in terms of $\tan\beta$. At tree level, the result is:
\begin{align}
b&=\frac{m_{H_1}^2-m_{H_2}^2+m_Z^2\cos(2\beta)}{\tan\beta-\cot\beta}\label{binbeta}\\
|\mu|^2&=\frac{1}{2}\left(b(\cot\beta+\tan\beta)-m_{H_1}^2-m_{H_2}^2\right)\label{muinbeta}
\end{align}
A complex phase of $\mu$ would introduce large $CP$-violating effects that are not observed in nature, so we demand $\mu$ to be real. That means we 
can eliminate $b$ and $\mu$ in favour of $\tan\beta$ and the sign of $\mu$. For phenomenological purposes, these variables are more convenient, since particle masses are closely related to $\tan\beta$.

In the Standard Model, three of the four degrees of freedom of the Higgs doublet were `eaten' by the gauge bosons and one physical Higgs boson was left. In the MSSM there are two complex Higgs doublets, so a total of eight degrees of freedom. The gauge bosons still absorb three of them, so we now end up with five Higgs bosons: $h^0,H^0,H^+,H^-$ and $A^0$. By convention $h^0$ is lighter than $H^0$ and $A^0$ is the only $CP$-odd state.

Due to the mixing terms, the mass eigenstates are considerably different from the gauge eigenstates. Diagonalizing the mass matrix gives the following masses at tree level \cite{martin2006}:
\begin{align}
m_{A_0}^2&=\frac{2b}{\sin(2\beta)}\label{A0mass}\\
m_{h^0,H^0}^2&=\frac{1}{2}\left(m_{A^0}^2+m_Z^2\mp\sqrt{(m_{A^0}^2-m_Z^2)^2+4m_Z^2m_{A^0}^2\sin^2(2\beta)}\right)\label{h0mass}\\
m_{H^\pm}^2&=m_{A^0}^2+m_W^2\label{Hpmmass}
\end{align}

This result indicates a low $h^0$ mass. In fact $m_{h^0}\leq m_Z|\cos(2\beta)|$ \cite{martin2006}, which is excluded experimentally. Due to large radiative corrections from especially tops and stops, the actual $h^0$ mass is larger, but it cannot exceed 135~GeV \cite{Heinemeyer:1998np}, which means it has to be found at the LHC if supersymmetry exists. The other Higgs bosons are much heavier and nearly degenerate in mass.

\subsection{Minimal Supergravity}\label{s:msugra}

We now have a model with spontaneous symmetry breaking and mass differences, but we do have a problem. We introduced supersymmetry at the cost of a single new parameter. Including the soft breaking terms leaves us with over a hundred free parameters. That means our theory has lost all predictive power. Fortunately, experimental limits put stringent bounds on many of them. 

\begin{wrapfigure}[10]{l}{50mm}\centering{
\begin{fmfchar*}(40,18)
\fmfforce{(0,.3h)}{in}
\fmfforce{(w,.3h)}{out}
\fmf{plain,label=$\mu^-$}{in,mu}
\fmf{gaugino,label=$\tilde B$}{mu,e}
\fmf{plain,label=$e^-$}{e,out}
\fmf{phantom_crossed, left, tension=0}{mu,e}
\fmf{dashes, left, tension=0,label=$\tilde\mu_R\quad\tilde e_R$}{mu,e}
\fmffreeze
\fmfforce{(.61w,.58h)}{v2}
\fmfrightn{gamma}{2}
\fmf{phantom}{v2,gamma1}
\fmf{photon,label=$\gamma$,label.side=right}{v2,gamma2}
\fmfdot{mu,e,v2}
\end{fmfchar*}\caption{Contribution of $\mathbf{m}_{\tilde{\bar e}}^2$ to $\mu^-\to e^-\gamma$}\label{f:mutoegamma}}\end{wrapfigure}
\noindent First look at the soft-breaking slepton mass terms. The mass term corresponding to the off-diagonal element $(\mathbf{m}_{\tilde{\bar{e}}}^2)_{12}$ couples a smuon to a selectron. This contributes to the branching ratio of $\mu^-\to e^-\gamma$ through the diagram in figure~\ref{f:mutoegamma}, where the smuon-selectron coupling is denoted by a cross. Because of the experimental limit on this branching ratio \cite{Brooks:1999pu}, $(\mathbf{m}_{\tilde{\bar{e}}}^2)_{12}$ has to be very small. The other off-diagonal terms are restricted by $\tau^-\to\mu^-\gamma$ and $\tau^-\to e^-\gamma$, although experimental limits on these branching ratios are weaker. Similar flavour mixing processes limit the off-diagonal elements of $\mathbf{m}_{\tilde L}^2$ and $\mathbf{a}_e$. This is straightforward for $\mathbf{m}_{\tilde L}^2$, but for $\mathbf{a}_e$ the argument is more complicated. The off-diagonal trilinear couplings combine with the Higgs vev to yield mass terms that couple sfermions of different flavours to each other. Thus the strength of the interaction depends on both the Higgs vev and the trilinear coupling, which makes limits depend on the other soft breaking parameters as well. Yet the general conclusion is that the off-diagonal elements have to be small.

\begin{wrapfigure}[13]{r}{50mm}
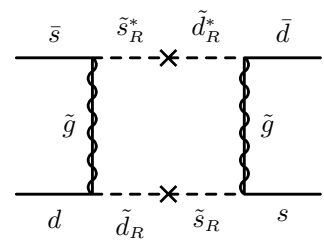
\centering{
\begin{fmfchar*}(40,30)
\fmfforce{(0,.8h)}{in1}
\fmfforce{(0,.2h)}{in2}
\fmfforce{(w,.8h)}{out1}
\fmfforce{(w,.2h)}{out2}
\fmf{plain,label=$\bar{s}$,label.side=left}{in1,v1}
\fmf{dashes,label=$\tilde s^*_R$,label.side=left}{v1,cross1}
\fmf{dashes,label=$\tilde d^*_R$,label.side=left}{cross1,v2}
\fmf{plain,label=$\bar d$,label.side=left}{v2,out1}
\fmf{plain,label=$d$}{in2,u1}
\fmf{dashes,label=$\tilde d_R$}{u1,cross2}
\fmf{dashes,label=$\tilde s_R$}{cross2,u2}
\fmf{plain,label=$s$}{u2,out2}
\fmffreeze
\fmf{phantom_crossed}{v1,v2}
\fmf{phantom_crossed}{u1,u2}
\fmf{gaugino,label=$\tilde g$}{v1,u1}
\fmf{gaugino,label=$\tilde g$,label.side=left}{v2,u2}
\end{fmfchar*}
\caption{Contribution of $\mathbf{m}_{\tilde{\bar d}}^2$ to $K^0\leftrightarrow\bar{K}^0$ mixing}\label{f:K-mixing}}\end{wrapfigure}

Constraints on the off-diagonal matrix elements in the squark sector come from experimental limits on meson mixing \cite{Yao:2006px}. Figure~\ref{f:K-mixing} shows an example of a diagram that contributes to kaon mixing that together with similar diagrams for $B_d^0\leftrightarrow\bar{B_d^0}$, $B_s^0\leftrightarrow\bar{B_s^0}$ and $D^0\leftrightarrow\bar{D^0}$ mixing put limits on other off-diagonal terms of $\mathbf{m}_{\tilde{\bar d},\tilde{\bar u},\tilde Q}^2$ and $\mathbf{a}_{u,d}$. Just like in the slepton case the limit on the triple scalar coupling comes from combining it with the Higgs vev. 

Finally, soft-breaking terms with a complex phase result in $CP$-violation \cite{Dugan:1984qf}, which we can avoid by requiring soft-breaking parameters to be real. Experimental bounds on flavour changing neutral currents put constraints on mass differences between flavours, so the breaking matrices are not only diagonal, their components must have the same order of magnitude \cite{Misiak:1997ei}. The experimental constraints on breaking parameters are summarized in table~\ref{t:softbreakinglimits}.

\begin{table}[!h]
\centering{
\begin{tabular}{|c|l|l|}
\hline
Breaking term & Constraint & For instance constrained by\\
\hline
$\mathbf{m}_{\tilde{\bar{e}}}^2,\mathbf{m}_{\tilde L}^2,\mathbf{a}_e$&Small off-diagonal elements&$\mu^-\to e^-\gamma$\\
$\mathbf{m}_{\tilde{\bar{u}}}^2,\mathbf{m}_{\tilde Q}^2,\mathbf{a}_u$&Small off-diagonal elements&$D^0\leftrightarrow\bar{D}^0$ mixing\\
$\mathbf{m}_{\tilde{\bar{d}}}^2,\mathbf{m}_{\tilde Q}^2,\mathbf{a}_d$&Small off-diagonal elements&$K^0\leftrightarrow\bar{K}^0$ mixing\\
&Complex phases have to be small&$CP$-violation\\
\hline
\end{tabular}
\caption[Experimental constraints on soft supersymmetry breaking terms]{Some experimental constraints on the form of soft supersymmetry breaking parameters. The last column mentions the most important sources of these constraints.}\label{t:softbreakinglimits}}
\end{table}
These constraints lead to soft supersymmetry breaking universality, which is the hypothesis that all mass matrices are proportional to the unit matrix, that the triple scalar couplings are proportional to the Yukawa matrices and that breaking parameters introduce no complex phases. These assumptions are the basis of minimal supergravity (mSUGRA), one of the most widely used models in supersymmetry. It assumes that breaking occurs through a coupling to gravity in its simplest form. Since gravity is flavour-blind, this justifies the assumption that the breaking mass matrices are proportional to the unit matrix. It also assumes unification at a high energy scale (see appendix~\ref{app:gut}). This leaves just four (and a half) free parameters: $M_0,\Mhalf,A_0,\tan\beta$ and the sign of $\mu$, where:
\begin{equation}
\Mhalf=M_1=M_2=M_3\label{Mhalf}
\end{equation}
\begin{equation}
M_0^2\mathbf{1}=\mathbf{m}_{\tilde Q}^2=\mathbf{m}_{\tilde L}^2=\mathbf{m}_{\tilde{\bar u}}^2=\mathbf{m}_{\tilde{\bar d}}^2=\mathbf{m}_{\tilde{\bar e}}^2\label{M0}
\end{equation}
\begin{equation}
M_0^2=m_{H_1}^2=m_{H_2}^2\label{M0Higgs}
\end{equation}
\begin{equation}
\mathbf{a}_u=A_0\mathbf{y}_u\quad\mathbf{a}_d=A_0\mathbf{y}_d\quad\mathbf{a}_e=A_0\mathbf{y}_e\label{A0}
\end{equation}
mSUGRA involves two assumptions that are weakly motivated. Even in the Standard Model we know of small nonzero parameters, such as the off-diagonal components and the $CP$-violating phase in the quark mixing matrix. So the case for strict soft supersymmetry breaking universality is not very strong. Yet even though it might not be exact, the constraints in table~\ref{t:softbreakinglimits} show that soft supersymmetry breaking universality is a good approximation.

The second assumption is unification. The MSSM is consistent with the minimal GUT (see page~\pageref{susyanddarkmatter}), but in general the theoretical motivation for unification is mainly aesthetical. In fact, the most important reason for assuming unification is a practical one. As long as no signs of supersymmetry have been found, the main purpose of models is to reveal general phenomenological implications. It is conceivable that dropping the requirement of unification yields a completely different phenomenology, but it is not very likely. Assuming unification reduces the complexity of the problem, but still allows a broad range of phenomenological scenarios.

In short, the most important property of mSUGRA is its predictive power. A theory with too many parameters is simply impractical, so we need the additional assumptions to make supersymmetry a workable theory.

\subsection{Renormalization Group Equations}

Knowing the soft breaking parameters at a high energy scale does not tell us anything about the phenomenology we could observe at the LHC. We first need to translate these parameters to their values at low energy. We know from section~\ref{s:RGE} how to use RGEs for that. In principle all parameters appearing in the supersymmetric Lagrangian are subject to RGEs. 

Obtaining the RGEs for supersymmetry is not an easy task. Firstly because of the enormous amount of possible interactions and loop corrections, but also because dimensional regularization poses a problem in supersymmetry. Remember from section~\ref{s:renormalizationsm} that a regularization scheme should not break any symmetries of a theory. However, the number of degrees of freedom within a supermultiplet needs to match. That means that the number of degrees of freedom of an on-shell Weyl fermion needs to be equal to the number of degrees of freedom of both an on-shell complex scalar and an on-shell massless gauge boson. A change in dimension does not affect scalars, but an on-shell massless gauge boson in $N-\epsilon$ dimensions has $N-2-\epsilon$ degrees of freedom. So in dimensional regularization, the number of degrees of freedom within supermultiplets does not match, which means supersymmetry is violated. A more fitting renormalization prescription is dimensional reduction (DRED) \cite{Capper:1979ns}\cite{Siegel:1979wq}. The basic concept is the same as in dimensional regularization, but the vector index of the gauge boson fields runs over 4 dimensions instead of $4-2\epsilon$. That ensures that the number of degrees of freedom within supermultiplets match. The usual renormalization procedure used in DRED is modified minimal subtraction ($\overline{DR}$).

The technical details of supersymmetric renormalization are complicated, but life is made easier by non-renormalization theorems. These state that chiral supermultiplets do not require mass renormalization \cite{Goity:1983aw}. As a result the RGEs for the superpotential have a simple form. Unfortunately, the soft breaking parameters from section~\ref{s:softbreakingterms} need to be renormalized as well, which leads to an enormous list of coupled differential equations.
 
This raises questions about the validity of soft supersymmetry breaking universality. This hypothesis is based on low-energy observations, but how do we know that for instance the diagonality of matrices stays intact after RG evolution? Fortunately, it turns out that many properties of the theory are RG-invariant \cite{Jack:1995gm}, including the form of the mass matrices and triple scalar couplings dictated by soft supersymmetry breaking universality.

Appendix~\ref{app:susyrge} contains the one-loop RGEs for softly broken supersymmetry. These RGEs are the dictionary between the unified theory at a high energy scale and the low-energy masses and couplings. A specific set of values for the high-energy parameters $M_0,\Mhalf,A_0,\tan\beta$ and $\mbox{sign}(\mu)$ are the boundary conditions for the RGEs and thus affect low-energy behaviour. Although some properties of the theory stay intact, low-energy phenomenology depends heavily on the boundary conditions.

\begin{figure}[!h]
\centering{
\includegraphics[width=9.5cm]{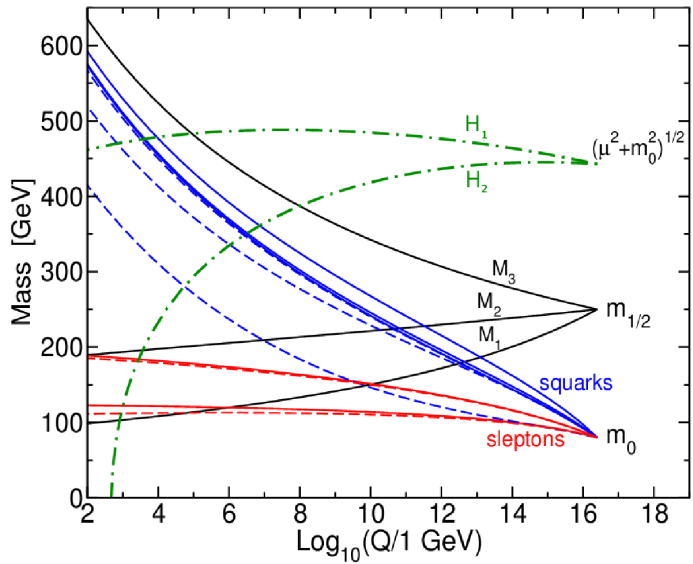}
\caption[Renormalization group evolution of the mSUGRA masses]{RG evolution of the mSUGRA masses for $m_0=80$~GeV, $m_{1/2}=250$~GeV, $A_0=-500$~GeV, $ \tan\beta=10$ and $\mu>0$ from \cite{martin2006}. Although the masses start out unified, the RGEs result in a non-degenerate mass spectrum at low energies. The large value of the top quark mass drives $|\mu|^2+m_{H_2}^2$ to small values, ensuring spontaneous symmetry breaking in large parts of mSUGRA parameter space.}\label{f:topmass}}
\end{figure}

The RGEs explain how spontaneous symmetry breaking can occur in a unified theory. The conditions for spontaneous symmetry breaking \eq{spontaneoussymmetrybreakinglowerlimit} and \eq{spontaneoussymmetrybreakingupperlimit} cannot be satisfied simultaneously if the two Higgs breaking masses are equal, which is assumed in mSUGRA. However, radiative corrections change the value of the Higgs breaking parameters through the RGEs. Equation \eq{RGEHiggs2} for the mass parameter $m_{H_2}^2$ has a positive contribution from the top quark mass through the parameter $X_t$. That means that $m_{H_2}^2$ decreases as the energy decreases. Since the top quark mass is large, this can drive $|\mu|^2+m_{H_2}^2$ to very small and even negative values at low energy, as shown in figure~\ref{f:topmass}.

\end{fmffile}

\begin{fmffile}{feynmethods}
\fmfset{dot_size}{1thick}
\fmfset{curly_len}{1.5mm}
\fmfset{arrow_len}{3mm}
\fmfpen{thin}

\fmfcmd{%
vardef cross_bar (expr p, len, ang) =
 ((-len/2,0)--(len/2,0))
 rotated (ang + angle direction length(p)/2 of p)
 shifted point length(p)/2 of p
enddef;
style_def phantom_crossed expr p=
 ccutdraw cross_bar (p, 3mm, 45);
 ccutdraw cross_bar (p, 3mm, -45);
enddef;
style_def gaugino expr p =
 cdraw (wiggly p);
 cdraw p;
enddef;
}

\section{Methods}\label{s:methods}

At this point we have a broken supersymmetric theory and we have limited the number of free parameters by restricting ourselves to mSUGRA. It is time to put theory to practice and discuss the implications for collider experiments, in particular the LHC. Predictions for the LHC are obtained in three steps.

The first step is to apply RG evolution to calculate low-energy masses and branching ratios\footnote{The branching ratio is the fraction of a certain type of particle that decays through a specific decay mode.} for a particular set of high-energy parameters. The RGEs are coupled differential equations that cannot be solved analytically. Low-energy phenomenology is complicated further because mixing angles and couplings depend on the high-scale parameters as well. Therefore the RGEs are solved numerically.

The second step is simulating a collision. In principle the phenomenology is fixed by masses, mixing angles and branching ratios, but the sheer number of possible decay chains makes it necessary to do a Monte Carlo simulation of collisions. The final step is to do a detector simulation. This is beyond the scope of this research.

This section is dedicated to explaining in more detail how to go from theory to experimental predictions. For each step, we first discuss the methods used in the field in general and then introduce the methods and programs specifically used in this research.

\subsection{Obtaining a Low-Energy Spectrum}

As mentioned before, obtaining a low-energy spectrum with the RGEs is done numerically. There are several programs available that do a two-loop calculation. They all have slightly different assumptions and precisions at various points in the calculation. For instance, some use dimensional regularization, others dimensional reduction. As a result, different programs yield slightly different results.
\begin{table}[!h]
\centering{
\begin{tabular}{|c|llll|c|llll|}
\hline
& SPheno & Isasugra & SuSpect & Softsusy&& SPheno & Isasugra & SuSpect & Softsusy\\
\hline
$h^0$&	110.38&	109.99&	110.12&	110.73&$\tilde e_L$&	231.12&	232.29&	229.01&	230.68\\
$H^0$&	517.01&	512.97&	513.62&	516.39&$\tilde e_R$&	156.87&	154.70&	155.18&	157.52\\
$A^0$&	516.24&	508.89&	512.84&	516.39&$\tilde \nu_e$&	217.64&	216.83&	215.78&	217.43\\
$H^+$&	522.71&	518.29&	519.36&	522.86&$\tilde \mu_L$&	231.13&	232.29&	229.01&	230.68\\
$\tilde d_L$&	675.18&	670.41&	666.64&	668.67&$\tilde \mu_R$&	156.85&	154.70&	155.18&	157.52\\
$\tilde d_R$&	648.48&	643.64&	640.51&	641.54&$\tilde \nu_\mu$&	217.64&	216.83&	215.78&	217.43\\
$\tilde u_L$&	670.83&	665.49&	662.15&	662.73&$\tilde\tau_1$&	151.53&	151.50&	149.86&	152.12\\
$\tilde u_R$&	649.47&	644.81&	641.57&	646.92&$\tilde\tau_2$&	232.76&	232.33&	230.72&	232.34\\
$\tilde s_L$&	675.18&	670.41&	666.64&	668.67&$\tilde\nu_\tau$&	216.96&	214.64&	215.13&	216.75\\
$\tilde s_R$&	648.47&	643.64&	640.51&	641.54&$\tilde g$&	719.96&	719.82&	716.83&	720.01\\
$\tilde c_L$&	670.84&	665.49&	662.15&	662.73&$\tilde\chi^{10}$&	118.33&	118.69&	118.66&	117.96\\
$\tilde c_R$&	649.46&	644.81&	641.57&	646.92&$\tilde\chi^{20}$&	222.90&	222.78&	222.98&	222.97\\
$\tilde b_1$&	606.22&	606.43&	603.53&	601.25&$\tilde\chi^{30}$&	466.63&	456.47&	462.00&	468.66\\
$\tilde b_2$&	647.94&	642.25&	640.34&	639.16&$\tilde\chi^{40}$&	481.64&	475.08&	478.94&	482.48\\
$\tilde t_1$&	452.36&	440.25&	446.58&	449.11&$\tilde\chi^{1+}$&	222.66&	222.83&	222.52&	224.56\\
$\tilde t_2$&	677.34&	670.96&	672.03&	673.33&$\tilde\chi^{2+}$&	481.89&	472.91&	478.30&	478.52\\
\hline
\end{tabular}
\caption[Low-energy mass spectrum results for different programs.]{Mass spectrum in~GeV for the high-scale parameters $M_0=100$~GeV, $\Mhalf=300$~GeV, $A_0=-300$~GeV, $\tan\beta=6,\mu>0$ computed by SPheno 2.2.3 \cite{Porod:2003um}, ISASUGRA 7.78 \cite{Paige:2003mg}, SuSpect 2.34 \cite{Djouadi:2002ze} and SOFTSUSY 2.0.17 \cite{Allanach:2001kg}. For this point in parameter space, the differences are smaller than 3\%.}\label{t:masscompare}}
\end{table}
A comparison of the mass spectra in~GeV for a generic point in mSUGRA parameter space is shown in table~\ref{t:masscompare}.
For this set of high-energy parameters, the differences are smaller than 3\%. The differences tend to increase for more exotic regions of the parameter space, but in general the results of different programs are in reasonable agreement. A systematic review of these programs can be found in \cite{Allanach:2002pz}.
\newline

For this research, SPheno 2.2.3 was used to calculate masses, couplings, branching ratios and decay widths of supersymmetric particles. The advantage of SPheno is that it is the only program that calculates both masses and branching ratios and uses $\overline{DR}$ as regularization scheme. To easily scan parameter space, a C++ program was written to generate an input file for specific high-energy parameters and then call SPheno to generate a spectrum file. The resulting mass spectrum is discussed in section~\ref{s:massspectrum}. Occasionally ISASUGRA linked with ISATOOLS is used to calculate the dark matter relic density.

\subsection{Event Simulation}\label{s:pythia}

The next step is to run a Monte Carlo simulation of proton-proton collisions at 14 TeV. There are Monte Carlo generators available that can do this. In general these calculations are done to leading order only.
\newline

In this research, the event generator Pythia 6 \cite{Sjostrand:2006za} was used through an interface within the ROOT framework \cite{rootsite}. It automatically reads in the mass spectrum and branching ratios generated by SPheno. For each point in parameter space, 20.000 SUSY events were generated. These are events where supersymmetric or Higgs particles are produced in the initial interaction.

The Pythia event record contains information on the particles that were produced in the event, for instance their identity, their origin (mothers) and decay products (daughters) and their momenta. Part of a sample event summary can be found in Appendix~\ref{app:pythia}. Figure \ref{f:decaychain} shows a graphical representation of the most important particles in this event.

\begin{figure}[!h]
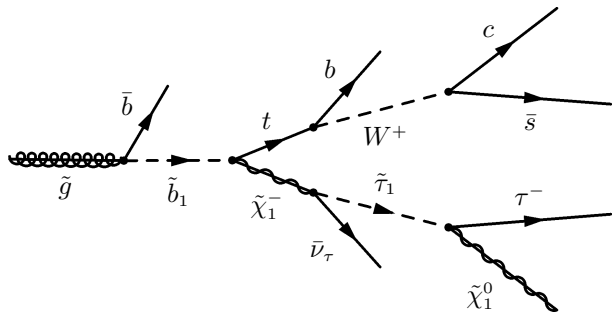

\centering{
\begin{fmfchar*}(80,40)
\fmfleft{in}
\fmfright{o1,o2,o4,o5}
\fmftop{p1,o7,p3,p5}
\fmfbottom{p2,o8,p4,p6}
\fmf{phantom}{in,p1}
\fmf{phantom}{in,p2}
\fmf{phantom,tension=0.01}{v10,p3}
\fmf{phantom,tension=0.01}{v10,p4}
\fmf{plain,tension=6,label=$\tilde g$}{in,v}
\fmf{dashes_arrow,tension=6,label=$\tilde b_1$,label.side=right}{v,v10}
\fmf{fermion,tension=0.3,label=$\bar b$,label.side=left}{v,v11}
\fmf{phantom,tension=1}{v,v12}
\fmf{gaugino,tension=4,label=$\tilde\chi_1^-$,label.side=right}{v10,v20}
\fmf{fermion,tension=4,label=$t$,label.side=left}{v10,v21}
\fmf{phantom,tension=0.3}{v11,o7}
\fmf{phantom,tension=0.3}{v12,o8}
\fmf{dashes_arrow,tension=2,label=$\tilde\tau_1$,label.side=left}{v20,v30}
\fmf{fermion,tension=0.8,label=$\bar\nu_\tau$,label.side=right}{v20,v31}
\fmf{dashes,tension=2,label=$W^+$,label.side=right}{v21,v32}
\fmf{fermion,tension=0.8,label=$b$,label.side=left}{v21,v33}
\fmf{gaugino,label=$\tilde\chi_1^0$,label.side=right}{v30,o1}
\fmf{fermion,label=$\tau^-$,label.side=left}{v30,o2}
\fmf{phantom,tension=1.5}{v31,p4}
\fmf{fermion,label=$\bar s$,label.side=right}{v32,o4}
\fmf{fermion,label=$c$,label.side=left}{v32,o5}
\fmf{phantom,tension=1.5}{v33,p3}
\fmfdot{v,v10,v20,v21,v30,v32}
\fmffreeze
\fmfi{curly}{vpath (__in,__v) shifted (thin*(0,1))}
\fmf{phantom}{v10,p5}
\fmf{phantom}{v10,p6}
\end{fmfchar*}
\caption[Part of a possible decay chain of a SUSY event.]{Part of a possible decay chain of a SUSY event. The notation is further explained in section~\ref{s:massspectrum}. The only stable particles in this chain are $\tilde\chi_1^0$ and $\bar\nu_\tau$. The strange quark will form a Kaon in the hadronization chain. All other particles will decay inside the detector.}\label{f:decaychain}}
\end{figure}
A typical event record starts with a short summary of the event, where the most important particles in the event are listed with their mothers and momenta. As can be seen in Appendix~\ref{app:pythia}, these particles have `status code' 21, and are usually repeated later in the event record. After this summary, all particles produced in the event are listed. They are sorted such that particles with the same mother are grouped together.

Not the entire event record was saved, since for instance hadronization chains are not interesting for this research. The following particles were saved for analysis:
\begin{itemize}
\setlength{\itemsep}{2pt}
\setlength{\parskip}{0pt}
\setlength{\parsep}{0pt}
\item Particles in the descriptive part of the event record (status code 21).
\item Supersymmetric particles and their daughters.
\item Taus and their daughters.
\item Tops and their daughters.
\item Bottoms
\item Final state leptons (leptons with status code 1).
\end{itemize}
Most particles in the event record shown in Appendix~\ref{app:pythia} are not saved by this procedure. Mothers of the remaining particles are adjusted to ensure consistency of the event record. If the actual mother of a particle was not saved, the mother was changed to be the last particle in the decay chain that was saved. This reduces the length of the event record from over a thousand lines to less than fifty lines, while retaining the important information. This makes it possible to save events and run a new analysis without having to repeat the event simulation.

In the analysis, not all final state leptons should be taken into account. Many electrons and muons are produced in jets and we can be certain that they cannot be distinguished from the jet. Leptons whose mother is a supersymmetric particle or a $W$ or $Z$ boson are listed in the descriptive part of the event record. Thus when determining the number of final state leptons, only those electrons and muons are counted whose mother either has status code 21 or is a tau. This is not entirely correct, but for an actual prediction of the number of observed final state leptons produced by SUSY events we would need to perform a detector simulation.

The events contain information about the relative quantities in which supersymmetric particles are produced. The number of times a particular supersymmetric particle is produced is determined by multiplying the number of times it is produced per SUSY event by the total cross section for supersymmetric particles. This cross section is different for every point in parameter space and is calculated by Pythia during the event simulation.
\newline

\textsc{Supersymmetric Particle Production.} By only selecting SUSY events, the possibility that Standard Model particles from the initial interaction at the LHC produce supersymmetric particles later in the decay chain is neglected. This is a reasonable assumption since only those Standard Model particles whose mass is large enough to decay to two supersymmetric particles can produce secondary supersymmetric particles. As an extra check, supersymmetric particles that were not produced in the first interaction and do not have a supersymmetric particle as their mother were counted in all events. The highest number occurred, as was to be expected, for a low-mass scenario and was $3.6\cdot10^{-3}$ particle per SUSY event, which is indeed negligible since two particles are produced in the initial interaction. In fact, this number is an overestimate, since it also counts the supersymmetric particles produced by Higgs particles, which are included in the initial interaction.
\newline

\textsc{Instabilities.} Although Pythia is widely used, a program of such complexity is not flawless. For a number of points in mSUGRA parameter space, internal consistency checks fail regularly. Also, Pythia occasionally generates negative cross sections or daughters with masses that exceed the energy of their mother. These errors are most likely due to phase space effects in the Monte Carlo generator and are sometimes dealt with by trying a different probability distribution as the basis of the Monte Carlo simulations. This was not done in this research. Instead, inconsistent events are not saved and if their number reaches 1\% of the total number of events for a particular point in mSUGRA parameter space, this point is not taken into account in the analysis.
\newline

\textsc{Detector Simulation.} For a full analysis, generating events is not sufficient. Due to the complicated design of detectors and event reconstruction algorithms, the only way to obtain a proper prediction is by including a detector simulation. As mentioned before, this is not done in this research, which means no direct predictions for the LHC are obtained.
\newline

\textsc{K-factors.} The total cross section calculated by Pythia does not include radiative corrections. Especially in the QCD sector, next to leading order contributions can be 30\% or more of the original cross section. This can be corrected using $K$-factors, with which the cross section of a particular process is multiplied. $K$-factors for the benchmark points (see section~\ref{s:benchmark}) have been calculated, but for most parts of parameter space they are not readily available. In addition, it is difficult to properly implement them in an event simulation, since that requires changing every single branching ratio. Therefore $K$-factors are not taken into account in this research. The $K$-factors for the overall supersymmetry cross sections for different benchmark points have the same order of magnitude \cite{cscnote}, so the general phenomenological implications derived in this research should hold.

\subsection{Benchmark Points}\label{s:benchmark}

Calculating the low-energy spectrum for a given set of high-energy parameters takes less than a minute, but generating thousands of events easily takes an hour. Doing a detector simulation would take even longer. Clearly it is impossible to scan the entire mSUGRA parameter space this way. Some low-energy observables limit the possible parameter space, but it is still too big to do a full simulation on a reasonably fine grid.

The conventional solution is to use benchmark points. For these points a full simulation is performed to find general predictions of supersymmetric phenomenology. Table~\ref{t:susypoints} lists the benchmark points used by ATLAS \cite{cscnote}. The SU5 points are not used in recent simulations. The names of the points reflect the different regions of parameter space that are not excluded by low-energy observables \cite{allanach2002}.
\begin{table}[!h]
\centering{
\begin{tabular}{|l|ccccc|}
\hline
Point & $M_0$ (GeV) & $\Mhalf$ (GeV) & $A_0$ (GeV) & $\tan\beta$ & sgn($\mu$)\\
\hline
Coannihilation (SU1) & 70 & 350 & 0 & 10 & +\\
Focus Point (SU2) & 3550 & 300 & 0 & 10 & +\\
Bulk (SU3) & 100 & 300 & -300 & 6 & +\\
Low Mass (SU4) & 200 & 160 & -400 & 10 & +\\
Scan (SU5.1) & 130 & 600 & 0 & 10 & +\\
Scan (SU5.2) & 250 & 600 & 0 & 10 & +\\
Scan (SU5.3) & 500 & 600 & 0 & 10 & + \\
Funnel (SU6) & 320 & 375 & 0 & 50 & + \\
Coannihilation (SU8.1) & 210 & 360 & 0 & 40 & + \\
Coannihilation (SU8.2) & 215 & 360 & 0 & 40 & + \\
Coannihilation (SU8.3) & 225 & 360 & 0 & 40 & +\\
\hline
\end{tabular}
\caption[Benchmark Points]{Benchmark points used by the ATLAS collaboration \cite{cscnote}.}\label{t:susypoints}}
\end{table}

The necessity of the benchmark points is clear, but their foundation in theory is not very strong. Firstly, table~\ref{t:masscompare} shows that the mass spectrum of different programs can be different for the same point in parameter space. Event simulations use the low-energy spectrum, which means their result depends not only on the high-energy parameters, but also on (the version of) the program used to do the RG evolution. What happens in practice is that one program is chosen to do the RG evolution and everyone works with the resulting low-energy spectrum. If a different program or version is used, the high-scale parameters are adjusted to reproduce the original spectrum as well as possible. Of course there is no guarantee that this is possible, since it is quite improbable that a different treatment of the RGEs yields unification of masses and couplings starting from the same low-energy spectrum.

Although this procedure might seem strange, the motivation for GUT was not very strong to begin with. Also, an important reason to assume unification was to have a workable theory. If letting go of this assumption at this point makes the theory more workable, there is no good reason to require strict unification anymore. It might not be very appealing from a theoretical point of view, but it is the only way results can be compared in a consistent manner. As long as there is no proper theory for supersymmetry breaking, the only approach is a practical one.

Unfortunately, there is a more serious drawback to the benchmark points. The mSUGRA parameter space is enormous and there is a good chance that certain phenomenological scenarios are not included in the analysis if only these eleven points are used. The main goal of this research is to look beyond the benchmark points and see if there are points with completely different phenomenology that are not currently studied in preparation for the LHC. This is done in section~\ref{s:regions}. Another part is to investigate the stability of the benchmark points. What happens to the predictions if the high-scale parameters are changed by a small amount? This is discussed in section~\ref{s:benchmarkresult}.

\subsection{Focus on Low Masses}

When studying supersymmetry in preparation for the LHC, one has to consider a number of things. In order to detect signs of supersymmetry, one wants to study particles that are produced with reasonable cross sections and can easily be distinguished from the Standard Model background. Of course, `reasonable' and `easily' are relative terms, since the Standard Model background at 14 TeV collisions is enormous and particle identification is difficult at a hadron collider. Missing energy would be a particularly good signal for supersymmetry, since the undetectable LSP is stable, but it will be difficult to accurately measure missing energy at the LHC. These factors make it very important to find as many distinguishable signals for supersymmetry as possible.

In some parts of the parameter space, supersymmetry is almost impossible to detect. The simplest example is a point where all supersymmetric particles have a very high mass. In that case the cross section at the LHC is very low and could even be zero due to insufficient energy. Therefore this research will focus on supersymmetry with relatively low masses.

\subsection{Uncertainties}

In the previous sections, several sources of errors in this research were mentioned. The most important one is the fact that all calculations are at leading order only. As a result, the predicted numbers of different particles are not correct. However, this research focuses on qualitative effects. Whether or not a decay is allowed does not depend on next to leading order effects in the collision or decay. Thus exact error calculation does not add significantly to this analysis and is not included. Statistical errors are shown for results of Pythia simulations, but are usually too small to be visible in graphs.

\end{fmffile}

\section{The Mass Spectrum}\label{s:massspectrum}

In this section we discuss the low-energy mass spectrum that follows from the RGEs. We start with the general properties and then investigate the influence of different high-energy parameters on the low-energy spectrum using SPheno. In the second part, we limit ourselves to scenarios with $M_0$ and $\Mhalf$ smaller than 600~GeV.

Most characteristics are explained with the one-loop RGEs (Appendix~\ref{app:susyrge}). Since the masses of particles and antiparticles are equal, only one of them is mentioned. Sfermions from the first two generations are given the labels $L$ and $R$, denoting the helicity of their fermionic partners. For the third generation this identification is not possible due to mixing effects, so the labels 1 and 2 are used. 

\subsection{General Properties}\label{s:massgeneralproperties}

The low-energy mass spectrum follows from the RGEs. However, the interaction eigenstates are usually not the mass eigenstates, so mixing plays an important role in the spectrum.
\newline

\textsc{The Higgs Sector.} The Higgs sector was discussed in section~\ref{s:susyhiggsmech}. After diagonalization of the mass matrix, the lightest Higgs has a mass smaller than 135~GeV and the heavy Higgses are nearly degenerate. The lightest Higgs is always subject to large radiative corrections. When the couplings to the Higgses are strong, as is the case for large values of $\tan\beta$ or $|A_0|$, loop corrections become so important that for the heavy Higgses the tree-level equations from section~\ref{s:susyhiggsmech} do not hold either.
\newline

\textsc{The Gaugino Sector.} In the gaugino sector, the gluino is heavier than the other gauginos. This is a consequence of equation \eq{RGE-gaugino}, where $M_3$ has a negative slope, while $M_1$ and $M_2$ have a positive slope. So as the energy decreases, the gluino mass $M_3$ becomes larger, while the other gaugino masses become smaller. However, this is complicated by mixing. The gluinos $\tilde g$ form a colour octet and cannot mix with any of the other gauginos, but the $U(1)\times SU(2)$ gauginos mix with each other \emph{and} with the Higgsinos. The resulting mass eigenstates are called `charginos' if they are charged and `neutralinos' if they are neutral. The neutralino mass matrix is a $4\times4$ matrix that contains a number of terms:
\begin{itemize}
\item Two diagonal entries $M_1$ and $M_2$ that are subject to the RGEs \eq{RGE-gaugino}.
\item Relatively small off-diagonal elements due to the Higgs-Higgsino-gaugino coupling of the form $g\langle H^0_i\rangle\tilde h\tilde\chi$ where $\langle H^0_i\rangle$ ($i=1,2$) is the vev of the Higgs field that couples to the Higgsino $\tilde h$ and the gaugino $\tilde\chi$ with interaction strength $g$. These terms come from the interactions between the gauge and chiral supermultiplets and give rise to mass terms after the Higgs fields obtain a vev.
\item An off-diagonal term $-\mu$ coming from the superpotential coupling between the two Higgsinos.
\end{itemize}
The chargino mass matrices are $2\times2$ matrices with diagonal entries $M_2$ and $\mu$ coming from the Winos and the Higgsinos respectively. They also have off-diagonal entries that come from the Higgs-Higgsino-gaugino coupling. After the mass matrices are diagonalized, the charginos and neutralinos are denoted as $\tilde\chi_1^\pm,\tilde\chi_2^\pm,\tilde\chi_1^0,\tilde\chi_2^0,\tilde\chi_3^0$ and $\tilde\chi_4^0$, where by definition the lower index is higher for an eigenstate with higher mass and the upper index is the charge.
\newpage
Now we can distinguish two cases. In the most common scenario, the neutralino $\tilde\chi_1^0$ is Bino-like, $\tilde\chi_2^0$ and $\tilde\chi_1^\pm$ $SU(2)$ gaugino-like, and $\tilde\chi_3^0$, $\tilde\chi_4^0$ and $\tilde\chi_2^\pm$ Higgsino-like. The states $\tilde\chi_2^0$ and $\tilde\chi_1^\pm$ are nearly degenerate, as are $\tilde\chi_4^0$ and $\tilde\chi_2^\pm$. The state $\tilde\chi_3^0$ has a somewhat smaller mass than $\tilde\chi_4^0$. Because of their $\mu$ dependence, the mass spectra of the (Higgsino-like) neutralinos and charginos share some of the characteristics of the Higgs mass spectra.

The second case occurs if the soft-breaking scalar mass $M_0$ has the same order of magnitude as the electroweak symmetry breaking scale $\mu$ (a few TeV). In that case, the Higgsino-like neutralinos and charginos have a lower mass than the other gauginos and the LSP is Higgsino-like. This results in very different phenomenology, but we will not discuss this scenario in detail for reasons that are explained in section~\ref{s:phenomenology}.
\newline

\textsc{The Sfermion Sector.} Because mSUGRA is constructed to conserve flavour, mixing between sfermions of different flavours can be neglected. However, left-handed and right-handed sfermions with the same flavour can interact with each other so these states \emph{are} subject to mixing. After the Higgs fields obtain a vev, the $2\times2$ sfermion mass matrices are composed of several components:
\begin{itemize}
\item Diagonal soft-breaking sfermion mass terms that are subject to RGEs.
\item A relatively small diagonal mass term coming from the $D$ terms in the Lagrangian.
\item Diagonal terms coming from the $F$ terms that have the form $|y|^2\langle H^0_i\rangle^2\tilde f^2$ or $|y|^2\langle H^0_i\rangle^2\tilde{\bar f}^2$, where $\langle H_i^0\rangle$ is the vev of the Higgs field that couples to the sfermion $\tilde f$.
\item Terms of the form $-\mu y \tilde f\tilde{\bar f} \langle H^0_i\rangle$ that also come from the $F$ terms and that are proportional to $\cos\beta$ for up-type squarks and to $\sin\beta$ for down-type squarks. These are off-diagonal terms in the mass matrix that are proportional to the Yukawa couplings.
\item Terms that contain the soft-breaking trilinear couplings and that are proportional to the Yukawa couplings. These too are off-diagonal contributions to the mass matrix, but unlike the off-diagonal terms coming from the $F$ terms, they are proportional to $\sin\beta$ for up-type sfermions and to $\cos\beta$ for down-type sfermions.
\end{itemize}
The contributions containing Yukawa couplings are particularly important for the third generation, so mixing effects are much stronger there. Yet several properties of the sfermion sector follow directly from the RGEs, without including mixing effects. Firstly, sfermions from the first two generations that are in the same representation are nearly degenerate in mass, since they have the same RGEs. This can be seen in for instance equations \eq{RGEQl} and \eq{RGEe}. Secondly, because of the large value of the gluino mass, the terms proportional to $M_3$ are the main contribution to the RGEs of the light squarks \eq{RGEQl}, \eq{RGEu} and \eq{RGEd}. As a result, the up-type and the down-type squarks of the first two generations are also nearly degenerate. Thirdly, if mixing can be neglected, sleptons are lighter than squarks. This can be seen most easily by comparing RGEs \eq{RGEu} and \eq{RGEe}:
\[\beta_{m^2_{\tilde{\bar u}}}-\beta_{m^2_{\tilde{\bar e}}}=\frac{8}{3}g_1^2M_1^2-\frac{32}{3}g_3^2M_3^2<0\]
This inequality certainly holds at the unification scale, where $g_1=g_3$ and $M_1=M_3$. For lower energies, it follows from equation \eq{RGE-gaugecoupling} that $g_3>g_1$ and from equation \eq{RGE-gaugino} that $M_3>M_1$, so the difference becomes larger. As a result, the slope of the squark mass is always more negative than the slope of the slepton mass, which means that the squark mass increases faster as the energy decreases. Similar arguments hold for the other sfermions, although they are complicated by additional terms in the RGEs.
\newpage
Within the squark sector, the stop and sbottom are generally lighter than squarks from the first two generations due to the $|y_t|^2X_t$ and $|y_b|^2X_b$ terms in the RGEs \eq{RGEt} and \eq{RGEb}:
\[\beta_{m^2_{\tilde{\bar t}}}-\beta_{m^2_{\tilde{\bar u}}}=4|y_t|^2X_t>0\]
Similar equations hold for the down-type squarks and the sleptons, so the slope of the masses of the first two generations is more negative than the slope of the third generation masses. That means that the masses of the first two generations increase faster as the energy decreases. The mass difference is the largest for the up-type squarks, because of the high value of the top Yukawa coupling. In the slepton sector, the effect is negligible, because the tau Yukawa coupling is much smaller.

Finally, the masses of the left-handed states are different from the masses of the right-handed states because the RGEs of the left-handed doublets have an additional $SU(2)$ gaugino mass contribution compared to the right-handed singlets. For instance, comparing the RGEs \eq{RGEQl} and \eq{RGEu} or \eq{RGELl} and \eq{RGEe}:
\begin{align}
\beta_{m_{\tilde Q_\ell}^2}-\beta_{m_{\tilde{\bar u}}^2}&=2g_1^2|M_1|^2-6g_2^2|M_2|^2<0\\
\beta_{m_{\tilde L_\ell}^2}-\beta_{m_{\tilde{\bar e}}^2}&=\frac{18}{5}g_1^2|M_1|^2-6g_2^2|M_2|^2<0\label{left-rightsplitting}
\end{align}
It follows from unification and the RGEs for the gauge couplings \eq{RGE-gaugecoupling} and the gaugino masses \eq{RGE-gaugino}, that this is true for all energies. As a result, left-handed sfermions are heavier than right-handed sfermions.

For the first two generations, the off-diagonal entries in the mass matrices can be neglected. This is not the case for the third generation due to the large Yukawa couplings. For the stop, sbottom and stau, diagonalizing the mass matrix gives a heavy and a light mass eigenstate. The splitting is most important for the stop, because of the large top Yukawa coupling. The bottom Yukawa coupling is large enough to cause considerable mixing as well. In addition, the top Yukawa coupling affects $A_b$ through the RGE \eq{RGEAb} and $A_b$ contributes to the off-diagonal elements. The tau Yukawa coupling is smaller and $A_\tau$ is not affected by $y_t$, so for the stau the mass splitting only becomes significant for very large or small $\tan\beta$. 
\newline

\textsc{Running Masses.} RGEs evolve masses to a certain energy scale, for instance the scale of the $Z$-mass. However, masses are still scale-dependent. A more sensible quantity than the mass at the $Z$-scale is the pole mass, which is $m_{ph}(m_{ph})$ in the notation of section \ref{s:RGE}. This is indeed what is calculated by SPheno. As a result, masses for different particles are given at different energy scales. These effects are typically small, but they can slightly change particle masses in ways that do not immediately follow from the RGEs.
\newline

Summarizing this section, we conclude that supersymmetric masses are determined by two building blocks. The general mass hierarchy follows from the RGEs. The mass splitting follows from the relative importance of the off-diagonal terms in the mass matrix, which is also determined by the RGEs. These building blocks are the basis of the discussion in the following sections.

\newpage
\subsection{Soft-Breaking Scalar Mass}\label{s:M0explanation}

\begin{figure}[!h]
\centering{
\includegraphics[width=\linewidth]{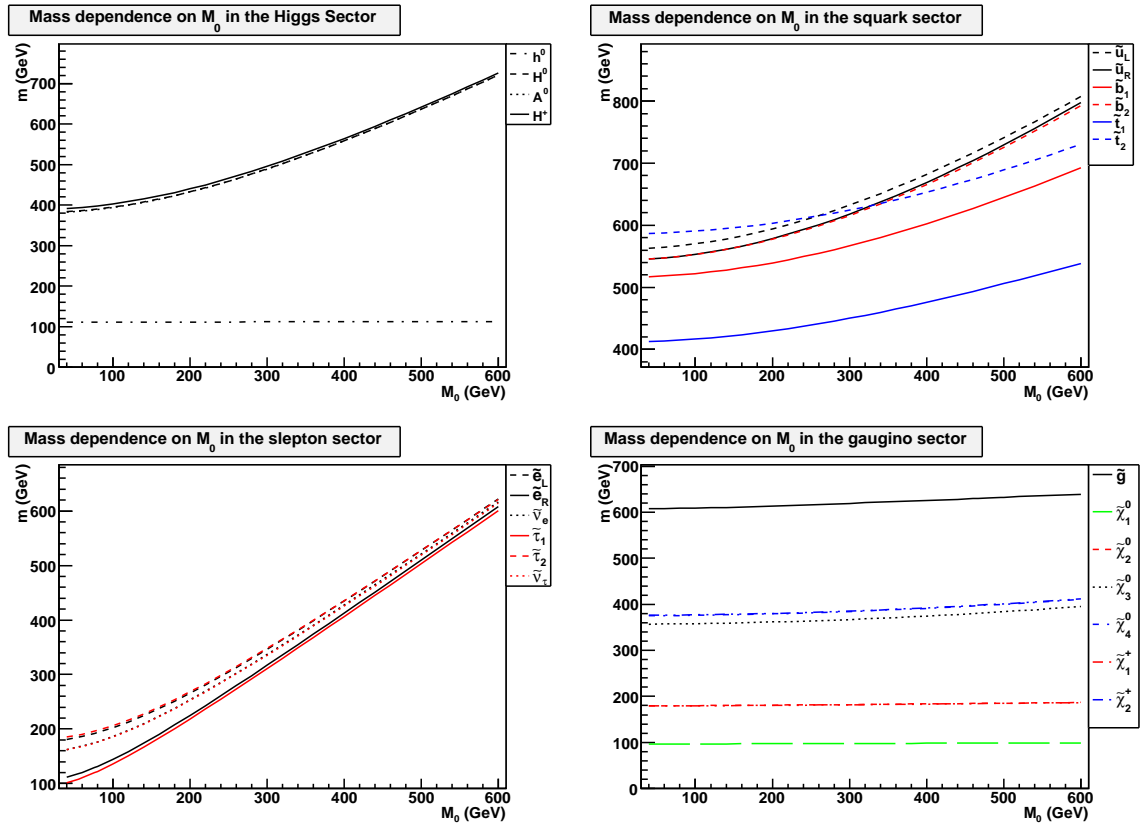}
\caption[Supersymmetric mass dependence on $M_0$]{Low-energy mass dependence on $M_0$ calculated with SPheno. All other parameters are kept constant at $\Mhalf=250$~GeV, $A_0=0$~GeV, $\tan\beta=10,\mu>0$.}\label{f:massM0}}
\end{figure}

Soft-breaking scalar masses occur in the RGEs of the Higgs fields and the third generation sleptons. Figure~\ref{f:massM0} shows how the low-energy mass spectrum changes as $M_0$ is varied with all other parameters kept constant. As mentioned before, the discussion is limited to $M_0,\Mhalf\leq600$~GeV. Also, not all particles from the first two generations are plotted, since they are nearly degenerate.

The general characteristics of the mass spectrum explained in section~\ref{s:massgeneralproperties} can easily be recognised in figure~\ref{f:massM0}. The mass splittings for third generation squarks can be seen clearly. For the much lighter sleptons, the splitting between left-handed and right-handed particles is more important for low $M_0$. This is caused by the gaugino mass terms in the RGEs. If the scalar masses get smaller, which is the case for low $M_0$, the relative effect of the gaugino masses in equation \eq{left-rightsplitting} is larger. This does not play a role in the squark sector, since even for low $M_0$ the squark masses are much larger than $M_2$.

As one could expect from the RGEs, the scalar masses are affected the most by a change in $M_0$. The shape of the graphs is explained by the relative importance of $M_0$. The RGEs for the scalar masses in Appendix~\ref{app:susyrge} are for the masses squared. To obtain the linear mass parameters plotted in figure~\ref{f:massM0}, we essentially have to take the square root of a quadratic function in $M_0$. If the terms containing $M_0$ are the leading contributions, which is the case for the slepton sector for large $M_0$, this yields an almost linear function. If other contributions are more important, such as $M_3$ in the squark sector, the result behaves more like a quadratic function. We can also see the quadratic behaviour for small $M_0$ in the slepton sector.

The masses of the gauginos and the lightest Higgs hardly change with $M_0$. For the Higgs this is because of equation \eq{h0mass}, which keeps the value low. The mass difference over the entire range of $M_0$ in figure~\ref{f:massM0} is only 1~GeV. The gaugino masses hardly depend on $M_0$ because they are affected through higher-order corrections only.

\subsection{Soft-Breaking Gaugino Mass}\label{s:Mhalfexplanation}

\begin{figure}[!h]
\centering{
\includegraphics[width=\linewidth]{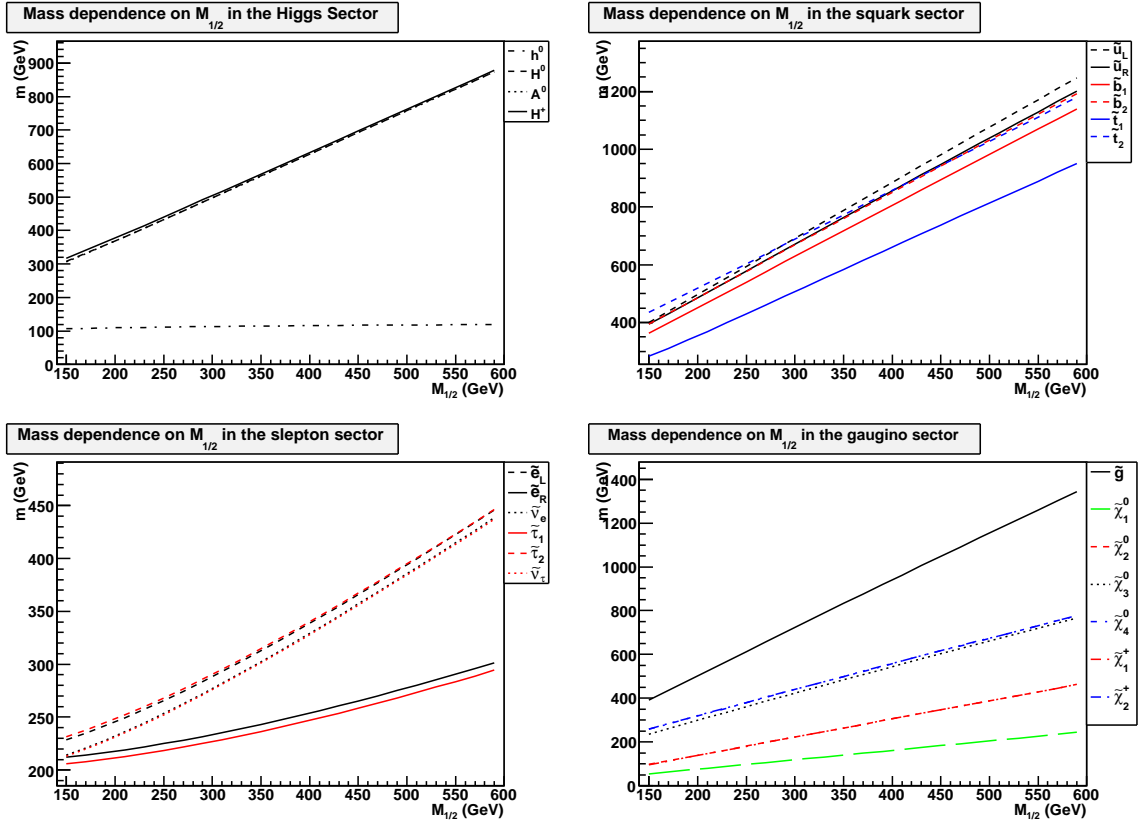}
\caption[Supersymmetric mass dependence on $\Mhalf$]{Low-energy mass dependence on $\Mhalf$ calculated with SPheno. All other parameters are kept constant at $M_0=200$~GeV, $A_0=0$~GeV, $\tan\beta=10,\mu>0$.}\label{f:massMhalf}}
\end{figure}

Gaugino masses occur in the RGEs of all soft-breaking terms. So changing the unification value $\Mhalf$ of the soft-breaking gaugino mass affects all masses, as can be seen in figure~\ref{f:massMhalf}.

The RGEs for the gaugino sector \eq{RGE-gaugino} are relatively simple. They are linear and depend only on the gauge couplings. Since the RGEs of gauge couplings \eq{RGE-gaugecoupling} do not depend on any of the other running masses or couplings, the relation between the low-energy gaugino mass and $\Mhalf$ is linear.

The RGEs of all other mass terms contain the gaugino masses as a $-|M_i|^2$ contribution. This means that if $\Mhalf$ is larger, the slope of these mass terms is more negative, so they grow more rapidly with decreasing energy. This leads to a higher mass for larger $\Mhalf$. The squarks have a stronger dependence on $\Mhalf$ than the sleptons, since their RGEs depend on the gluino mass $M_3$. From equation \eq{RGE-gaugino} and figure~\ref{f:massMhalf} it is clear that $M_3$ has a stronger dependence on $\Mhalf$ than the other gauginos, which results in a stronger $\Mhalf$ dependence in the squark sector.

This also affects other characteristics of the graphs. Firstly, it explains the difference in shape between the graphs of the slepton sector and those of the squark sector. As mentioned in section~\ref{s:M0explanation}, the square root of a quadratic function yields a quadratic behaviour if the influence is small and linear behaviour if the influence is large. Thus the dependence in the squark sector is practically linear, while the slepton curves bend for low $\Mhalf$.

The splitting between left-handed and right-handed sleptons is more important for higher $\Mhalf$, which can be explained by the relative contribution of $M_2$ to the slepton mass RGEs that was mentioned in section~\ref{s:M0explanation}. As $\Mhalf$ gets larger, so does $M_2$. This makes the slope of the left-handed slepton masses more negative, so the splitting increases, as follows from equation \eq{left-rightsplitting}. In the squark sector, this effect does not play a role, since the squarks are much heavier than $M_2$. As a result, the mass splitting caused by $M_2$ is small compared to the other contributions in the RGEs (compare section~\ref{s:M0explanation}).

Although the effect is small compared to the changes in the other masses, $\Mhalf$ has a larger influence on the mass of the lightest Higgs than $M_0$. Over the range of $\Mhalf$ plotted in figure~\ref{f:massMhalf}, the mass of the lightest Higgs increases by as much as 10~GeV. In fact, for these parameters, the lightest Higgs crosses the LEP Standard Model Higgs bound\footnote{The famous LEP limit on the Higgs mass is that the Standard Model Higgs cannot be lighter than 114.4~GeV at 95\% confidence level \cite{Barate:2003sz}. For the MSSM Higgs, this limit is actually 92.8~GeV. But since the light Higgs in mSUGRA models usually resembles the Standard Model Higgs, the exclusion limit for most points in mSUGRA parameter space is actually closer to 114.4~GeV.} for $\Mhalf\approx470$~GeV.

\subsection{Higgs Vacuum Expectation Value}

\begin{figure}[!h]
\centering{
\includegraphics[width=\linewidth]{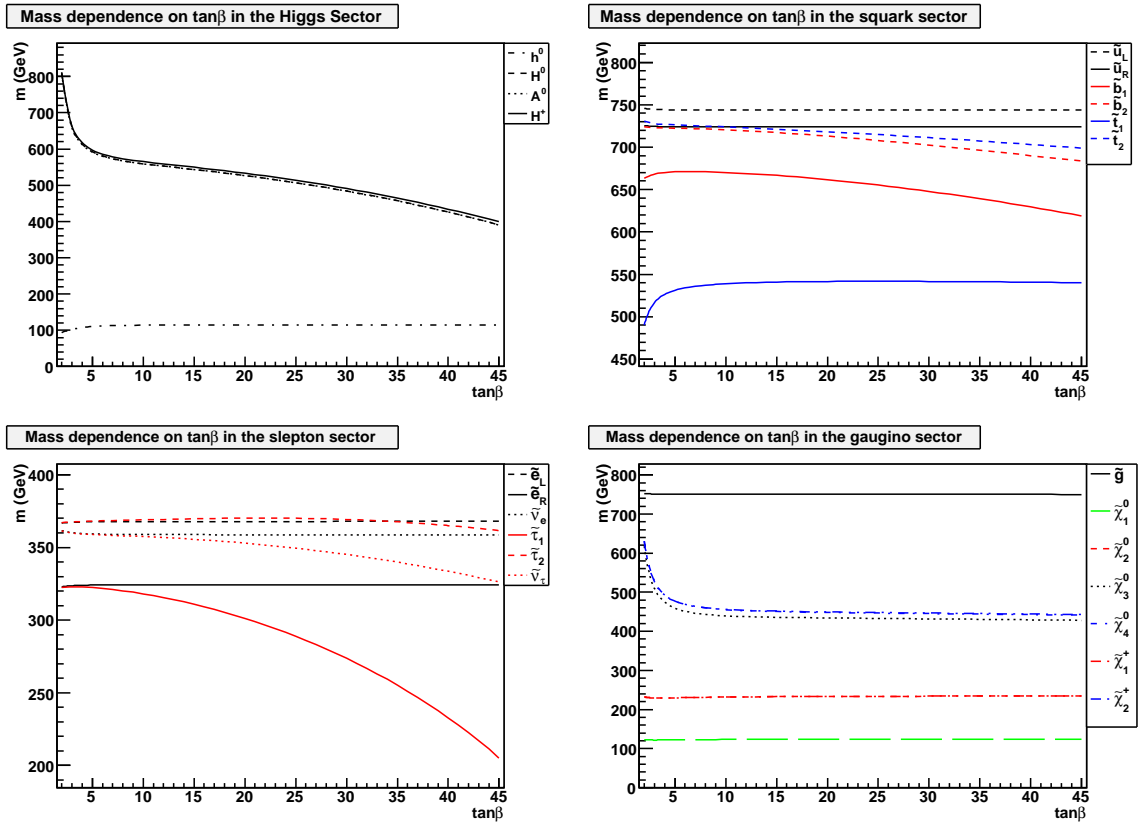}
\caption[Supersymmetric mass dependence on $\tan\beta$]{Low-energy mass dependence on $\tan\beta$ calculated with SPheno. All other parameters are kept constant at $M_0=300$~GeV, $\Mhalf=310$~GeV, $A_0=0$~GeV, $\mu>0$}\label{f:massbeta}}
\end{figure}

The ratio of the Higgs vevs influences the masses of particles through their coupling to the Higgs fields. The dependence on $\tan\beta$ is shown in figure~\ref{f:massbeta}.

As mentioned in section~\ref{s:susyhiggsmech}, the masses of the heavy Higgses depend on $m_{A_0}^2$ and are nearly degenerate. From equations \eq{binbeta} and \eq{A0mass}, the sharp rise in the heavy Higgs masses for low $\tan\beta$ ($\tan\beta<5$) can be understood, since this is the behaviour of $b/\sin(2\beta)$. Even though radiative corrections are enormous, the drop in the $h^0$ mass for low $\tan\beta$ follows from equation \eq{h0mass} as well. However, based on the same equations, one would expect the masses of the heavy Higgses to remain constant for high $\tan\beta$, which is clearly not the case. The reason for this is that equation \eq{binbeta} holds for the tree-level Higgs potential only. For large $\tan\beta$, loop corrections from the bottom and sbottom loops that have a $\sin\beta$ dependence (see section~\ref{s:massgeneralproperties}) become increasingly important. This drives the Higgs mass down for large $\tan\beta$. The lower Higgs mass also affects the masses of the third generation sleptons through loop corrections.

The off-diagonal terms in the third generation slepton mass matrices are important for large and small $\tan\beta$, since they depend on $\cos\beta$ and $\sin\beta$. As mentioned in section~\ref{s:massgeneralproperties}, the mass splitting for the stau is most important for large $\tan\beta$, although in some parts of parameter space it occurs for low $\tan\beta$ as well. For the stop the small $\tan\beta$ effect is quite strong. In some exotic parts of parameter space, the lightest stop can even become the LSP, for instance at $M_0 = 140$~GeV, $\Mhalf = 150$~GeV, $A_0 = -600$~GeV, $\tan\beta = 4, \mu >0$. A large $\tan\beta$ has a particularly strong effect on the stau mass and can easily drive it below the mass of the lightest neutralino in parts of parameter space. If the LSP is not a neutral particle, it would have been observed long ago, so these points are excluded. The exact splitting and the extent to which sfermion masses are driven down depends on the other mSUGRA parameters as well as on $\tan\beta$.

The gauginos are mostly unaffected by a change in $\tan\beta$. An exception is the low $\tan\beta$ behaviour of the Higgsino-like neutralinos and charginos. This is caused by the $\beta$-dependence of the off-diagonal terms in the mass matrix.

\subsection{Soft-Breaking Trilinear Coupling}\label{s:A0massexplanation}

\begin{figure}[!h]
\centering{
\includegraphics[width=\linewidth]{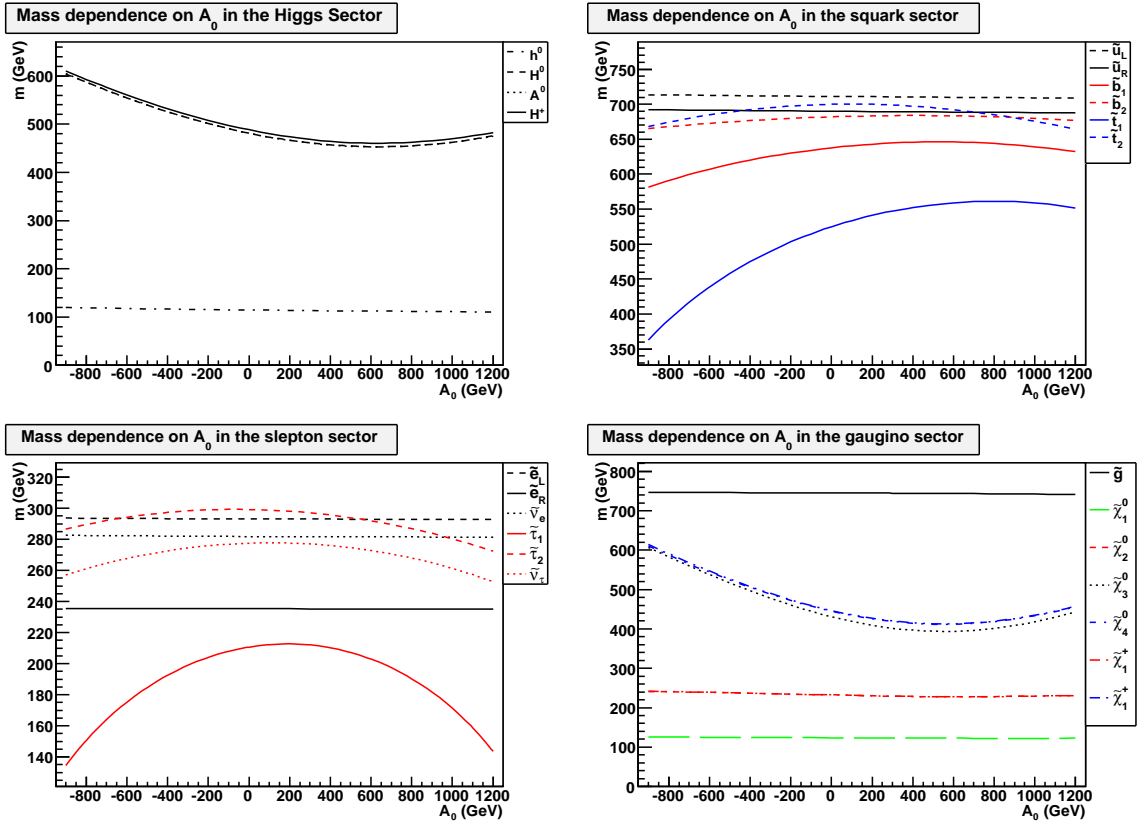}
\caption[Supersymmetric mass dependence on $A_0$]{Low-energy mass dependence on $A_0$ calculated with SPheno. All other parameters are kept constant at $M_0=200$~GeV, $\Mhalf=310$~GeV, $\tan\beta=20,\mu>0$}\label{f:massA0}}
\end{figure}

The unification value of the trilinear couplings $A_0$ is perhaps the most complicated parameter in mSUGRA. It always occurs in combination with a Yukawa coupling. The parameter $A_0$ is not directly a boundary condition for the RGEs of any of the masses, but the trilinear couplings occur in the RGEs of all third generation sfermions as well as the Higgses. On top of that, they occur in the off-diagonal elements in the sfermion mass matrices, which are important for the third generation as well. The $A_0$ dependence of supersymmetric masses is shown in figure~\ref{f:massA0}.

Of the two effects that change the sfermion masses through the parameter $A_0$, we first discuss the off-diagonal terms in the sfermion mass matrices. Diagonalizing the mass matrix yields the square root of a quadratic function in $A_0$. Since the influence of $A_0$ is small compared to the other parameters, this results in an approximate quadratic dependence on $A_0$. This effect explains the strong mass splitting for large $|A_0|$ as well as the $-A_0^2$ dependence of the light third generation sfermions. It follows from section \ref{s:massgeneralproperties} that for the up-type sfermions this effect is especially important for large $\tan\beta$, while for down-type sfermions it is significant for small $\tan\beta$. Due to the large value of the top Yukawa coupling, the stop mass is always affected.

For the heavy third generation sfermions one would predict an $A_0^2$ dependence based on this argument. In contrast, figure~\ref{f:massA0} shows a slight $-A_0^2$ dependence. This is caused by the second effect: the RGEs in Appendix~\ref{app:susyrge}. The trilinear couplings also affect the third family \emph{diagonal} sfermion mass terms through the $X_{t,b,\tau}$ terms in the RGEs. If $A_0^2$ increases, the slopes of the third generation soft-breaking sfermion masses increase as well, so the masses at low energy decrease. This gives rise to an overall $-A_0^2$ dependence of the sfermion masses, which has to be added to the splitting effect of the off-diagonal terms.

Of course, the trilinear couplings themselves are subject to RGEs as well, so it is not actually $A_0$ that occurs quadratically in the equations. This is the reason that the maxima do not occur at $A_0=0$~GeV. The actual position of a maximum is determined by all values of that particular trilinear coupling over the entire range of the RGEs. Thus they are different for the different sfermions, and also change if other breaking parameters are varied. The maxima of all graphs are shifted to more positive values of $A_0$, but not all are shifted by the same amount.

Based on the RGEs one would expect the Higgs sector to behave in approximately the same way as the sfermion sector, but figure~\ref{f:massA0} tells a different story. Most likely this is due to radiative corrections, but it cannot be explained by simple first-order arguments and only follows from a full calculation such as done by SPheno.

Some of these loop corrections have an important effect on the mass of the lightest Higgs. The $h^0$ mass becomes larger as $A_0$ gets more negative and changes by as much as 10~GeV over the range shown in figure~\ref{f:massA0}. This is quite a strong effect and is caused by corrections of the trilinear couplings to the Higgs quartic interaction \cite{Okada:1990gg}, which affects the Higgs mass through loop corrections.

The masses of most gauginos are hardly affected by $A_0$, as can be expected from the RGEs and the mass matrices. The influence on the Higgsino-like neutralinos and charginos is much larger. They behave very much like the masses of the Higgses, because they are subject to the same RGEs and interactions.

\subsection{The Superpotential Higgs Mass}

The sign of $\mu$ does not have a major effect on the mass spectrum. It causes a small shift in the masses, but the effect is negligible compared to the other parameters. This is the reason that most research focuses on $\mu>0$. The parameter space for $\mu<0$ also seems to be more restricted than for $\mu>0$, for instance because the stau usually becomes the LSP for lower $\tan\beta$. Although it is possible that nature is described by supersymmetry with $\mu<0$, this scenario does not seem to introduce new phenomenology, so as long as no signs of supersymmetry are found, it is more useful to focus research on $\mu>0$.

\subsection{Rules of Thumb}

The mSUGRA mass spectrum depends in a complicated way on the boundary conditions at the unification scale, but there are a few observations that hold for most parts of mSUGRA parameter space.
\begin{itemize}
\item The lightest stop, sbottom and stau are lighter than their counterparts of the first two generations.
\item Squarks are heavier than sleptons.
\item Right-handed sfermions are lighter than left-handed sfermions.
\item The gluino is the heaviest gaugino. Unless the soft-breaking scalar mass becomes as large as the electroweak symmetry breaking scale, the lightest neutralino is Bino-like, while $\tilde\chi_2^0$ and $\tilde\chi_1^\pm$ are SU(2) gaugino-like and nearly degenerate in mass. The heaviest neutralinos and charginos are then Higgsino-like and $\tilde\chi_4^0$ and $\tilde\chi_2^\pm$ are nearly degenerate in mass.
\end{itemize}
The masses depend on the mSUGRA parameters roughly in the following way:
\begin{itemize}
\item If $M_0$ increases, all masses rise considerably except the lightest Higgs and the gauginos, which are nearly constant.
\item All masses increase with increasing $\Mhalf$.
\item The masses of the gauginos and the first two generations of sfermions hardly depend on $\tan\beta$.
\item The mass of the lightest Higgs is constant for large $\tan\beta$, but drops considerably if $\tan\beta$ becomes small.
\item For small $\tan\beta$, the heavy Higgs and Higgsino masses rise sharply with decreasing $\tan\beta$. The masses of third generation sfermions decrease for lower $\tan\beta$ depending on the other mSUGRA parameters.
\item For large $\tan\beta$, the heavy Higgses and the third generation sfermions become lighter. The mass splitting of the third generation sfermions, especially the $\tilde\tau_1$ and $\tilde\tau_2$, is more important for large $\tan\beta$ as well.
\item Sfermions of the third generation have an approximate $-A_0^2$ dependence with a maximum that is shifted to positive values of $A_0$.
\item The mass of the lightest Higgs is higher for lower $A_0$.
\item The sign of $\mu$ has little effect compared to the other parameters.
\end{itemize}
These conclusions are only valid within the mSUGRA framework. mSUGRA was constructed on the hypotheses of soft supersymmetry-breaking universality and unification, which leaves little freedom for extreme mixing angles or large differences in couplings between different points in parameter space. In the general MSSM, and even more so in extensions of the MSSM, the effects of couplings and off-diagonal terms can be much larger.

Although the mass spectrum by itself does not fix phenomenology, it does give important clues. The mass spectrum determines which decays are kinematically allowed, and thereby fixes the phenomenology to a large extent. Also, lighter particles are more likely to be produced at the LHC than heavier particles if the couplings have the same magnitude. So the mass spectrum also has important implications for the total supersymmetric cross section and thus for detecting supersymmetry.
\section{Phenomenology}\label{s:phenomenology}

If we do not want to restrict ourselves to benchmark points, we have essentially two ways to deal with the mSUGRA parameter space. The first is, putting it crudely, to simulate events for an enormous amount of points and see if the result of any of these points happens to coincide with measurements at the LHC. The second is a more systematic approach that we will develop in this section.

We expect the phenomenology at a collider to be determined by three ingredients. The first is the type of particles that is produced in the initial interaction. This is determined largely by the couplings: particles that couple more strongly to the particles in the beam are produced more abundantly. In case of the LHC, the beam particles are protons, so we expect the supersymmetric QCD sector to be important.

The second ingredient is the supersymmetric cross section. This determines the amount of supersymmetric particles that is produced at the LHC. The total supersymmetric cross section depends on the masses of the particles produced the most in the first interaction. If these particles are light, they are produced more abundantly relative to the Standard Model background, which results in a better distinguishable signal at the LHC. Therefore, low-mass points in the mSUGRA parameter space have the best discovery potential in the first years of the LHC.

Finally, the phenomenology at the LHC depends on the decay chains of supersymmetric particles. These are determined by the supersymmetric mass spectrum that was discussed in section~\ref{s:massspectrum}. Particles with a lower mass are produced more often in the decay chain than heavier particles. In fact, heavy particles might not be produced at all.

The following sections address these issues. Section~\ref{s:firstint} is dedicated to studying which particles are produced in the inital interaction. In section~\ref{s:crosssection} the consequences for the total supersymmetric cross sections are discussed. Then regions in parameter space with similar phenomenology are identified in section~\ref{s:regions}. Finally, stability and representativeness of benchmark points is discussed in section~\ref{s:benchmarkresult}. In our discussion we will limit ourselves to points with relatively low masses to ensure a reasonable cross section for supersymmetry.

\subsection{Particle Production}\label{s:firstint}

As mentioned in section~\ref{s:pythia}, only supersymmetric and Higgs particles are selected to be produced in the initial interaction of the Pythia simulation. Our first step is to determine which particles are produced in the initial interaction for generic points in mSUGRA parameter space. After this we discuss parts of parameter space that have a different behaviour. 

\begin{figure}[!h]
\centering{
\includegraphics[width=0.8\linewidth]{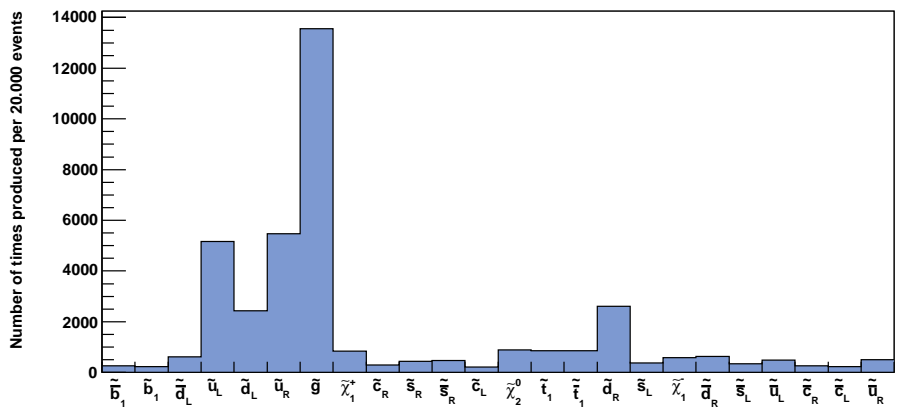}
\caption[Supersymmetric particles produced in the initial interaction for SU3]{The number of times a given particle was produced in the initial interaction of 20.000 SUSY events for benchmark point SU3. Only the particles that are produced in at least 1\% of the events are shown.}\label{f:firstint}}
\end{figure}
Figure~\ref{f:firstint} shows the particles that are produced in the initial interactions of 20.000 SUSY events for the benchmark point SU3. Only the particles that are produced in the initial interaction of at least 1\% of the events are shown. A complete graph containing all initial interactions can be found in Appendix~\ref{app-firstint}. 

Due to the strength of the QCD coupling, the initial interaction is dominated by QCD processes. This is not true if the gluino and all squarks are too heavy to be produced. But since such points yield extremely low cross sections, we will not investigate this case.

\subsubsection{The Effects of the Parton Distribution Function}\label{s:pdf}

\begin{figure}[!h]
\centering{
\includegraphics[width=90mm]{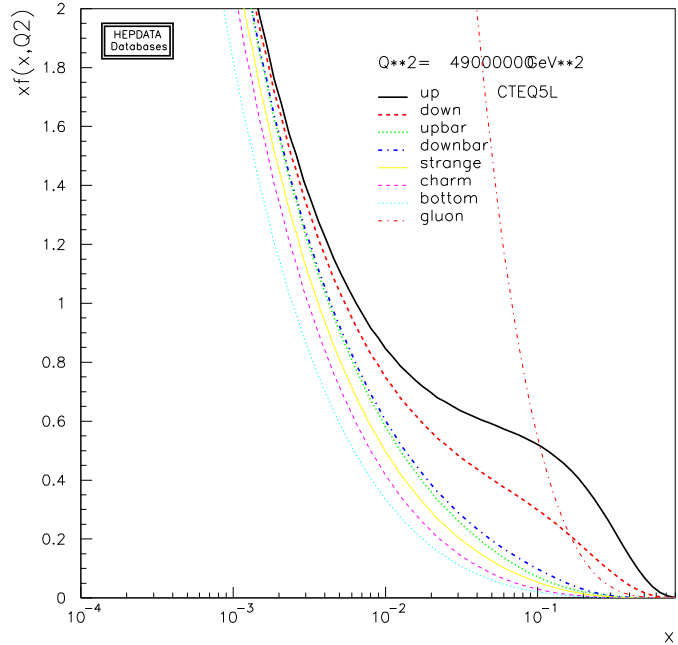}
\caption[Parton distribution function of a 7 TeV proton]{Parton distribution function of a 7 TeV proton from \cite{durhampdfs}.}\label{f:pdf}}
\end{figure}

Figure~\ref{f:firstint} shows that gluinos are produced the most, followed by up squarks and then down squarks. Furthermore, more positive than negative particles are produced. This is explained by the parton distribution function of a 7 TeV proton shown in figure~\ref{f:pdf}.

Let us use the rough estimate that supersymmetric particles have a mass of at least 100~GeV. Then partons need to have at least 1.4\% of the momentum fraction of a 7 TeV proton to produce supersymmetric particles. In fact, it is more likely that they have a higher momentum fraction, since cross sections are typically small close to a kinematic threshold. Figure~\ref{f:pdf} shows that for such high momentum fractions the content of a high-energy proton resembles that of a low-energy proton. In other words, partons with high momentum fractions are dominantly valence quarks or gluons. The couplings predicted by mSUGRA models do not have strong flavour-changing components, so up squarks and down squarks are produced more often than other flavoured particles, usually in combination with a gluino.

\begin{figure}[!ht]
\includegraphics[width=0.5\linewidth]{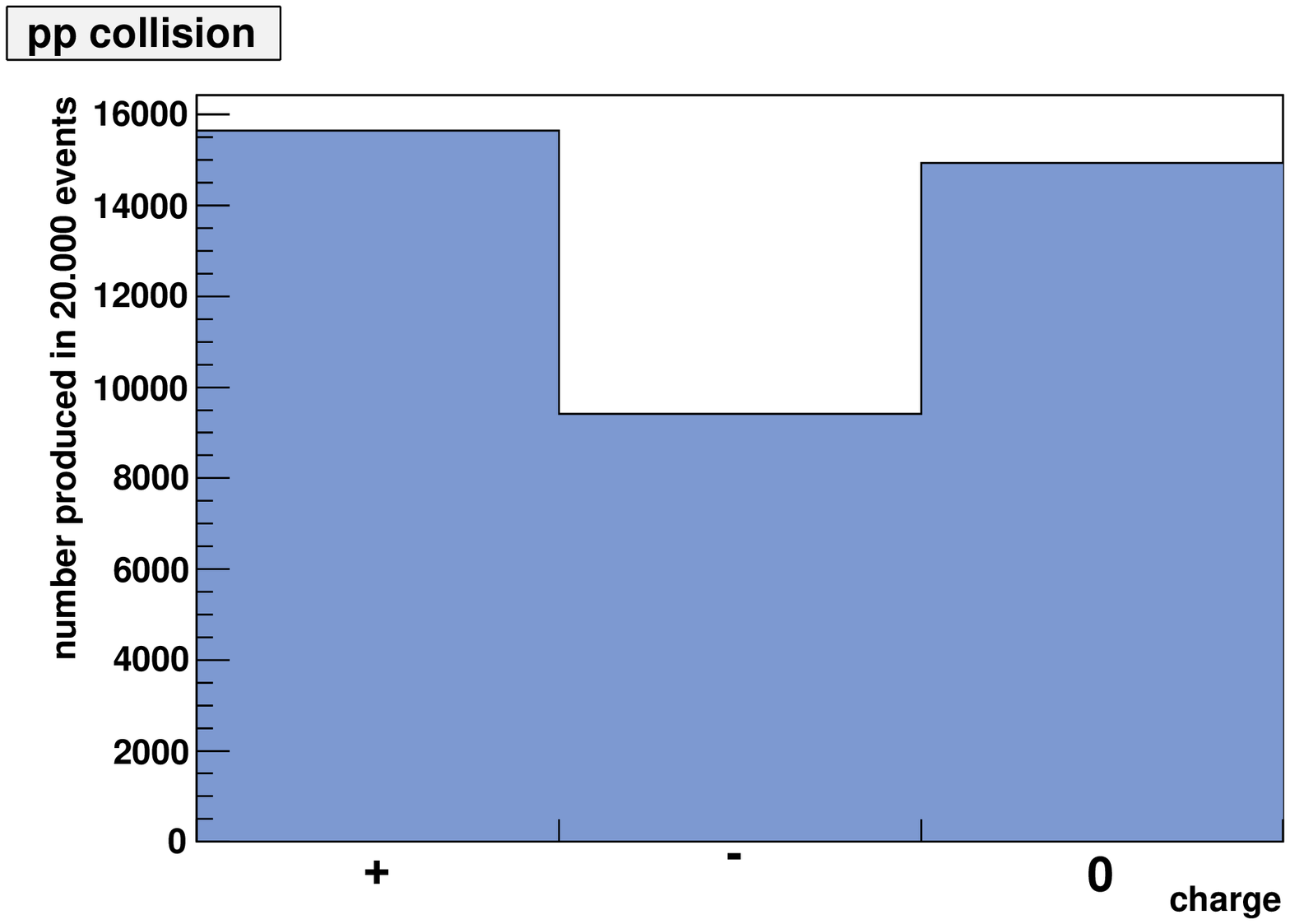}
\includegraphics[width=0.5\linewidth]{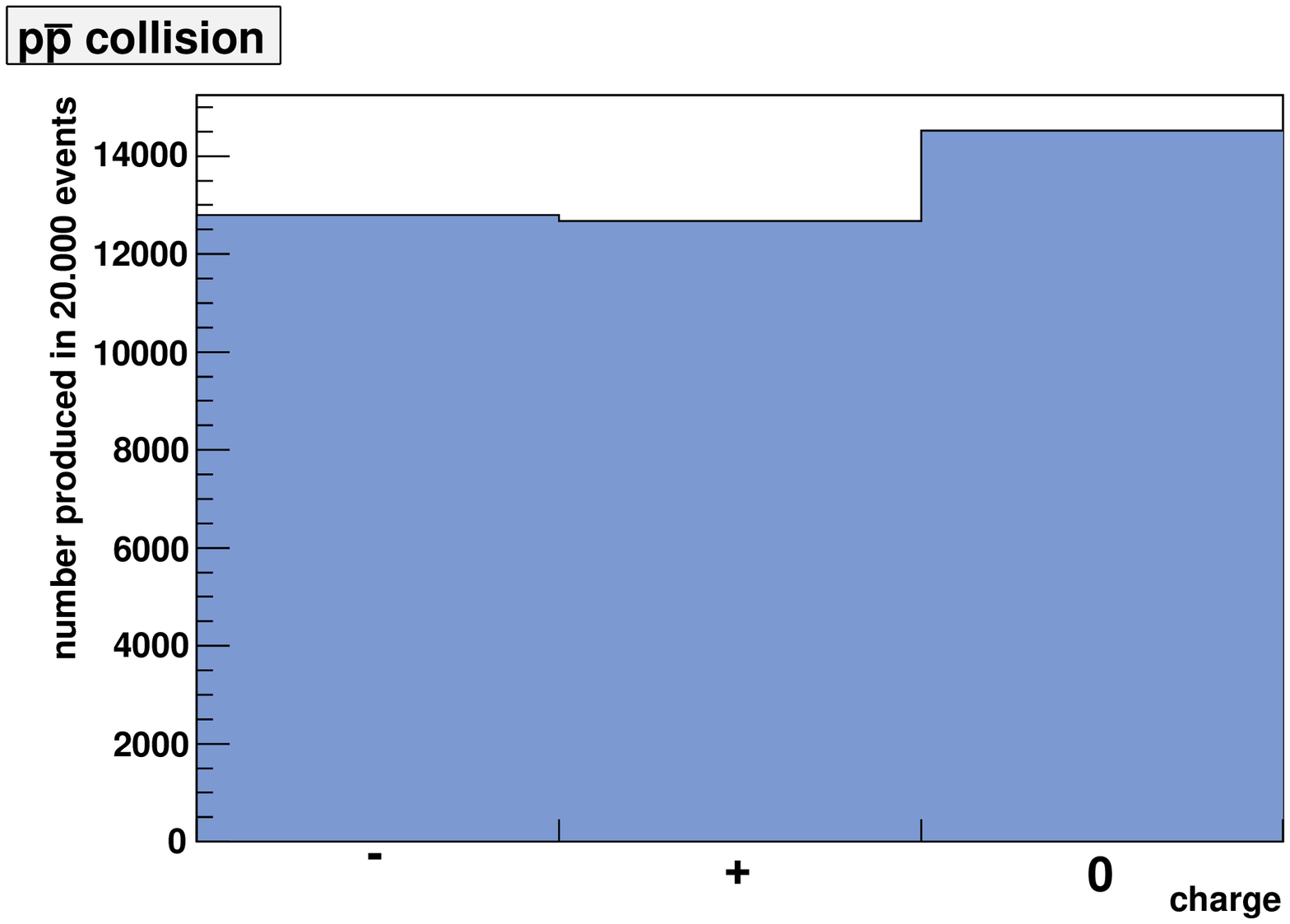}
\caption[Charge of supersymmetric particles from a $pp$ collision and a $p\bar p$ collision]{The number of supersymmetric particles of a certain charge produced in the initial interaction in case of a proton-proton collision and a proton-antiproton collision at the same energy. These numbers are based on 20.000 SUSY events for benchmark point SU3.}\label{f:chargedist}
\end{figure}

The role of the proton content can be verified by simulating a proton-antiproton collision with the same parameters. On average this should yield an equal number of positive and negative particles. Figure~\ref{f:chargedist} shows the number of positive, negative and neutral particles produced in the initial interaction for both a $pp$ collision and a $p\bar p$ collision.
\newpage
Figure~\ref{f:chargedist} confirms that the abundance of positively charged particles is caused by the content of the proton. This means that this effect is not specific to supersymmetry. From the parton distribution function in figure~\ref{f:pdf}, one would expect this behaviour for any high-mass particle at the LHC.

In short, gluinos are produced the most in the initial interaction, followed by up squarks and then down squarks. This is caused by the parton distribution function of the proton shown in figure~\ref{f:pdf} and is true for most points in mSUGRA parameter space that have a reasonably high cross section. There are a few exceptions to this general conclusion. Firstly stops can be an important contribution to the total cross section if the mass of the lightest stop is small. Alternatively, if gluino masses are very large while squark masses are small, or the other way around, the initial interaction could yield different results.

\subsubsection{Light Stops}\label{s:firstint-lightstops}

\begin{figure}[!h]
\centering{
\includegraphics[width=0.6\linewidth]{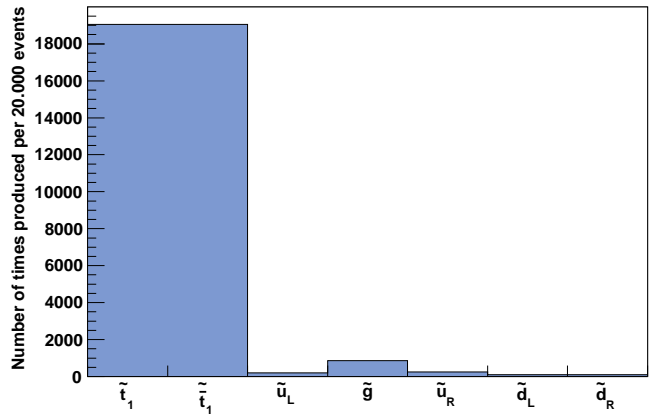}
\caption[Supersymmetric particles produced in the initial interaction in case of a low stop mass]{The number of times a given particle is produced in the initial interaction of 20.000 SUSY events for the point $M_0=200$~GeV, $\Mhalf=200$~GeV, $A_0=-800$~GeV, $\tan\beta=15,\mu>0$, with a stop mass of 87~GeV. Only the particles that are produced in at least 0.5\% of the events are shown.}\label{f:firstint-lightstop}}
\end{figure}

The only particle that has QCD interactions and yet can have quite a small mass is the stop. In section \ref{s:massspectrum} we saw that the stop mass can easily become very low for large values of $|A_0|$, especially if $\tan\beta$ is large. For extreme points where the stop almost becomes the LSP, stop production can be the dominant process in the initial interaction, as can be seen in figure \ref{f:firstint-lightstop}.

Large negative values of $A_0$ that can yield such light stops are particularly interesting because the mass of the lightest Higgs becomes large (see section~\ref{s:A0massexplanation}).

\subsubsection{Heavy Gluinos}

If the soft-breaking gaugino mass is large while the soft-breaking scalar mass is small, the first interaction produces mainly squarks and very little gluinos. However, this situation cannot be taken to extremes for two reasons.

Firstly, we know from section~\ref{s:massspectrum} that the only parameter that has a large effect on the gluino mass is $\Mhalf$. However, scalar masses are affected by $\Mhalf$ as well. So within the framework of mSUGRA, heavy gluinos automatically imply rather heavy scalar masses.

Secondly, if the gluino is heavy, so are the other gauginos. If $\Mhalf$ is too large compared to $M_0$, the LSP is a charged scalar particle, usually the lightest stau. In that case we would have observed it already, so this scenario is excluded. In other words, $\Mhalf$ cannot become extremely large unless $M_0$ becomes large as well, in which case the cross section drops.

\begin{figure}[!h]
\centering{
\includegraphics[width=0.9\linewidth]{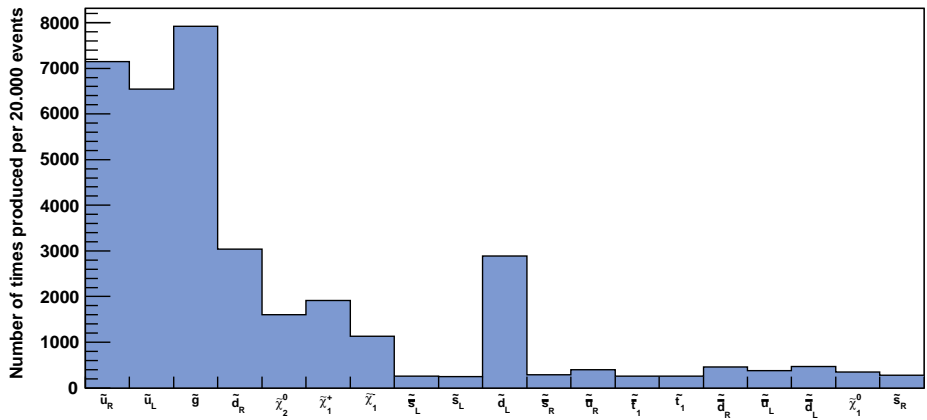}
\caption[Supersymmetric particles produced in the initial interaction in case of a heavy gluino]{Particles produced most abundantly in the initial interaction for the point $M_0=120$~GeV, $\Mhalf=600$~GeV, $A_0=0$~GeV, $\tan\beta=10,\mu>0$, which has a gluino mass of 1.4~TeV and relatively light scalars. Only particles produced in at least 1\% of the initial interactions is shown. Clearly the gluino production is suppressed compared to figure~\ref{f:firstint}}\label{f:firstint-heavygluino}}
\end{figure}

To investigate if this effect can be important nonetheless, we take a point in mSUGRA parameter space with large gaugino masses and relatively small scalar masses. Due to the coupling strength, gluinos are dominantly produced by processes that involve gluons. Thus judging from the parton distribution function in figure~\ref{f:pdf}, one expects the gluino contribution to drop if the gluino is heavier than 1~TeV. So we choose a value for $\Mhalf$ that ensures the gluino is heavier than 1~TeV and find a value for $M_0$ such that the scalar masses are just large enough for the LSP to be a neutralino. The particles produced in the initial interaction of such a point are shown in figure~\ref{f:firstint-heavygluino}.

The supersymmetric cross section for this point is 0.43~pb, which is a factor 500 smaller than the supersymmetric cross section for the low-mass benchmark point SU4. The chance of discovering such high-mass supersymmetry in the first years of the LHC is small. In addition, although gluino production is suppressed, it is still significant even in this extreme scenario. The initial interaction yields slightly different results, but that will not result in radically different collider phenomenology and we will not treat this case separately.

\subsubsection{Heavy Scalars}\label{s:focuspointregion}

The last exception is a large scalar mass. Constraints from low-energy observables for this scenario are weak. The most important constraint on supersymmetric masses comes from the dark matter relic density. Unfortunately this constraint has a loophole if the LSP has a large Higgsino component \cite{allanach2002}, which is the case if $M_0$ is a few TeV (see section~\ref{s:massgeneralproperties}).

However, supersymmetric collider phenomenology is rather straightforward in this case. Because the scalars are much heavier than the gauginos, they are hardly ever produced. Thus supersymmetry is effectively limited to the gaugino-sector. The gluinos decay through three-body decays involving two Standard Model particles and a gaugino. The neutralinos and charginos usually decay to a lighter gaugino and a $W,Z$ or Higgs boson. At a collider, the most likely way to recognise this scenario is by missing energy signals combined with the signals from the Standard Model bosons. We do not discuss this part of parameter space in depth, because the supersymmetric phenomenology does not need further discussion. It is an interesting region to study in the context of discovery potential, but that is not the goal of this research.
\newline

Summarizing the conclusions of this section, the particles that are produced most abundantly in the initial interaction are $\tilde g,\tilde u$ and $\tilde d$ and in some cases $\tilde t_1$. In general gluinos are produced the most. Within the low-mass region, the relative numbers of supersymmetric particles produced in the initial interaction can be taken as constant, with the exception of the stops. As a result, the effect of the initial interaction can almost completely be absorbed in the total supersymmetric cross section. This is a single number for a particular point in parameter space and is calculated by Pythia during the event simulation. Only stop production has to be taken into account separately.

\subsection{The Supersymmetric Cross Section}\label{s:crosssection}

We concluded in section~\ref{s:firstint} that the most important processes in the initial interaction are the production of gluinos, up or down squarks, and in some cases the production of two light stops. If we treat stop production separately and only take moderate values of $M_0$ and $\Mhalf$ into account, the gluino is involved in all processes. As a result, the total supersymmetric cross section heavily depends on the gluino mass. The scalar masses also influence the total cross section, but their effect is not as large as that of the gluino.

\begin{figure}[!h]
\centering{
\includegraphics[width=\linewidth]{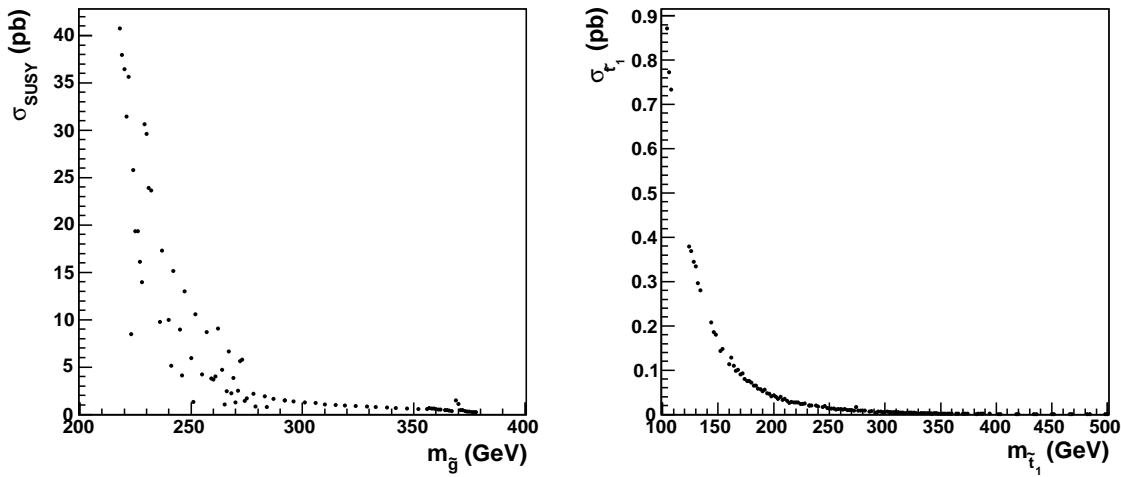}
\caption[The supersymmetric cross section]{The total supersymmetric cross section and the cross section for stop production for many values of the high-scale parameters.}\label{f:crosssection}}
\end{figure}

The stop cross section depends on the stop mass and follows the gluon parton distribution function, since gluons are the leading contribution to stop production. Both the total supersymmetric cross section and the cross section for stop production are shown in figure~\ref{f:crosssection}.

These plots contain the cross sections for many different values of the high scale parameters: $M_0,\Mhalf,A_0$ and $\tan\beta$ are all varied. The total cross section clearly depends on the gluino mass. The scatter in the graph for the total cross section is caused by the difference in scalar masses. The total cross section is not influenced by the changes in couplings that are associated with varying $A_0$ and $\tan\beta$. As expected, the stop production rises considerably for low stop mass. For relatively high gluino masses, it can even become the main contribution to the cross section. For instance the small increase in the total cross section in figure \ref{f:crosssection} near $m_{\tilde g}=370$~GeV is caused by stop production.

In short, the cross section generally increases as the gluino mass decreases. The scalar masses have a smaller effect on the total cross section, but it is still significant, particularly in case of a small stop mass. In general the particles produced in the first interaction do not have much effect on phenomenology for moderate values of $M_0$ and $\Mhalf$. The only exception is the case of a stop mass smaller than approximately 200 GeV. If the gluino mass is large, stop production can be the leading contribution to the cross section in that case.

\subsection{Phenomenological Regions}\label{s:regions}

With the exception of stops, the changes in the type of particles produced in the first interaction are negligible over the part of parameter space we investigate. In this section we take a closer look at the last ingredient that determines the phenomenology: the mass spectrum. We limit ourselves to $M_0,\Mhalf\leq600$~GeV to ensure reasonable cross sections and interesting phenomenology.

The goal of this section is to predict phenomenology directly from the mass spectrum. This has several advantages. Firstly, as mentioned in section~\ref{s:methods}, doing a complete event simulation takes much more time than generating a mass spectrum. Secondly, a systematic approach gives more insight in supersymmetric phenomenology. Finally, event generators have become so complex, that they are practically black boxes. The results from an RG evolution program are checked more easily, for instance using the results from chapter~\ref{s:massspectrum}. In this section we will not take the total supersymmetric cross section into account, but focus on how the phenomenology follows from the relative masses of the particles. Graphs that include the effects of the supersymmetric cross section can be found in Appendix \ref{app:phenocross}.

The first step is to identify which particles are particularly important for the phenomenology that could be measured at the LHC. We first consider the points in parameter space where the lightest stop has a mass of at least 200 GeV. In section~\ref{s:firstint} we concluded that in that case the particles produced most abundantly in the first interaction are gluinos, up squarks and down squarks. Referring to section~\ref{s:massspectrum}, we see that these are the heaviest particles in the spectrum. That means that supersymmetric phenomenology is not determined by the beginning of the decay chain, since the particles produced in the initial interaction can decay to almost all other supersymmetric particles. The masses of the initial particles do affect the cross section, but that is a constant factor for a particular point in parameter space and mainly influences the discovery potential, not the phenomenology within supersymmetry. 

If the particles produced in the initial interaction are not the determining factor for the phenomenology, how about their decay products? Gluinos decay to a squark and a quark, where decays to the lightest stops and sbottoms are preferred because of their low mass. Up and down squarks decay to up and down quarks because mixing between families is negligible. These decays also produce charginos and neutralinos. 

This suggests that the most important factor in determining the phenomenology is the spectrum of the low-mass particles, which determines the possible decay chains. Therefore we will study the lightest particles in sets of particles that have similar couplings. In other words: the lightest particle of the squark sector, the lightest particle in the slepton sector, and the next to lightest particles in the gaugino sector, since the lightest gaugino is the stable LSP. From section~\ref{s:massspectrum} it follows that these particles are $\tilde t_1, \tilde\tau_1, \tilde\chi_2^0$ and $\tilde\chi_1^\pm$. We discuss the properties of these particles and the phenomenological implications in more detail in the following sections.
\newpage
\noindent The analyses can be divided in three steps
\begin{enumerate}
\setlength{\itemsep}{2pt}
\setlength{\parskip}{0pt}
\setlength{\parsep}{0pt}
\item The leading production and decay modes of a specific particle are determined. The results are listed in Appendix~\ref{app:processes}.
\item Based on the mass spectrum obtained with SPheno, different phenomenological regions in the mSUGRA parameter space are identified. The effects of stop production in the initial interaction are taken into acount as well.
\item The validity of these regions and the resulting predictions is checked with an event simulation using Pythia.
\end{enumerate}
Decays that are not kinematically allowed anywhere in the low-mass region of mSUGRA parameter space are not considered. These decay modes are suppressed, since they are only accessible to off-shell particles.

In all graphs, particles and antiparticles are treated as one variable. For instance, the total number of light staus is a shorthand for the number of light staus plus the number of light anti-staus. This ensures that we do not have to correct for the surplus of positive particles produced in the initial interaction when comparing charged particles.

\subsubsection{Phase Space Effects}\label{s:phasespace}

If a decay is kinematically allowed, it is not automatically important. Whether or not it gives a significant contribution depends on the coupling and the available phase space. If the mass of the mother particle barely exceeds the mass of the daughters, a decay is typically suppressed due to phase space effects. As more phase space is available, the decay will become more important. In other words, the boundary of a phenomenological region is the point where a decay begins to contribute, not the point where it is the most important.

As a result, small phenomenological regions do not yield a new type of phenomenology, since they resemble their neigbhouring regions. Small phenomenological regions arise when the total mass of the daughters in two separate decay modes of a particular particle is almost equal. We will see several examples of this situation.

\subsubsection{The Light Stau}\label{s:stau}

\begin{figure}[!p]
\centering{
\includegraphics[width=\linewidth]{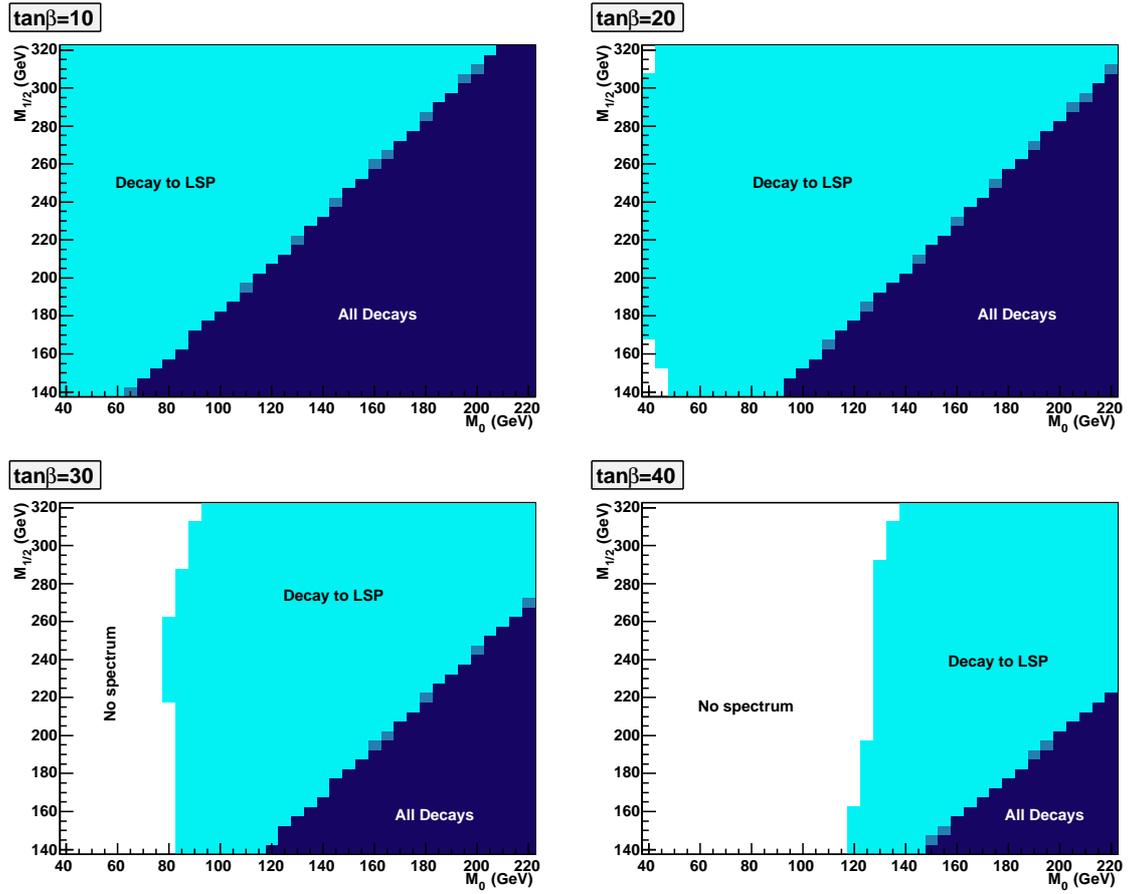}
\caption[Phenomenological regions based on stau decay]{Phenomenological regions based on the decays of light staus in the $M_0,\Mhalf$ plane. In all plots $A_0=0$~GeV and $\mu>0$. In the dark blue region, the lightest stau is heavier than the next to lightest gauginos, while in the cyan region it is lighter.}\label{f:staudecay-regions}}
\end{figure}

The decay modes of light staus are limited, since the tau-lepton quantum number is conserved. The possible decays are listed in table~\ref{t:staudecay}. Whether or not these decays are allowed depends on the specific point in parameter space. The decay to the lightest neutralino is always allowed for otherwise the light stau would be stable and it would have been observed already. We distinguish three regions in parameter space that are shown in figure~\ref{f:staudecay-regions}.
\begin{itemize}
\item If the lightest stau is lighter than the lightest charginos, it is likely to be produced quite abundantly because of its low mass. It can only decay to the LSP and a $\tau$, so supersymmetry in this region of parameter space produces many taus.
\item If the lightest stau is sufficiently heavy, it can decay to $\tilde\chi_2^0$, $\tilde\chi_1^\pm$ and the LSP. Since decays to charginos produce neutrinos rather than taus, one would expect the number of taus to go down in this region. However, as we will see later, a second effect decreases the number of taus even more.
\item In the small band where $m_{\tilde\chi_1^\pm}<m_{\tilde\tau}<m_{\tilde\chi_2^0}+m_{\tau}$, the lightest stau can decay to a chargino and to the LSP, but the decay to $\tilde\chi_2^0$ is not kinematically allowed. However, the phase space for the decay to charginos is small and off-shell staus can decay to a $\tilde\chi_2^0$. As explained in section \ref{s:phasespace}, such a transition region yields no new phenomenology.
\end{itemize}
In the white regions in figure~\ref{f:staudecay-regions} no proper spectrum was created. This is either because electroweak symmetry breaking does not occur or because the LSP is not neutral. In both cases the spectrum can immediately be discarded.

As for production, light staus can be produced in several ways, which are listed in table~\ref{t:staufrom}. The dominant production processes are those involving $\tilde\chi_1^\pm$ and $\tilde\chi_2^0$. These particles occur more abundantly than the other possible mothers because they are produced in many decays.

\begin{figure}[!h]
\centering{
\includegraphics[width=0.9\linewidth]{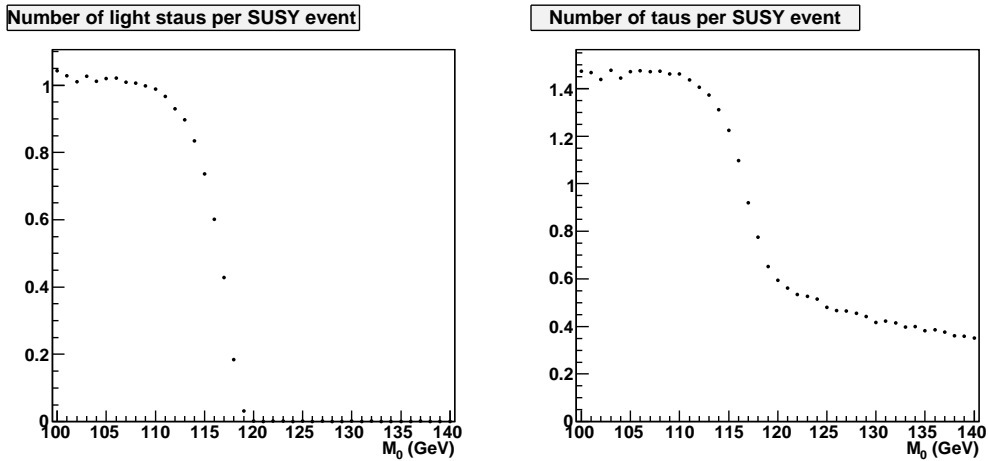}
\caption[A jump in the number of staus and taus due to kinematics]{The number of light staus and taus produced per SUSY event depending on $M_0$ with $\Mhalf=205$~GeV, $A_0=0$~GeV, $\tan\beta=10,\mu>0$. The number of staus and taus drops when the stau becomes heavier than the light gauginos. A graph that includes the cross section can be found in figure~\ref{f:staujump-cross}.}\label{f:staujump}}
\end{figure}

This has important consequences for the number of taus produced by supersymmetric processes in different parts of parameter space. As mentioned earlier, fewer staus decay to taus if the stau is heavier than the lightest charginos. However, a more important reason for the number of taus to drop in this region is that fewer staus are produced: if a stau can decay to a chargino, a chargino cannot decay to a stau. This is confirmed by an event simulation shown in figure~\ref{f:staujump}.
There is a clear drop in the number of staus if the stau is heavier than the light gauginos. Although taus can also come from other particles than staus (for instance from $W$ bosons or gauginos decaying through a three-body decay), this drop is also visible in the number of taus. A comparison with figure~\ref{f:staudecay-regions} shows that the point where the number of staus and taus drops coincides with the point in parameter space where the stau becomes heavier than the light charginos.

\subsubsection{The Light Gauginos}

The gauginos $\tilde\chi_1^\pm$ and $\tilde\chi_2^0$ couple to almost all particles, just like the Standard Model $W$ and $Z$ bosons. As a result there are many possible production and decay channels. Fortunately we can bring some order to chaos using the general mass hierarchy of the mass spectrum.

As mentioned in section~\ref{s:massspectrum}, the squarks and gluinos are heavier than the other supersymmetric particles. The only squark that can be lighter than the next to lightest gauginos is the stop, but the combined mass of the stop and the top is so high that this decay channel is not open. That means that the squarks and the heavier gauginos can only play a role in the production of $\tilde\chi_1^\pm$ and $\tilde\chi_2^0$ and not in the decay. Still, all sleptons can be part of either the production mechanism or the decay chain depending on the high-energy parameters. The possible decays are listed in table~\ref{t:gauginodecay}.

Because squarks are produced abundantly in both the initial interaction and through gluino decay, they are the main source of charginos and neutralinos. Thus, unless the stop is light, one would expect the number of light gauginos per SUSY event to be quite stable over parameter space. If the stop becomes light, the number of gauginos is affected. This is discussed in section \ref{s:phenA0}.

\begin{figure}[!h]
\centering{
\includegraphics[width=\linewidth]{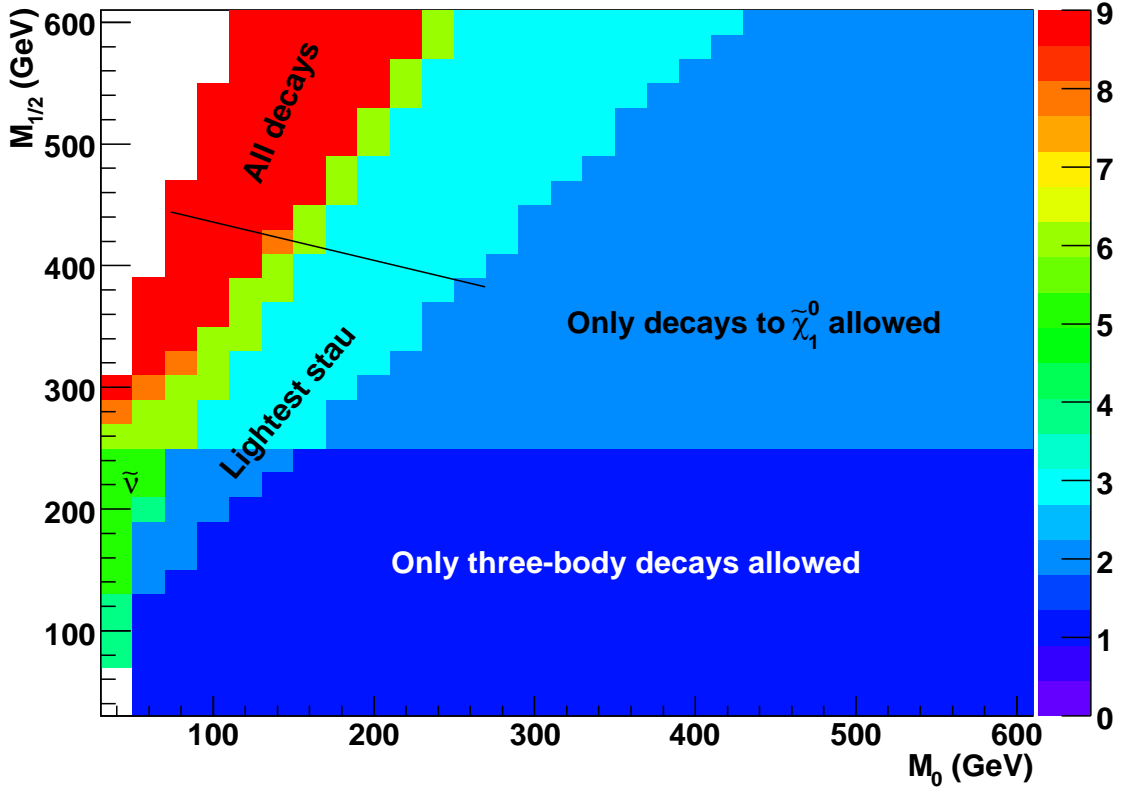}
\caption[Phenomenological regions based on $\tilde\chi_1^\pm$ decay]{Possible decay modes of the light charginos in different parts of parameter space with $A_0=0$~GeV, $\tan\beta=10,\mu>0$. The labels indicate which of the decay modes listed in table~\ref{t:gauginodecay} are kinematically allowed in a certain region. The colour code indicates how many of the decay modes are kinematically allowed. Three-body decays are always allowed.}\label{f:chargino-region10}}
\end{figure}

\begin{figure}[!h]
\centering{
\includegraphics[width=\linewidth]{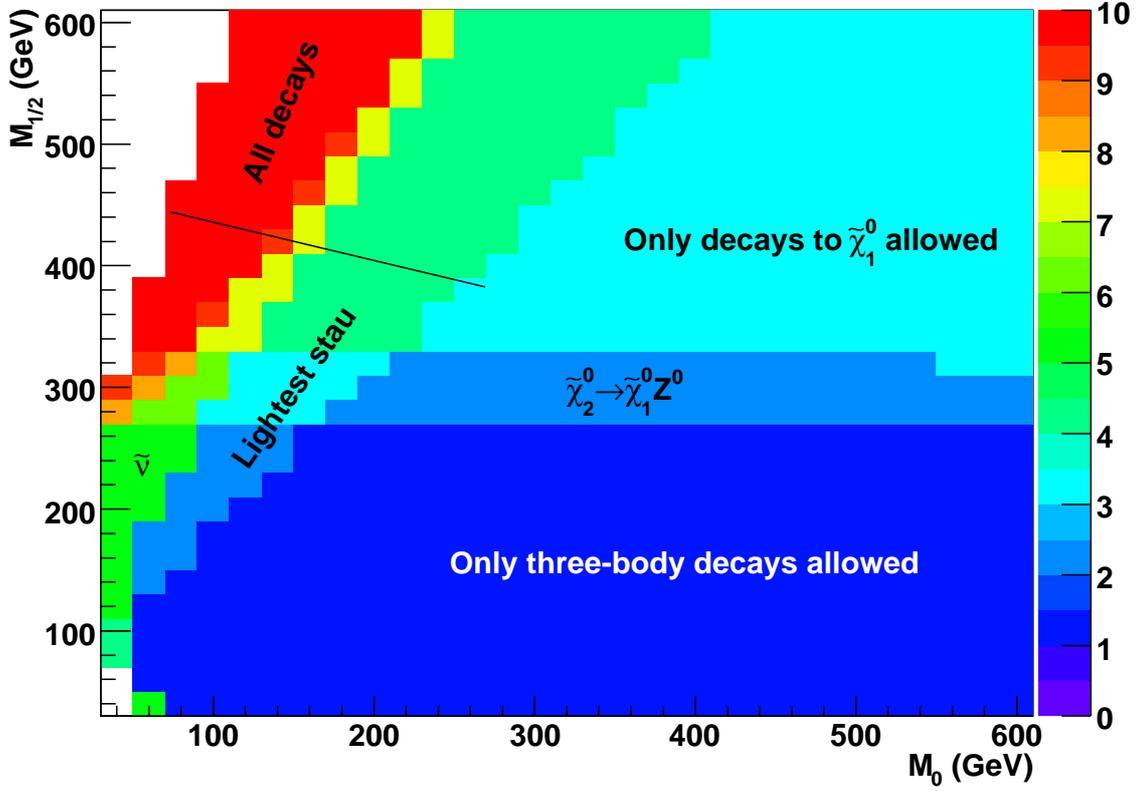}
\caption[Phenomenological regions based on $\tilde\chi_2^0$ decay]{Possible decay modes of $\tilde\chi_2^0$ in different parts of parameter space with $A_0=0$~GeV, $\tan\beta=10,\mu>0$. The labels indicate which of the decay modes listed in table~\ref{t:gauginodecay} are kinematically allowed in a certain region. The colour code indicates how many of the decay modes are kinematically allowed. Three-body decays are always allowed.}\label{f:neutralino-region10}}
\end{figure}

\begin{figure}[!h]
\centering{
\includegraphics[width=\linewidth]{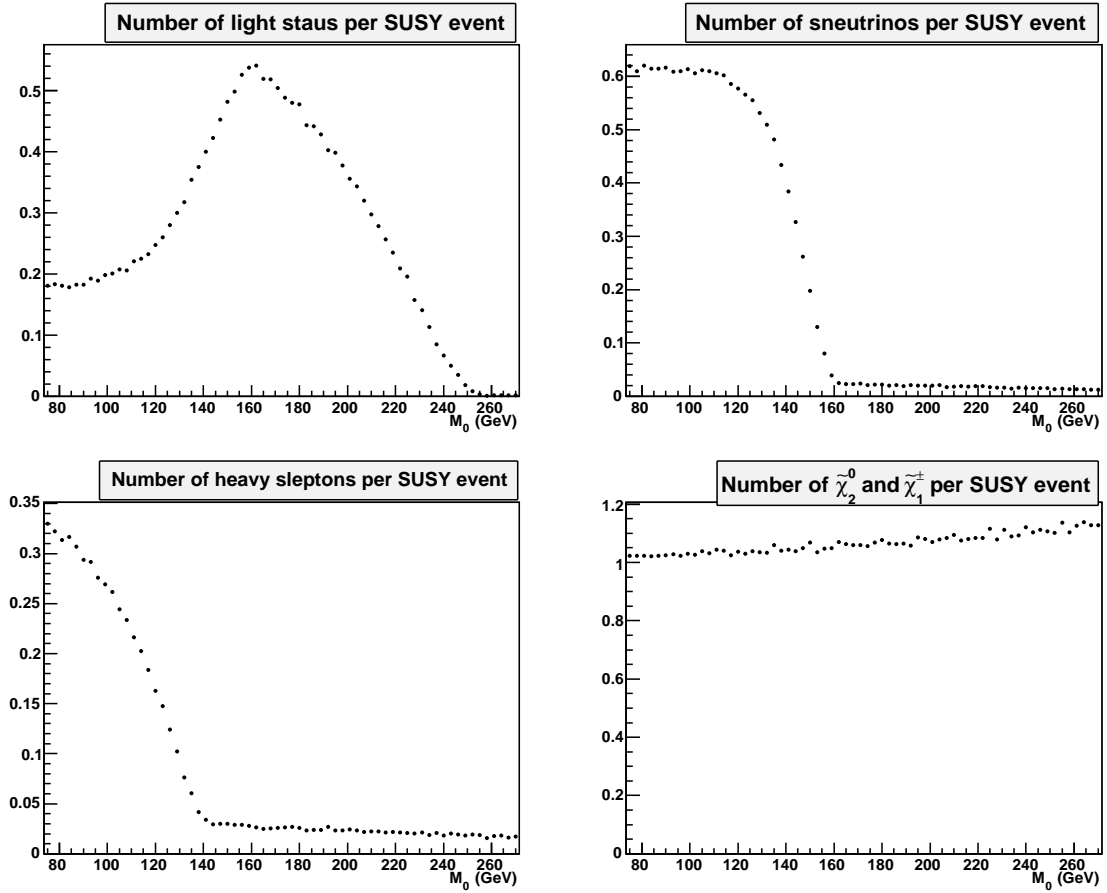}
\caption[The effects of gaugino decay channels on slepton production]{The number of times a particular particle is produced per event along the path through parameter space indicated in figures~\ref{f:chargino-region10} and~\ref{f:neutralino-region10}. Both $M_0$ and $\Mhalf$ are varied, $A_0=0$~GeV, $\tan\beta=10,\mu>0$. The kinks in the graphs exactly coincide with the boundaries of the phenomenological regions. Graphs that include cross section effects can be found in figure \ref{f:gauginodecay-cross}.}\label{f:gauginodecay}}
\end{figure}

We distinguish regions in parameter space in the same way as in section~\ref{s:stau}. The graphs are a bit more colourful than for the staus due to the larger number of possible decay modes, but the approximate mass degeneracies in the sfermion sector limit the number of regions. Figure~\ref{f:chargino-region10} shows the number of possible decay modes of a chargino in the $M_0,\Mhalf$ plane. In the white region the spectrum is discarded because the LSP is not neutral. Plots for the individual decay channels from table~\ref{t:gauginodecay} can be found in Appendix~\ref{app:charginodecay}. The black line is for later reference. We can recognise several regions in parameter space based on the mass spectrum.
\begin{itemize}
\item In the dark blue region, only three-body decays are kinematically allowed. These are decays such as $\tilde\chi_1^+\to\tilde\chi_1^0 u\bar d$ that are suppressed in the rest of the parameter space.
\item In the large light blue region, the decays $\tilde\chi_1^\pm\to\tilde\chi_1^0W^\pm$ are allowed, but charginos cannot decay to any other supersymmetric particles than the LSP.
\item In the cyan region, $m_{\tilde\tau_1}<m_{\tilde\chi_1^\pm}$. This region extends to lower values of $\Mhalf$, so for instance at $M_0=90$~GeV and $\Mhalf=200$~GeV the decay to the light stau is kinematically allowed, even though the decay to the LSP and a $W^\pm$ is not.
\item In the green region, the sneutrinos are lighter than the light charginos as well, making three additional decays kinematically possible. The small mass differences between the sneutrino species and between the Standard Model leptons can usually be neglected due to the phase space effects explained in \ref{s:phasespace}.
\item In the red region, all decays from table~\ref{t:gauginodecay} are kinematically allowed. In the small orange region, the decay to the heavy stau is not kinematically accessible, but this region is too small to yield a new phenomenology (see section \ref{s:phasespace}).
\end{itemize}
As can be seen in table~\ref{t:gauginodecay}, the neutralino $\tilde\chi_2^0$, has even more possible decay modes than the light charginos. Figure~\ref{f:neutralino-region10} shows the number of decay modes that are kinematically allowed for different regions. Many decays involve particles whose masses are (nearly) degenerate, so we can distinguish six main regions.
\begin{itemize}
\item In the dark blue region, only three-body decays such as $\chi_2^0\to\chi_1^0 u\bar u$ are kinematically allowed.
\item In the light blue region just above it, the decay to the LSP and the $Z^0$ is allowed. Three body decays are suppressed, so almost all $\tilde\chi_2^0$ will decay to a $Z$ boson and the LSP.
\item In the large cyan region, the decay $\tilde\chi_2^0\to\tilde\chi_1^0 h^0$ is allowed as well. Since the coupling to the Higgs is stronger than the coupling to the $Z$ boson, this is the dominant decay mode.
\item In the dark-green area, the next to lightest neutralino is heavier than the light stau, so the decay to the light stau is allowed as well. This region extends to the cyan and light blue region with the same label, although the decays to the LSP and the lightest Higgs or $Z$ boson are not kinematically allowed there.
\item In the yellow and light-green region, the sneutrinos are lighter than $\tilde\chi_2^0$, making those decays kinematically possible. Once again, the mass differences are negligible due to the phase space effects explained in section \ref{s:phasespace}.
\item In the red region, all decays listed in table~\ref{t:gauginodecay} are kinematically allowed. The orange region, where the decay to the heavy stau is not allowed, is too small to yield a new phenomenology.
\end{itemize}
To test if these regions indeed yield different phenomenology, 20.000 SUSY events were generated with Pythia along the line in figures~\ref{f:chargino-region10} and~\ref{f:neutralino-region10}. The results are shown in figure~\ref{f:gauginodecay}.

\begin{figure}[!h]
\centering{
\includegraphics[width=0.5\linewidth]{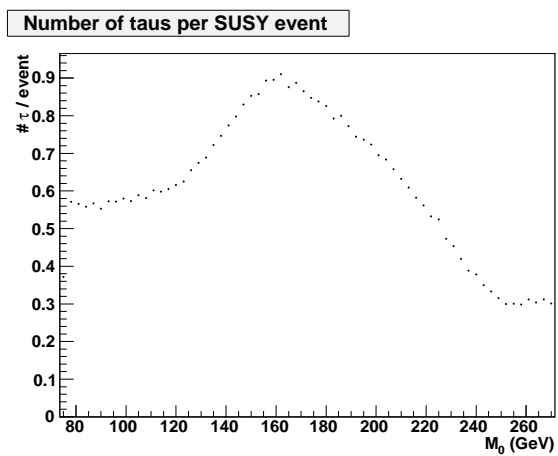}
\caption[The effects of gaugino decay channels on tau production]{The number of times a tau is produced per SUSY event along the path through parameter space indicated in figures~\ref{f:chargino-region10} and~\ref{f:neutralino-region10}. A graph including cross section effects can be found in \ref{f:decayjump-totaltau-cross}}\label{f:decayjump-totaltau}}
\end{figure}

Figure~\ref{f:gauginodecay} shows the expected characteristics of the different phenomenological regions. Firstly, the number of light gauginos rises slightly as $M_0$ increases, since that makes gauginos lighter compared to the rest of the spectrum. Secondly, we can clearly see when a boundary of a phenomenological region is crossed. Reading the graphs from high to low $M_0$, we can see exactly when decays become kinematically accessible. For $M_0$~\raisebox{0.2ex}{$\scriptstyle\lesssim$}~$250$~GeV, the decay to the light stau is kinematically allowed and the number of staus increases. Below $M_0\approx160$~GeV, the decays to sneutrinos become kinematically allowed as well. This can be seen in a rise in the number of sneutrinos, but also in a drop of the number of staus. At $M_0\approx140$~GeV, the remaining decays become kinematically accessible and we see the number of heavy sleptons increase. As more decay channels are allowed, a smaller fraction of the charginos and neutralinos decays to staus and the number of staus decreases.
\newpage
The number of staus has a direct effect on the number of taus, as can be seen in figure~\ref{f:decayjump-totaltau}, which shows the number of taus produced per SUSY event.

\begin{figure}[!h]
\centering{
\includegraphics[width=\linewidth]{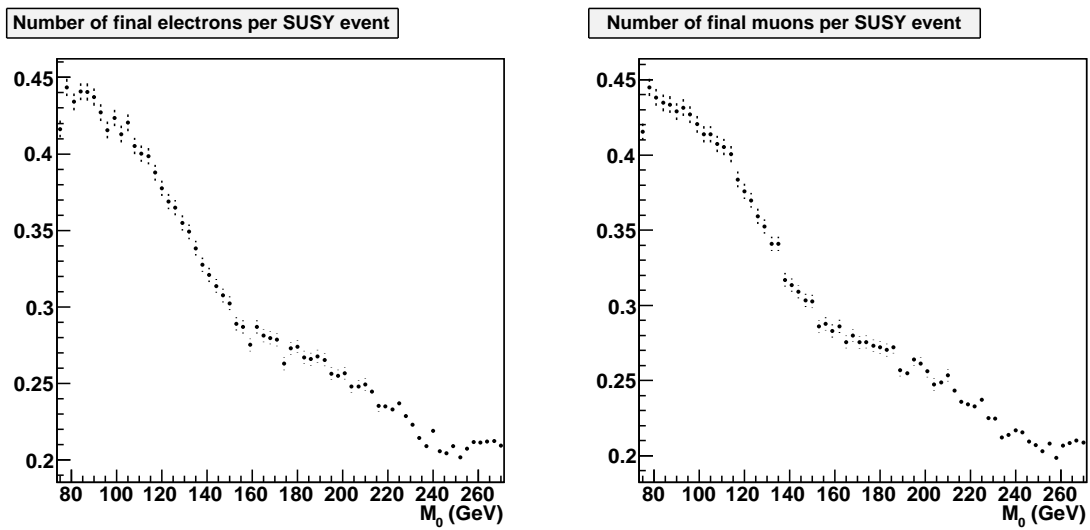}
\caption[The effects of gaugino decay channels on the number of final leptons]{The number of final state muons and electrons at points along the path through parameter space indicated in figures~\ref{f:chargino-region10} and~\ref{f:neutralino-region10}. Both $M_0$ and $\Mhalf$ are varied, $A_0=0$~GeV, $\tan\beta=10,\mu>0$. Graphs that include cross section effects can be found in figure \ref{f:gauginodecay-finalpart-cross}.}\label{f:gauginodecay-finalpart}}
\end{figure}

The effects can also be seen in the number of final muons and electrons, as can be seen in figure~\ref{f:gauginodecay-finalpart}. The number of final muons and electrons is determined using the procedure described in section \ref{s:pythia}.
There are two kinks in both graphs. One occurs when the decay to the stau becomes kinematically accessible. Since the stau can be produced by light gauginos, it can only decay to the LSP and a tau at this point in parameter space. Approximately 35\% of the taus decay to electrons or muons, which can be detected relatively easily. The other kink is at the point where the chargino decay to the sneutrinos is kinematically accessible. A chargino decaying to an electron (muon) sneutrino also produces an electron (muon), since lepton quantum numbers are conserved. This decay replaces part of the decays to staus, so more electrons and muons are produced. Of course neutralinos decaying to heavy sleptons also produce the associated leptons, which adds to the total number. The change in the number of final muons and electrons is not as pronounced as the change in the number of taus, so in order to determine the high-scale parameters, tau identification is necessary.

\begin{figure}[!h]
\centering{
\includegraphics[width=\linewidth]{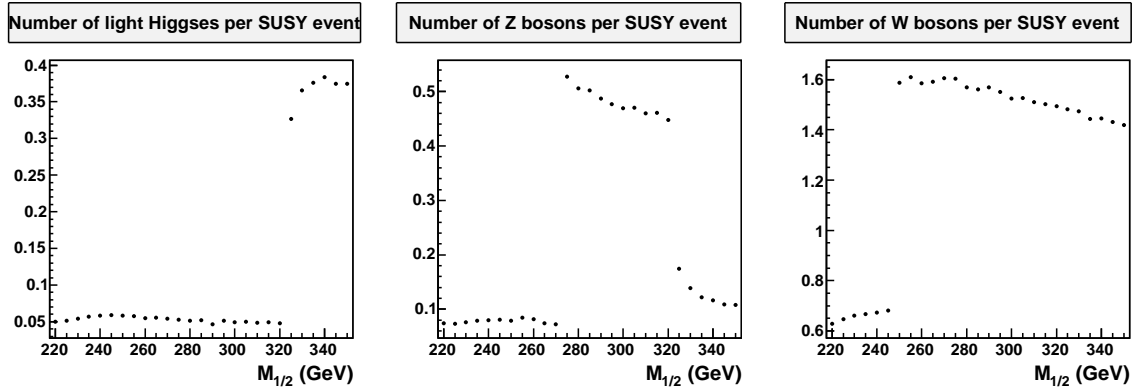}
\caption[The effects of gaugino decay channels on the number of Higgs, $Z$ and $W$ bosons]{The number of $h^0$, $Z^0$ and $W^\pm$ bosons per event depending on $\Mhalf$ with $M_0=300$~GeV, $A_0=0$~GeV, $\tan\beta=10$ and $\mu>0$}\label{f:gauginodecay-higgs}}
\end{figure}

Finally, from figures~\ref{f:chargino-region10} and~\ref{f:neutralino-region10} we see that there is a part of parameter space where the light gauginos cannot decay to any sleptons. There are distinct phenomenological regions here too. Figure~\ref{f:gauginodecay-higgs} shows the number of $h^0$, $Z$ and $W$ bosons produced per SUSY event.
There is a jump in the number of $Z$ bosons at the point where the decay $\tilde\chi_2^0\to\tilde\chi_1^0Z^0$ becomes kinematically allowed. Once the decay $\tilde\chi_2^0\to\tilde\chi_1^0h^0$ is allowed as well, it almost completely replaces the first decay due to the stronger coupling to the Higgs boson. This results in an increased number of Higgs bosons and a corresponding drop in the number of $Z$ bosons. The number of $W$ bosons produced only has one jump, which corresponds to the point where the chargino decay to the $W$ boson becomes kinematically allowed. If supersymmetry is in this part of the parameter space, the high-scale parameters can be determined by studying the number of $Z$, $W$ and Higgs bosons.

\subsubsection{The Light Stop}\label{s:phenostops}

The possible production and decay channels of stops are limited because in general the mixing between different generations is negligible. The production processes of stops are listen in table~\ref{t:stopproduction} and the decay channels in table~\ref{t:stopdecay}. The large mass of the top quark is an important constraint on which decays are kinematically accessible.

\begin{figure}[!h]
\centering{
\includegraphics[width=\linewidth]{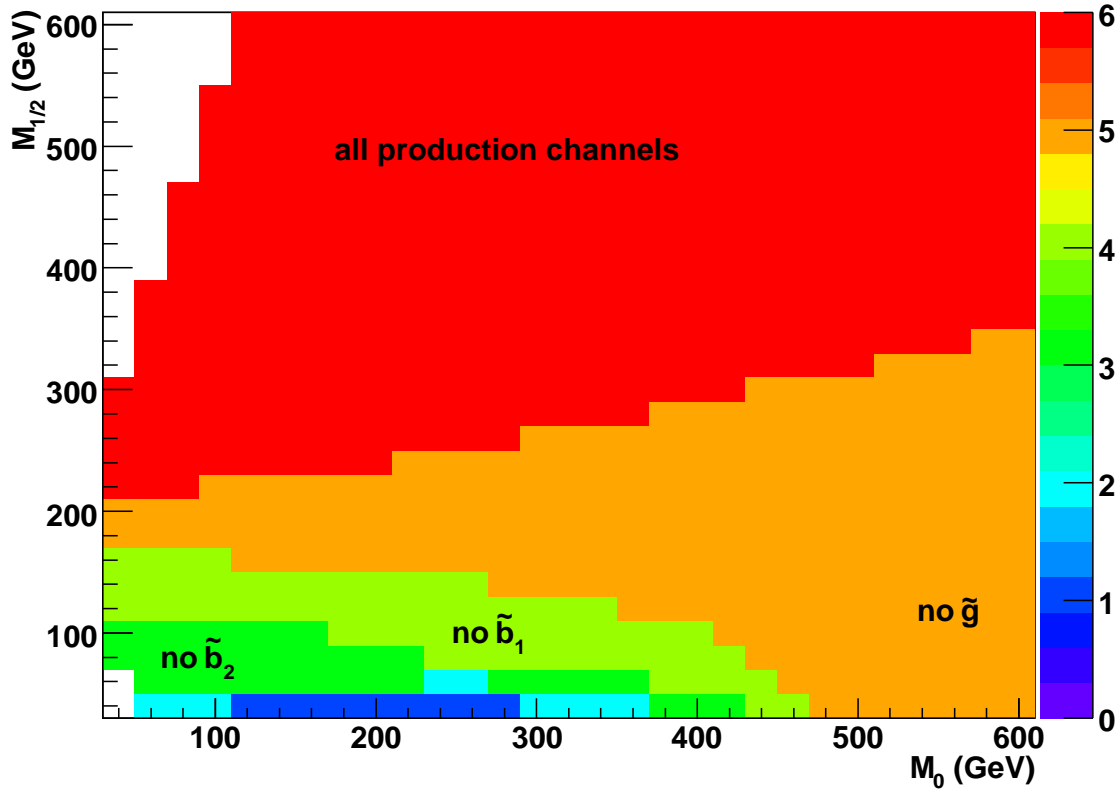}
\caption[Phenomenological regions in parameter space based on stop production]{Phenomenological regions based on stop production for $A_0=0$~GeV, $\tan\beta=10,\mu>0$. The labels indicate which production modes from table~\ref{t:stopproduction} are not allowed in a specific part of parameter space. The colour code indicates how many of the decay modes are kinematically allowed. The stop can always be produced in the initial interaction.}\label{f:stopproduction-region10}}
\end{figure}

The number of stops and their decay modes affect the number of $t$ and $b$ quarks measured at the LHC. As mentioned in section \ref{s:firstint-lightstops}, light stops can also be produced abundantly in the initial interaction, but we will discuss that in section \ref{s:phenA0}. Figure~\ref{f:stopproduction-region10} shows the phenomenological regions based on stop production for $A_0=0$~GeV, $\tan\beta=10,\mu>0$, where stop production in the first interaction does not play a large role. Plots for the individual decays can be found in Appendix~\ref{app:stopprocesses}.
\newpage

We can distinguish several regions in parameter space.
\begin{itemize}
\item In the red region, all production processes listed in table~\ref{t:stopproduction} are kinematically allowed.
\item In the orange region, the gluino is not heavy enough to decay to a stop and a top. Stops can still be produced by sbottom decay.
\item In the light-green region, the decay of the lighter sbottom to a stop and a $W$ is not kinematically allowed either.
\item In the dark-green region, the heavy sbottom is too light to decay to stops, so only the production processes involving heavy stops are kinematically accessible.
\item In the cyan region, the heavy stop cannot decay to a light stop and a Higgs.
\item In the dark blue region, none of the processes from table \ref{t:stopproduction} are allowed.
\end{itemize}
The phenomenological regions based on the subsequent decay of the light stop are shown in figure~\ref{f:stopdecay-region10}. Plots for individual decays can be found in Appendix~\ref{app:stopprocesses}. In this part of parameter space, the decay to the lightest chargino and a $b$ quark is always allowed. This is not necessarily the case for other values of $A_0$.
\newpage
\begin{figure}[!h]
\centering{
\includegraphics[width=\linewidth]{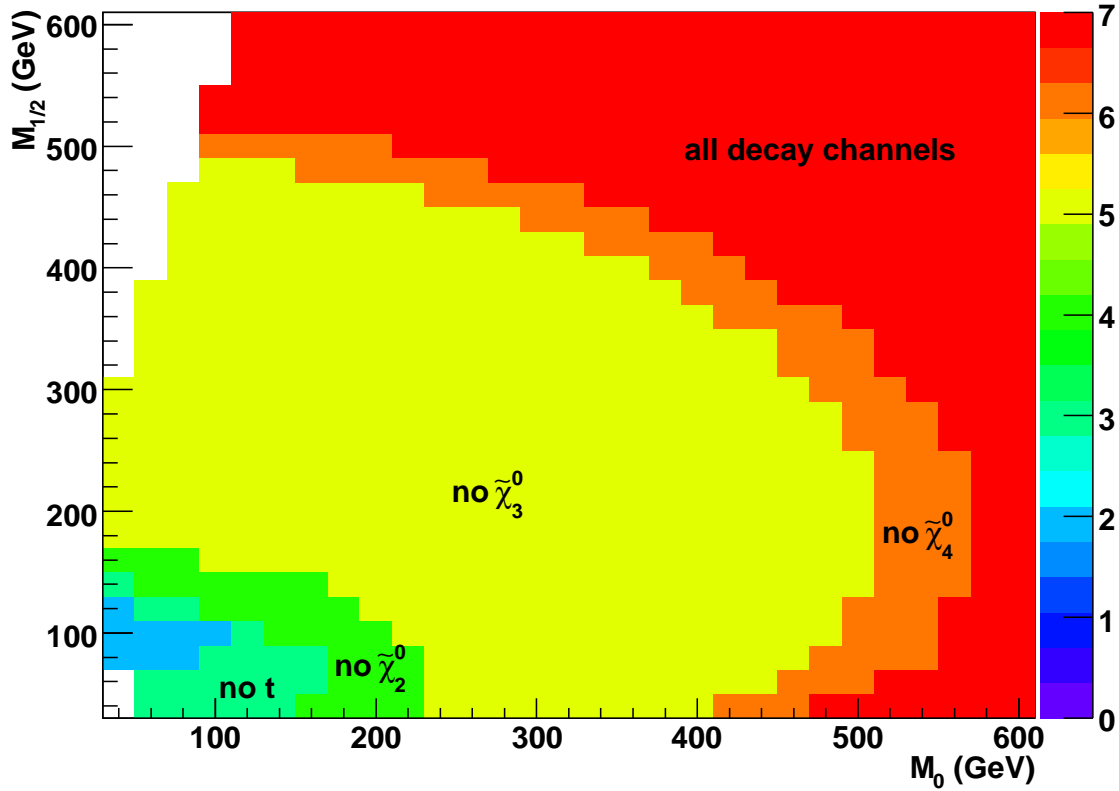}
\caption[Phenomenological regions based on stop decay]{Phenomenological regions in parameter space based on stop decay for $A_0=0$~GeV, $\tan\beta=10,\mu>0$. The labels indicate which decay modes from table~\ref{t:stopdecay} are not allowed in that part of parameter space. The colour code indicates how many of the decay modes are kinematically allowed. The suppressed decay $\tilde t_1\to\tilde\chi_1^0c$ is always kinematically accessible.}\label{f:stopdecay-region10}}
\end{figure}

\begin{figure}[!h]
\centering{
\includegraphics[width=\linewidth]{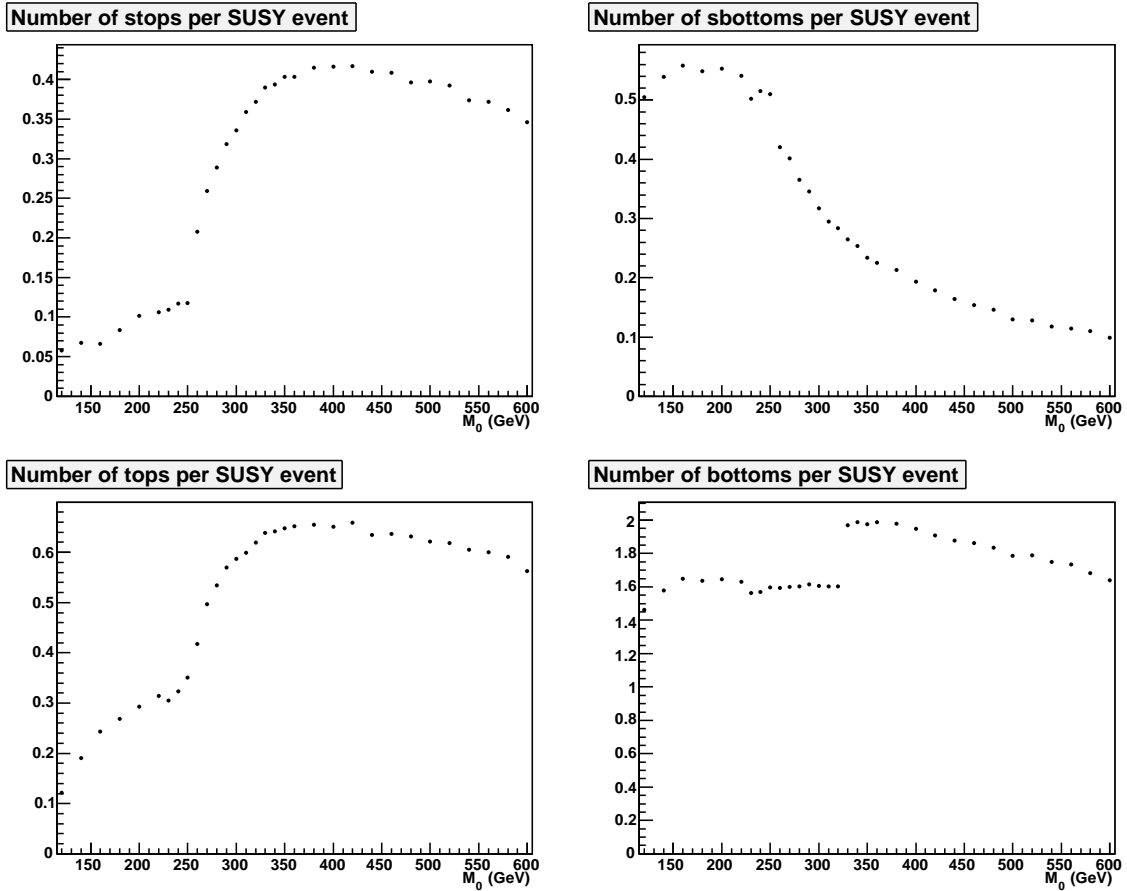}
\caption[The effect of phenomenological regions on stops, sbottoms, tops and bottoms]{The number of stops, sbottoms, tops and bottoms per SUSY event along the line $M_0=\Mhalf$ for $A_0=0$~GeV, $\tan\beta=10,\mu>0$.}
\label{f:stopjump-pythia}}
\end{figure}

We can identify several regions that have different phenomenology based on the mass spectrum.
\begin{itemize}
\item In the blue region, only the decay to $\tilde\chi_1^\pm$ and a $b$ quark is kinematically allowed.
\item In the dark-green region, the decay to the heavy chargino and a $b$ quark is allowed as well, but the light stop cannot decay to a $t$ quark.
\item In the light-green region, the decay to the LSP and a $t$ quark is kinematically allowed, but decays to other neutralinos are not.
\item In the yellow region the stop can also decay to $\tilde\chi_2^0$.
\item In the orange region the decay to $\tilde\chi_3^0$ is kinematically accessible as well.
\item In the red region, all decays listed in table~\ref{t:stopdecay} are kinematically allowed.
\end{itemize}
\newpage
Figure~\ref{f:stopjump-pythia} shows the number of stops, sbottoms, tops and bottoms produced on the line $M_0=\Mhalf,A_0=0$~GeV, $\tan\beta=10,\mu>0$.
Near $M_0\approx160$~GeV, a small kink in the number of stops is visible. In figure~\ref{f:stopproduction-region10} we can see that this is the point in parameter space where the decay of the light sbottom to a stop and a $W$ boson becomes kinematically accessible. A more significant kink in both the stop and the sbottom production is at $M_0\approx250$~GeV. From figure~\ref{f:stopproduction-region10} we can see that this is the point where the decay $\tilde g\to\tilde t_1\bar t$ is kinematically allowed. This results in more stops and tops and less sbottoms, since the decay to the stop replaces part of the decay to the sbottom. The number of $b$ quarks remains constant since most tops decay to $b$ quarks, so they end up in both plots.

However, there is a sudden jump in the number of $b$ quarks at $M_0=330$~GeV. This is caused by the effect from figure~\ref{f:gauginodecay-higgs}. Looking at figure~\ref{f:neutralino-region10}, we see that the decay $\tilde\chi_2^0\to\tilde\chi_1^0h^0$ is kinematically accessible for $\Mhalf>330$~GeV. Since the channel $h^0\to b\bar b$ is one of the most important decay channels of the light Higgs, this results in a higher number of $b$ quarks.
\newpage
\subsubsection{Small Stop Mass}\label{s:phenA0}

For large negative values of $A_0$, which are interesting because of the high mass of the lightest Higgs, the stop mass can become quite small, especially for extreme values of $\tan\beta$. To show what happens if we vary $A_0$ and $\tan\beta$, we first discuss the phenomenological regions based on production and decay channels of stops in the $A_0,\tan\beta$ plane. After this we include the stop cross section in the analysis.

The phenomenological regions based on stop decay in the $A_0,\tan\beta$ plane for $M_0=\Mhalf=200$~GeV are shown in figure \ref{f:stopdecayA0beta}. Plots for the individual processes can be found in Appendix \ref{app:stopprocesses}.

\begin{figure}[!h]
\centering{
\includegraphics[width=\linewidth]{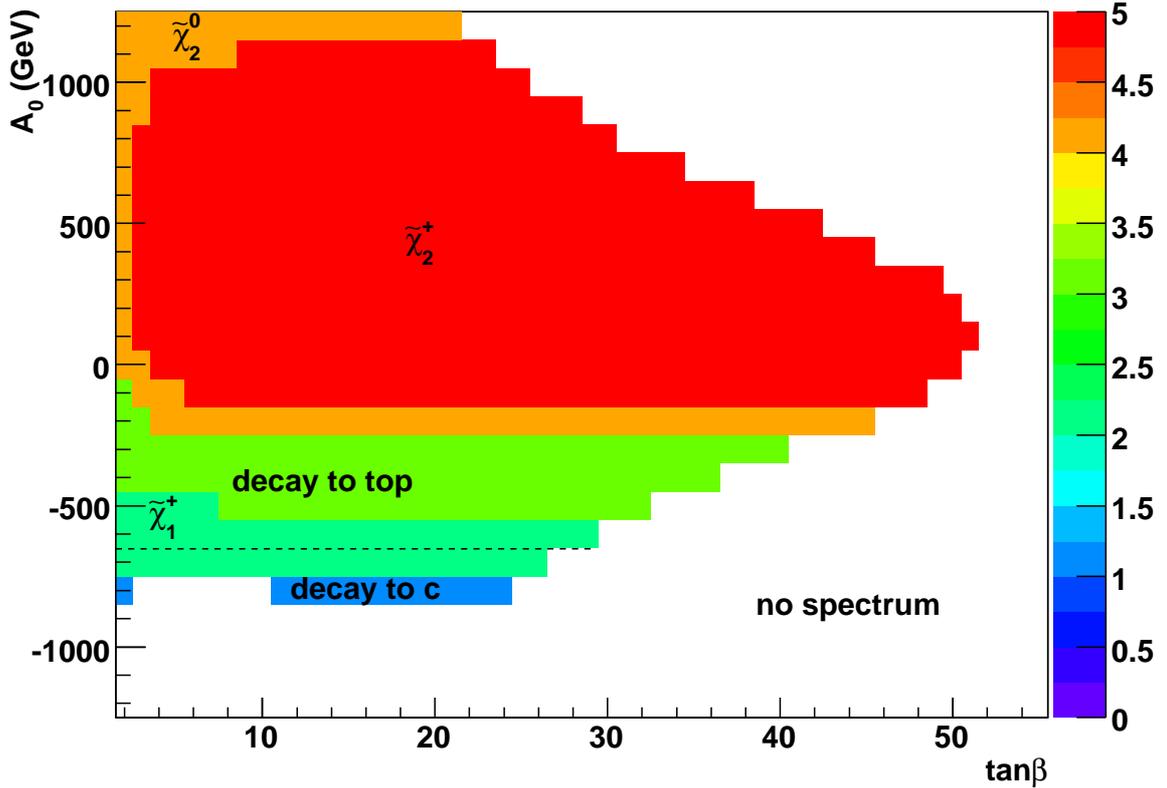}
\caption[Phenomenological regions based on stop decay in the $A_0,\tan\beta$ plane]{Phenomenological regions in parameter space based on stop decay for $M_0=\Mhalf=200$~GeV and $\mu>0$. The labels indicate which decay modes from table~\ref{t:stopdecay} are allowed.}\label{f:stopdecayA0beta}}
\end{figure}
\begin{figure}[!h]
\centering{
\includegraphics[width=\linewidth]{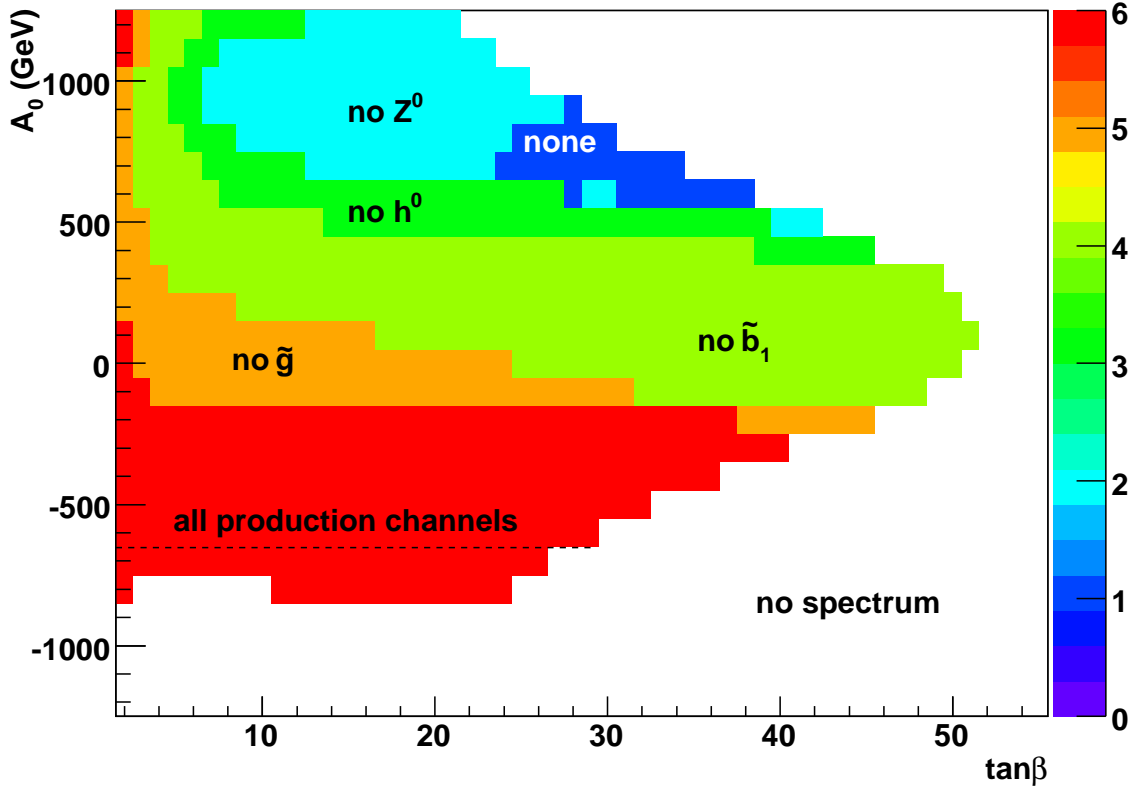}
\caption[Phenomenological regions based on stop production in the $A_0,\tan\beta$ plane]{Phenomenological regions in parameter space based on stop production for $M_0=\Mhalf=200$~GeV and $\mu>0$. The labels indicate which production modes from table~\ref{t:stopproduction} are not allowed in that part of parameter space.}\label{f:stopproductionA0beta}}
\end{figure}

For these values of $M_0$ and $\Mhalf$, the decays to $\tilde\chi_{3,4}^0$ are never allowed. Most of the other decays are familiar from figure \ref{f:stopdecay-region10}. An exception is the blue region. Here the stop is so light that even the decay $\tilde t_1\to\tilde\chi_1^+ b$ is not allowed. Because all other decay channels are suppressed in this region, stops decay through the flavour-changing process $\tilde t_1\to\tilde\chi_1^0c$.
\newpage
Figure \ref{f:stopproductionA0beta} shows a similar plot for the production of stops. Plots from the individual processes can be found in Appendix \ref{app:stopprocesses}.

We can distinguish several phenomenological regions based on the mass spectrum.
\begin{itemize}
\item In the red region, the lightest stop is one of the lightest supersymmetric particles and all production processes from table \ref{t:stopproduction} are possible.
\item In the orange region, the decay $\tilde g\to\tilde t_1\bar t$ is not allowed due to the large top mass.
\item In the light-green region, the light sbottom cannot decay to a stop and a $W$. Figure \ref{f:massA0} shows why: the mass difference between the light stop and the light sbottom decreases for increasing $A_0$ until the stop mass has reached its maximum. This region extends to the small orange strip for large $\tan\beta$, where the gluino decay is kinematically allowed.
\item In the dark-green region, the mass difference between the light and the heavy stop is too small for the decay $\tilde t_2\to\tilde t_1h^0$.
\item In the cyan region, the decay $\tilde t_2\to\tilde t_1Z^0$ is not kinematically allowed either. The only production process from table \ref{t:stopproduction} that is allowed here is the one involving the heavy sbottom.
\item In the dark blue region, this last process is not kinematically allowed either and stops can only be produced in the initial interaction.
\end{itemize}
Now it is time to discuss how the stop cross section fits into this. From figure \ref{f:crosssection} we know that the stop cross section rises considerably when $m_{\tilde t_1}$~\raisebox{0.2ex}{$\scriptstyle\lesssim$}~$200$~GeV. This is the case below the dotted lines in figures \ref{f:stopdecayA0beta} and \ref{f:stopproductionA0beta}. Many stops are produced in this region. If it is kinematically allowed, they decay to $b$ quarks. In the region where the stop cross section is the dominant contribution to the supersymmetric cross section, even the decay to $b$ quarks is not kinematically accessible, so stops decay to a $c$ quark and the LSP.

A low stop mass also influences the gaugino sector. If the decay $\tilde t_1\to\tilde\chi_2^0t$ is not kinematically allowed, the number of $\tilde\chi_2^0$ decreases. At the same time the number of $\tilde\chi_1^\pm$ increases, since the decay $\tilde t_1\to\tilde\chi_1^\pm b$ replaces part of the decays to $\tilde\chi_2^0$. The number of charginos rises even more once no decays to tops are allowed. If only decays to $c$ quarks are allowed, the number of $\tilde\chi_2^0$ and $\tilde\chi_1^\pm$ drops dramatically. In that case supersymmetric phenomenology is essentially limited to stops and the lightest neutralino.

\begin{figure}[!h]
\centering{
\includegraphics[width=\linewidth]{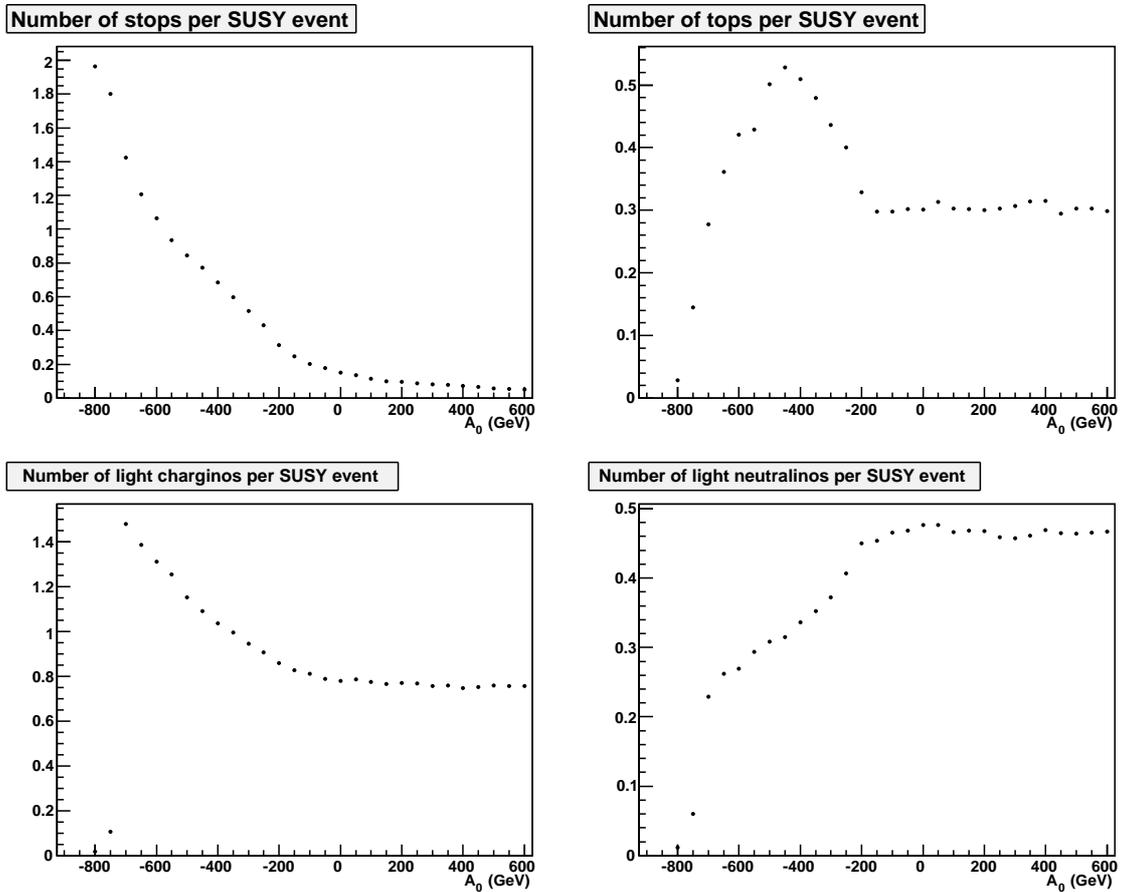}
\caption[The number of stops, tops, $\tilde\chi_1^\pm$ and $\tilde\chi_2^0$ produced per SUSY event depending on $A_0$]{The number of stops, tops, $\tilde\chi_1^\pm$ and $\tilde\chi_2^0$ produced per SUSY event depending on $A_0$ with $M_0=\Mhalf=200$~GeV, $\tan\beta=15,\mu>0$. For large negative values of $A_0$, the stop mass is small, so many stops are produced.}\label{f:pythiadecayjumps-stopA0}}
\end{figure}

This is confirmed by an event simulation over the line $M_0=\Mhalf=200$~GeV, $\tan\beta=15,\mu>0$ shown in figure \ref{f:pythiadecayjumps-stopA0}.

For positive values of $A_0$, most stops are produced by decays of light sbottoms. The number of stops increases considerably at $A_0\approx-200$~GeV when they can be produced by gluinos. That decay produces top quarks as well so the number of top quarks rises. Once the stop can no longer decay to a top, the number of tops decreases. At $A_0\approx-700$~GeV, stop production becomes an important process in the initial interaction and the number of stops per SUSY event rises even more. Naturally, the number of $t$ quarks drops dramatically at that point because they are no longer produced by the decay $\tilde g\to\tilde t_1t$.

The gauginos also behave as expected. As the stop gets lighter, the number of charginos per SUSY event first increases, while the number of neutralinos decreases. When the stop can only decay to a $c$ quark very few gauginos are produced. Of course a drop in gauginos affects stau phenomenology. But since charginos and neutralinos have the opposite behaviour, the effect is small unless the stop mass is so small that only the decay to the $c$ quark is allowed.
\newline

\begin{figure}[!h]
\centering{
\includegraphics[width=\linewidth]{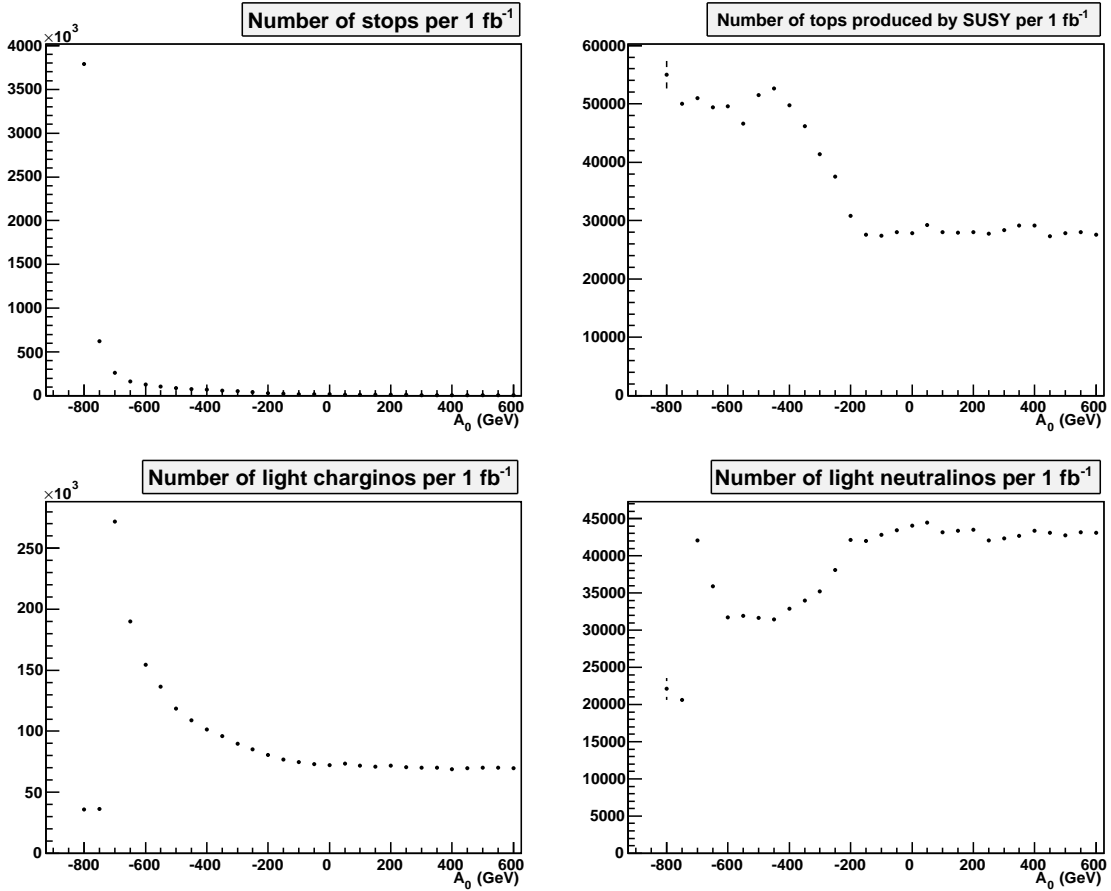}
\caption[The number of stops, tops, $\tilde\chi_1^\pm$ and $\tilde\chi_2^0$ produced per fb$^{-1}$ of luminosity depending on $A_0$.]{The number of stops, tops, $\tilde\chi_1^\pm$ and $\tilde\chi_2^0$ produced per fb$^{-1}$ of luminosity depending on $A_0$. For all points $M_0=\Mhalf=200$~GeV, $\tan\beta=15,\mu>0$. For large negative values of $A_0$, the stop mass is so small that the total cross section rises considerably. This changes the shape of the graphs compared to figure \ref{f:pythiadecayjumps-stopA0}}\label{f:pythiadecayjumpscrosssection-stopA0}}
\end{figure}

These graphs and the conclusions derived from them hold for the number of stops produced per SUSY event. However, we know from section \ref{s:crosssection} that a small stop mass strongly influences the stop cross section. To obtain a prediction for the number of stops produced at the LHC for different points in parameter space, we have to take the cross section into account as well. Figure \ref{f:pythiadecayjumpscrosssection-stopA0} shows the total number of stops, tops, charginos and neutralinos produced per 1 fb$^{-1}$ of luminosity. The shape of these graphs are considerably different from the graphs in figure \ref{f:pythiadecayjumps-stopA0} because of the increase in the total supersymmetric cross section that is caused by the small stop mass.

Of all other graphs presented in this section, a version where the cross section is taken into account can be found in Appendix \ref{app:phenocross}.
\newline

The parameters $A_0$ and $\tan\beta$ influence the couplings as well as the masses. However, the effect of the couplings on supersymmetric phenomenology is small. The reason is that $A_0$ and $\tan\beta$ only influence interactions that involve the Higgses. The heavy Higgses are usually too heavy to be produced in large amounts at the LHC, while the light Higgs is too light to decay to supersymmetric particles. Thus, $A_0$ and $\tan\beta$ only affect the supersymmetric phenomenology through their effect on the mass spectrum. The change in couplings will most likely affect Higgs phenomenology, but that is beyond the scope of this research. In Appendix \ref{app:otherbeta} the phenomenological regions based on the decays of the light gauginos for $\tan\beta=30$ are discussed. The regions have a different form, but the tools developed in this section still apply, confirming that supersymmetric phenomenology is dominantly determined by the mass spectrum.

\subsection{Benchmark Points}\label{s:benchmarkresult}

We have now developed a way to study mSUGRA parameter space based on phenomenology. An important question is in which phenomenological regions the benchmark points are located and if they are representative for the mSUGRA parameter space. In table~\ref{t:susypoints} we see that almost all the benchmark points have a moderate value of $A_0$. The only exception is the SU4 benchmark point with $A_0=-400$~GeV. We discuss each of the benchmark points and their phenomenology. We also investigate how the phenomenology changes around benchmark points. For brevity, not all these aspects will be discussed in detail for all benchmark points.

\subsubsection{Benchmark Points and Dark Matter}

The strongest experimental bound on high-mass supersymmetry is the dark matter relic density. If $R$-parity is conserved, the LSP contributes to the total amount of dark matter in the universe. This can result in a dark matter relic density that is many times higher than the measured value of $\Omega_mh^2=0.1277^{+0.0080}_{-0.0079}$ \cite{Spergel:2006hy}.

The supersymmetric contribution to the total amount of dark matter in the universe is equal to the number of LSPs times the mass of the LSP. Thus there are two ways to obtain a low dark matter relic density. The first is to have a light LSP, which is another argument for relatively low-mass supersymmetry. However, most points in mSUGRA parameter space that yield an LSP that is light enough are ruled out by other experimental bounds, such as the LEP bound on the Higgs mass. The second option is to decrease the number of LSPs. In the high-density early universe, supersymmetric particles could annihilate into Standard Model particles. Since two supersymmetric particles are involved in the initial state, this decreases the number of supersymmetric particles without violating $R$-parity. Thus the number of LSPs depends on the thermodynamics in the early universe and on the annihilation cross section that is determined by the supersymmetric theory. In general only annihilation of two LSPs plays a role. However, there are several regions where other processes are important that enhance supersymmetric annihilation processes. The benchmark points are chosen in these regions. An insightful description of the calculation of the dark matter relic density can be found in \cite{Griest:1990kh}. 

ISASUGRA linked with ISATOOLS was used to calculate the dark matter relic densities mentioned in this section. The SU3 benchmark point, the SU5 benchmark points and two of the SU8 benchmark points have dark matter relic densities that exceed the present experimental value. The SU5 benchmark points are not studied anymore due to their high dark matter relic density, so we will not discuss them here either.

\newpage
\subsubsection{Coannihilation Benchmark Points SU1 and SU8}\label{s:coannihilation}

Coannihilation regions exist because of a loophole in the calculation of the total dark matter relic density. If the mass of the light stau is close to the mass of the LSP, the decay of the stau to the LSP and a tau is suppressed. As a result, the LSP and the light stau coexist in the early universe. The cross section for stau-neutralino interactions is much higher than the cross section for neutralino-neutralino interactions, so this leads to higher annihilation rates of supersymmetric particles and thus a lower dark matter relic density.

If the mass difference between the stau and the LSP is too small, the dark matter relic density is many times smaller than the measured value. These points in parameter space are not excluded, since it is possible that there are other sources of dark matter, but they are not the most interesting points. The mass difference that yields a correct dark matter relic density depends on the mass of the LSP.

For benchmark point SU1, SPheno yields a mass difference of $\delta m=8.0$~GeV, while for benchmark point SU8.1 it gives $\delta m=8.8$~GeV. The other SU8 benchmark points have dark matter relic densities that exceed the recent experimental value and since all SU8 benchmark points are in the same phenomenological region, we only discuss the SU8.1 benchmark point.

Coannihilation lines were determined by a recursive algorithm that finds a value for $\Mhalf$ for a given value of $M_0$ such that $\delta m-0.1$~GeV~$<m_{\tilde \tau_1}-m_{\tilde\chi_1^0}<\delta m+0.1$~GeV. If no integer value of $\Mhalf$ can be found, the algorithm allows a smaller or larger mass difference. However, the mass difference is not allowed to exceed $2\delta m$. This procedure does not yield a constant dark matter relic density, but it is a reasonable approximation of a coannihilation line for a limited mass range. The points used for the SU1 and the SU8 coannihilation lines and their total supersymmetric cross sections are listed in table~\ref{t:coannihilationpoints}. All points in table \ref{t:coannihilationpoints} have an acceptable dark matter relic density. The lowest-mass points for both coannihilation lines have a dark matter relic density that is approximately a factor 10 too low, but as mentioned before this does not necessarily exclude them.
\begin{table}[!h]
\centering{
\begin{tabular}{|c|c|c|c|c|c|}
\hline
\multicolumn{3}{|c|}{SU1 ($\tan\beta=10$)}&\multicolumn{3}{c|}{SU8.1 ($\tan\beta=40$)}\\
\hline
$M_0$&$\Mhalf$&$\sigma$ (pb)&$M_0$&$\Mhalf$&$\sigma$ (pb)\\
\hline
40 & 221 & $74.78$ &180 & 274 & $23.16$\\
45 & 241 & $48.86$ &185 & 288 & $18.31$ \\
50 & 262 & $32.81$ &190 & 302 & $14.52$ \\
55 & 285 & $21.61$ &195 & 316 & $11.55$ \\
60 & 306 & $15.24$ &200 & 331 & $9.187$ \\
65 & 328 & $10.74$ &205 & 345 & $7.445$ \\
70 & 350 & $7.774$ &210 & 360 & $6.007$\\
75 & 372 & $5.657$ &215 & 375 & $4.893$\\
80 & 394 & $4.203$ &220 & 389 & $4.054$ \\
85 & 415 & $3.232$ &225 & 404 & $3.351$ \\
90 & 437 & $2.443$ &230 & 419 & $2.762$ \\
95 & 460 & $1.857$ &235 & 433 & $2.299$ \\
100 & 480 & $1.456$ &240 & 448 & $1.914$ \\
\hline
\end{tabular}
\caption[Points in coannihilation regions and their cross sections]{Points used for the coannihilation lines and their cross sections. All masses are in GeV.}\label{t:coannihilationpoints}}
\end{table}

\begin{figure}[!h]
\includegraphics[width=0.5\linewidth]{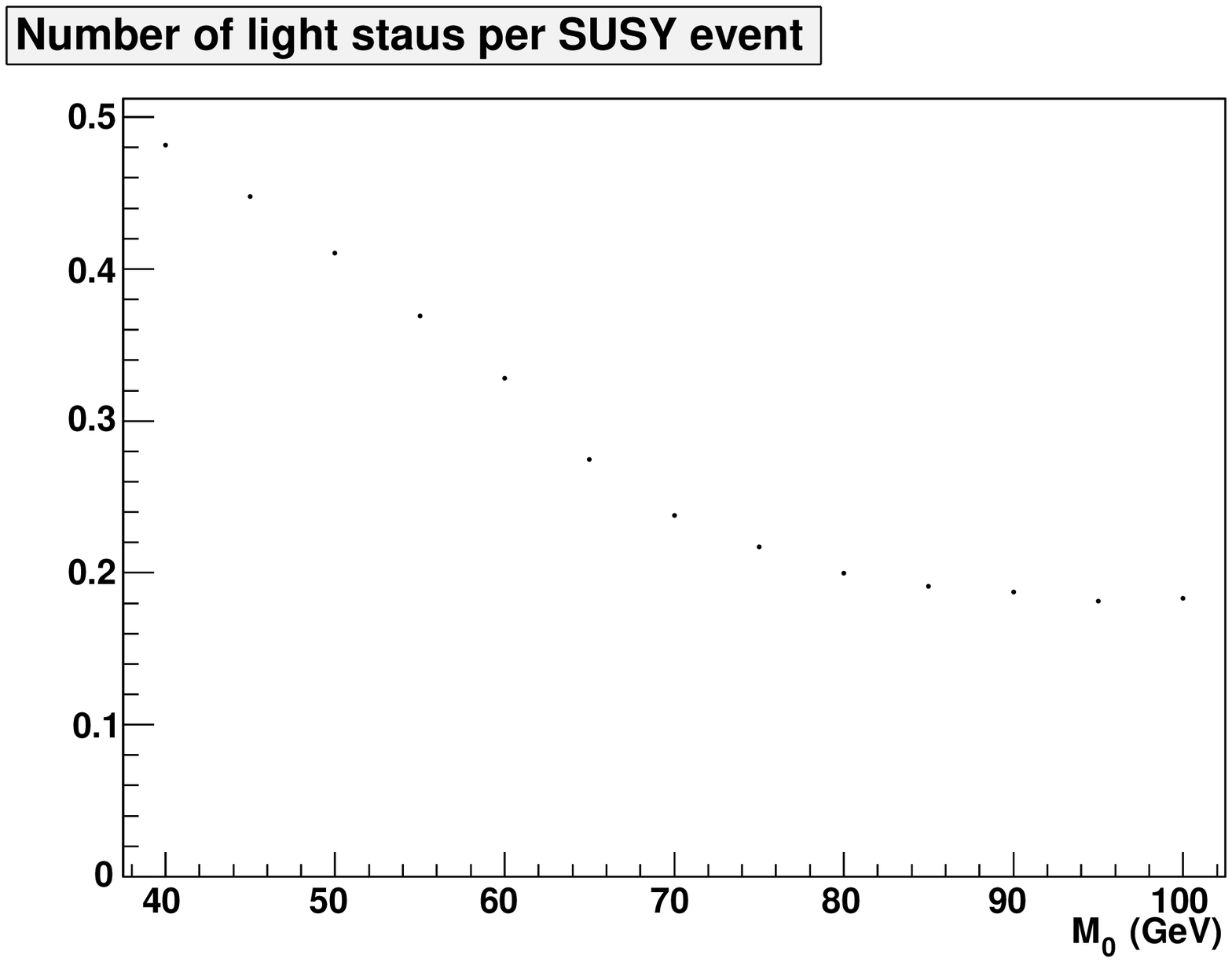}
\includegraphics[width=0.5\linewidth]{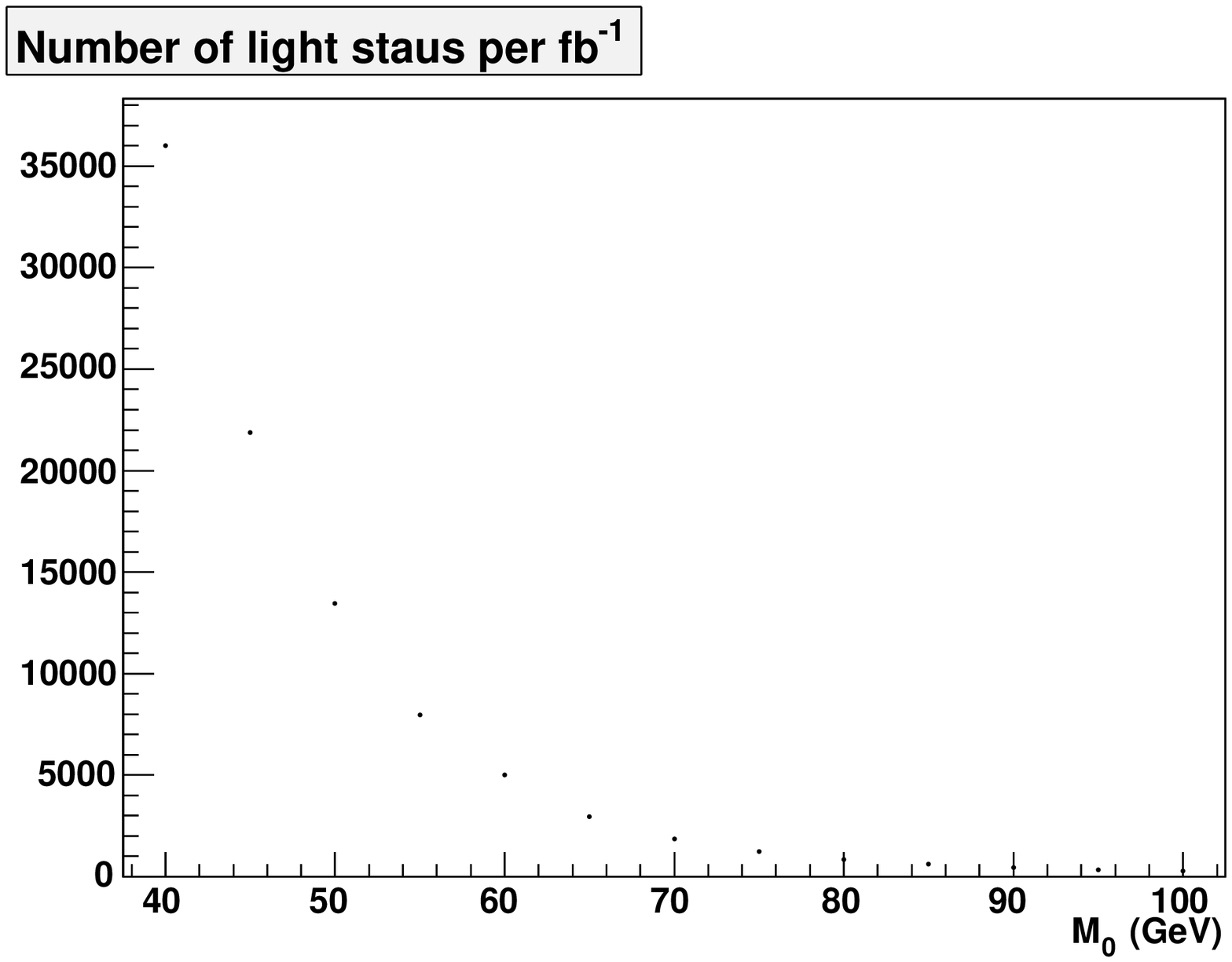}
\caption[Stau production along an SU1 coannihilation line]{Staus produced per SUSY event and per fb$^{-1}$ of luminosity along the SU1 coannihilation line from table \ref{t:coannihilationpoints}.}\label{f:SU1staus}
\end{figure}

\begin{figure}[!h]
\includegraphics[width=0.5\linewidth]{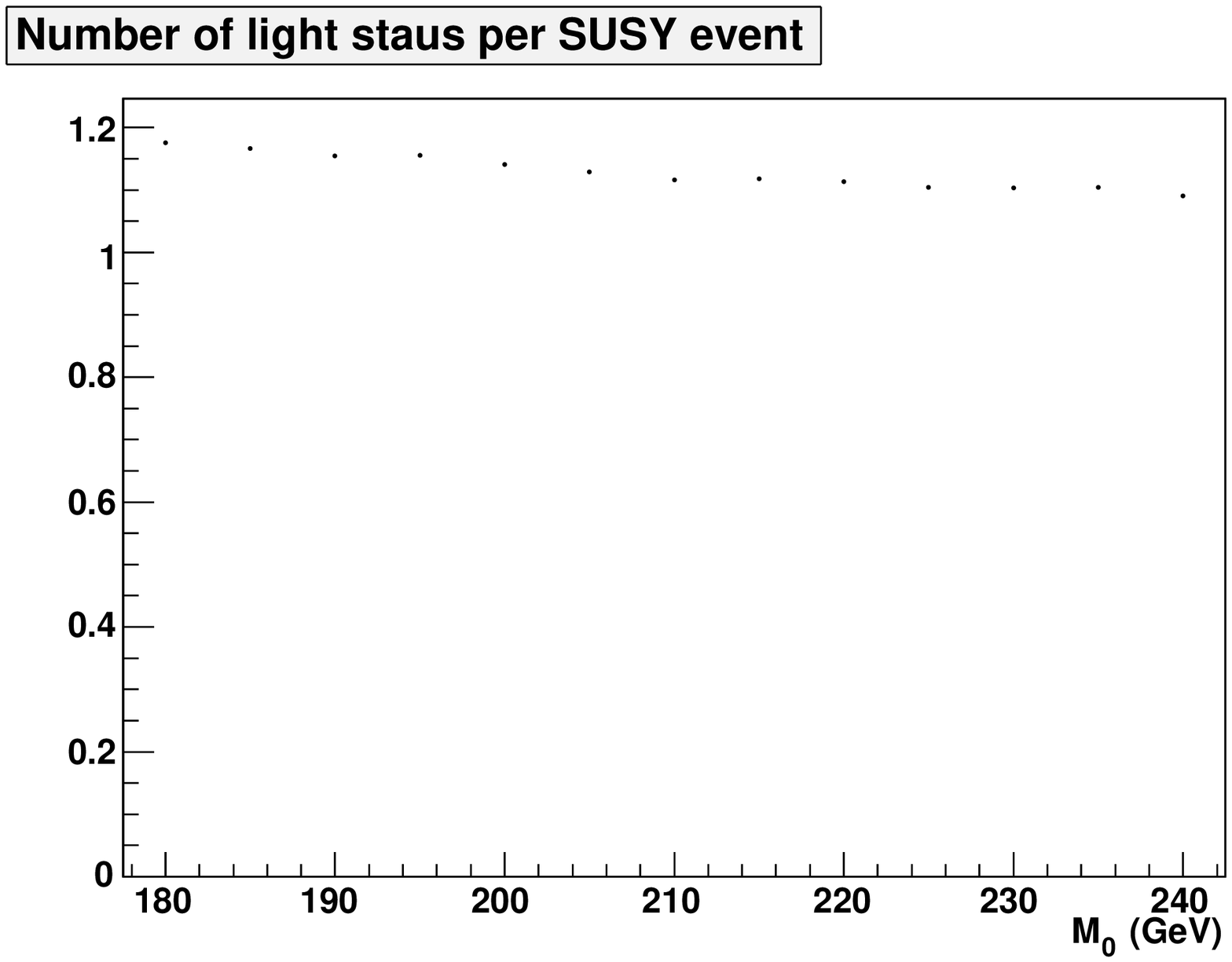}
\includegraphics[width=0.5\linewidth]{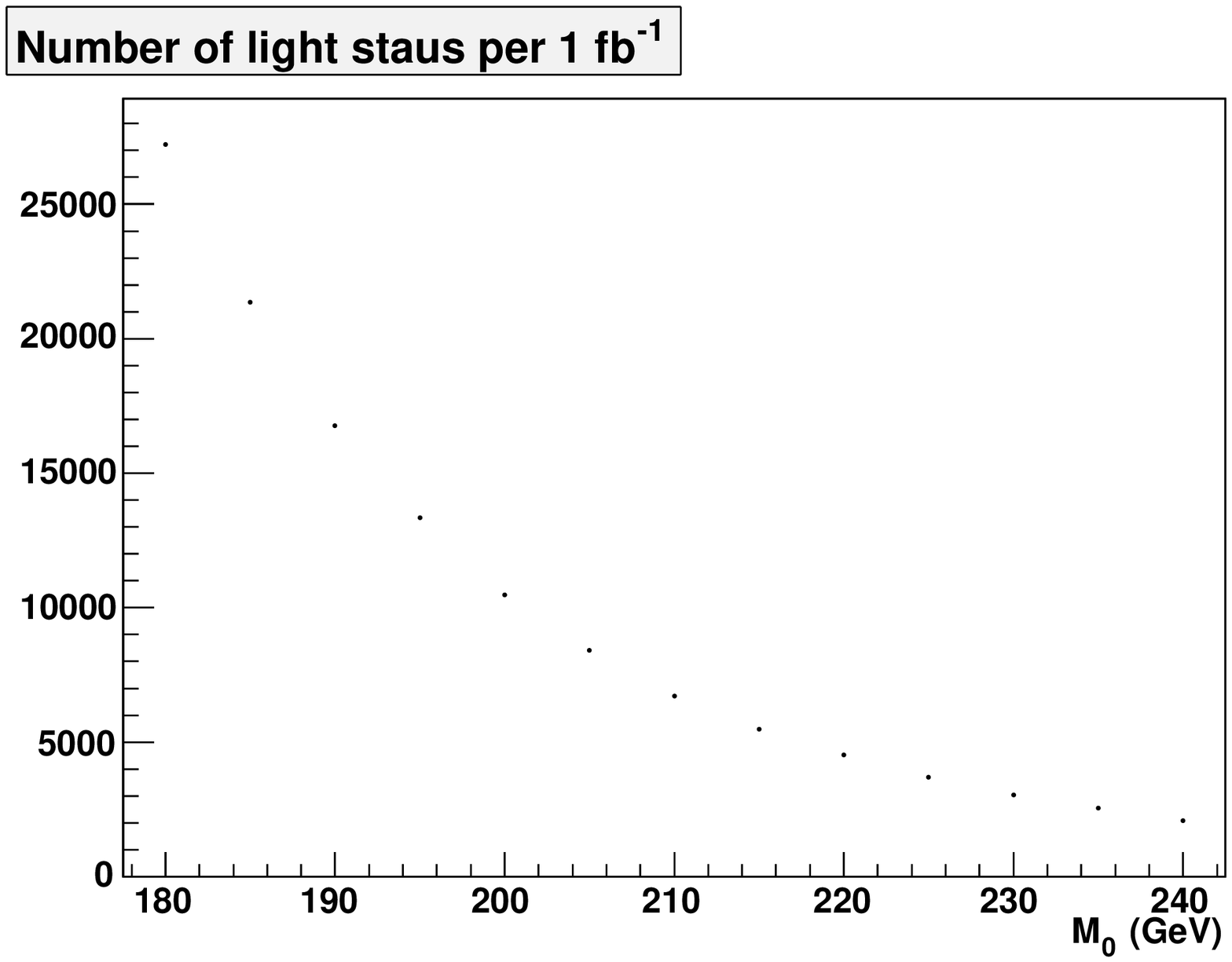}
\caption[Stau production along and SU8.1 coannihilation line]{Staus produced per SUSY event and per fb$^{-1}$ of luminosity along the SU8 coannihilation line from table \ref{t:coannihilationpoints}. The number of staus per SUSY event is almost constant, so the change is completely due to the higher cross section that is associated with lower $\Mhalf$.}\label{f:SU8staus}
\end{figure}

Figure~\ref{f:SU1staus} shows the changes in the number of staus along the coannihilation line belonging to the SU1 benchmark point and figure~\ref{f:SU8staus} shows the changes along the coannihilation line belonging to the SU8 benchmark point. 
\newpage

The total supersymmetric cross section plays an important role for both coannihilation lines, but there are some important differences. The SU1 benchmark point is in a phenomenological region where all decays from table \ref{t:gauginodecay} are allowed. In contrast, for the SU8 benchmark points only decays to light staus and the LSP are kinematically accessible. This results in a higher number of staus per SUSY event for the SU8.1 benchmark point. However, this effect is partly cancelled by cross section effects.

The second difference is that the number of staus per SUSY event along the SU1 coannihilation line changes considerably, while it is nearly constant for the SU8.1 benchmark point. The reason can once again be found in the phenomenological regions. In the language of figures \ref{f:chargino-region10} and \ref{f:neutralino-region10}, the SU1 coannihilation line essentially follows the boundary of the white and the red region. Thus it moves further away from the green and yellow region as the mass increases. In other words, for higher masses more phase space is available for $\tilde\chi_2^0$ and $\tilde\chi_1^\pm$ decays to sneutrinos and heavy sleptons. Thus the number of staus decreases.

This effect does not play a role for the SU8.1 coannihilation line, since all points on this line are in the stau-dominated phenomenological region. For the low masses along the SU8.1 coannihilation line, the gaugino-decays to the Higgs boson is no longer kinematically accessible, which results in a slightly higher number of staus.

\subsubsection{Focus Point Benchmark Point SU2}\label{s:SU2}

\begin{figure}[!h]
\centering{
\includegraphics[width=\linewidth]{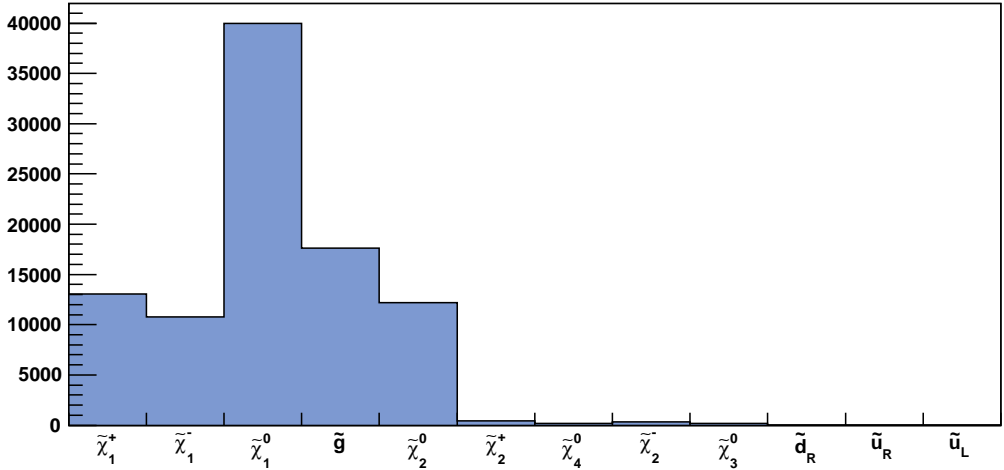}
\caption[Particles produced in SU2]{The number of times a particle was produced in 20.000 SUSY events of benchmark point SU2.}\label{f:focuspointfirstint}
}
\end{figure}

The focus point region is the region where the LSP has a large Higgsino component that enhances its annihilation. As mentioned in section \ref{s:massgeneralproperties}, this is the case for very large values of $M_0$. As a result, the gaugino masses are much smaller than the scalar masses. The phenomenology of this region was briefly discussed in section \ref{s:focuspointregion}. Figure~\ref{f:focuspointfirstint} shows all supersymmetric particles produced in the initial interactions and decay chains of 20.000 SUSY events. 

Because of conservation of $R$-parity, the LSP is produced twice per event, which leads to a total of 40.000 $\tilde\chi_1^0$ particles. As predicted in section \ref{s:firstint}, the number of scalars produced is negligible. This type of phenomenology is representative for the entire focus point region, since as mentioned in section \ref{s:focuspointregion}, the phenomenology is dominated by gauginos.

\subsubsection{Bulk Benchmark Point SU3}

Benchmark point SU3 is quite close to the border of two phenomenological regions based on gaugino decays. Decays to sneutrinos are allowed, but a small change in parameters increases the phase space for this decay considerably. As a result, a small variation around the SU3 benchmark point yields very different results for the number of staus and taus. The SU3 benchmark point is stable for stop phenomenology. As can be seen in table \ref{t:masscompare}, the stop mass is well above the critical value of 200~GeV and no decays become (im)possible by a small change in parameters.

\begin{figure}[!h]
\includegraphics[width=0.5\linewidth]{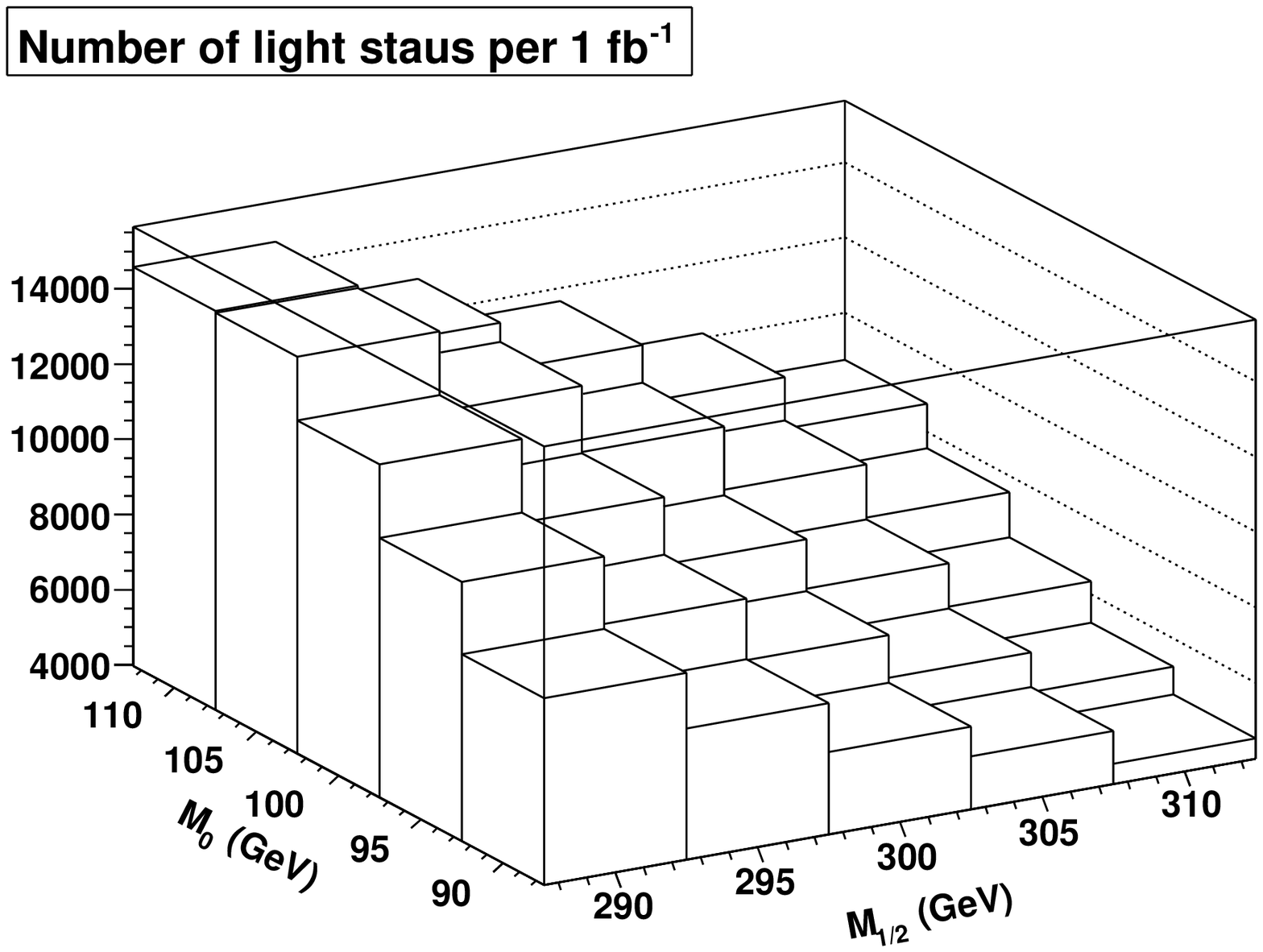}
\includegraphics[width=0.5\linewidth]{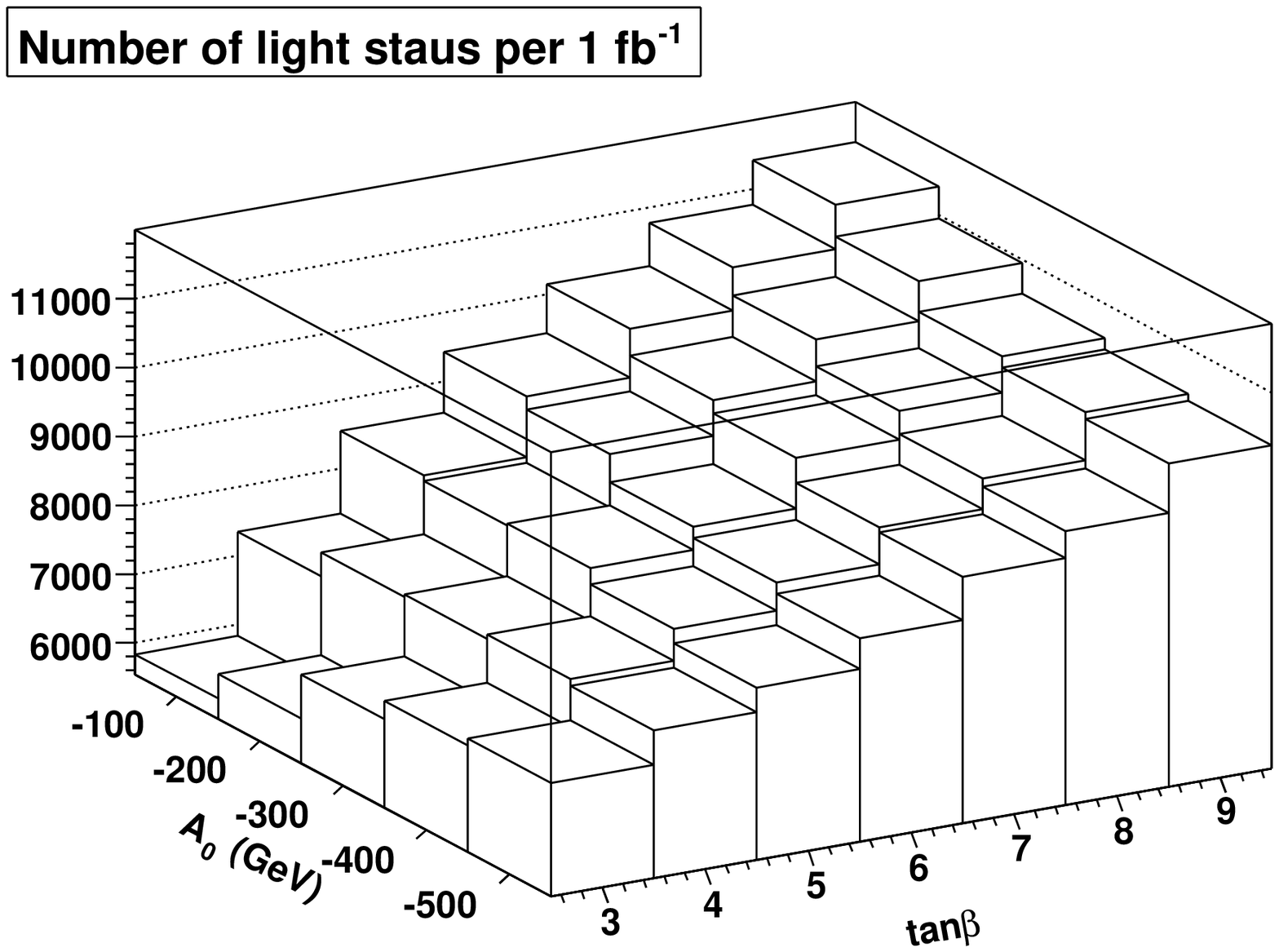}
\caption[Number of light staus on two grids around SU3]{The number of light staus produced by supersymmetric processes per 1 fb$^{-1}$ of luminosity for two grids around the SU3 benchmark point.}\label{f:SU3-totalstau}
\end{figure}

Figure~\ref{f:SU3-totalstau} shows the total number of staus produced in 1 fb$^{-1}$ of 14~TeV proton-proton collisions for two grids around benchmark point SU3, one in the parameters $M_0$ and $\Mhalf$, and one in the parameters $A_0$ and $\tan\beta$. The total supersymmetric cross section calculated by Pythia has been taken into account.
\newpage

\begin{figure}[!h]
\includegraphics[width=0.5\linewidth]{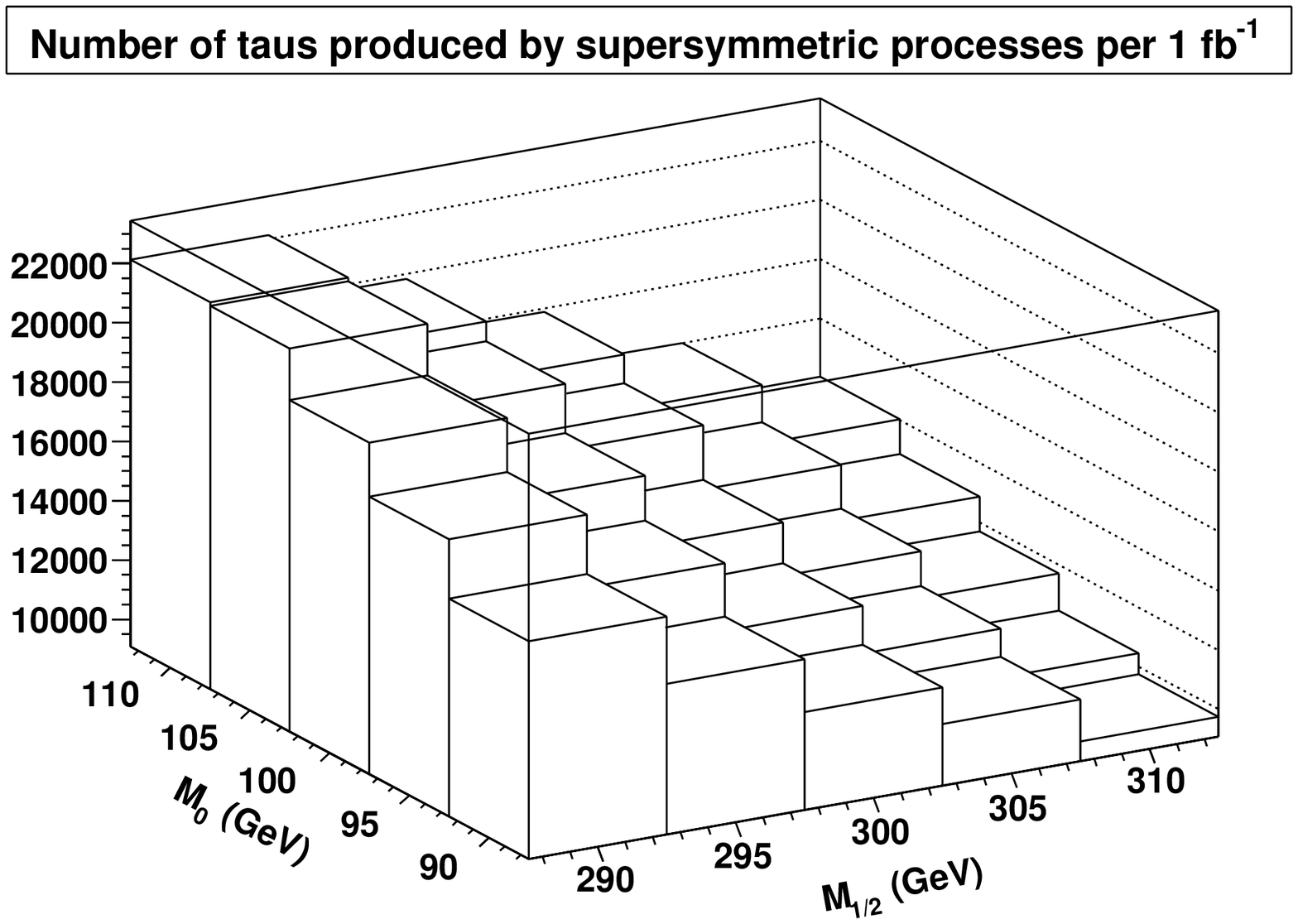}
\includegraphics[width=0.5\linewidth]{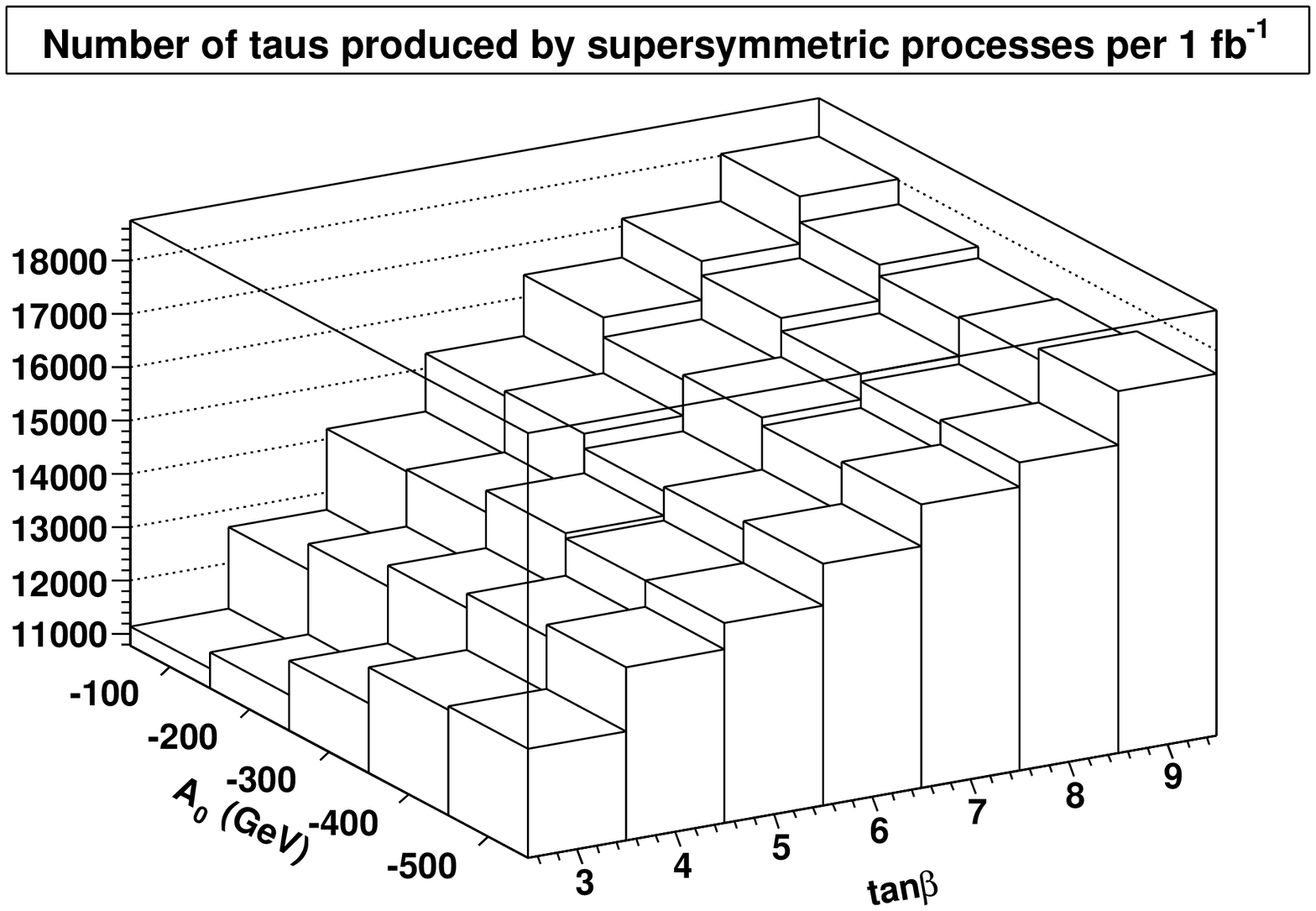}
\caption[Number of taus on two grids around SU3]{Number of taus produced per 1 fb$^{-1}$ of luminosity for two grids around the SU3 benchmark point.}\label{f:SU3-totaltau}
\end{figure}

Even though $M_0$ and $\Mhalf$ are varied by only 20~GeV, the number of light staus produced changes by more than a factor of 3. Varying $A_0$ and $\tan\beta$ can double the number of staus produced. This results in a similar change in the number of taus produced, which is shown in figure~\ref{f:SU3-totaltau}.

\begin{figure}[!h]
\includegraphics[width=0.5\linewidth]{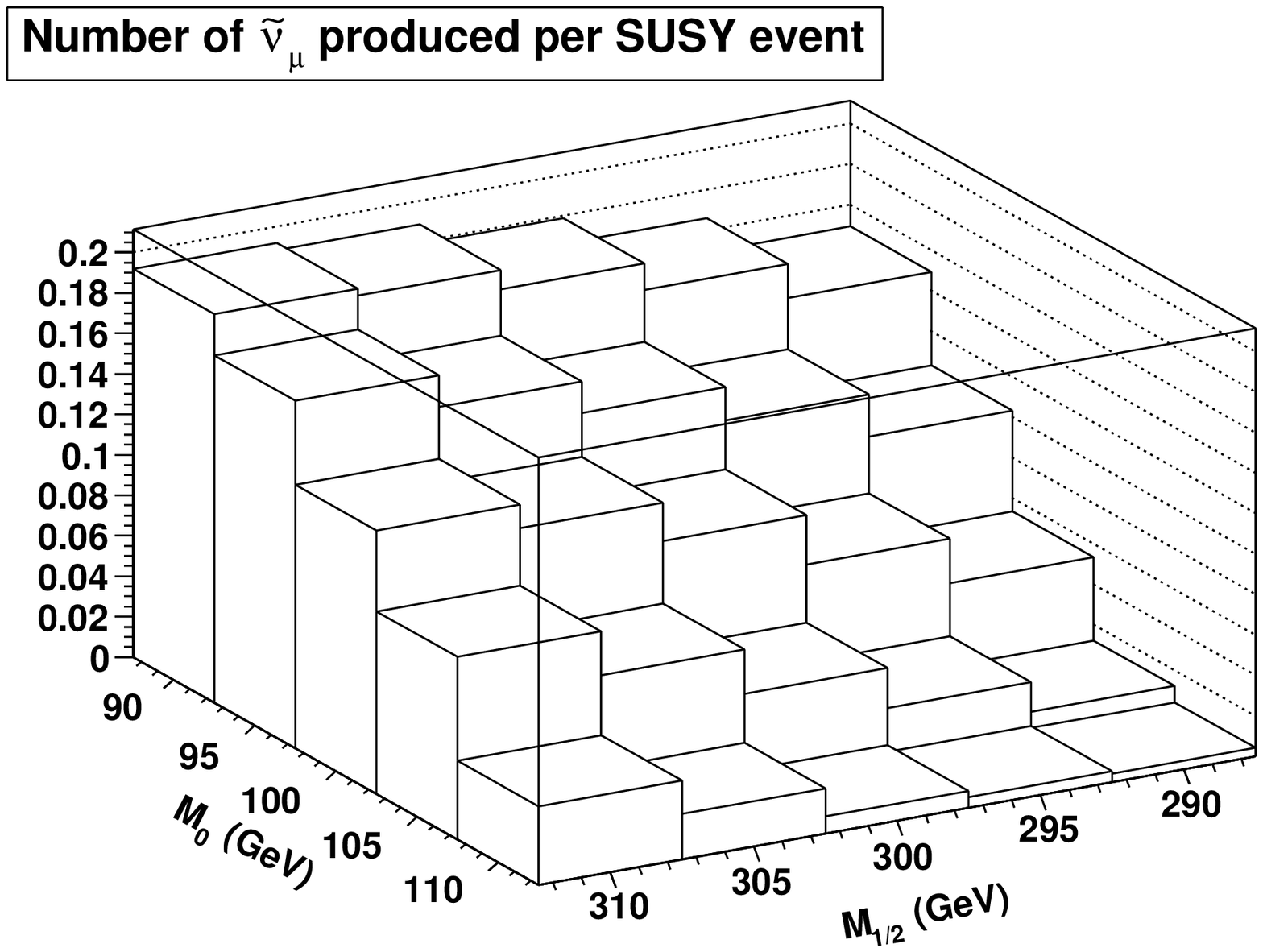}
\includegraphics[width=0.5\linewidth]{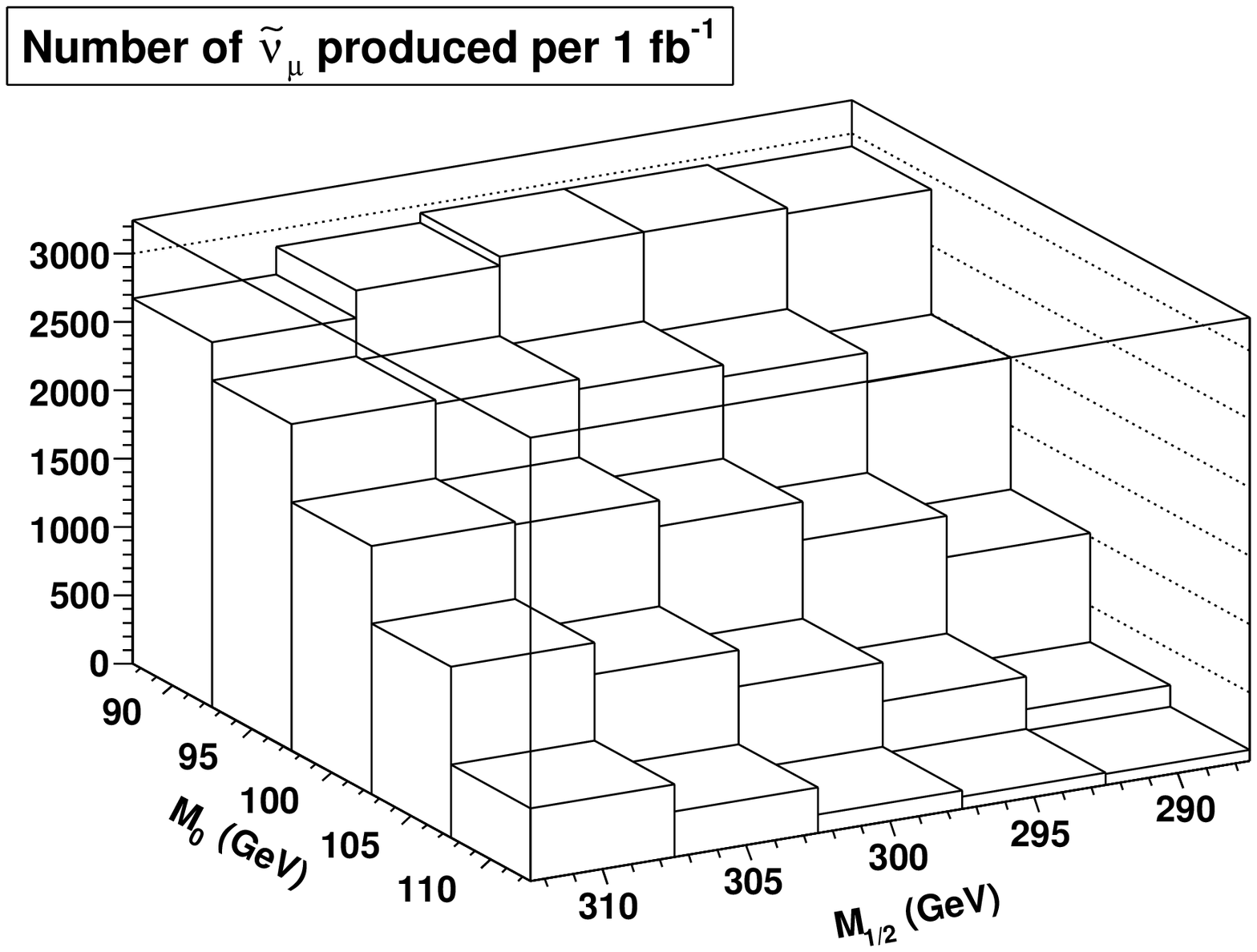}
\caption[Number of muon-sneutrinos on a grid around SU3]{Number of $\tilde\nu_\mu$'s produced per SUSY event and per 1 fb$^{-1}$ of luminosity for a grid around the SU3 benchmark point.}\label{f:SU3-numu}
\end{figure}

There are two reasons for this change. The first is a cross section effect: lower masses lead to higher cross sections. The second reason is that the gaugino-decay to sneutrinos and heavy sleptons becomes more important for lower $M_0$ or higher $\Mhalf$. This can also be seen in figure~\ref{f:SU3-numu}, where the number of muon-sneutrinos for an $M_0,\Mhalf$ grid around SU3 is plotted. Note that the axes are rotated relative to the other graphs for readability of figure \ref{f:SU3-numu}.

\begin{figure}[!h]
\includegraphics[width=0.5\linewidth]{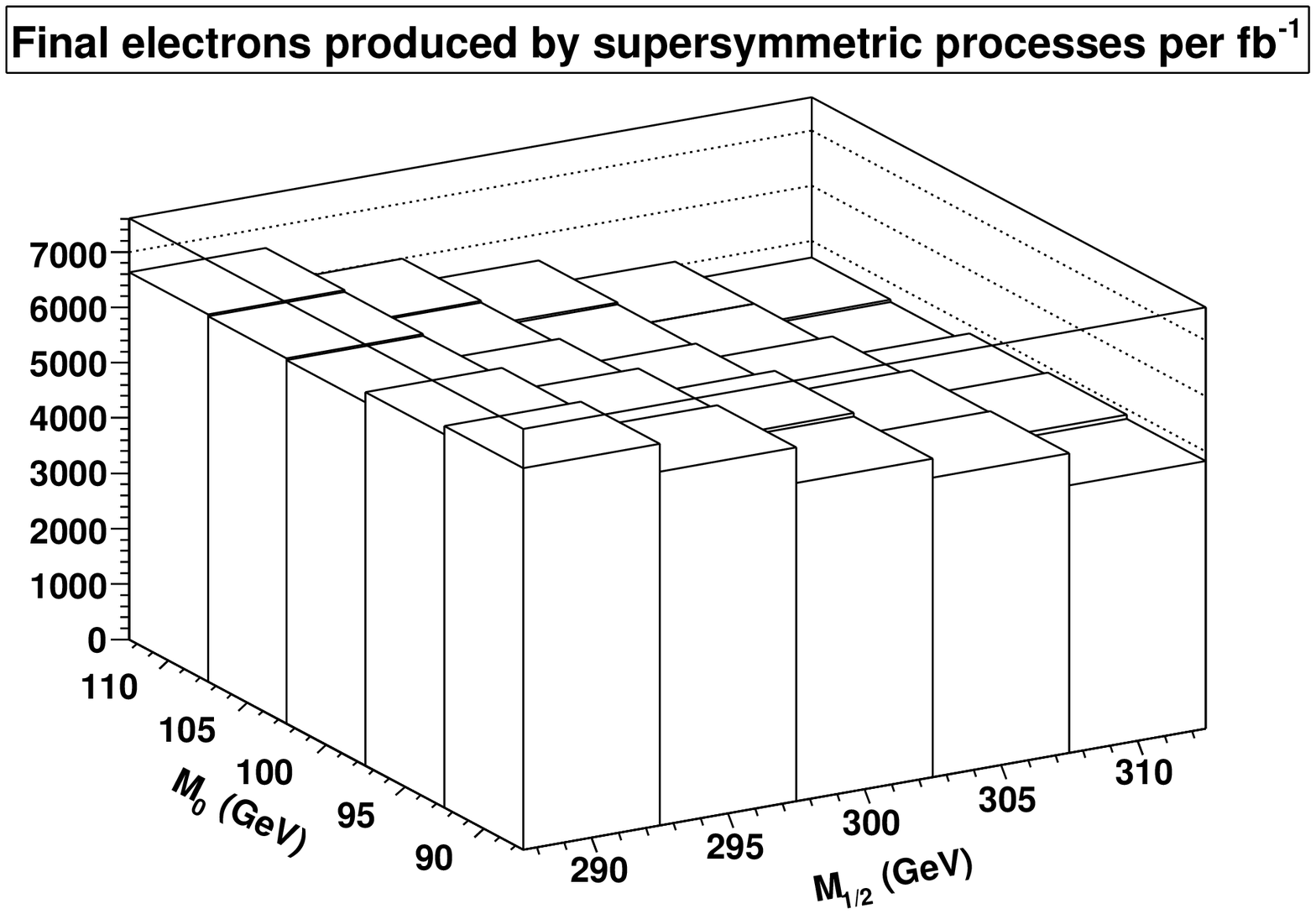}
\includegraphics[width=0.5\linewidth]{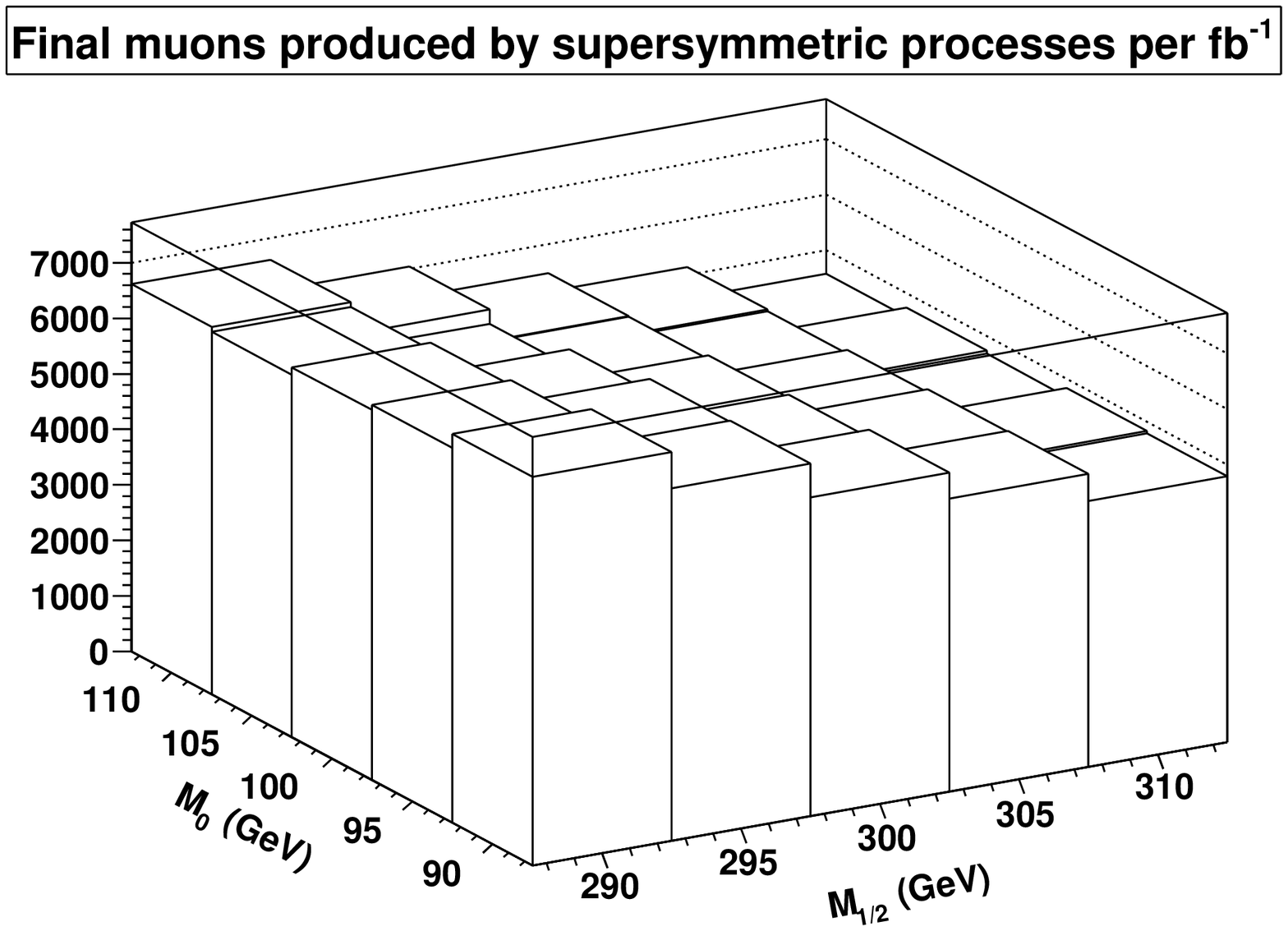}
\caption[Number of final electrons and muons on a grid around SU3]{The number of final electrons and muons produced by supersymmetric processes per 1~fb$^{-1}$ of luminosity around the SU3 benchmark point.}\label{f:SU3-M0Mhalf-final}
\end{figure}

Figure \ref{f:SU3-numu} shows that the effect of the supersymmetric phenomenology is more important than the effect of the cross section in this case. If $\Mhalf$ is changed, the two contributions (cross section and phenomenology) have the opposite effect on the number of sneutrinos. Near the boundary of a phenomenological region, phenomenology usually wins.
\newpage
The effect of such small variations in $M_0$ and $\Mhalf$ is less visible in the number of final state muons and electrons. Figure~\ref{f:SU3-M0Mhalf-final} shows the number of final state electrons and muons produced by supersymmetric processes per 1 fb$^{-1}$ of luminosity.

As we already saw in figure~\ref{f:gauginodecay-finalpart}, the number of final muons and electrons depends significantly less on the point in parameter space than the number of taus. The small changes in figure \ref{f:SU3-M0Mhalf-final} are almost entirely due to the supersymmetric cross section. Since many points in the mSUGRA parameter space have the same supersymmetric cross section yet different tau phenomenology, the identification of taus is vital to determine the soft breaking parameters if supersymmetry is found at the LHC.

\subsubsection{Funnel Benchmark Point SU6}

Funnel regions are another loophole to lower the dark matter relic density. If the mass of the lightest neutralino is of the order of half the mass of the $Z$ boson or half the mass of a Higgs boson, its annihilation is enhanced, because it takes place near a resonance\footnote{In literature, this resonance is often referred to as a pole. This terminology is not entirely correct, since the resonance does not contain a singularity for real momenta (it has a finite width).} in the cross section. This results in more rapid annihilation and thus a lower dark matter relic density. Annihilation is also enhanced if the mass of the LSP is half the mass of a neutral resonance such as the $J/\Psi$. Given the large masses of supersymmetric particles, the most interesting annihilation processes are those involving the neutral heavy Higgses.

If we neglect the kinetic energy of the neutralinos, annihilation takes place at the peak of the resonance if $m_{H^0}=2m_{\tilde\chi_1^0}$. In that case the predicted dark matter relic density is considerably lower than the experimental value. For the SU6 benchmark point, the mass difference calculated by SPheno is $m_{H^0}-2m_{\tilde\chi_1^0}=69$~GeV. The width of $H^0$ in benchmark point SU6 is only 12~GeV, so the difference is quite large. That means that this mechanism works for a large range of Higgs and neutralino masses.

Yet in most parts of parameter space the Higgs mass is much larger than twice the LSP mass. From section \ref{s:massspectrum} we know that the only high-scale parameter that affects the neutralino mass is $\Mhalf$, and $\Mhalf$ affects the Higgs mass as well. Thus this mechanism can only play a role if the heavy Higgses are relatively light. In practice this means that the funnel region is important only for high values of $\tan\beta$ and relatively low values of $M_0$.

Benchmark point SU6 is chosen to be in such a funnel region. The SU6 benchmark point is in the center of the phenomenological region where the light gauginos are heavier than staus, but lighter than all other sfermions. Thus staus are produced abundantly and the number of staus is stable under a small variation of the parameters. The light stop can decay through almost all decay channels from table \ref{t:stopdecay} except $\tilde t_1\to\tilde\chi_{3,4}^0\bar t$. It can be produced by a gluino and by the heavy sbottom, but not by the light sbottom. This region corresponds to the small orange strip for high $\tan\beta$ in figure \ref{f:stopproductionA0beta}, although the high-energy parameters are different.

For high $\tan\beta$, the phenomenological region where all decays from table \ref{t:gauginodecay} are allowed is quite small, so production of heavy sleptons and sneutrinos is suppressed by phase space effects. However, the funnel region does contain several phenomenological regions, since it extends to parts of parameter space where the $\tilde\chi_1^\pm$ and $\tilde\chi_2^0$ cannot decay to staus. Therefore the SU6 benchmark point is not representative for the entire funnel region.

\subsubsection{Low-Mass Benchmark Point SU4}

\begin{figure}[!h]
\centering{
\includegraphics[width=0.8\linewidth]{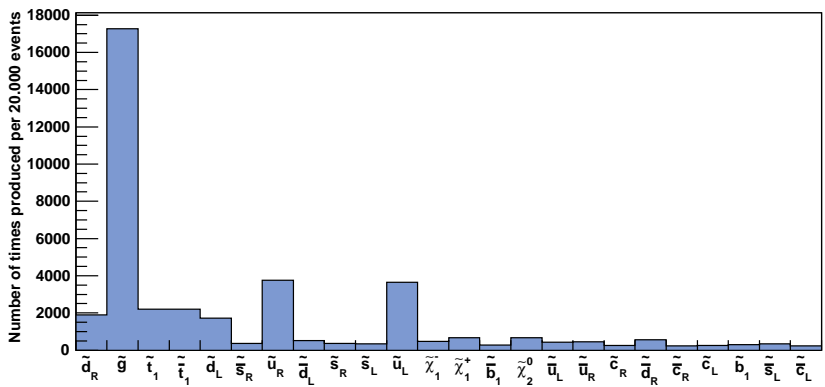}
\caption[Supersymmetric particles produced in the first interaction for SU4]{The number of times a given particle was produced in the first interaction of 20.000 SUSY events for benchmark point SU4. Only the particles that are produced in at least 1\% of the events are shown.}\label{f:SU4-firstint}}
\end{figure}

Benchmark point SU4 is the only low-mass point and thus has a much higher supersymmetric cross section than the other benchmark points. Due to the large negative value of $A_0$, the stop mass is low. However, with a stop mass of 209.8~GeV the dominant process in the first interaction is still gluino production. This can be seen in figure~\ref{f:SU4-firstint}, which shows the particles produced most abundantly in the first interaction in the SU4 benchmark point. Only the particles produced in at least 1\% of the events are shown. 

Gluinos are still the most important contribution, but the contribution from stop production is not negligible. The light stop is too light to decay to a top, but tops are produced in approximately one third of the SUSY events through the decay $\tilde g\to\tilde t_1 \bar t$ and through sbottom decays. The number of staus and taus is very low (only a few staus per fb$^{-1}$), since benchmark point SU4 is in a region where the light gauginos can only decay through three-body decays. Thus few sleptons are produced. The SU4 benchmark point is quite stable under a change in parameters, but it is not representative for the entire low-mass region. As we can see in figures \ref{f:chargino-region10} and \ref{f:neutralino-region10}, there are also low-mass points where $\tilde\chi_1^\pm$ and $\tilde\chi_2^0$ can decay to sleptons. For instance, the point $M_0=80$~GeV, $\Mhalf=170$~GeV, $A_0=-400$~GeV, $\beta=10,\mu>0$ is comparable to the benchmark point SU4 except that almost $4\times10^5$ staus are produced per 1 fb$^{-1}$ of luminosity.

\subsubsection{Representativeness of the Benchmark Points}\label{s:repbenchmark}

An important question is if the benchmark points are representative for the mSUGRA parameter space. In other words: if we have benchmark points for all phenomenological regions. The answer is no, but this is not always a problem. Some of the effects of the phenomenology are counteracted by the cross section, as can be seen by comparing figures \ref{f:pythiadecayjumps-stopA0} and \ref{f:pythiadecayjumpscrosssection-stopA0}. It might not be necessary to have benchmark points in all phenomenological regions to have a representative set of points. And for practical reasons we want to keep the number of benchmark points as low as possible.

Another reason not to study a particular phenomenological region in detail is that it is excluded by low-energy observables such as the dark matter relic density. However, one has to be extremely careful with such exclusion limits, since the dark matter relic density can change considerably if the requirement of unification is dropped \cite{Griest:1991gu}. Table \ref{t:benchmarkpheno} summarizes the stau and stop phenomenology of the benchmark points that are studied in the most recent ATLAS note \cite{cscnote}.
\begin{table}[!h]
\centering{
\begin{tabular}{|l|l|l|}
\hline
Point & Slepton/lepton phenomenology & Stop phenomenology\\
\hline
SU1&many sleptons, all flavours&many stops\\
SU2& few sleptons, leptons from Standard Model gauge bosons & few stops (few scalars)\\
SU3&many sleptons, particularly staus&many stops\\
SU4& few sleptons, leptons from three-body decays & many stops\\
SU6&dominated by staus&many stops\\
SU8&dominated by staus&many stops\\
\hline
\end{tabular}
\caption[Phenomenology of the benchmark points.]{The phenomenology of the benchmark points. With the exception of benchmark point SU2, stops can be produced by gluino decay in all benchmark points. All SU8 points are in the same phenomenological region, so they are not listed separately. The SU5 benchmark points are not studied anymore, so they are not listed either.}\label{t:benchmarkpheno}}
\end{table}

Not all phenomenological regions are addressed by the benchmark points. Firstly, there are no benchmark points in the region where the decay to the Higgs is the dominant decay mode of $\tilde\chi_2^0$. As we can see in figure \ref{f:gauginodecay-higgs}, this scenario yields a significant number of light Higgses and thus $b$ quarks. The original ATLAS benchmark points contained such a point (SU5.3), but this point is not studied anymore due to its high dark matter relic density. An example of a point with the required phenomenology is the point $M_0=420$~GeV, $\Mhalf=350$~GeV, $A_0=0$~GeV, $\tan\beta=50,\mu>0$. This point has a total supersymmetric cross section of 5.2 pb, which is comparable to the cross section of benchmark point SU2. Since it is within the funnel region, it also has an acceptable dark matter relic density.

Secondly, all benchmark points with the exception of SU2 are in regions where stops can be produced by gluino decay. As mentioned in section \ref{s:SU2} the phenomenology of the SU2 benchmark point is dominated by gauginos. However, there are points where the decay $\tilde g\to\tilde t_1\bar t$ is not kimenatically allowed while the decay $\tilde g\to\tilde b_1\bar b$ is accessible. It is an interesting question whether it is possible to distinguish this case at the LHC.

Given the LEP bounds on the Higgs mass, it is very likely that the decay $\tilde b_1\to\tilde t_1W^-$ is allowed\footnote{The LEP bounds on the Standard Model Higgs mass makes it unlikely that $M_0$ and $\Mhalf$ both have a low value. From figure \ref{f:stopproduction-region10} we can see that this is exactly the region where the decay $\tilde b_1\to\tilde t_1W^-$ is not allowed. Thus it is unlikely that a light sbottom cannot decay to a light stop.}. The final states in both cases are quite similar:
\begin{itemize}
\item $\tilde g\to\tilde t_1\bar t\to\tilde\chi_1^+bW^-\bar b$ or $\tilde g\to\tilde t_1\bar t\to\tilde\chi_1^0tW^-\bar b\to\tilde\chi_1^0bW^+W^-\bar b$
\item $\tilde g\to\tilde{\bar b}_1 b\to \tilde\chi_1^+\bar tb\to\tilde\chi_1^+\bar bW^- b$ or $\tilde g\to\tilde b_1\bar b\to\tilde t_1W^-\bar b\to \tilde\chi_1^+bW^-\bar b$ or $\tilde\chi_1^0tW^-\bar b\to\tilde\chi_1^0bW^+W^-\bar b$
\end{itemize}
However, the kinematics are different, so the jets should yield a different invariant mass. It is difficult to distinguish these cases at the LHC, but it might be possible. An example of such a point is $M_0=420$~GeV, $\Mhalf=250$~GeV, $A_0=0$~GeV, $\tan\beta=50,\mu>0$, which is in the funnel region and has a supersymmetric cross section of 23 pb. 

Finally, another type of benchmark point that would be interesting to study in the context of stop phenomenology is a point with large negative $A_0$. At the moment there is no such benchmark point. There are three reasons to study this part of parameter space in more detail. Firstly, by not having a benchmark point with relatively low stop mass, we have a gap in the possible phenomenology. The second and third reason come from the low-energy constraints on parameter space. We saw in section \ref{s:A0massexplanation} that low values of $A_0$ yield a relatively high mass for the lowest Higgs, which is interesting in the light of the LEP Higgs mass limit. In addition, if the stop is so light that it can only decay through the channel $\tilde t_1\to\tilde\chi_1^0c$, it is stable enough to drive the dark matter relic density down through coannihilation with the LSP. In other words: there is a stop coannihilation region. This is a part of parameter space that is not studied in detail at the moment.
\newline

\noindent In short, we propose three additional benchmark points:
\begin{itemize}
\item A point such as $M_0=420$~GeV, $\Mhalf=350$~GeV, $A_0=0$, $\tan\beta=50,\mu>0$, where the dominant decay mode of $\tilde\chi_2^0$ is the decay $\tilde\chi_2^0\to\tilde\chi_1^0h^0$.
\item A point such as $M_0=420$~GeV, $\Mhalf=250$~GeV, $A_0=0$~GeV, $\tan\beta=50,\mu>0$, where the stop cannot be produced by a gluino, but the sbottom can.
\item Points with $A_0\ll0$~GeV. Specifically, points in the stop coannihilation region, such as the point $M_0=160$~GeV, $\Mhalf=320$~GeV, $A_0=-1200$~GeV, $\tan\beta=10,\mu>0$, where $\tilde t_1$ is lighter than $\tilde\chi_1^\pm$ and can only decay to the LSP and a $c$ quark.
\end{itemize}
There are other regions that are not addressed by the benchmark points in table \ref{t:susypoints}, such as the scenario that the stop cannot decay to a top or the scenario that the main gluino decay mode is $\tilde g\to\tilde t_1c$. These would be interesting to study as well. However, the reason to use benchmark points was to make supersymmetric parameter space manageable. Thus while preparing for the LHC it is more important to limit the benchmark points to a set that shows the most important phenomenological scenarios than it is to cover every remote corner of parameter space. 

\subsection{Using Phenomenological Regions}

The predictive power of the phenomenological regions introduced in the previous sections is striking, but they are not a substitute for a full analysis. Decay chains can be quite complex and one can easily overlook an important contribution. Take for example figure \ref{f:stopjump-pythia}. The jump in the $b$ quark production cannot be explained by stop and sbottom phenomenology without taking gaugino decays into account as well.

Also, phenomenological regions do not predict the absolute number of particles produced. They can only be used to predict how the number of specific particles per SUSY event changes if the high-energy parameters are varied. To do a complete analysis, one has to take the entire decay chain into account.

However, phenomenological regions can tell us where to start looking. Thus, in combination with a full event simulation, they are a powerful tool. For instance, if we are interested in a particular particle, phenomenological regions are a quick way to estimate the range of possibilities. Simply list the possible processes that produce the particle of interest and determine which ones are the most important. Then list the possible decay modes of the most important mothers and determine where in the parameter space these decays are kinematically allowed. After that an event simulation is needed to obtain the actual numbers. For a check on the validity of phenomenological regions, relatively few events (1000 should suffice to see the most important changes) should be generated along the following lines through parameter space:
\begin{itemize}
\item Lines crossing phenomenological regions, to confirm the expected changes.
\item A line along which the results are expected to remain constant to confirm that the regions are not unexpectedly affected by other processes, such as production processes of the mothers. 
\end{itemize}
No detector simulation is necessary for these checks. If the phenomenological regions pass these tests, one representative point for every region can be chosen for more elaborate simulations. These points represent all possibilities for the phenomenology of the particle of interest.

As far as the benchmark points are concerned, phenomenological regions show that the cross section of a benchmark point is not very important, since many points with similar phenomenology but totally different cross section can be found. Also, the benchmark points do not represent all possible phenomenologies (see section \ref{s:repbenchmark}).

While preparing for the LHC, phenomenological regions can be used to identify the possible signals of supersymmetry. However, they are more interesting once the LHC is running. If no supersymmetry is found, phenomenological regions combined with the predicted supersymmetric cross section can be used to exclude parts of parameter space. If supersymmetry is found at the LHC, phenomenological regions can be used to quickly identify the interesting parts of parameter space. In short, phenomenological regions are an efficient tool to exclude parts of the mSUGRA parameter space so research can focus on the interesting regions.

\section{Conclusion and Outlook}

In this research we developed a method to predict supersymmetric phenomenology at the LHC from the mass spectrum in the context of mSUGRA. While preparing for the LHC, this method can be used to find the possible phenomenological scenarios. Once the LHC is collecting data, phenomenological regions can be used to exclude parts of parameter space and identify the interesting regions.

Within the low-mass mSUGRA parameter space, phenomenology is determined by the total supersymmetric cross section, the stop cross section and the phenomenology within supersymmetry. Cross sections are relatively simple functions that depend on the masses of the supersymmetric particles involved in the production process. However, the phenomenology within supersymmetry can produce sudden jumps in the number of particles near the border of a phenomenological  region. These effects are particularly strong in the slepton sector.

Phenomenology can vary considerably over different phenomenological regions. For example, decays of light gauginos give rise to a slepton-dominated region, a Higgs-dominated region, a gauge boson-dominated region and a region that is characterized by three-body decays. If supersymmetry is in the slepton-dominated region, the identification of taus will be vital to determine the supersymmetric phenomenology more precisely. For other regions the effect of the cross section can be more important. For instance, the Higgs-dominated region spans an enormous part of parameter space. Within this region, cross section effects are more important than changes in supersymmetric phenomenology.

In the SUSY QCD sector phenomenological regions are generally larger than in the slepton sector. As a result, the supersymmetric cross section plays an important role. There are still points that have the same cross section yet different phenomenology. However, most phenomenological regions based on stop and sbottom phenomenology can only be distinguished by studying top and bottom kinematics, which is a very difficult task.
\newline

There are many interesting questions left about supersymmetry, some of which are very important for the interpretation of the results at the LHC. Firstly, in preparation for the LHC, detector simulations have to be performed for a representative set of points in the supersymmetric parameter space. In addition to the existing benchmark points, it would be interesting to study the representative points proposed in section \ref{s:repbenchmark} in more detail.

Secondly, to obtain actual exclusion limits from LHC data, next to leading order calculations are needed as well. Implementing loop corrections in Monte Carlo simulations is quite difficult, but it is necessary for accurate predictions. Next to leading order corrections do not change the shape of the phenomenological regions, but they do affect the number of supersymmetric particles that are produced. Since these corrections can be as large as roughly 100\%, they have a large effect on exclusion limits.

Finally, this research is limited to mSUGRA. An important question is if the phenomenology of mSUGRA is representative for the supersymmetric parameter space. Can the method developed in this research be applied to other breaking models, the MSSM or perhaps even a general supersymmetric extension of the Standard Model?

Most likely it is possible to generalize the methods presented in this research, although in particular the general MSSM is more complex than mSUGRA. For instance, the mass spectrum can be very different if the requirement of unification is dropped. This affects not only the phenomenology within supersymmetry, it can also change which particles are produced in the initial interaction. In the most general extension, with more than one supersymmetry generator, there are probably too many possibilities to make any sensible predictions.

In all generalizations it is important to include the low-energy observables in the analysis, since these put severe restrictions on the mass spectrum and the possible phenomenology. Research into these observables is ongoing, particularly in the areas of the dark matter relic density and the loop corrections that contribute to flavour changing processes. Fortunately, LHC Higgs searches will put a strong restriction on the supersymmetric mass spectrum. In the meantime, it might be interesting to find a quick way to put more accurate limits on the parameter space using the Higgs branching ratios as well as the mass. This would yield a more accurate bound than the LEP bound on the Higgs mass that is currently used in many analyses.

There are many other things that are interesting to investigate. However, as far as phenomenology is concerned, the representativeness of the benchmark points is probably the most important issue, both in the context of mSUGRA and beyond. This is an interesting field of research from both a theoretical and an experimental point of view. For experiments, the biggest challenge after finding signs of supersymmetry will be to identify the type of phenomenology and thus the interesting part of supersymmetric parameter space. It is vital to have a representative set of benchmark points to identify the phenomenological region.

To obtain a good set of benchmark points, one has to answer a more theoretically motivated question. Is mSUGRA representative for the parameter space of the general MSSM? Answering this question while taking into account the low-energy parameters will truly show the range of possible phenomenologies we could expect at the LHC.

\appendix
\begin{fmffile}{feynrules}
\fmfset{dot_size}{1thick}
\fmfset{curly_len}{1.5mm}
\fmfset{arrow_len}{3mm}

\section{Feynman Rules}\label{feynmanrules}

This appendix lists the relevant Feynman rules used in this thesis.

\subsection[$\phi^3$ theory]{\boldmath$\phi^3$ theory}

\begin{align}
\mbox{Scalar propagator:}\qquad\parbox{15mm}{\begin{fmfgraph*}(12,5)
\fmfleft{q1}
\fmfright{q2}
\fmf{dashes,label=$p$}{q1,q2}
\fmfdot{q1}
\fmfdot{q2}
\end{fmfgraph*}} & = \frac{i}{p^2-m^2+i\varepsilon}\label{phi^3prop}\\
\mbox{Trilinear scalar interaction:}\qquad\parbox{11mm}{\begin{fmfgraph}(8,10)
\fmfleft{q1}
\fmfright{q2,q3}
\fmf{dashes}{q1,v}
\fmf{dashes}{v,q2}
\fmf{dashes}{v,q3}
\fmfdot{v}
\end{fmfgraph}} & = -ig\label{phi^3int}
\end{align}

\subsection{Quantum Electrodynamics}

With $Q$ the charge fraction, $e$ the charge of the positron and $\slashed{p}=p_\mu\gamma^\mu$, the Feynman rules are:
\begin{align}
\mbox{Fermion propagator:}\qquad\parbox{17mm}{\begin{fmfgraph*}(12,5)
\fmfleft{q1}
\fmfright{q2}
\fmf{fermion,label=$p$}{q1,q2}
\fmfdot{q1}
\fmfdot{q2}
\end{fmfgraph*}} & = i\,\frac{\slashed{p}+m}{p^2-m^2+i\varepsilon}\label{fermionprop}\\
\mbox{Photon propagator:}\qquad\parbox{17mm}{\begin{fmfgraph*}(12,5)
\fmfleft{q1}
\fmfright{q2}
\fmf{photon,label=$p$}{q1,q2}
\fmflabel{$\mu$}{q1}
\fmflabel{$\nu$}{q2}
\fmfdot{q1}
\fmfdot{q2}
\end{fmfgraph*}} & = i\,\frac{-g_{\mu\nu}}{p^2+i\varepsilon}\label{photonprop}\\
\mbox{Electromagnetic interaction:}\qquad\parbox{17mm}{\begin{fmfgraph*}(10,10)
\fmfleft{q1,q2}
\fmfright{q3}
\fmf{fermion}{q1,v}
\fmf{fermion}{v,q2}
\fmf{photon}{v,q3}
\fmflabel{$\mu$}{q3}
\fmfdot{v}
\end{fmfgraph*}} & = iQe\gamma^\mu \label{photonfermionfermionint}
\end{align}

\subsection{Higgs sector}

\begin{align}
\mbox{Scalar propagator:}\qquad\parbox{15mm}{\begin{fmfgraph*}(12,5)
\fmfleft{q1}
\fmfright{q2}
\fmf{dashes,label=$p$}{q1,q2}
\fmfdot{q1}
\fmfdot{q2}
\end{fmfgraph*}} & =\frac{i}{p^2-m^2+i\varepsilon}\label{higgsprop}\\
\mbox{Yukawa interaction:}\qquad\parbox{12mm}{\begin{fmfgraph}(8,8)
\fmfleft{q3}
\fmfright{q1,q2}
\fmf{fermion}{q1,v}
\fmf{fermion}{v,q2}
\fmf{dashes}{v,q3}
\fmfdot{v}
\end{fmfgraph}} & =-iy\label{higgsfermionfermionint}\\
\mbox{Quartic scalar interaction:}\qquad\parbox{12mm}{\begin{fmfgraph}(8,8)
\fmfleft{i1,i2}
\fmfright{o1,o2}
\fmf{dashes}{v,i1}
\fmf{dashes}{v,i2}
\fmf{dashes}{v,o1}
\fmf{dashes}{v,o2}
\fmfdot{v}
\end{fmfgraph}} & = -i\lambda_h\label{higgsfourpointint}
\end{align}

\end{fmffile}

\section{Grand Unified Theories}\label{app:gut}

If we look at the Standard Model, we have three forces. We could in fact say the Standard Model is not one theory, but a combination of three distinct theories. It would be more appealing to have one Grand Unified Theory (GUT). A natural way to do this would be by embedding $SU(3)\times SU(2)\times U(1)$ in a larger symmetry group \cite{Slansky:1981yr}. This larger symmetry would have to be broken, since we do not observe it in nature, but it could be a good symmetry at some high energy scale. Ideally the breaking mechanism should explain the value of masses and coupling constants at low energies.

As shown in section \ref{s:RGE}, RGEs predict running coupling constants. If there is an energy scale where the three coupling constants get the same value, that would be a good candidate for the unification scale. A number of assumptions has to be made to arrive at this energy, especially concerning the non-existence of particles in the energy range between $100$ GeV and $10^{15}$ GeV, referred to as the `desert', since such particles would affect the RGEs. Furthermore, although we know the normalization of the non-Abelian gauge couplings because of their self-interactions, we do not know the relative normalization for the $U(1)$-coupling, which means we can choose it. Obviously, any three lines can always be made to cross each other in one point if we can scale one of them with an arbitrary factor.

However, if the Standard Model gauge group is embedded in a larger symmetry group, the relative normalization of the $U(1)$ coupling is determined by the normalization of the larger group. When the idea of unification was first explored, the existing data hinted at unification with a scaling factor for the $U(1)$ coupling that could be explained by group theory. Meanwhile, measurements from LEP have convincingly ruled out the possibility of the Standard Model couplings converging to a single point \cite{Amaldi:1991cn} in the minimal GUT.

Unification is not ruled out if there are particles in the desert, which is the case for supersymmetry. The extra particles contribute to the loop corrections and change the RGEs. If they have masses in the TeV range, the coupling constants can converge to a unification point. To our present knowledge supersymmetry is consistent with unification \cite{Dimopoulos:1981yj}.

The smallest group in which the Standard Model gauge group can be embedded is $SU(5)$ \cite{Georgi:1974sy}. Actually, $SU(5)$ does not quite contain $SU(3)\times SU(2)\times U(1)$ but rather $S(U(3)\times U(2))$, which is a slightly different group. It does have the same Lie-algebra, which ensures most of the phenemenology is unchanged. The most important difference is that $S(U(3)\times U(2))$ predicts charge quantization \cite{schellekensbtsm}, so it is probably a better description of nature than $SU(3)\times SU(2)\times U(1)$.

Now consider the matrix representation of $SU(5)$. The part that contains the Standard Model can be written as a matrix of the form:
\[\left(\begin{array}{cc}U_3&0\\0&U_2\end{array}\right)\]
where $U(2)$ and $U(3)$ are unitary matrices of dimension $2$ and $3$ with $\det(U_3)\det(U_2)=1$. This constraint corresponds to a single $U(1)$ phase freedom and it follows that the corresponding representations (compare the notation to Table \ref{standardmodelparticles}) are given by:
\[(3,1,-\frac{1}{3}q)\quad\quad(1,2,\frac{1}{2}q)\]
where $q$ is the scaling factor that is determined by the normalization of the vector representation of $SU(N)$:
\begin{equation}\Tr(T^aT^b)=\frac{1}{2}\delta^{ab}\label{groupnormalization}\end{equation}
Since the normalization of $SU(5)$, $SU(3)$ and $SU(2)$ is given by equation \eq{groupnormalization}, this fixes the $U(1)$ scaling factor:
\begin{equation}
\frac{1}{2}=3\times\left(-\frac{q}{3}\right)^2+2\times\left(\frac{q}{2}\right)^2=\frac{5}{6}q^2\Rightarrow q=\sqrt{\frac{3}{5}}
\end{equation}
This factor is usually absorbed in the definition of the coupling constant.

\section{Renormalization Group Equations of the MSSM}\label{app:susyrge}

The one-loop\footnote{The two-loop equations can be found in for instance \cite{martin1994}, \cite{Yamada:1994id} and \cite{Jack:1994kd}} RGEs for the parameters of the MSSM given below are based on \cite{martin1994}, \cite{Bertolini:1990if} and \cite{falck1985}. In this appendix we will first give the matrix equations for the general case and then derive the leading order RGEs for mSUGRA. The following notation is used for the different parameters:
\begin{table}[!h]
\begin{tabular}{ll}
$g_i$ & Gauge couplings\\
$M_i$ & Soft-breaking gaugino mass parameters\\
$\mathbf{y}_{u,d,e}$ & Yukawa matrix for up-type and down-type squarks as well as for charged sleptons\\
$\mathbf{m}_{\tilde Q,\tilde{\bar u},\tilde{\bar d},\tilde L,\tilde{\bar e}}^2$ & Soft-breaking sfermion mass matrices\\
$m_{H_1,H_2}^2$&Soft-breaking Higgs mass parameters\\
$\mathbf{a}_{u,d,e}$ & Soft-breaking trilinear terms\\
$b$ & Soft-breaking Higgs mixing parameter\\
$\mu$ & Superpotential $\mu$ parameter
\end{tabular}
\end{table}

The $U(1)$ hypercharge is normalized according to the $SU(5)$ GUT (see appendix \ref{app:gut}). Indices are used to distinguish breaking terms belonging to different particles and are not summed over. The following shorthand notation is used for parameters that enter the RGEs through common combinations of second Casimir invariants:
\begin{align*}
b_i&=\left(\frac{33}{5},1,-3\right)&c_i'&=\left(\frac{7}{15},3,\frac{16}{3}\right)\\
c_i&=\left(\frac{13}{15},3,\frac{16}{3}\right)&c_i''&=\left(\frac{9}{5},3,0\right)
\end{align*}
Another useful combination is:
\begin{equation}\label{S-U1trace}
\mathcal{S}=m_{H_2}^2-m_{H_1}^2+\Tr\left[\mathbf{m}_{\tilde Q}^2-\mathbf{m}_{\tilde L}^2-2\mathbf{m}_{\tilde{\bar u}}^2+\mathbf{m}_{\tilde{\bar d}}^2+\mathbf{m}_{\tilde{\bar e}}^2\right]
\end{equation}
The $\beta$-function of a variable $v$ is defined as:
\begin{equation}
\beta_v=16\pi^2\frac{dv}{dt}
\end{equation}
where compared to the notation of section \ref{s:RGE}, an extra factor $16\pi^2$ is absorbed in the definition of $\beta$ and $t=\ln(\mu/\mu_0)$ with $\mu_0$ the GUT scale that is used to make the argument of the logarithm dimensionless. The beta functions for the supersymmetric parameters are:
{\allowdisplaybreaks
\begin{align}
\beta_{g_i}=&b_ig_i^3\label{RGE-gaugecoupling}\\
\beta_{M_i}=&2b_ig_i^2M_i\label{RGE-gaugino}\\
\beta_{\mathbf{y}_u}=&\mathbf{y}_u\Big(3\Tr(\mathbf{y}_u\mathbf{y}_u^\dagger)+3\mathbf{y}_u^\dagger\mathbf{y}_u+\mathbf{y}_d^\dagger\mathbf{y}_d-\sum_ic_ig_i^2\Big)\\
\beta_{\mathbf{y}_d}=&\mathbf{y}_d\Big(\Tr(3\mathbf{y}_d\mathbf{y}_d^\dagger+\mathbf{y}_e\mathbf{y}_e^\dagger)+3\mathbf{y}_d^\dagger\mathbf{y}_d+\mathbf{y}_u^\dagger\mathbf{y}_u-\sum_ic_i'g_i^2\Big)\\
\beta_{\mathbf{y}_e}=&\mathbf{y}_e\Big(\Tr(3\mathbf{y}_d\mathbf{y}_d^\dagger+\mathbf{y}_e\mathbf{y}_e^\dagger)+3\mathbf{y}_e^\dagger\mathbf{y}_e-\sum_ic_i''g_i^2\Big)\\
\beta_{\mathbf{a}_u}=&-\sum_ic_ig_i^2\mathbf{a}_u+2\sum_ic_ig_i^2M_i\mathbf{y}_u+4\mathbf{y}_u\mathbf{y}_u^\dagger\mathbf{a}_u+6\Tr(\mathbf{a}_u\mathbf{y}_u^\dagger)\mathbf{y}_u\\
&+5\mathbf{a}_u\mathbf{y}_u^\dagger\mathbf{y}_u+3\Tr(\mathbf{y}_u\mathbf{y}_u^\dagger)\mathbf{a}_u+2\mathbf{y}_u\mathbf{y}_d^\dagger\mathbf{a}_d+\mathbf{a}_u\mathbf{y}_d^\dagger\mathbf{y}_d\nonumber\\
\beta_{\mathbf{a}_d}=&-\sum_ic_i'g_i^2\mathbf{a}_d+2\sum_ic_i'g_i^2M_i\mathbf{y}_d+4\mathbf{y}_d\mathbf{y}_d^\dagger\mathbf{a}_d+6\Tr(\mathbf{a}_d\mathbf{y}_d^\dagger)\mathbf{y}_d+5\mathbf{a}_d\mathbf{y}_d^\dagger\mathbf{y}_d\\
&+3\Tr(\mathbf{y}_d\mathbf{y}_d^\dagger)\mathbf{a}_d
+2\mathbf{y}_d\mathbf{y}_u^\dagger\mathbf{a}_u+\mathbf{a}_d\mathbf{y}_u^\dagger\mathbf{y}_u+2\Tr(\mathbf{a}_e\mathbf{y}_e^\dagger)\mathbf{y}_d+\Tr(\mathbf{y}_e\mathbf{y}_e^\dagger)\mathbf{a}_d\nonumber\\
\beta_{\mathbf{a}_e}=&-\sum_ic_i''g_i^2\mathbf{a}_e+2\sum_ic_i''g_i^2M_i\mathbf{y}_e+4\mathbf{y}_e\mathbf{y}_e^\dagger\mathbf{a}_e+2\Tr(\mathbf{a}_e\mathbf{y}_e^\dagger)\mathbf{y}_e\\
&+5\mathbf{a}_e\mathbf{y}_e^\dagger\mathbf{y}_e+\Tr(\mathbf{y}_e\mathbf{y}_e^\dagger)\mathbf{a}_e+6\Tr(\mathbf{a}_d\mathbf{y}_d^\dagger)\mathbf{y}_e+3\Tr(\mathbf{y}_d\mathbf{y}_d^\dagger)\mathbf{a}_e\nonumber\\
\beta_{b}=&\left(\frac{6}{5}g_1^2M_1+6g_2^2M_2+\Tr(6\mathbf{y}_u^\dagger\mathbf{a}_u+6\mathbf{y}_d^\dagger\mathbf{a}_d+2\mathbf{y}_e^\dagger\mathbf{a}_e)\right)\mu\\
&+\left(-\frac{3}{5}g_1^2-3g_2^2+\Tr(3\mathbf{y}_u\mathbf{y}_u^\dagger+3\mathbf{y}_d\mathbf{y}_d^\dagger+\mathbf{y}_e\mathbf{y}_e^\dagger)\right)b\nonumber\\
\beta_{\mu}=&\mu\Big(-\frac{3}{5}g_1^2-3g_2^2+\Tr(3\mathbf{y}_u\mathbf{y}_u^\dagger+3\mathbf{y}_d\mathbf{y}_d^\dagger+\mathbf{y}_e\mathbf{y}_e^\dagger)\Big)\\
\beta_{m_{H_1}^2}=&-\frac{6}{5}g_1^2|M_1|^2-6g_2^2|M_2|^2+\Tr\left(6(m_{H_1}^2+\mathbf{m}_{\tilde Q}^2)\mathbf{y}_d^\dagger\mathbf{y}_d+2(m_{H_1}^2+\mathbf{m}_{\tilde L}^2)\mathbf{y}_e^\dagger\mathbf{y}_e\right)\\
&+\Tr\left(6\mathbf{y}_d^\dagger\mathbf{m}_{\tilde{\bar d}}^2\mathbf{y}_d+2\mathbf{y}_e^\dagger\mathbf{m}_{\tilde{\bar e}}^2\mathbf{y}_e+6\mathbf{a}_d^\dagger\mathbf{a}_d+2\mathbf{a}_e^\dagger\mathbf{a}_e\right)-\frac{3}{5}g_1^2\mathcal{S}\nonumber\\
\beta_{m_{H_2}^2}=&-\frac{6}{5}g_1^2|M_1|^2-6g_2^2|M_2|^2+6\Tr\left((m_{H_2}^2+\mathbf{m}_{\tilde Q}^2)\mathbf{y}_u^\dagger\mathbf{y}_u+\mathbf{y}_u^\dagger\mathbf{m}_{\tilde{\bar u}}^2\mathbf{y}_u+\mathbf{a}_u^\dagger\mathbf{a}_u\right)+\frac{3}{5}g_1^2\mathcal{S}\\
\beta_{\mathbf{m}_{\tilde Q}^2}=&-\frac{2}{15}g_1^2|M_1|^2-6g_2^2|M_2|^2-\frac{32}{3}g_3^2|M_3|^2+2\mathbf{y}_u^\dagger\mathbf{m}_{\tilde{\bar u}}^2\mathbf{y}_u+2\mathbf{y}_d^\dagger\mathbf{m}_{\tilde{\bar d}}^2\mathbf{y}_d+2\mathbf{a}_u^\dagger\mathbf{a}_u+2\mathbf{a}_d^\dagger\mathbf{a}_d\\
&+(\mathbf{m}_{\tilde Q}^2+2m_{H_2}^2)\mathbf{y}_u^\dagger\mathbf{y}_u+(\mathbf{m}_{\tilde Q}^2+2m_{H_1}^2)\mathbf{y}_d^\dagger\mathbf{y}_d+(\mathbf{y}_u^\dagger\mathbf{y}_u+\mathbf{y}_d^\dagger\mathbf{y}_d)\mathbf{m}_{\tilde Q}^2+\frac{1}{5}g_1^2\mathcal{S}\nonumber\\
\beta_{\mathbf{m}_{\tilde{\bar u}}^2}=&-\frac{32}{15}g_1^2|M_1|^2-\frac{32}{3}g_3^2|M_3|^2+(2\mathbf{m}_{\tilde{\bar u}}^2+4m_{H_2}^2)\mathbf{y}_u\mathbf{y}_u^\dagger\\
&+4\mathbf{y}_u\mathbf{m}_{\tilde Q}^2\mathbf{y}_u^\dagger+2\mathbf{y}_u\mathbf{y}_u^\dagger\mathbf{m}_{\tilde{\bar u}}^2+4\mathbf{a}_u\mathbf{a}_u^\dagger-\frac{4}{5}g_1^2\mathcal{S}\nonumber\\
\beta_{\mathbf{m}_{\tilde{\bar d}}^2}=&-\frac{8}{15}g_1^2|M_1|^2-\frac{32}{3}g_3^2|M_3|^2+(2\mathbf{m}_{\tilde{\bar d}}^2+4m_{H_1}^2)\mathbf{y}_d\mathbf{y}_d^\dagger\\
&+4\mathbf{y}_d\mathbf{m}_{\tilde Q}^2\mathbf{y}_d^\dagger+2\mathbf{y}_d\mathbf{y}_d^\dagger\mathbf{m}_{\tilde{\bar d}}^2+4\mathbf{a}_d\mathbf{a}_d^\dagger+\frac{2}{5}g_1^2\mathcal{S}\\
\beta_{\mathbf{m}_{\tilde L}^2}=&-\frac{6}{5}g_1^2|M_1|^2-6g_2^2|M_2|^2+(\mathbf{m}_{\tilde L}^2+2m_{H_1}^2)\mathbf{y}_e^\dagger\mathbf{y}_e+2\mathbf{y}_e^\dagger\mathbf{m}_{\tilde{\bar e}}^2\mathbf{y}_e+\mathbf{y}_e^\dagger\mathbf{y}_e\mathbf{m}_{\tilde L}^2+2\mathbf{a}_e^\dagger\mathbf{a}_e-\frac{3}{5}g_1^2\mathcal{S}\\
\beta_{\mathbf{m}_{\tilde{\bar e}}^2}=&-\frac{24}{5}g_1^2|M_1|^2+(2\mathbf{m}_{\tilde{\bar e}}^2+4m_{H_1}^2)\mathbf{y}_e\mathbf{y}_e^\dagger+4\mathbf{y}_e\mathbf{m}_{\tilde L}^2\mathbf{y}_e^\dagger+2\mathbf{y}_e\mathbf{y}_e^\dagger\mathbf{m}_{\tilde{\bar e}}^2+4\mathbf{a}_e\mathbf{a}_e^\dagger+\frac{6}{5}g_1^2\mathcal{S}\label{RGE-leptonR}
\end{align}}

Since these equations are not very enlightening, we give the leading order contributions for the case of mSUGRA (section \ref{s:msugra}). We use the following approximations:
\begin{itemize}
\item The soft-breaking mass matrices are assumed to be diagonal.
\item The trilinear couplings are taken proportional to the Yukawa couplings: $\mathbf{a}_u=A_u\mathbf{y}_u$, $\mathbf{a}_d=A_d\mathbf{y}_d$ and $\mathbf{a}_e=A_e\mathbf{y}_e$.
\item Since the corresponding Standard Model particles are so much heavier, only the Yukawa couplings of the third generation $y_t,y_b,y_\tau$ are considered significant.
\item If all masses in equation \eq{S-U1trace} are equal, then $\mathcal{S}=0$. From equation \eq{S-U1trace} and the $\beta$-functions it follows that $d\mathcal{S}/dt\propto\mathcal{S}$. That means that if the masses are unified at some point, $\mathcal{S}$ will vanish at any energy scale. Since mSUGRA assumes this unification, we take $\mathcal{S}=0$.
\end{itemize}
Due to the approximation that only the Yukawa couplings of the third generation are significant, the result for the heaviest generation is different from the two lighter generations. If the difference is not clear from the names of the particles, the light generations will be denoted by an $\ell$ and the heavy generations by an $\mathpzc{h}$. For instance $\tilde Q_\mathpzc{h}$ denotes the top-bottom squark doublet. To obtain the RGEs for the trilinear couplings $A_i$, we have to separate them from the Yukawa couplings. For instance, for the stop trilinear coupling, we get:
\begin{equation}
\left[\frac{d\mathbf{a}_u}{dt}\right]_{tt}=\frac{dA_ty_t}{dt}=\frac{dA_t}{dt}y_t+\frac{dy_t}{dt}A_t\label{partialdertrilinear}\end{equation}
Finally, the following combinations will be useful:
\begin{align}
X_t&=m_{\tilde Q_\mathpzc{h}}^2+m_{\tilde{\bar t}}^2+m_{H_2}^2+A_t^2\label{X_t}\\
X_b&=m_{\tilde Q_\mathpzc{h}}^2+m_{\tilde{\bar b}}^2+m_{H_1}^2+A_b^2\label{X_b}\\
X_\tau&=m_{\tilde L_\mathpzc{h}}^2+m_{\tilde{\bar\tau}}^2+m_{H_1}^2+A_\tau^2\label{X_tau}
\end{align}
Then the result is:
\allowdisplaybreaks{
\begin{align}
\beta_{y_t}&\approx y_t\Big(6|y_t|^2+|y_b|^2-\sum_i c_ig_i^2\Big)\label{RGEyt}\\
\beta_{y_{u,c}}&\approx y_{u,c}\Big(3|y_t|^2-\sum_i c_ig_i^2\Big)\label{RGEyu}\\
\beta_{y_b}&\approx y_b\Big(|y_t|^2+6|y_b|^2+|y_\tau|^2-\sum_i c_i'g_i^2\Big)\label{RGEyb}\\
\beta_{y_{d,s}}&\approx y_{d,s}\Big(3|y_b|^2+|y_\tau|^2-\sum_i c_i'g_i^2\Big)\label{RGEyd}\\
\beta_{y_\tau}&\approx y_\tau\Big(3|y_b|^2+4|y_\tau|^2-\sum_ic_i''g_i^2\Big)\label{RGEytau}\\
\beta_{y_{e,\mu}}&\approx y_{e,\mu}\Big(3|y_b|^2+|y_\tau|^2-\sum_ic_i''g_i^2\Big)\label{RGEye}\\
\beta_{A_t}&=2\sum_ic_ig_i^2M_i+12|y_t|^2A_t+2|y_b|^2A_b\label{RGEAt}\\
\beta_{A_{u,c}}&\approx2\sum_ic_ig_i^2M_i+6|y_t|^2A_t\label{RGEAu}\\
\beta_{A_b}&\approx2\sum_ic_i'g_i^2M_i+12|y_b|^2A_b+2|y_t|^2A_t+2|y_\tau|^2A_\tau\label{RGEAb}\\
\beta_{A_{d,s}}&\approx2\sum_ic_i'g_i^2M_i+6|y_b|^2A_b+2|y_\tau|^2A_\tau\label{RGEAd}\\
\beta_{A_\tau}&\approx2\sum_ic_i''g_i^2M_i+6|y_b|^2A_b+8|y_\tau|^2A_\tau\label{RGEAtau}\\
\beta_{A_{e,\mu}}&\approx2\sum_ic_i''g_i^2M_i+6|y_b|^2A_b+2|y_\tau|^2A_\tau\label{RGEAe}\\
\beta_{b}&\approx \mu\Big(\frac{6}{5}g_1^2M_1+6g_2^2M_2+6|y_t|^2A_t+6|y_b|^2A_b+2|y_\tau|^2A_\tau\Big)\label{RGEB}\\
&+b\Big(-\frac{3}{5}g_1^2-3g_2^2+3|y_t|^2+3|y_b|^2+|y_\tau|^2\Big)\nonumber\\
\beta_{\mu}&\approx\mu\Big(-\frac{3}{5}g_1^2-3g_2^2+3|y_t|^2+3|y_b|^2+|y_\tau|^2\Big)\label{RGEmu}\\
\beta_{m_{H_1}^2}&\approx-\frac{6}{5}g_1^2|M_1|^2-6g_2^2|M_2|^2+6|y_b|^2X_b+2|y_\tau|^2X_\tau\label{RGEHiggs1}\\
\beta_{m_{H_2}^2}&\approx-\frac{6}{5}g_1^2|M_1|^2-6g_2^2|M_2|^2+6|y_t|^2X_t\label{RGEHiggs2}\\
\beta_{m_{\tilde Q_\mathpzc{h}}^2}&\approx-\frac{2}{15}g_1^2|M_1|^2-6g_2^2|M_2|^2-\frac{32}{3}g_3^2|M_3|^2+2|y_t|^2X_t+2|y_b|^2X_b\label{RGEQh}\\
\beta_{m_{\tilde Q_\ell}^2}&\approx-\frac{2}{15}g_1^2|M_1|^2-6g_2^2|M_2|^2-\frac{32}{3}g_3^2|M_3|^2\label{RGEQl}\\
\beta_{m_{\tilde{\bar t}}^2}&\approx-\frac{32}{15}g_1^2|M_1|^2-\frac{32}{3}g_3^2|M_3|^2+4|y_t|^2X_t\label{RGEt}\\
\beta_{m_{\tilde{\bar u},\tilde{\bar c}}^2}&\approx-\frac{32}{15}g_1^2|M_1|^2-\frac{32}{3}g_3^2|M_3|^2\label{RGEu}\\
\beta_{m_{\tilde{\bar b}}^2}&\approx-\frac{8}{15}g_1^2|M_1|^2-\frac{32}{3}g_3^2|M_3|^2+4|y_b|^2X_b\label{RGEb}\\
\beta_{m_{\tilde{\bar d},\tilde{\bar s}}^2}&\approx-\frac{8}{15}g_1^2|M_1|^2-\frac{32}{3}g_3^2|M_3|^2\label{RGEd}\\
\beta_{m_{\tilde L_\mathpzc{h}}^2}&\approx-\frac{6}{5}g_1^2|M_1|^2-6g_2^2|M_2|^2+2|y_\tau|^2X_\tau\label{RGELh}\\
\beta_{m_{\tilde L_\ell}^2}&\approx-\frac{6}{5}g_1^2|M_1|^2-6g_2^2|M_2|^2\label{RGELl}\\
\beta_{m_{\tilde{\bar \tau}}^2}&\approx-\frac{24}{5}g_1^2|M_1|^2+4|y_\tau|^2X_\tau\label{RGEtau}\\
\beta_{m_{\tilde{\bar e},\tilde{\bar\mu}}^2}&\approx-\frac{24}{5}g_1^2|M_1|^2\label{RGEe}
\end{align}}

\section{Pythia Event Record}\label{app:pythia}

An example of a summary of an event generated by Pythia. Only part of the event is shown, since the complete record in this case contains 1205 lines. The first column is the line number of the particle and the second column its name. The third line contains the particle's status code, that displays if a particle is part of the event summary (status code 21), has decayed (status code 11 and 12 in this record) or is part of the final state (status code 1). The letters `A' and `I' are for internal use of the program. The `KF' column contains the particle `PDG' code as defined by the Particle Data Group. The `orig' column contains the line number of the mother of the particle. The other columns are the momentum in $x$, $y$ and $z$ direction, the energy and the mass in GeV. For brevity, only the summary event listing is shown. The complete event record also contains other information on the particles such as the line numbers of daughters of particles.

\begin{verbatim}

                            Event listing (summary)

    I particle/jet KS     KF  orig    p_x      p_y      p_z       E        m

    1 !p+!         21    2212    0    0.000    0.000 7000.000 7000.000    0.938
    2 !p+!         21    2212    0    0.000    0.000-7000.000 7000.000    0.938
 ==============================================================================
    3 !d!          21       1    1   -1.430    0.413 4410.288 4410.288    0.000
    4 !d!          21       1    2    1.139    0.838-2920.576 2920.576    0.000
    5 !g!          21      21    3   28.978 -132.907  991.344 1000.633    0.000
    6 !g!          21      21    4  516.567   50.851 -585.399  782.381    0.000
    7 !~g!         21 1000021    0  519.815   12.491  529.544 1044.688  735.249
    8 !~g!         21 1000021    0   25.730  -94.548 -123.600  738.326  721.282
    9 !~b_1!       21 1000005    7  330.757   50.445  322.928  762.027  603.704
   10 !bbar!       21      -5    7  189.058  -37.954  206.617  282.661    4.800
   11 !~b_2bar!    21-2000005    8   -5.926 -147.829 -103.563  673.015  648.332
   12 !b!          21       5    8   31.657   53.282  -20.036   65.311    4.800
   13 !~chi_1-!    21-1000024    9   97.658  153.547  -32.593  289.915  223.326
   14 !t!          21       6    9  233.324 -103.206  355.786  472.316  177.220
   15 !~chi_10!    21 1000022   11 -282.270   13.942 -163.130  347.120  118.366
   16 !bbar!       21      -5   11  277.228 -159.834   59.355  325.497    4.800
   17 !~tau_1-!    21 1000015   13    7.085   97.212  -11.949  181.272  152.369
   18 !nu_taubar!  21     -16   13   90.572   56.336  -20.644  108.643    0.000
   19 !W+!         21      24   14  236.125 -122.559  281.406  395.259   79.145
   20 !b!          21       5   14   -2.802   19.353   74.380   77.057    4.800
   21 !~chi_10!    21 1000022   17   18.788   95.970  -33.043  157.052  118.366
   22 !tau-!       21      15   17  -11.702    1.241   21.094   24.219    1.777
   23 !sbar!       21      -3   19  180.995  -83.323  157.519  253.996    0.500
   24 !c!          21       4   19   53.498  -38.327  122.209  138.810    1.500
 ==============================================================================
   25 (~chi_1-)    11-1000024   13   97.658  153.547  -32.593  289.915  223.326
   26 ~chi_10       1 1000022   15 -270.976   12.999 -156.868  334.985  118.366
   27 (~tau_1-)    11 1000015   17    7.085   97.212  -11.949  181.272  152.369
   28 nu_taubar     1     -16   18   90.572   56.336  -20.644  108.643    0.000
   29 (W+)         11      24   19  234.493 -121.650  279.728  392.806   79.145
   30 ~chi_10       1 1000022   21   18.788   95.970  -33.043  157.052  118.366
   31 (tau-)       11      15   22  -11.702    1.241   21.094   24.219    1.777
   32 (s)       A  12       3    4  -59.584  -29.032 -146.611  160.898    0.500
   33 (g)       I  12      21    4  -32.513  -19.990  -85.737   93.848    0.000
   34 (g)       I  12      21    4  -10.387   -3.936  -25.949   28.227    0.000
   35 (g)       I  12      21    4   -0.597    0.293   -0.485    0.824    0.000
  169 nu_mubar      1     -14   31   -3.416    0.871    6.407    7.313    0.000
  170 mu-           1      13   31   -4.481    0.548    8.877    9.960    0.106
  171 nu_tau        1      16   31   -3.805   -0.177    5.809    6.947    0.000
  172 (string)     11      92   32 -104.154  239.546 3444.034 4265.239 2502.532
  173 (K*bar0)     11    -313  172  -18.651   -9.271  -46.248   50.729    0.874
  174 (pi0)        11     111  172   -6.162   -2.771  -15.625   17.023    0.135
  175 (rho-)       11    -213  172  -20.572  -11.217  -51.591   56.668    0.753
  176 (K*+)        11     323  172  -17.056   -9.103  -43.189   47.328    0.932
  177 (Kbar0)      11    -311  172  -13.460   -6.293  -32.427   35.673    0.498
  178 pi-           1    -211  172   -3.335   -1.460   -8.023    8.811    0.140
  179 (rho+)       11     213  172  -13.550   -8.094  -37.453   40.652    0.845
  180 (rho-)       11    -213  172   -5.265   -3.005  -12.388   13.815    0.805
  181 pi+           1     211  172   -2.236   -0.912   -5.890    6.367    0.140
  182 pi-           1    -211  172    0.053    0.005   -0.105    0.183    0.140
  183 (pi0)        11     111  172   -2.262   -0.931   -5.372    5.904    0.135
  184 pi+           1     211  172   -0.312    0.207   -0.957    1.037    0.140
  185 (eta)        11     221  172   -0.007   -0.447   -1.226    1.415    0.547
  186 (Lambda0)    11    3122  172    0.722   -0.063   -6.301    6.440    1.116
  187 (Sigmabar-)  11   -3222  172    0.870    0.584   -5.863    6.074    1.189
  188 (omega)      11     223  172    1.004    0.237   -6.381    6.511    0.779
  189 pi+           1     211  172    0.875   -0.910   -7.929    8.030    0.140
  190 (pi0)        11     111  172    1.024   -0.270   -3.165    3.340    0.135
  191 (K*0)        11     313  172    2.387    0.768  -15.735   15.967    1.043
  192 (K*bar0)     11    -313  172    2.464   -0.593  -14.186   14.434    0.824
  193 (pi0)        11     111  172    0.887    0.519   -1.754    2.037    0.135
  194 (rho0)       11     113  172    0.479    0.401   -0.576    1.149    0.773
 1188 K-            1    -321 1150   -0.200    0.572    1.585    1.767    0.494
 1189 pi+           1     211 1150   -0.106    0.126    2.176    2.187    0.140
 1190 gamma         1      22 1154    4.757   -2.839   10.941   12.263    0.000
 1191 gamma         1      22 1154    0.040   -0.020    0.066    0.080    0.000
 1192 gamma         1      22 1172   10.222   -1.995   10.844   15.035    0.000
 1193 gamma         1      22 1172   10.993   -2.050   11.793   16.252    0.000
 1194 gamma         1      22 1174    5.870   -0.939    6.197    8.587    0.000
 1195 gamma         1      22 1174    5.351   -0.813    5.818    7.946    0.000
 1196 (K_S0)       11     310 1175    1.898    3.716   -1.139    4.354    0.498
 1197 (K_S0)       11     310 1186   -0.765    0.634    2.108    2.383    0.498
 1198 pi+           1     211 1196    0.661    1.053   -0.194    1.266    0.140
 1199 pi-           1    -211 1196    1.237    2.663   -0.946    3.088    0.140
 1200 (pi0)        11     111 1197   -0.066    0.196    0.282    0.375    0.135
 1201 (pi0)        11     111 1197   -0.699    0.438    1.826    2.008    0.135
 1202 gamma         1      22 1200   -0.006    0.002   -0.014    0.016    0.000
 1203 gamma         1      22 1200   -0.061    0.194    0.297    0.360    0.000
 1204 gamma         1      22 1201   -0.683    0.422    1.811    1.981    0.000
 1205 gamma         1      22 1201   -0.016    0.016    0.015    0.027    0.000
 ==============================================================================
                   sum:  2.00          0.00     0.00     0.00 14000.00 14000.00
		   \end{verbatim}

\section{Particles Produced in the First Interaction}\label{app-firstint}

These are all the particles produced in the first interaction of 20.000 SUSY events of benchmark point SU3 on a logarithmic scale.
\begin{figure}[!h]
\centering{
\includegraphics[angle=90,width=0.88\linewidth]{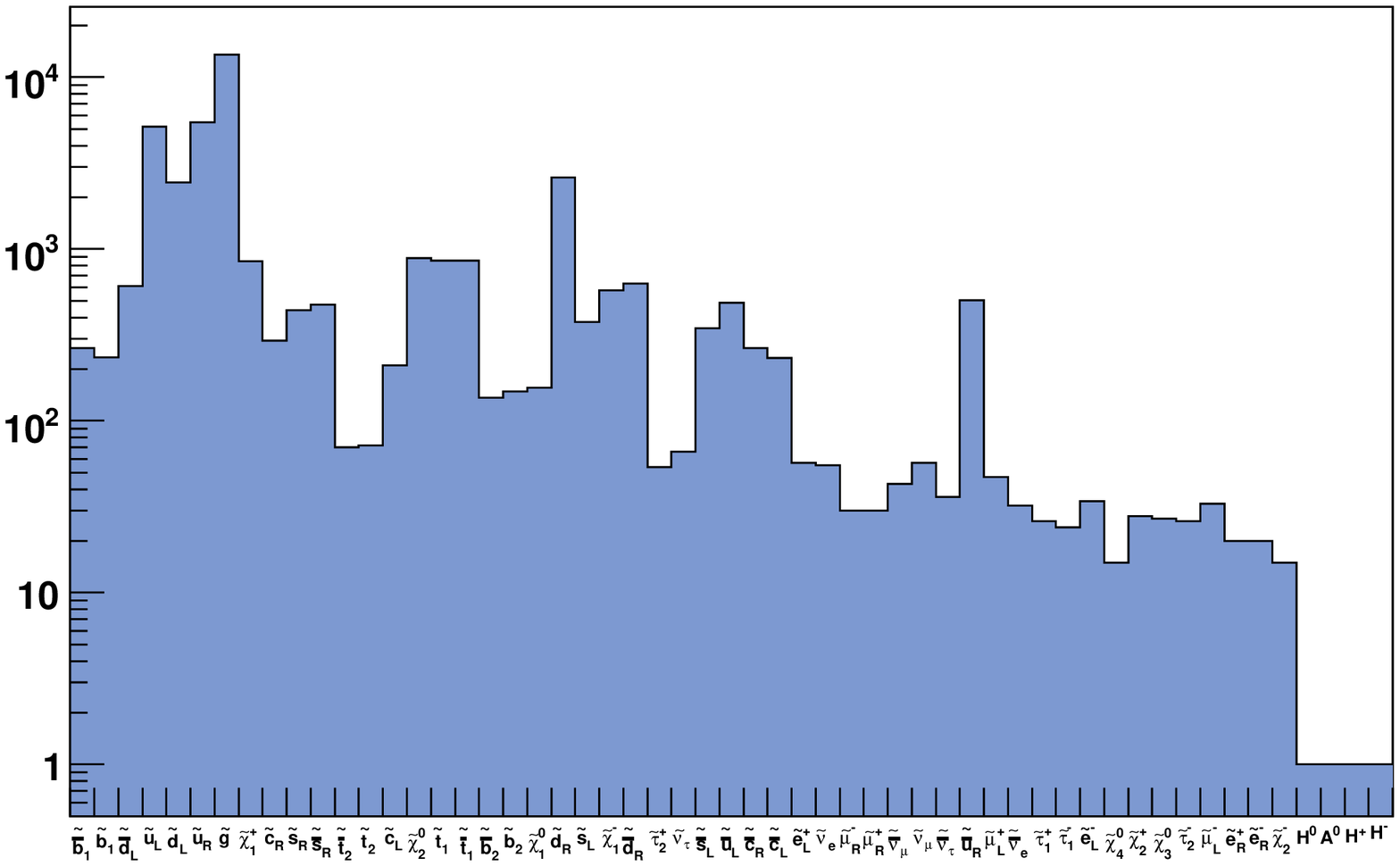}}
\end{figure}
\section{Production and Decay Modes}\label{app:processes}

This appendix lists possible decays and production processes of the supersymmetric particles studied in section \ref{s:phenomenology}. Of course all particles can be produced in the first interaction as well. 

The decay modes listed are the most important contribution for most points in mSUGRA parameter space. In some parts of the parameter space none of these decays are allowed. In that case particles decay through channels that are suppressed in other points of parameter space. An example is the three-body decay $\tilde\chi_2^0\to\tilde\chi_1^0u\bar u$. Other decays mediated by loop corrections can also become important if the normal decay modes are not kinematically accessible. For instance, $\tilde t_1\to\tilde\chi_1^0 c$ is the most important decay mode for the stop if the stop cannot decay to a $b$ quark. Such decays are not listed here. Possible decays of off-shell particles ($\gamma$, $Z^0$ or Higgs radiation) are not listed either.

Finally, there are other decays that are suppressed but occasionally occur. An example is the decay $\tilde\chi_2^0\to\tilde e_R^+e^-$, which would not be allowed if $\tilde\chi_2^0$ were a pure $SU(2)$ gaugino, which only couples to left-handed sfermions. However, $\tilde\chi_2^0$ is only approximately an $SU(2)$ gaugino, since it contains Bino and Higgsino components as well. Also, due to mixing the actual sfermion mass eigenstate contains a small left-handed component. Since these effects are usually small, we have not listed the decays. The only exception is the third generation sfermions, where mixing is so strong that it is impossible to define left-handed and right-handed states.

\begin{table}[!h]
\begin{minipage}[t]{0.3\linewidth}
\centering{
$\begin{array}{|l|}
\hline
\tilde\tau_1^-\to\tilde\chi_1^-\nu_\tau\\
\tilde\tau_1^+\to\tilde\chi_1^+\bar\nu_\tau\\
\tilde\tau_1^\pm\to\tilde\chi_i^0\tau^\pm\\
\hline
\end{array}$
\caption{Possible decays of light staus}\label{t:staudecay}}
\end{minipage}
\hspace{0.1\linewidth}
\begin{minipage}[t]{0.6\linewidth}
\centering{
$\begin{array}{|l|l|l|l|}
\hline
\tilde\chi_i^+ \to \tilde\tau_1^+ \nu_\tau &
\tilde\chi_i^- \to \tilde\tau_1^- \bar\nu_\tau&
\tilde\chi_i^0 \to \tilde\tau_1^\pm \tau^\mp&
A^0 \to \tilde\tau_1^\pm \tilde\tau_2^\mp\\
H^0 \to \tilde\tau_1^\pm \tilde\tau_1^\mp&
H^0 \to \tilde\tau_1^\pm \tilde\tau_2^\mp&
H^+ \to \tilde\nu_\tau \tilde\tau_1^+&
H^- \to \tilde{\bar\nu}_\tau \tilde\tau_1^-\\
\tilde\tau_2^\pm \to \tilde\tau_1^\pm Z&
\tilde\tau_2^\pm \to \tilde\tau_1^\pm h^0&
\tilde\nu_\tau \to \tilde\tau_1^- W^+&
\tilde{\bar\nu}_\tau \to \tilde\tau_1^+ W^-\\
\hline
\end{array}$
\caption{Possible production processes for light staus.}\label{t:staufrom}}
\end{minipage}
\end{table}

\begin{table}[!h]
\centering{
$\begin{array}{|l|l|l|l|}
\hline
\tilde\chi_1^+\to\tilde e_L^+\nu_e&
\tilde\chi_1^-\to\tilde e_L^-\bar\nu_e&
\tilde\chi_2^0\to\tilde e_L^\mp e^\pm&
\tilde\chi_2^0\to\tilde \mu_L^\mp \mu^\pm\\
\tilde\chi_1^+\to\tilde\nu_e e^+&
\tilde\chi_1^-\to\tilde{\bar\nu}_e e^-&
\tilde\chi_2^0\to\tilde\nu_e\bar\nu_e&
\tilde\chi_2^0\to\tilde{\bar\nu}_e\nu_e\\
\tilde\chi_1^+\to\tilde \mu_L^+\nu_\mu&
\tilde\chi_1^-\to\tilde \mu_L^-\bar\nu_\mu&
\tilde\chi_2^0\to\tilde\nu_\mu\bar\nu_\mu&
\tilde\chi_2^0\to\tilde{\bar\nu}_\mu\nu_\mu\\
\tilde\chi_1^+\to\tilde\nu_\mu \mu^+&
\tilde\chi_1^-\to\tilde{\bar\nu}_\mu \mu^-&
\tilde\chi_2^0\to\tilde\nu_\tau\bar\nu_\tau&
\tilde\chi_2^0\to\tilde{\bar\nu}_\tau\nu_\tau\\
\tilde\chi_1^+\to\tilde\tau_1^+\nu_\tau&
\tilde\chi_1^-\to\tilde\tau_1^-\bar\nu_\tau&
\tilde\chi_2^0\to\tilde\tau_1^\mp\tau^\pm&
\tilde\chi_2^0\to\tilde\tau_2^\mp\tau^\pm\\
\tilde\chi_1^+\to\tilde\tau_2^+\nu_\tau&
\tilde\chi_1^-\to\tilde\tau_2^-\bar\nu_\tau&
\tilde\chi_2^0\to\tilde\chi_1^0 Z^0&
\tilde\chi_2^0\to\tilde\chi_1^0 h^0\\
\tilde\chi_1^+\to\tilde\nu_\tau \tau^+&
\tilde\chi_1^-\to\tilde{\bar\nu}_\tau \tau^-&
&
\\
\tilde\chi_1^+\to\tilde\chi_1^0 W^+&
\tilde\chi_1^-\to\tilde\chi_1^0 W^-&
&\\\hline
\end{array}$
\caption{Possible decays of $\tilde\chi_1^\pm$ and $\tilde\chi_2^0$}\label{t:gauginodecay}}
\end{table}

\begin{table}[!h]
\begin{minipage}[b]{0.5\linewidth}
\centering{
$\begin{array}{|l|}
\hline
\tilde t_2\to \tilde t_1 Z^0\\
\tilde t_2\to \tilde t_1 h^0\\
\tilde b_i\to\tilde t_1W^-\\
\tilde g\to\tilde t_1 \bar t\\
\hline
\end{array}$
\caption{Possible production processes for light stops}\label{t:stopproduction}}
\end{minipage}
\begin{minipage}[b]{0.5\linewidth}
\centering{
$\begin{array}{|l|}
\hline
\tilde t_1\to\tilde\chi_i^0t\\
\tilde t_1\to\tilde\chi_i^+b\\
\hline
\end{array}$
\caption{Possible decays of the light stop}\label{t:stopdecay}}
\end{minipage}
\end{table}

The production and decay channels for the anti-stop are not listed, since they follow trivially from the processes for the stop.

\section{Gaugino Decays}\label{app:charginodecay}

These are the individual plots on which figures \ref{f:chargino-region10} and \ref{f:neutralino-region10} are based. In the cyan region, a specific decay is kinematically allowed. In the dark-blue region it is not allowed. Similar decay channels of charginos and neutralinos are shown in the same plot. The kinematical boundaries in such combined plots are not exactly at the same point in parameter space, as can be seen in for instance the $\tilde\chi_2^0\to\tilde\chi_1^0Z^0$ and $\tilde\chi_1^+\to\tilde\chi_1^0W^+$ plot.
\begin{figure}[!h]
\includegraphics[width=\linewidth]{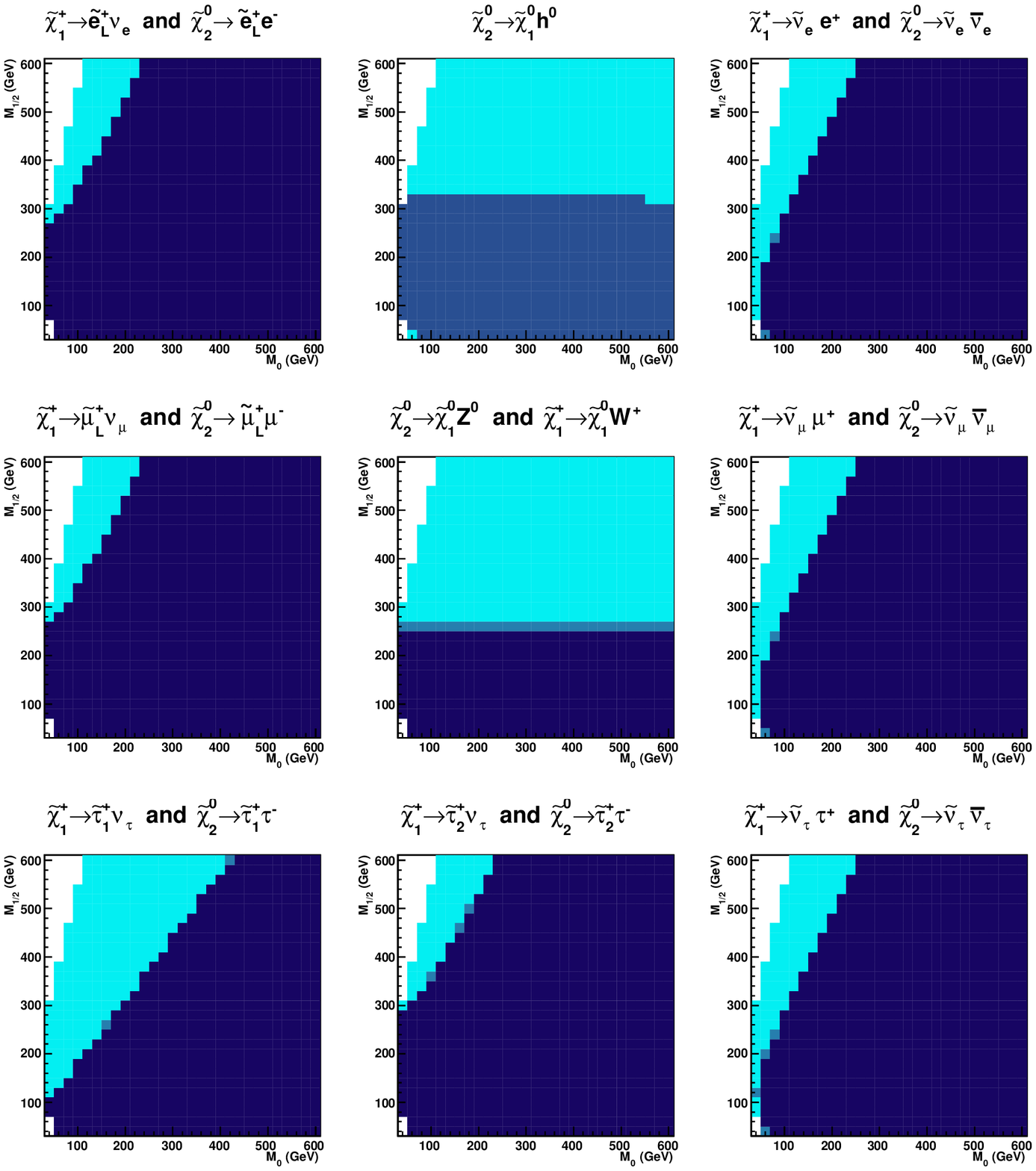}
\end{figure}
\section{Stop Production and Decay}\label{app:stopprocesses}

These are the individual plots for stop production and decay modes. In the cyan regions, a particular decay is allowed. In the dark blue regions it is not kinematically accessible.

\begin{figure}[!h]
\centering{
\includegraphics[width=0.89\linewidth]{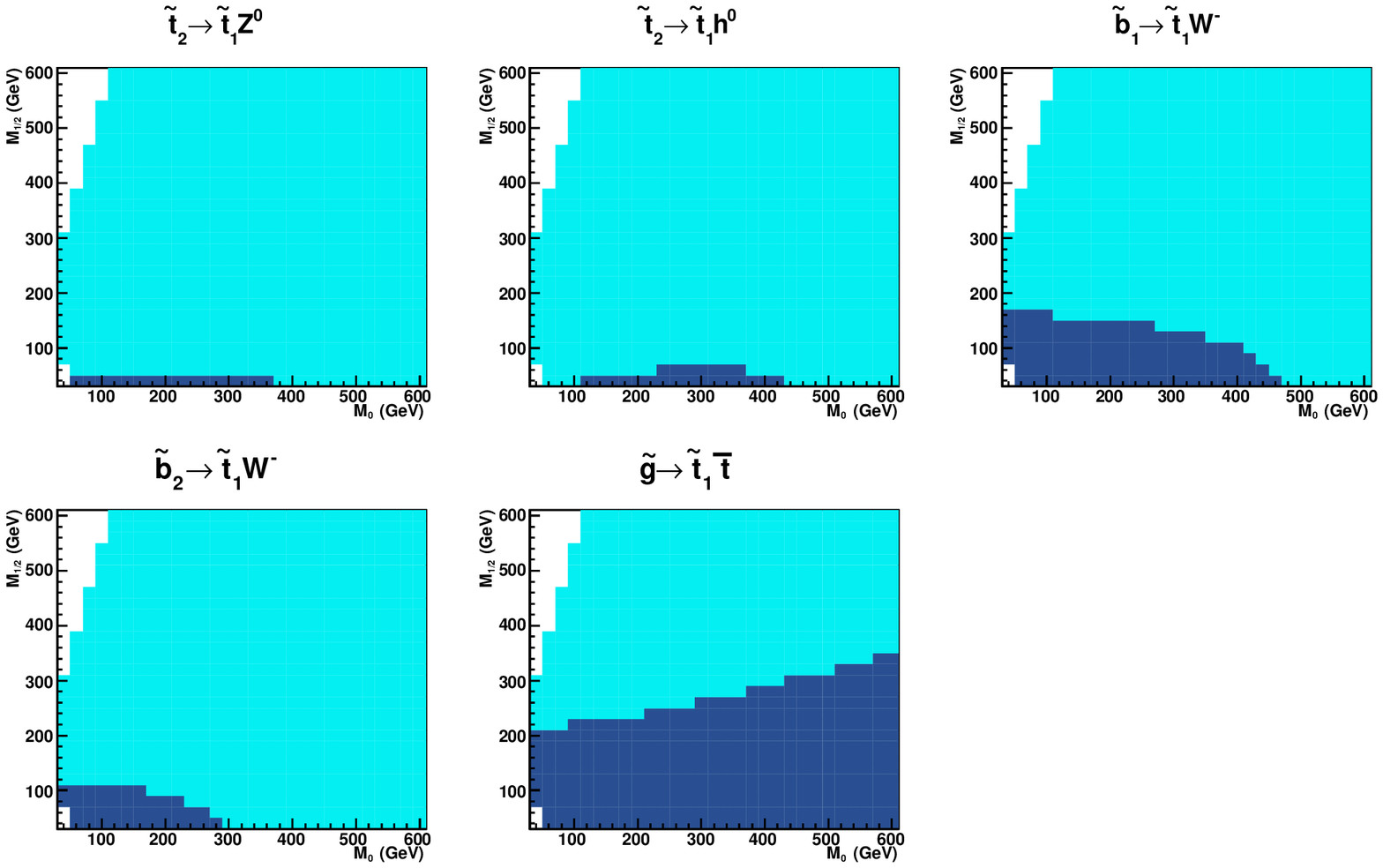}}
\end{figure}

\begin{figure}[!h]
\centering{
\includegraphics[width=0.89\linewidth]{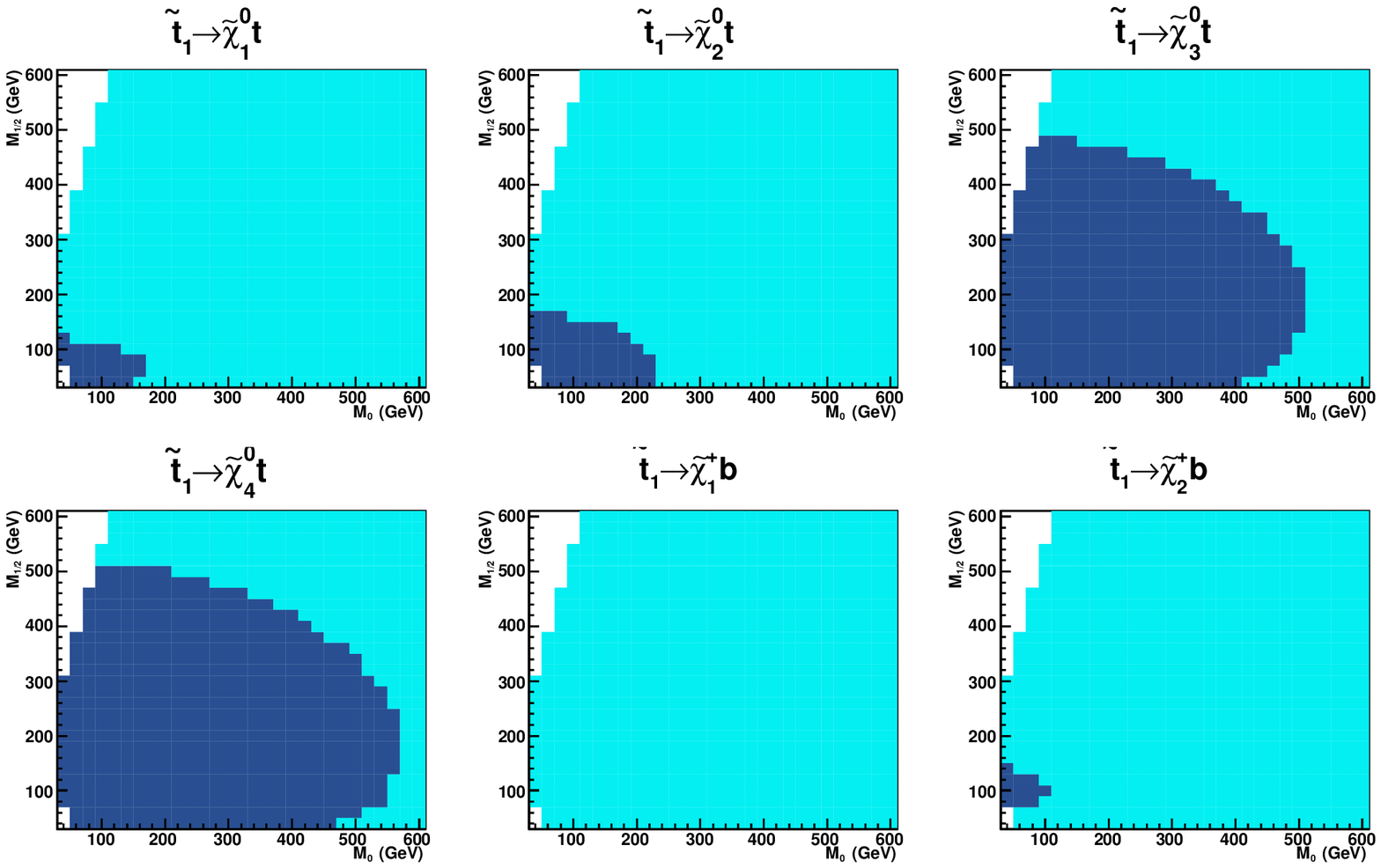}}
\end{figure}

\begin{figure}[!h]
\centering{
\includegraphics[width=0.9\linewidth]{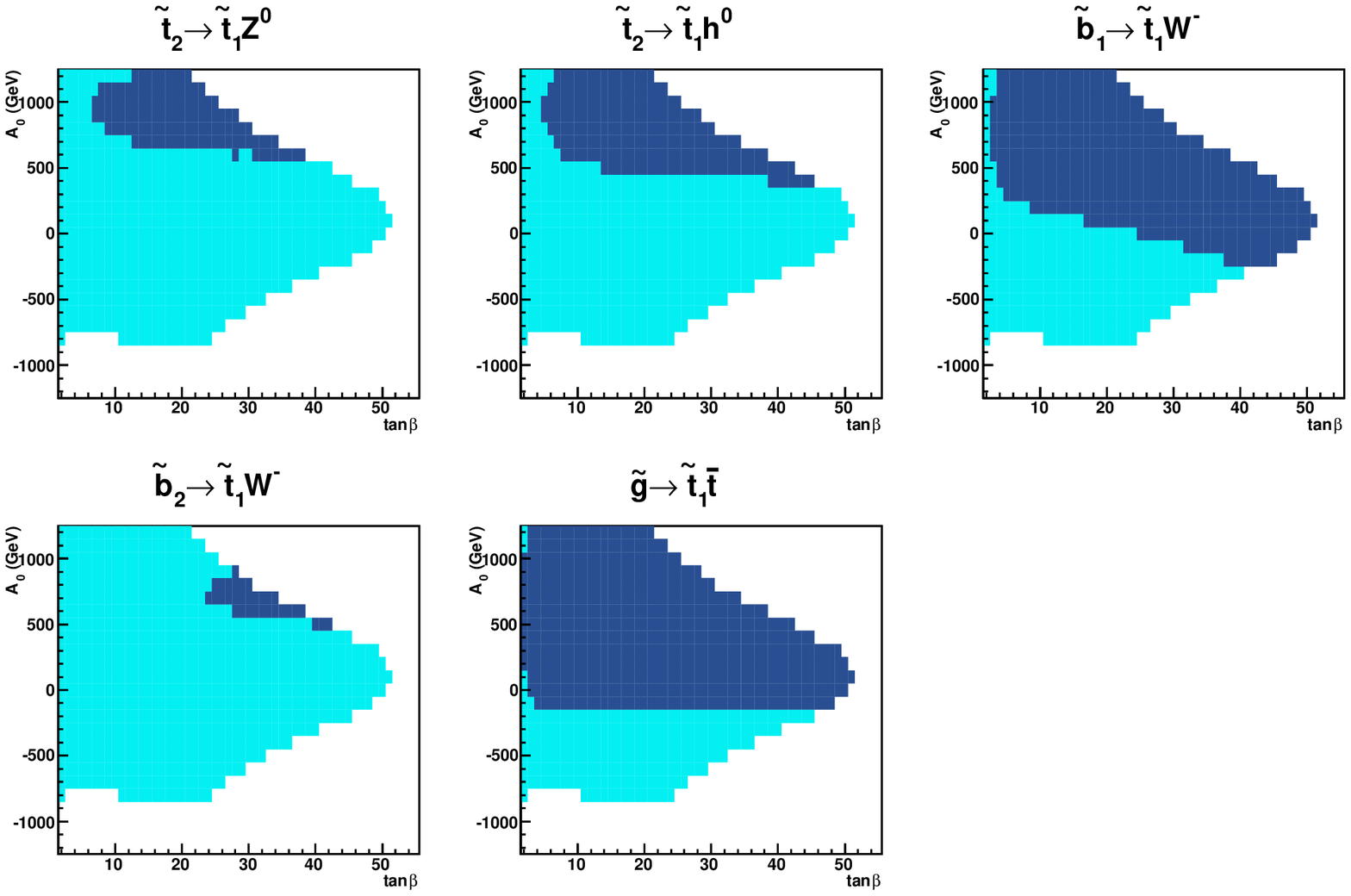}}
\end{figure}

\begin{figure}[!h]
\centering{
\includegraphics[width=0.9\linewidth]{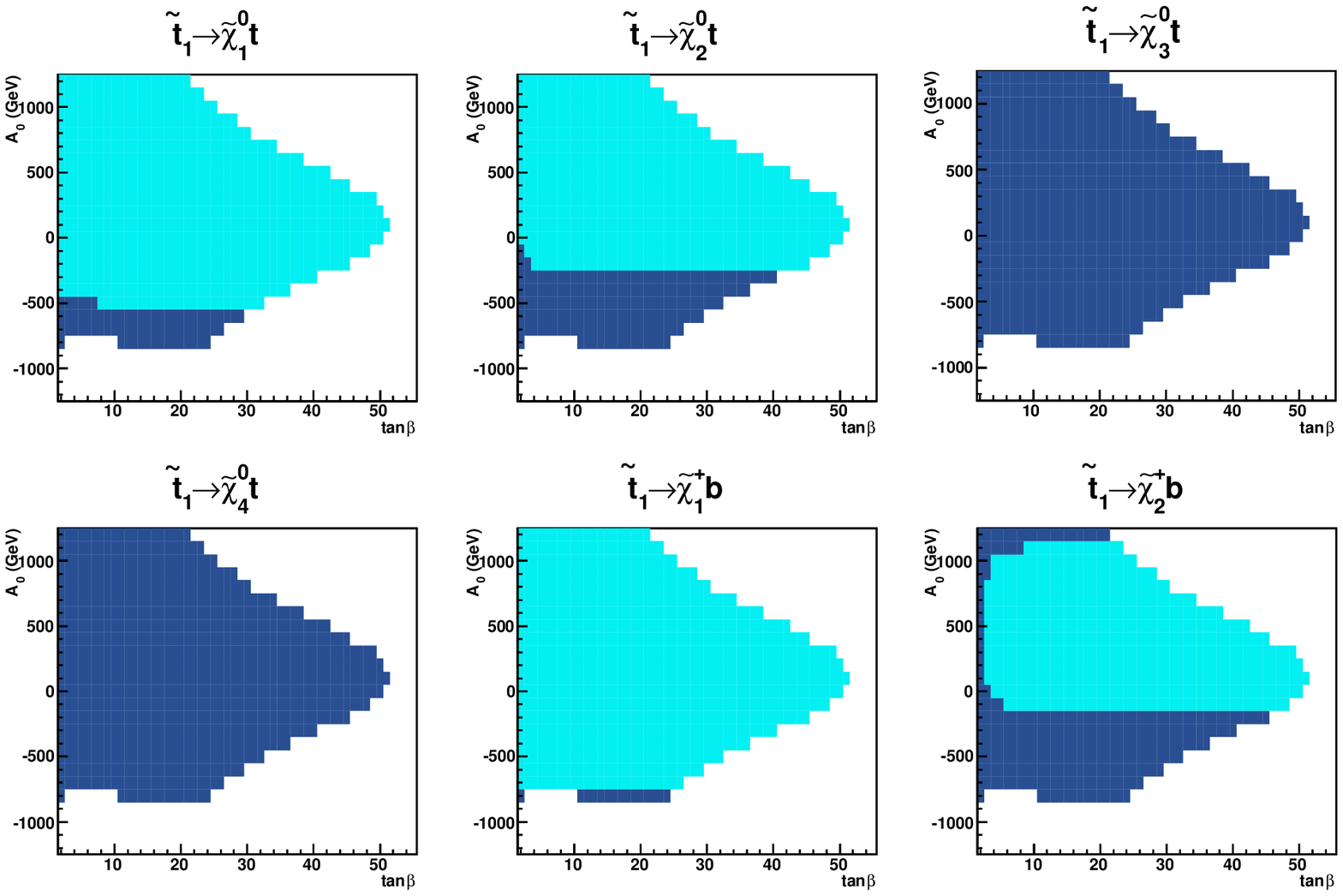}}
\end{figure}

\section{Supersymmetric Phenomenology with Cross Section Effects}\label{app:phenocross}

\begin{figure}[!h]
\centering{
\includegraphics[width=0.82\linewidth]{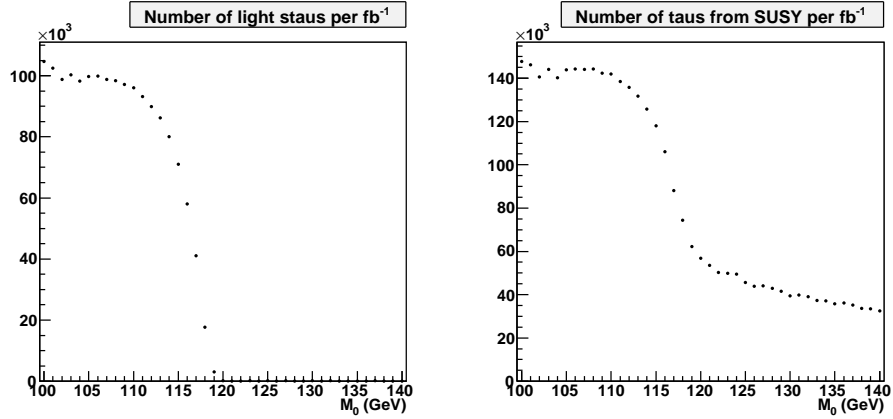}
\caption[Effects of phenomenological regions on stau production with the cross section included]{The number of light staus and taus produced by SUSY events per 1 fb$^{-1}$ of luminosity. This graph has the same specifications as the graph in figure \ref{f:staujump} except that the effects of the cross section are included.}\label{f:staujump-cross}}
\end{figure}

\begin{figure}[!h]
\centering{
\includegraphics[width=0.9\linewidth]{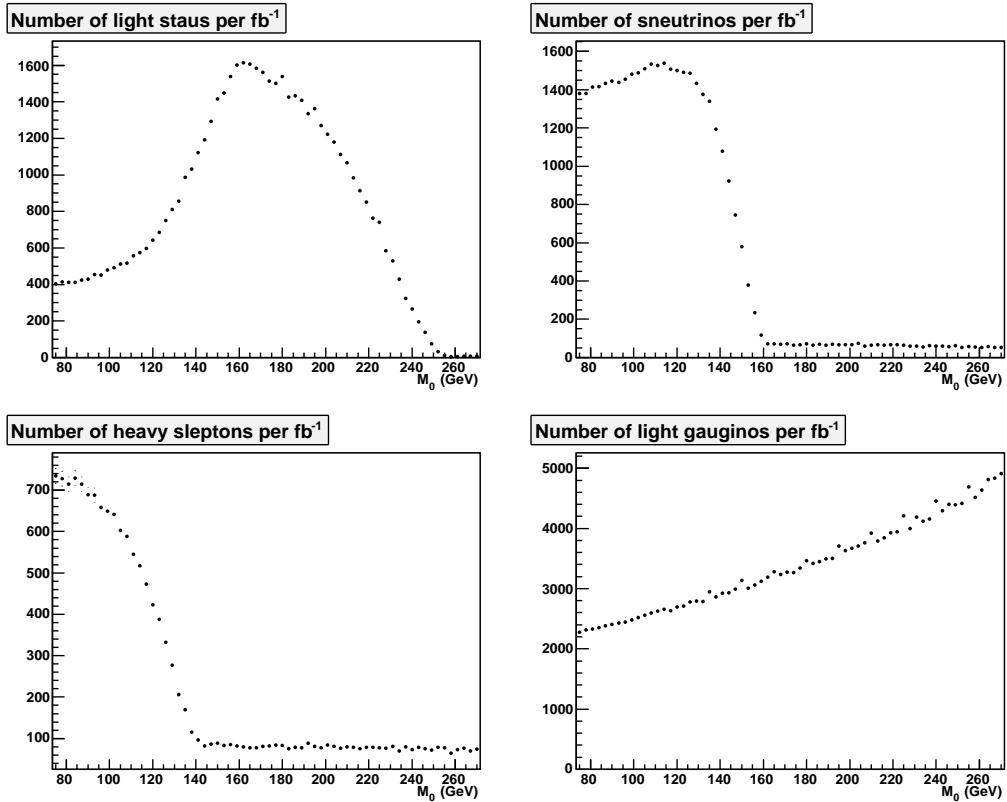}
\caption[Effects of phenomenological regions on slepton production with the cross section included]{The number of times $\tilde\chi_1^\pm,\tilde\chi_2^0$ and sleptons are produced per fb$^{-1}$ of luminosity. This graph has the same specifications as the graph in figure \ref{f:gauginodecay} except that the effects of the cross section are included.}\label{f:gauginodecay-cross}}
\end{figure}

\begin{figure}[!h]
\centering{
\includegraphics[width=0.5\linewidth]{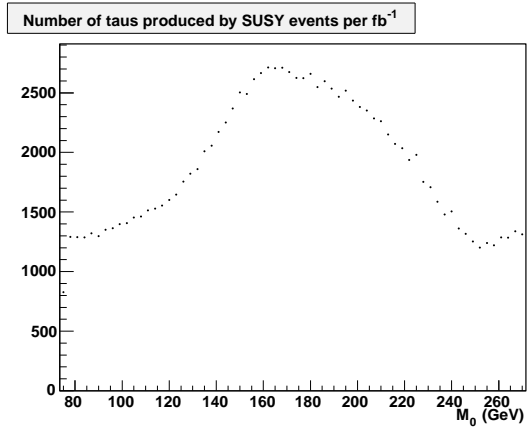}
\caption[Effects of phenomenological regions on tau production with the cross section included]{The number of times a stau is produced per fb$^{-1}$ of luminosity. This graph has the same specifications as the graph figure \ref{f:decayjump-totaltau} except that the effects of the cross section are included. So higher $M_0$ corresponds to lower $\Mhalf$. The higher cross section for low $\Mhalf$ changes the shape of the graph considerably.}\label{f:decayjump-totaltau-cross}}
\end{figure}

\begin{figure}[!h]
\centering{
\includegraphics[width=\linewidth]{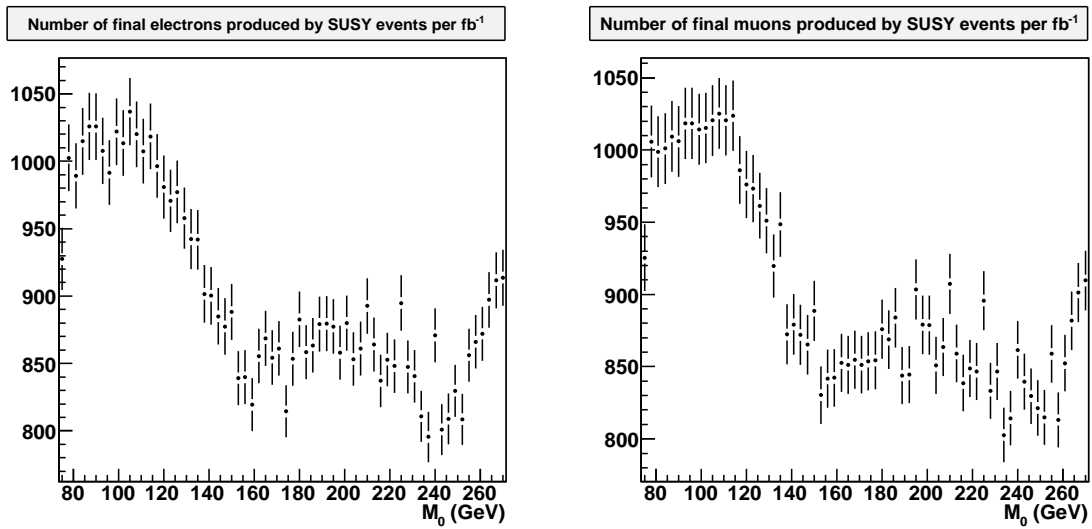}
\caption[Effects of phenomenological regions on final leptons with the cross section included]{The number of final muons and electrons per fb$^{-1}$ of luminosity. This graph has the same specifications as the graph from figure \ref{f:gauginodecay-finalpart} except that the effects of the cross section are included. So higher $M_0$ corresponds to lower $\Mhalf$. The higher cross section for low $\Mhalf$ makes the differences in the number of muons and electrons even smaller.}\label{f:gauginodecay-finalpart-cross}}
\end{figure}

\begin{figure}[!h]
\centering{
\includegraphics[width=\linewidth]{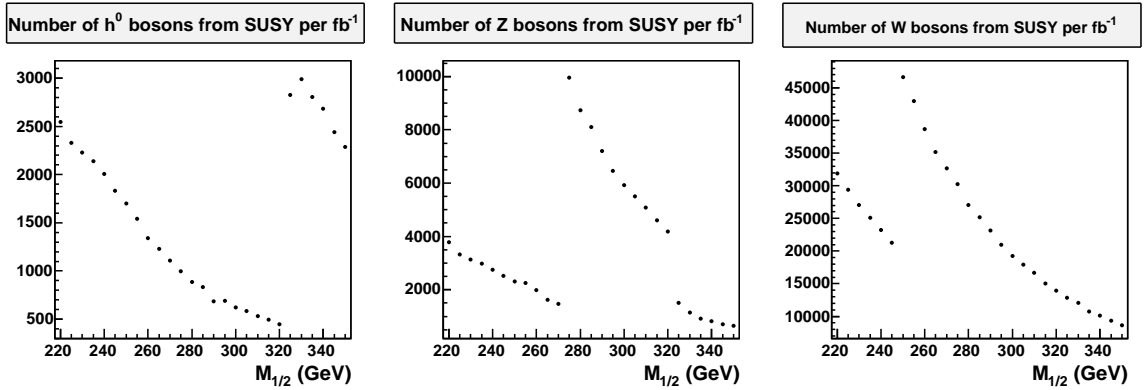}
\caption[Effects of phenomenological regions on $h^0$, $Z$ and $W$ production with the cross section included]{The number of $h^0$, $W$ and $Z$ bosons produced by SUSY events per 1 fb$^{-1}$ of luminosity. This graph has the same specifications as the graph in figure \ref{f:gauginodecay-higgs} except that the effects of the cross section are included. The jumps are still clearly visible, but it is difficult to determine the point in parameter space based on these numbers alone.}\label{f:gauginodecay-higgs-cross}}
\end{figure}

\begin{figure}[!h]
\centering{
\includegraphics[width=0.9\linewidth]{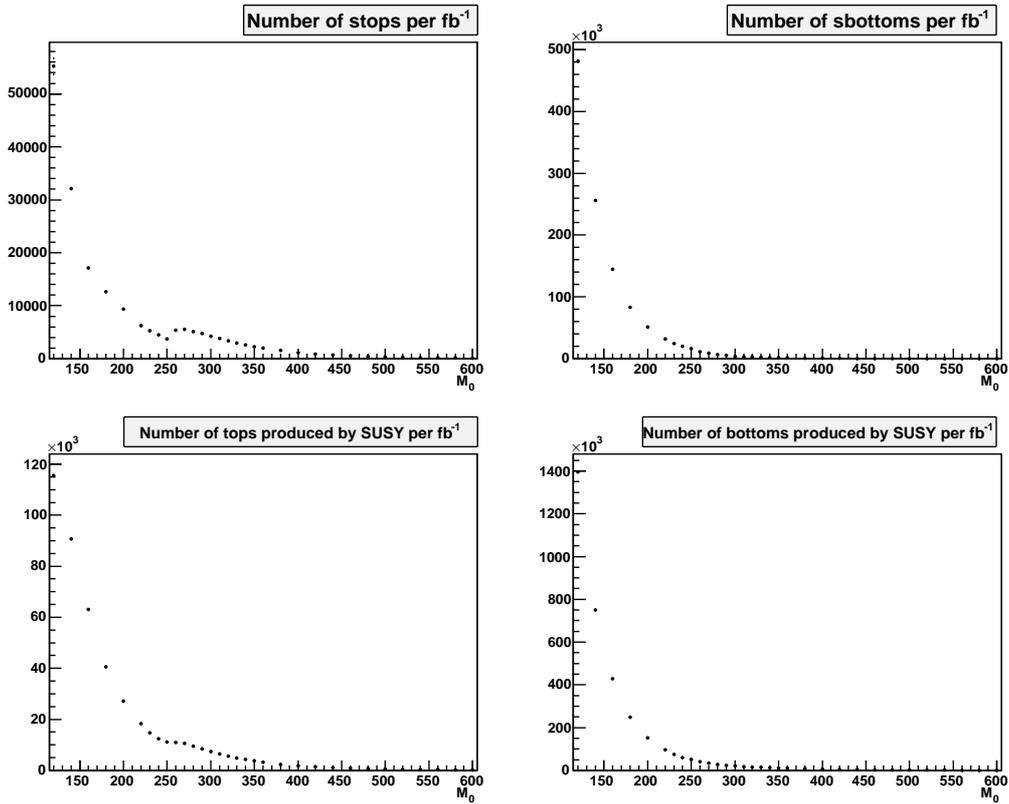}
\caption[Effects of phenomenological regions on the number of stops, sbottoms, tops and bottoms with the cross section included]{The number of stops, sbottoms, tops and bottoms produced by SUSY events per 1 fb$^{-1}$ of luminosity. This graph has the same specifications as the graph in figure \ref{f:stopjump-pythia} except that the effects of the cross section are included. Clearly the cross section is the main effect in this case. However, cross section effects along this line are particularly strong. It is possible to find a line through parameter space that crosses several phenomenological regions and does not span such a large range of cross sections.}
\label{f:stopjump-pythia-cross}}
\end{figure}

\section{Phenomenological Regions for Higher \boldmath$\tan\beta$\unboldmath}\label{app:otherbeta}

\begin{figure}[!h]
\includegraphics[width=0.5\linewidth]{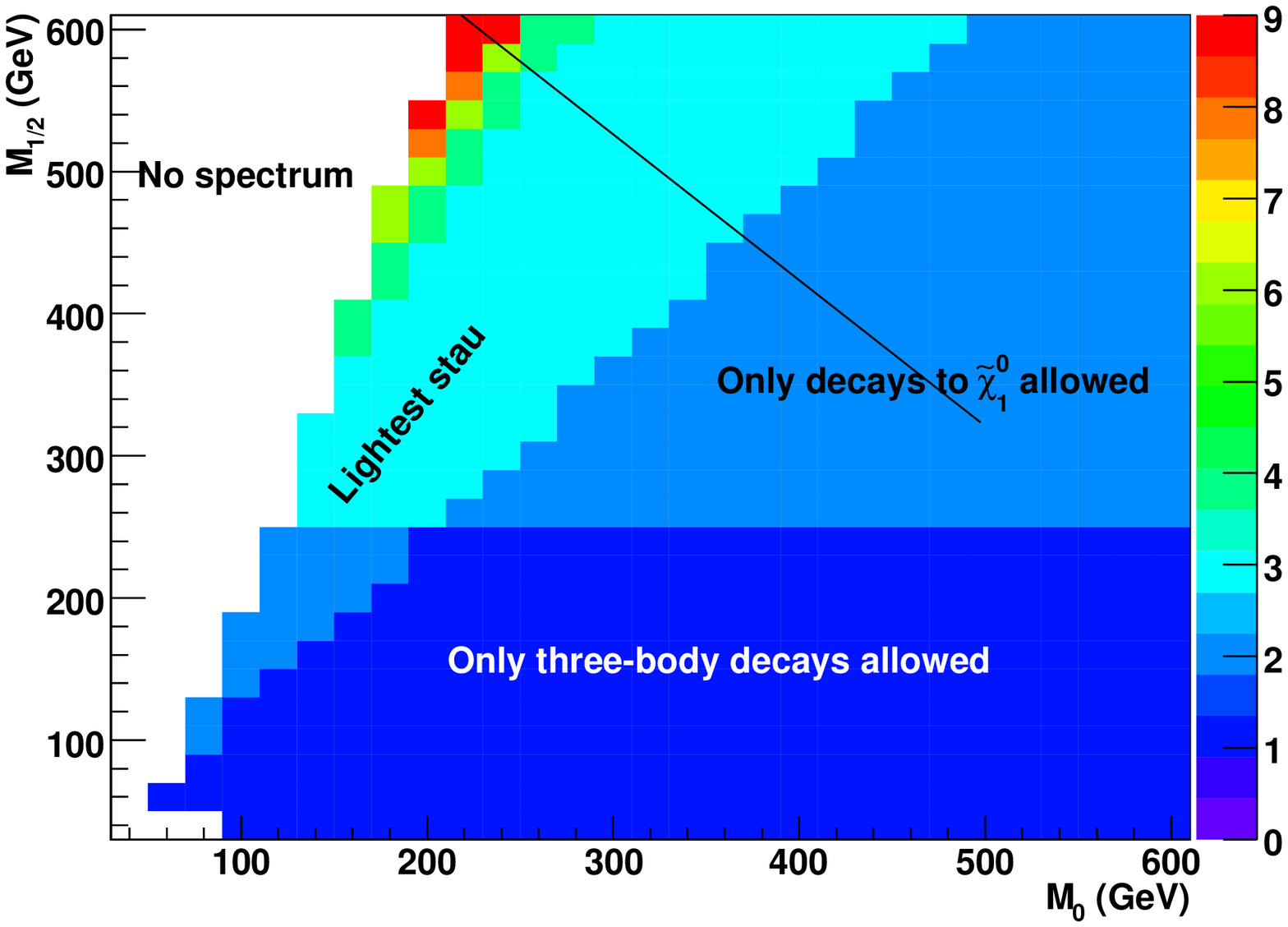}
\includegraphics[width=0.5\linewidth]{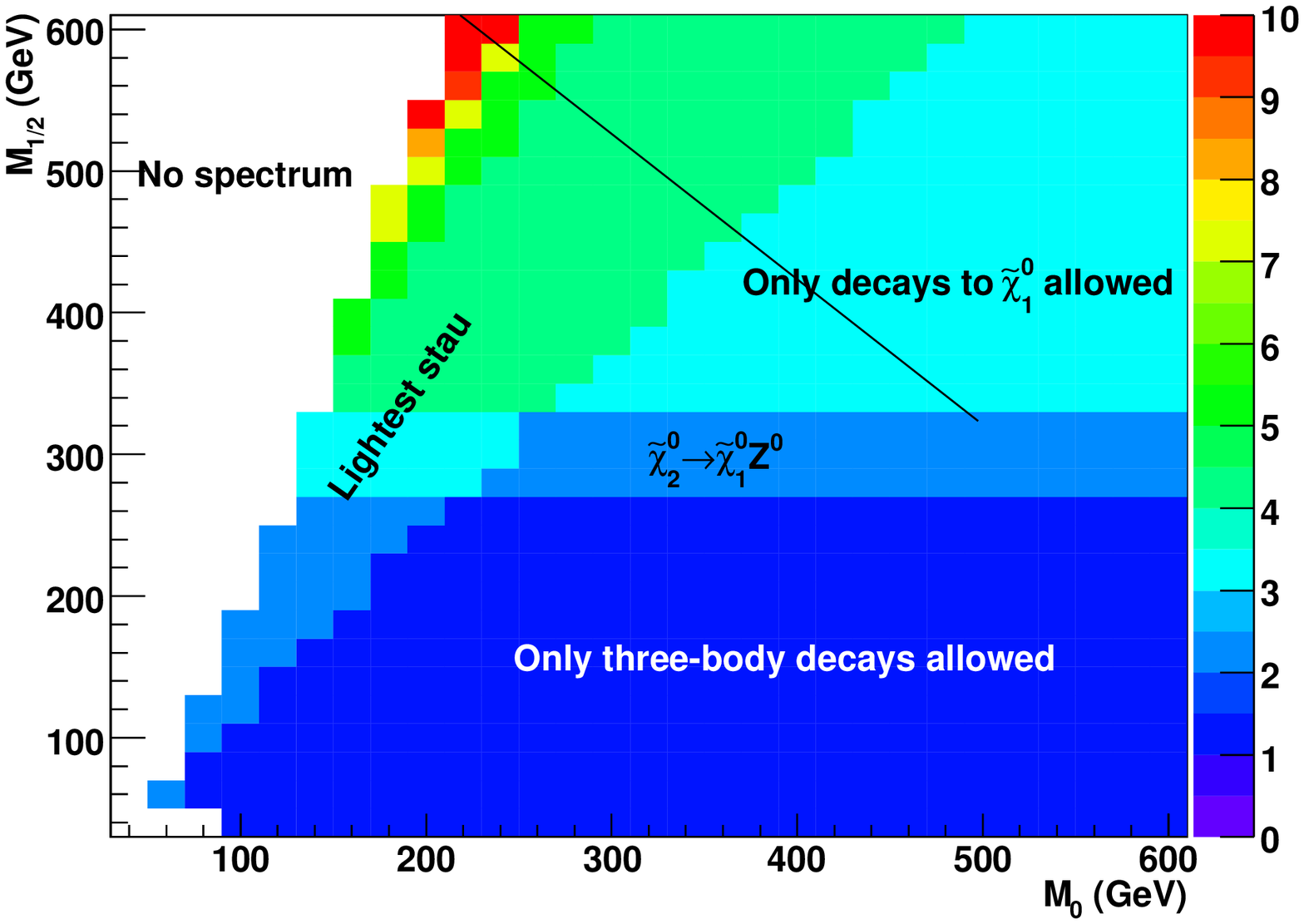}
\caption[Phenomenological regions based on chargino and neutralino decay for $\tan\beta=30$]{Phenomenological regions based on chargino decay (left) and neutralino decay (right) for $A_0=0$~GeV, $\tan\beta=30,\mu>0$}\label{f:appgauginoregion}
\end{figure}

To illustrate that the method introduced in section \ref{s:phenomenology} also works for higher $\tan\beta$, figure \ref{f:appgauginoregion} shows the phenomenological regions based on gaugino decays for $A_0=0$~GeV, $\tan\beta=30,\mu>0$. Comparing these plots to figures \ref{f:chargino-region10} and \ref{f:neutralino-region10}, shows that a change in $\tan\beta$ shifts the boundaries of phenomenological regions and thus changes their shape. Most of the characteristics are unchanged though. 

The spectrum is discarded in a larger part of parameter space, since large $\tan\beta$ results in a lower stau mass, so the lightest stau often ends up being the LSP. The spectrum in the lower left corner is rejected because the lightest stop is the LSP.



\begin{figure}[!h]
\centering{
\includegraphics[width=\linewidth]{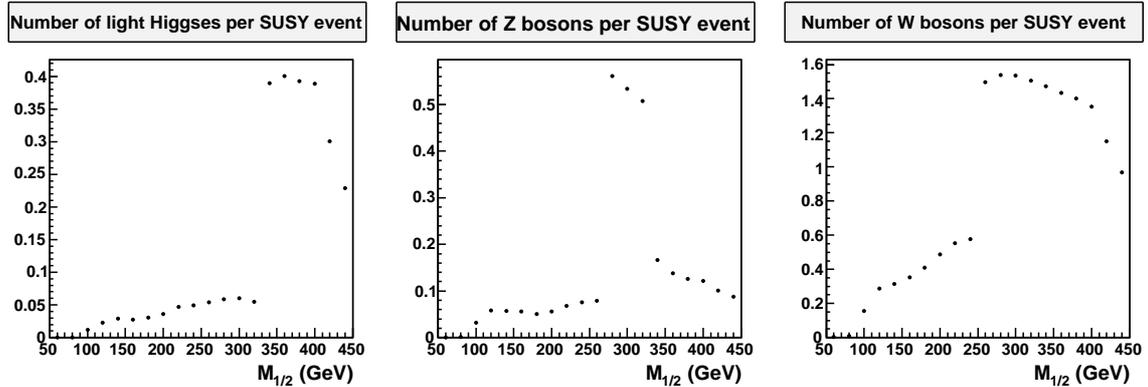}
\caption[The number of Higgs, $Z$ and $W$ bosons per SUSY event for high $\tan\beta$.]{The number of Higgs, $Z$ and $W$ bosons per SUSY event depending on $\Mhalf$ with $M_0=320$~GeV, $A_0=0$~GeV, $\tan\beta=30,\mu>0$. The decline in the number of Higgs and $W$ bosons for $\Mhalf$~\raisebox{0.2ex}{$\scriptstyle\gtrsim$}~$400$~GeV is due to the decay channel to the light stau that becomes kinematically accessible at this point.}\label{f:app-pythiahiggsjump}}
\end{figure}

Figure \ref{f:app-pythiahiggsjump} shows the number of Higgs, $Z$ and $W$ bosons produced per SUSY event along the line $M_0=320$~GeV in figure \ref{f:appgauginoregion}.

\begin{figure}[!h]
\centering{
\includegraphics[width=\linewidth]{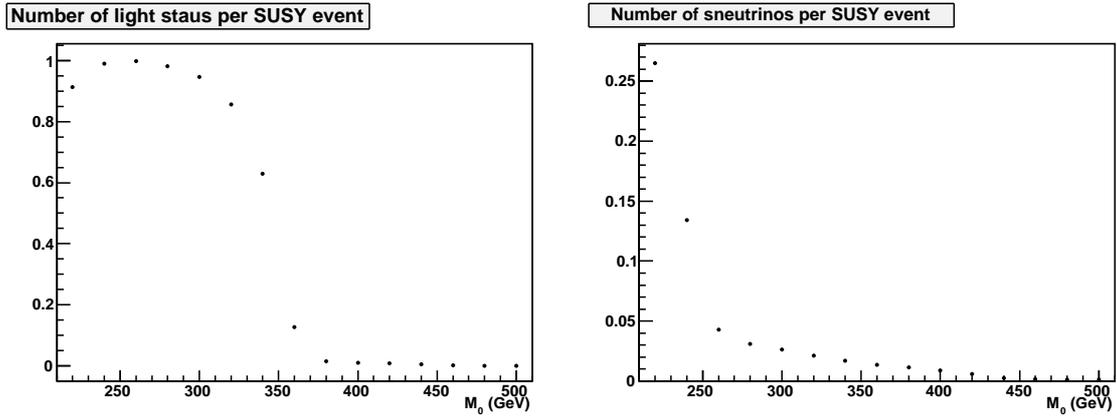}
\caption[The number of staus and sneutrinos per SUSY event for high $\tan\beta$.]{The number of staus and sneutrinos per SUSY event for high $\tan\beta$. These points are generated along the line indicated in figure \ref{f:appgauginoregion}. The jump in staus coincides with the boundary of the phenomenological region.}\label{f:app-pythiastaujump}}
\end{figure}

The jumps in the graphs exactly coincide with the boundaries of the phenomenological regions in figure \ref{f:appgauginoregion}. In figure \ref{f:app-pythiastaujump} the number of staus along the line indicated in figure \ref{f:appgauginoregion} is plotted. It shows that the jump in the number of staus occurs just as in the low $\tan\beta$ region. For $M_0$~\raisebox{0.2ex}{$\scriptstyle\lesssim$}~$250$~GeV the number of staus decreases because the sneutrino decay channel becomes kinematically allowed.

In the stop section (discussed in sections \ref{s:phenostops} and \ref{s:phenA0}), higher values of $\tan\beta$ can give rise to new phenomenological regions. For instance, for some points in parameter space the gluino cannot decay through any of the regular decay channels. It turns out that the decay mode $\tilde g\to\tilde t_1\bar c$ is more important than three body decays if it is kinematically allowed. For a complete description of stop phenomenology at high $\tan\beta$, one needs the phenomenological regions of stop production, gluino decay and sbottom decay, since the production and decay modes of these particles are closely related. We will not explicitly perform this analysis here, but simply summarize the result that phenomenological regions apply to this case as well.

\addcontentsline{toc}{section}{List of Figures}
\listoffigures
\addcontentsline{toc}{section}{List of Tables}
\listoftables
\newpage
\addcontentsline{toc}{section}{References}
\bibliographystyle{JHEP-2}
\bibliography{../literature/literature}

\end{document}